\documentclass[10pt]{article}
\usepackage{amsmath,amssymb,graphicx}
\usepackage{amsthm}

\batchmode

\newtheorem{defn}{Definition}[section]
\newtheorem{lem}{Lemma}[section]
\newtheorem{thm}{Theorem}[section]

\newcommand{\pr}{\noindent{\bf Proof}. }
\newcommand{\re}{\noindent{\bf Remark}. }
\newcommand{\res}{\noindent{\bf Remarks}. }

\newcommand{\pa}{\partial}
\newcommand{\one}{\cO(1)}

\newcommand{\const}{\textrm{const}}

\newcommand{\supp}{ \mathrm{ supp  }}
\newcommand{\hs}{ \hspace{1cm}}
\newcommand{\tr}{\textrm{ tr }}

\newcommand{\loc}{ \mathrm{loc}}
\newcommand{\Vol}{\mathrm{Vol}}
\newcommand{\nat}{\natural}
\newcommand{\tk}{\bbT^{-k}_{\sM +\sN -k}}

\newcommand{\be}{\begin{equation}}
\newcommand{\ee}{\end{equation}}
\newcommand{\bsp}{\begin{split}}
\newcommand{\esp}{\end{split}}

\newcommand{\bom}{\mbox{\boldmath $\Om$}}
\newcommand{\bla}{\mbox{\boldmath $\La$}}

\newcommand{\bpi}{\mbox{\boldmath $\Pi$}}

\newcommand{\ba}{\mathbf{a}}

\newcommand{\sM}{\mathsf{M}}
\newcommand{\sN}{\mathsf{N}}
\newcommand{\sfb}{\mathsf{b}}

\newcommand{\al}{\alpha}
\newcommand{\De}{\Delta}
\newcommand{\de}{\delta}
\newcommand{\ga}{\gamma}

\newcommand{\ka}{\kappa}
\newcommand{\La}{\Lambda}
\newcommand{\la}{\lambda}
\newcommand{\Om}{\Omega}
\newcommand{\om}{\omega}

\newcommand{\ep}{\epsilon}

\newcommand{\vep}{\varepsilon}

\newcommand{\cB}{{\cal B}}
\newcommand{\cC}{{\cal C}}
\newcommand{\cD}{{\cal D}}

\newcommand{\cO}{{\cal O}}

\newcommand{\cS}{{\cal S}}

\newcommand{\cR}{{\cal R}}
\newcommand{\cT}{{\cal T}}

\newcommand{\cN}{{\cal N}}
\newcommand{\cW}{{\cal W}}

\newcommand{\cL}{{\cal L}}
\newcommand{\cP}{{\cal P}}

\newcommand{\cZ}{{\cal Z}}

\newcommand{\bbR}{{\mathbb{R}}}
\newcommand{\bbZ}{{\mathbb{Z}}}
\newcommand{\bbC}{{\mathbb{C}}}

\newcommand{\bbT}{{\mathbb{T}}}

\newcommand{\bbA}{{\mathbb{A}}}
\newcommand{\bbB}{{\mathbb{B}}}

\begin{document}

\title{The Renormalization Group According to Balaban\\ II. Large Fields}
\author{ 
J. Dimock
\thanks{dimock@buffalo.edu}\\
Dept. of Mathematics \\
SUNY at Buffalo \\
Buffalo, NY 14260 }
\maketitle

\begin{abstract}
This is   an expository account of Balaban's approach to the renormalization group.  The method is 
illustrated with a treatment of  the the ultraviolet problem  for     the scalar $\phi^4$ 
model  on  toroidal lattice  in dimension  $d=3$.   In this second   paper we control   the  large   field contribution to the partition function\end{abstract}

%\tableofcontents

\section{Introduction}

This paper is an extension of  part I  \cite{Dim11}.   We   recall  the  general  setup.
We  are  studying the  $\phi^4$  field theory   on   a toroidal    lattice  of the form  
 \begin{equation}
\bbT_{\sM}^{-\sN}  =   ( L^{-\sN}\bbZ  / L^{\sM} \bbZ )^3  
\end{equation}
The  theory  is  scaled up to the unit lattice   $\bbT^0_{\sM  + \sN }$
and there   the  partition  function  has the form  
\begin{equation}
Z_{\sM,  \sN}  =  \int   \rho^{\sN}_0(\Phi)  d \Phi
\end{equation}
where  for   fields  $\Phi:   \bbT^0_{\sM  + \sN } \to  \bbR$
we have the density  
\begin{equation}  \label{den0}
 \rho^{\sN}_0(\Phi)  
=\exp \left( - S^{\sN}_0(\Phi)  - V^{\sN}_0(\Phi)  \right) 
\end{equation}
with 
\begin{equation}
\begin{split}
S^{\sN}_0(\Phi)  = &  \frac12  \|  \pa  \Phi \|^2  +  \frac12    \bar  \mu^{\sN}_0  \|  \Phi  \|^2 \\
V^{\sN}_0(\Phi)  = & \vep^{\sN}_0     \Vol( \bbT_{\sM+\sN} )   +  \frac12   \mu^{\sN}_0      \|  \Phi  \|^2     + \frac14   \la^{\sN}_0  \sum_x  \Phi^4 (x)  \\
\end{split}
\end{equation} 
and very small    coupling  constants   $\la^{\sN}_0 = L^{-\sN}  \la,  \mu^{\sN}_0  =  L^{-2\sN} \mu$,   etc.
The  superscript  $\sN$ is generally omitted  so we have  $\la_0,  \mu_0,  etc.  $.

Our goal  is  to show  that with  intelligent   choices of the counter terms   $ \vep^{\sN}_0 ,   \mu^{\sN}_0$    the partition function  $Z_{\sM,  \sN}  $  satisfies  stability bounds   which are uniform in the ultraviolet 
cutoff  $\sN$   and with  bulk   dependence on the volume parameter  $\sM$.    The  method  is  
the renormalization group  method  of  Balaban (\cite{Bal82a}  - \cite{Bal98c}).   In fact   our primary goal 
is  not the stability bounds,  which are interesting  but  not new,   but rather   the illustration of Balaban's method.

The first renormalization group operation is defined as follows.     We  create a new  density  defined
for  $\Phi_1:    \bbT^1_{\sM  + \sN } \to  \bbR$  by   
 \begin{equation}    \label{sundry}
\tilde  \rho_{1}(\Phi_1) = \const   \int      \exp \left(
- \frac12  \frac{a}{L^2}  \|\Phi_1- Q \Phi_0\|^2  \right) \rho_0( \Phi_0)   \    d\Phi_{0}
\end{equation}
Here  $Q$  averages over blocks  of linear size $L$.  Then  one scales back to a unit lattice  replacing $\Phi_1$ by $\Phi_{1,L}$
where    now   $\Phi_1:    \bbT^0_{\sM  + \sN-1 } \to  \bbR$
and    
\begin{equation}
\Phi_{1,L}(x)  =  L^{-1/2} \Phi(x/L)
\end{equation}
 Thus we define 
\begin{equation}
  \rho_{1}(\Phi_1)= \const  \   \tilde  \rho_{1}(\Phi_{1,L})
\end{equation}
 The constants are chosen to  preserve the integral:
 \begin{equation}
    \int   \rho_{1}(\Phi_1) d \Phi_1  =   \int \rho_0(\Phi_0) d \Phi_0
    \end{equation}

This operation is repeated.  However to control  the densities   $\rho_0, \rho_1,  \rho_2, \dots  $
that  are  generated  we need an  extensive  analysis at  each stage.   The  key  idea
is  to   analyze large and small  field regions  separately.    We  give a first taste here in a  somewhat  simplified version.

In  (\ref{sundry})     insert  under the integral  sign   
\begin{equation}
1  =  \sum_{\Om_1}  \zeta( \Om^c_1,  \Phi_0)  \chi( \Om_1,  \Phi_0) 
\end{equation}
Here  we have partitioned the lattice into  cubes   $\square$  of  linear size  $M= L^m$.  We  are 
summing  over   regions  $\Om_1$   which are unions of such  cubes.    The function  $ \chi( \Om_1,  \Phi_0) $
is  the characteristic function of the  set of fields which satisfy  some  small field conditions
  in $\Om_1$.  The conditions are  
\begin{equation}
|\Phi_1-Q\Phi_0|  \leq  p_0  \hs  |\pa \Phi_0|  \leq   p_0  \hs    | \Phi_0|  \leq   \la_0^{- \frac14} p_o 
\end{equation}
where        $p_0  =  ( - \log \la_0 )^p$  is actually rather large.
The function  $ \zeta( \Om^c_1,  \Phi_0,  \Phi_1)$  is the characteristic function of fields  which violate
at least one of these inequalities at some point   in each cube  in   $\Om_1^c$. 

The resulting integral  can now be written  (splitting the bonds across $\pa \Om_1$)
 \begin{equation}    \label{sundry2}
 \begin{split}
\tilde  \rho_{1}(\Phi_1) = &   \const  \sum_{\Om_1}  \int      d\Phi_{0, \Om_1^c} \zeta( \Om^c_1,  \Phi_0)  
 \exp \left(- \frac12    \frac{a}{L^2}   \|\Phi_1- Q \Phi_0\|^2_{\Om^c_1}  - S_0(\Om^c_1, \Phi_0)  - V_0(\Om^c_1,  \Phi_0)   \right)   \\
&  \left[  \int        d\Phi_{0, \Om_1} \  \chi( \Om_1,  \Phi_0) 
 \exp \left(- \frac12  \frac{a}{L^2}  \|\Phi_1- Q \Phi_0\|^2_{\Om_1}  - S_0(\Om_1, \Phi_0)  - V_0(\Om_1,  \Phi_0)   \right) \right] \\
\end{split}
\end{equation}

The  idea  is now  to  carry  out  a detailed   analysis of the small field integral over  $[ \dots ]  $   This involves   expanding  around  the field  $\Phi_0 $  which  minimizes the first two  terms in the
exponent.     This  generates  a new  action    $S_1( \Om_1,  \Phi_1) $  and a fluctuation integral.      Then  one   writes the    fluctuation   integral in  a local form,   which  means    doing a cluster expansion.
After   scaling   the fields  and the region $\Om_1$ we  have  a new  contribution to the density of the form   
\begin{equation}
[\cdots] =\exp \Big( -S_1( \Om_1, \Phi_1)  - V_1(\Om_1, \Phi_1 )   +   \sum_{X \subset  \Om_1} E_1(X, \Phi_1) \Big)
\end{equation}
The  first two terms are  similar to what we  started with but now with new coupling constants
$\la_1= L \la_0$,  $\mu_1 = L^2 \mu_0 + \dots$,  etc.  
The localized    functions    $  E(X, \Phi_1)$  are defined for polymers  $X$  (connected unions of $M$ cubes),  depend  only    on
 $ \Phi_1$ restricted to $X$,  and are  exponentially decaying in  $|X|_M$ (the number of $M$ cubes in $X$).    

Now consider the large field region $\Om_1^c$.
In a cube        $\square \subset   \Om^c_1$   at least one of     $ \exp (- a/2L^2  \|\Phi_1- Q \Phi_0\|^2_{\square})$ or        $\exp( - S_0(\square, \Phi_0))    $  
or $  \exp(  - V_0(\square ,  \Phi_0 )  )      $ is bounded by  $e^{- \cO(1) p_0^2} $  .     Thus   the   first exponential in  (\ref{sundry2})  is bounded by     $ e^{- \cO(1) p_0^2|\Om_1^c|_M}$.
This tiny factor is sufficient to  strongly   suppress the contribution  of   any large field region and     control the sum  over  $\Om_1$.

  Now  one    repeats  this  operation.    At   the  $k^{th}$   step  we  have  a tighter  definition of  small fields
  based   on  the new  larger    coupling constants  $\la_k = L^k \la_0$  and smaller parameters   $p_k= (-\log \la_k)^p$. 
  Correspondingly  one makes a new large/small split   $\Om_{k+1}$ in the current small field region  $\Om_k$.
   The overall  result  is a 
  sum  over   decreasing  regions  $\Om_1  \supset  \Om _2  \supset  \cdots    \supset  \Om_k$
with    an  explicit     leading  action  in the current small field region  $\Om_k$.  

  The main  issues are  the detailed analysis of  (1.)   the  small field region,  (2.) the large field regions, (3.) the coupling between them,  and  (4.)  the convergence of all the sums.    
 Point (1.)  was  considered in detail  in   the first paper.   Here  we  are concerned with points (2.) and (3.).     A third and final paper
 establishes  point  (4.)   and  completes the proof of the stability bound.

Although the broad outlines of the procedure  are due to Balaban,  especially   in   his  treatment of the linear  sigma model \cite{Bal95}  - \cite{Bal98c},   we  deviate  in many of the particulars.

Before  plunging into the details of the general problem   we  open  with an  analysis of the free effective actions
which are generated by this process
\bigskip

\noindent
\textbf{convention:}   Throughout the text  $\one$ stands for a constant independent of all other parameters.     Also  $C$  stands for  a constant
depending on  $L$,  but on no other parameters.   Both  $\one$ and $C$ can change from line to line.

\section{Localized  block averaging}  \label{two}

\subsection{block averaging}

\label{averaging}

First some definitions.    In any  of our lattices  $ \bbT^{-k}_{\sM + \sN -k}$   the centers  of  $L^n$-cubes   
are the points in the lattice    $\bbT^{-(k-n)}_{\sM + \sN -k}$.  For  a  region  $\Om \subset  \bbT^{-k}_{\sM + \sN -k}$,   
let  $\Om^{(n)}$  be the    centers   of   $L^n$-cubes  in    $\Om$,  thus we have $\Om^{(n)} = 
\Om  \cap    \bbT^{-(k-n)}_{\sM + \sN -k}$.

Let   $\Phi_0$  be a function  on  the initial   torus   $\bbT^{0}_{\sM +  \sN}$  and  $\rho_0( \Phi_0)$   an initial density. 
 Let  $\Om_1$  be  a union  of $LM$ blocks   in   $\bbT^{0}_{\sM +  \sN}$ for some large $M= L^m$.    If   $\Phi_{0, \Om_1}$ is the
 restriction of $\Phi_0$    to  $\Om_1$ and  $Q$ is averaging over  $L$-cubes,   then  $Q\Phi_{0, \Om_1}$ is a function on
 $\Om_1^{(1) }$  ( $=\Om_1  \cap    \bbT^1_{\sM +  \sN}$).  If     $\Phi_{1, \Om_1}$  is any other function  on $\Om^{(1)}_1$ we can form  
  \footnote{In  general  if    $f $  is  defined    on   $\Om \subset  \bbT^{-k}$  then 
  \[   \|  f \|^2_{\Om}    \equiv L^{-3k} \sum_{x \in \Om} | f(x) |^2  \equiv   L^{-3k} | f |^2_{\Om}
\]     }
  \begin{equation}
\|   \Phi_1- Q \Phi_0\|^2_{  \Om^{(1)}_1}   \equiv  L^3  \sum_{x  \in \Om^{(1)}_1} |\Phi_1(x)- Q \Phi_0(x)|^2  \equiv   L^3   |\Phi_1- Q \Phi_0|^2_{  \Om^{(1)}_1}   
 \end{equation}
  In the following we generally    write an expression like  $ \|   \Phi_1- Q \Phi_0\|^2_{  \Om^{(1)}_1}  $ 
 as just   $ \|   \Phi_1- Q \Phi_0\|^2_{  \Om_1}  $  and the expression   $ |\Phi_1- Q \Phi_0|^2_{  \Om^{(1)}_1}$
 as just  $  |\Phi_1- Q \Phi_0|^2_{\Om_1}   $.
 It is   understood that  the norm is to be evaluated on the intersection of
 the domain  of the fields with  $\Om_1$,  with the appropriate weighting.

 We  define block  averaging  in $\Om_1$       by 
           \begin{equation}    
\begin{split}     \label{fluff}
\tilde  \rho_{1,\Om_1}(  \Phi_{0, \Om^c_1},\Phi_{1, \Om_1}) 
= &  \cN_{aL,  \Om^{(1)}_1} ^{-1} \int      \exp \left(- \frac12  \frac{a}{L^2} \|   \Phi_1- Q \Phi_0\|^2_{  \Om_1}   \right) \rho_0( \Phi_0)  \    d\Phi_{0,\Om_1}  \\
&      =   \cN_{aL,  \Om^{(1)}_1} ^{-1}\int   
  \exp \left(- \frac12   aL  |\Phi_1- Q \Phi_0|^2_{  \Om_1}   \right)       \rho_0( \Phi_0)    \    d\Phi_{0,\Om_1}  \\
\end{split}
\end{equation}
where      
\be   \cN_{a, \Om} =  ( 2 \pi/a)^{|\Om|/2}   \hs    |\Om| =\textrm{   number of elements in } \Om
\ee   The preserves the integral:
\begin{equation}
\int  \tilde  \rho_{1,\Om_1}(  \Phi_{0, \Om^c_1},\Phi_{1, \Om_1})\ d \Phi_{0, \Om^c_1} d\Phi_{1, \Om_1}
=   \int  \rho_0( \Phi_0)  d \Phi_0
\end{equation}

 Next we    scale  back down to a unit lattice .     Replace  $\Om_1$  by   $L\Om_1$
 where now     $\Om_1$ is a   union of $M$-blocks in  $\bbT^{-1}_{\sM +  \sN-1}$  (i.e. length $M$, not $M$ sites),
    so that   $L\Om_1$ is a union of $LM$ blocks in  
  $\bbT^0_{\sM +  \sN}$. 
 Then replace    $\Phi_{1, L\Om_1}$,  a function  on  $(L \Om_1)^{(1)}=  L \Om_1  \cap   \bbT^1_{\sM + \sN}$,      by    $[\Phi_{1,L}]_{L \Om_1}=[\Phi_{1, \Om_1}]_L$  where now   $\Phi_{1, \Om_1}$  is a function  on  $\Om_1^{(1)} =   \Om_1  \cap   \bbT^0_{\sM + \sN-1}$.
 Also replace   $  [ \Phi_0]_{L\Om_1^c}$,  a function on   $L \Om_1^c \subset   \bbT^0_{\sM + \sN}  $  
  by  $ [ \phi_{ L}]_{L\Om_1^c} = [\phi_{ \Om^c_1}]_L$  where now  $\phi_{ \Om^c_1}$  is a function  on   
  $ \Om_1^c \subset   \bbT^{-1}_{\sM + \sN-1}  $.  
  Thus   the definition is   
\begin{equation}    \label{sub0}
\begin{split}
    \rho_{1,\Om_1}( \phi_{ \Om^c_1},\Phi_{1, \Om_1}) 
  =& \const \  \tilde  \rho_{1,L\Om_1}\Big( [\phi_{ \Om^c_1}]_L,[\Phi_{1, \Om_1}]_L\Big) \\
   \end{split}
  \end{equation}
  with the constant chosen to preserve the integral.
  
\begin{figure}[t] 
\begin{picture}(250,250)(-120,-20)
\thinlines
 \multiput(60,60)(0,10){9}{\line(1,0){80}}
\multiput(60,60)(10,0){9}{\line(0,1){80}}  
 \multiput(40,40)(0,20){7}{\line(1,0){120}}
 \multiput(40,40)(20,0){7}{\line(0,1){120}}  
\multiput(0,0)(0,40){6}{\line(1,0){200}}
 \multiput(0,0)(40,0){6}{\line(0,1){200}}  
\linethickness{2pt}
\put(80,80){\framebox(40,40)}
\put(60,60){\framebox(80,80)}
\put(40,40){\framebox(120,120)}
\put(0,0){\framebox(200,200)}
\put(105, 85){\text{$\Om^c_1$} }
\put(125, 65){\text{$\Om^c_2$} }
\put(145, 45){\text{$\Om^c_3$} }
\put(185, 5){\text{$\Om^c_4$} }
\put (-50,100){\text{$\dots \dots$}}
\put (230,100){\text{$\dots \dots$}}
\end{picture}
\caption{Nested regions  $\Om_1 \supset \Om_2  \supset  \Om_3     \supset  \Om_4$   \label{rumple} }
\end{figure}
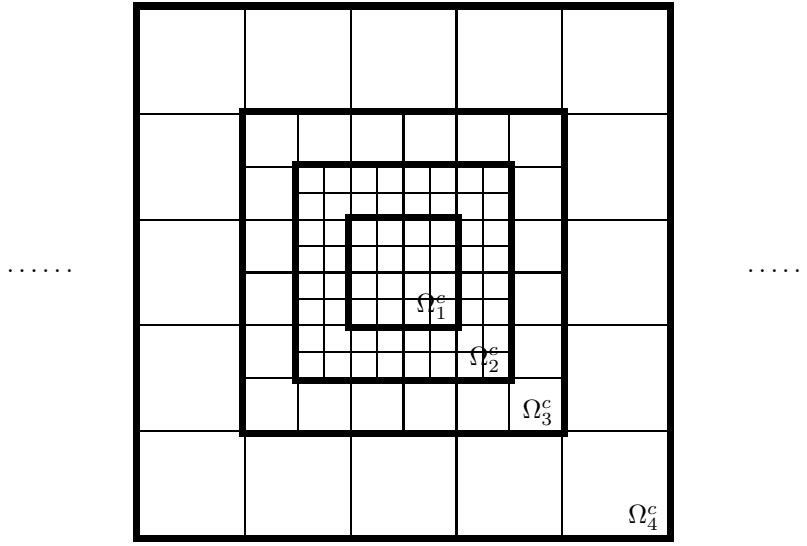

 We   iterate this procedure  always shrinking  the  region in which we are  averaging.  (Later we   impose  some conditions
 on this shrinking). After  $k$  steps  we will
 have a sequence  of   regions 
 \begin{equation}  \label{sundry3}
\bom  =    ( \Om_1, \Om_2,   \dots,  \Om_k)
 \end{equation}
 in     $\bbT^{-k}_{\sM + \sN -k}$  which satisfy 
 \begin{equation}
 \Om_1 \supset \Om_2 \supset  \cdots  \supset   \Om_k
 \end{equation}   
  The region      $\Om_j$ is  
a union  of     $L^{-(k-j)} M$  cubes.  We  also define
\begin{equation}
\de \Om_j  =  \Om_j - \Om_{j+1}    \hs   j=1,2, \dots,  k-1
\end{equation}
See figure \ref{rumple}  for an indication of how a piece of   this might look  in the case where  the complements  $\Om^c_j$  are
small   (the more likely case).

After $k$ steps we will have   
a density   $   \rho_{k, \bom}(  \phi_{\Om_1^c}, \Phi_{k,\bom})$  with the same integral.   Here  $\phi_{\Om_1^c}:  \Om_1^c \to  \bbR$
and    
 \begin{equation}  \label{honey}
\Phi_{k,\bom}  =  ( \Phi_{1, \de  \Om_1}  ,  \Phi_{2, \de  \Om_2} \dots,   \Phi_{k-1, \de    \Om_{k-1}},   \Phi_{k,  \Om_k  } )
\end{equation}
where  
\begin{equation}
\begin{split}
&\Phi_{j, \de \Om_j} :    \de \Om_j^{(j)}  \to   \bbR   \hs     j = 1, \dots,  k-1  \\
&  \Phi_{k,  \Om_k} :     \Om_k^{(k)}  \to   \bbR     \\
\end{split}
\end{equation}
Note that  $\de \Om_j^{(j)}  \subset   \bbT^{-(k-j)}_{\sM +\sN -k}$  and    $\ \Om_k^{(k)}  \subset   \bbT^0_{\sM +\sN -k}$.  
 The  fields   $\Phi_{k,\bom}$  can  also   be regarded as a single  function on 
\begin{equation}
\de       \Om^{(1)}_1 \cup  \de       \Om^{(2)}_2   \cup \cdots     \cup  \de      \Om^{(k-1)}_{k-1}  \cup        \Om^{(k)}_k  \  \subset   \bbT^{-k}_{\sM + \sN -k}
\end{equation}

The    next   step   is taken by introducing   by choosing  $\Om_{k+1}  \subset  \Om_k$   which is  a union 
of   $LM$ blocks  in  $\bbT^{-k}_{\sM + \sN -k}$.     There is  a    new  field   $\Phi_{k+1}:  \Om^{(k+1)}_{k+1} \to \bbR$
(   $\Om^{(k+1)}_{k+1}  =   \Om_{k+1} \cap   \bbT^{1}_{\sM + \sN -k}$)
Now define      \begin{equation}
   \begin{split}
    \bom^+  =& (\bom, \Om_{k+1}) =   (  \Om_1, \Om_2,   \dots,  \Om_k, \Om_{k+1})  \\
    \Phi_{k+1, \bom^+}  = &    ( \Phi_{1, \de  \Om_1}  , \dots,   \Phi_{k, \de \Om_{k}},   \Phi_{k+1,  \Om_{k+1}  } ) \\
    \end{split}
    \end{equation}
and
   \begin{equation}    
\begin{split}   
&\tilde  \rho_{k+1,  \bom^+  }(\phi_{\Om^c_1},   \Phi_{k+1,  \bom^+}  )\\
=  & \cN_{aL,   \Om^{(k+1)}_{k+1}}^{-1}     \int      \exp \left(
- \frac12  \frac{a}{L^2}  \|\Phi_{k+1}- Q \Phi_{k}\|^2_{  \Om_{k+1}}   \right)
 \rho_{k, \bom}(\phi_{\Om^c_1}, \Phi_{k, \bom} )   \
    d\Phi_{k,\Om_{k+1}}   \\
    = &  \cN_{aL,   \Om^{(k+1)}_{k+1}}^{-1}     \int      \exp \left(
- \frac12   aL  |\Phi_{k+1}- Q \Phi_{k}|^2_{  \Om_{k+1}}   \right)
 \rho_{k, \bom}(\phi_{\Om^c_1}, \Phi_{k, \bom} ) \
    d\Phi_{k,\Om_{k+1}}   \\
\end{split}
\end{equation}

Next  we   scale.  
 Replace   $\bom^+$  by  $L \bom^+$  where still   
$ \bom^+ =    (\Om_1,  \Om_2,  \dots, \Om_k,   \Om_{k+1})$  but now  
         $\Om_j$ is  a union  of  $L^{-(k+1-j)}M$  blocks in  $\bbT^{-k-1}_{\sM  + \sN -k-1}$.
 Replace     $\phi_{L\Om^c_1}$ by $[\phi_L]_{L\Om^c_1} =[\phi_{\Om^c_1}]_L $ where    $\phi$ is defined on  the new $\Om_1^c $.  For  $1 \leq  j  \leq  k$ replace    $\Phi_{j, L\de   \Om_j} $      by   
   $  [\Phi_{j,L} ]_{L \de  \Om_j}  =[\Phi_{j,\de   \Om_j}]_L $  where  now       $\Phi_{j, \de  \Om_j}$  is defined on   $\de  \Om_j^{(j)} = \de \Om_j  \cap  
   \bbT^{-k-j-1}_{\sM  + \sN -k-1})$,   and similarly replace   $\Phi_{k+1, L\Om_{k+1}} $  by 
     $ [\Phi_{k+1,L} ]_{L   \Om_{k+1}} =   [\Phi_{k+1,   \Om_{k+1}}]_L   $  where   now   $\Phi_{k+1}$
       is defined on   $\Om_{k+1}^{(k+1)}=  \Om_{k+1} \cap  \bbT^0_{\sM+\sN-k-1}$).
 With these changes we get   the      a function of  the new            
 $
\Phi_{k+1,  \bom^+}  =     ( \Phi_{1, \de  \Om_1}  , \dots,   \Phi_{k, \de \Om_k},   \Phi_{k+1, \Om_{k+1}}  )
 $ defined by   
\begin{equation}   \label{scale}
  \rho_{k+1,  \bom^+ }(\phi_{\Om^c_1},    \Phi_{k+1,  \bom^+}  )
  = \const \   \tilde  \rho_{k+1, L \bom^+  }\Big([\phi_{\Om^c_1}]_L,  [\Phi_{k+1,  \bom^+}]_L  \Big)   
\end{equation}

We  can also    compose the various averaging operators.
 For  any field on any lattice   define $Q_j \Phi  =   Q^j \Phi$.   This is averaging over cubes with
$L^j$  sites on a side.  Then  define  for  $\phi$  on  $\bbT^{-k}_{\sM + \sN  -k}$
\begin{equation}  \label{sunshine0}
Q_{k,\bom}  \phi =    \Big(  [  Q_1   \phi]_{\de \Om_1},  \dots,   [ Q_{k-1}   \phi]_{ \de    \Om_{k-1}}, [ Q_k   \phi]_{    \Om_k}   \Big)
\end{equation}
where   $ [  Q_j   \phi]_{\de   \Om_j}:  \de    \Om^{(j)}_j \to  \bbR$.   Let  $\Phi_{k, \bom}$  be any other field as in   (\ref{honey})  and
define  
 \begin{equation}
  \ba =  \ba^{(k)}  = \Big( a^{(k)}_1  ,  \dots ,     a^{(k)}_k  \Big)  \hs     a^{(k)}_j =   a_j   L^{2(k-j)}
  \end{equation}
Then we  can form  
\begin{equation}  \label{29}
   \|  \ba^{\frac12}(  \Phi_{k,\bom}  - Q_{k, \bom}  \phi ) \|^2 
\equiv  
    \sum_{j=1}^{k-1}  a^{(k)}_j \|  \Phi_{j}- Q_{j} \phi   \|^2_{ \de  \Om_j}    +  a_k \|  \Phi_k- Q_k \phi   \|^2_{   \Om_k}  
\end{equation}

    \begin{lem}  For some constant  $\cN_{k, \bom}$ 
      \begin{equation}  \label{mabel}
   \rho_{k,  \bom  }(\phi_{\Om_1^c},   \Phi_{k,  \bom}  )
    = \cN_{k, \bom}^{-1}  \int         \exp  \Big(   - \frac12   \|  \ba^{\frac12}(  \Phi_{k,\bom}  - Q_{k, \bom}  \phi ) \|^2      \Big)              \rho_0  ( \phi_{L^k}  )   d \phi_{\Om_1}
  \end{equation} 
  \end{lem}
  \bigskip

  \pr   The statement for  $k=1$ follows from  (\ref{fluff}),  (\ref{sub0}).
   Suppose it is true for  $k$.     Then    
     \begin{equation}    \label{sweet}
\begin{split}   
&\tilde  \rho_{k+1,  \bom^+  }(  \phi_{\Om^c_1} ,\Phi_{k+1,  \bom^+}  )\\
=  & \const   \int      \exp \Big(
- \frac12   \frac{a}{L^2}  \|\Phi_{k+1}- Q \Phi_{k}\|^2_{  \Om_{k+1}}  
   - \frac12   \|   (\ba^{(k)})^{\frac12}(  \Phi_{k, \bom}  - Q_{k, \bom}  \phi ) \|^2      \Big)              \rho_0  ( \phi_{L^k}  )   
    d\Phi_{k,\Om_{k+1}}   d \phi_{\Om_1}\\
\end{split}
\end{equation}
We  make the split   $\|  \Phi_k- Q_k \phi   \|^2_{   \Om_{k}}=\|  \Phi_k- Q_k \phi   \|^2_{   \Om_{k+1}}+ \|  \Phi_k- Q_k \phi   \|^2_{  \de  \Om_{k}}$.
To  evaluate the integral over  $\Phi_{k,\Om_{k+1}}$  we  expand  around the minimizer  in      $\Phi_{k,  \Om_{k+1}}$ of 
\begin{equation}  \label{stringy}
  \frac{a}{2L^2}  \|\Phi_{k+1}- Q \Phi_{k}\|^2_{ \Om_{k+1}}  +  \frac{a_k}{2} \|  \Phi_k- Q_k \phi   \|^2_{   \Om_{k+1}}
\end{equation}
  This is  a problem already  discussed in  part I on the whole torus.  The solution is the same here.  
  The   variational  equation  for     
   $\Phi_k$     is  
\begin{equation}  \label{cherry0}
\Big(a_k  +  \frac{a}{L^2}    Q^TQ  \Big)\Phi_k   =    a_k Q_{k} \phi    +\frac{a}{L^2}  Q^T  \Phi_{k+1}   
\end{equation}
The  solution is     
$\Psi_k=  \Psi_{k, \Om_{k+1}}  =        \Psi_{k, \Om_{k+1}} (\Phi_{k+1}, \phi)$   given  by   
\begin{equation}    \label{psik}
 \Psi_{k} =   Q_{k}  \phi  
- \frac{aL^{-2}}{a_k+ aL^{-2}}Q^T  Q_{k+1} \phi  +  \frac{aL^{-2}}{a_k+ aL^{-2}}Q^T  \Phi_{k+1}   
\end{equation}  
A  short calculation    shows  that  the value of  (\ref{stringy})  at the minimum is 
\begin{equation}        \label{stringy2}
   \frac{a}{2L^2}  \|\Phi_{k+1}- Q \Psi_{k}\|^2_{  \Om_{k+1}}  +  \frac{ a_k}{2} \|  \Psi_k- Q_k \phi   \|^2_{   \Om_{k+1}}  =
\frac { a_{k+1}}{2L^2} \| \Phi_{k+1}  - Q_{k+1}  \phi   \|^2_{\Om_{k+1}} 
\end{equation}  

Now   in  (\ref{sweet})    write    $\Phi_{k, \Om_{k+1}}  =  \Psi_{k, \Om_{k+1}} +  Z$  (all functions on the unit lattice   $\Om_{k+1}^{(k)}$) 
and integrate over  $Z$ instead of  $\Phi_{k, \Om_{k+1}}  $.
The terms with  no $Z's$  are  (\ref{stringy2}),  the terms linear in $Z$ vanish,  and the terms  quadratic in $Z$ when integrated over $Z$  
yield a constant.    Thus we have
  \begin{equation}    
\begin{split}   
&\tilde  \rho_{k+1,  \bom^+  }(  \phi_{\Om^c_1}, \Phi_{k+1,  \bom^+}  )\\ 
=\const     
  &    \int      \exp \Big(
- \frac {   a_{k+1}}{2L^2} \| \Phi_{k+1}  - Q_{k+1}  \phi   \|^2_{\Om_{k+1}}
 -  \frac{1}{2}  \sum_{j=1}^k  a^{(k)}_j \|  \Phi_{j}- Q_{j} \phi   \|^2_{ \de  \Om_j}       \Big)              \rho_0  ( \phi_{L^k}  )    d \phi_{\Om_1}  \\
\end{split}
\end{equation}
Scaling as  in   (\ref{scale})  yields 
 \begin{equation}
   \rho_{k+1,  \bom^+  }(\phi_{\Om_1^c},   \Phi_{k+1,  \bom^+}  )
    =  \const  \int         \exp  \Big(   - \frac12   \|  (\ba^{(k+1)})^{\frac12}(  \Phi_{k+1,\bom^+}  - Q_{k+1, \bom^+}  \phi ) \|^2      \Big)   
               \rho_0  ( \phi_{L^{k+1} } )   d \phi_{\Om_1}
  \end{equation} 
which is the result we  want.  The constant  $ \cN_{k, \bom}^{-1}$ can be  evaluated by integrating over all fields.

\subsection{free flow}

\label{freeflow}

Now  suppose we   start   with the free density  given by  (\ref{den0})  with  $V_0=0$. Written in scaled form  for    $\phi:  \bbT^{-k}_{\sM +\sN -k}  \to \bbR$ it is   
\begin{equation}    \label{tony1}
\rho_0( \phi_{L^k} )    =  \exp\Big(-  \frac12  <\phi, (- \De +   \bar  \mu_k)\phi>  \Big)
\end{equation}  
Thus we  wish to evaluate  
 \begin{equation}  \label{mabel2}
   \rho_{k,  \bom  }(\phi_{\Om_1^c},   \Phi_{k,  \bom}  )
    = \cN_{k, \bom}^{-1}  \int         \exp  \Big(   - \frac12   \|  \ba^{\frac12}(  \Phi_{k,\bom}  - Q_{k, \bom}  \phi ) \|^2    -  \frac12  <\phi, (- \De +   \bar  \mu_k)\phi>  \Big)
 \end{equation}

 The  analysis depends on the  decomposition 
  \begin{equation}  \label{tony2}
  \begin{split}
& \frac12  <\phi, (- \De +   \bar  \mu_k)\phi>\\
 = & \frac12  <\phi_{\Om^c_1}, [- \De +   \bar  \mu_k]_{\Om^c_1}\phi_{\Om^c_1}>  
    +    <\phi_{\Om_1}, [- \De]_{\Om_1,  \Om_1^c} \phi_{\Om^c_1}>  + \frac12<\phi_{\Om_1}, [- \De+   \bar  \mu_k]_{\Om_1}\phi_{\Om_1}> \\
    \end{split}
  \end{equation}  
  Here   $\phi_{\Om}$ is the restriction to  $\Om$,
   $ [- \De]_{\Om} \equiv 1_{\Om}[-\De] 1_{\Om}$ is the  Laplacian    with Dirichlet boundary conditions,
  and   $ [- \De]_{\Om,  \Om^c} \equiv     1_{\Om}[-\De] 1_{\Om^c}$.

  \begin{thm}  \label{sonnyboy}  Starting with the free density  
    after  $k$  steps   the density    has the form   
  \begin{equation}  \label{bombshell}
  \begin{split}
 &  \rho_{k, \bom}(  \phi_{\Om^c_1},   \Phi_{k, \bom}) \\
  =  & Z_{k, \bom}  \exp \Big  (- \frac{1}{2}  \|\ba^{1/2}   (\Phi_{k,\bom}- Q_{k, \bom} \phi )  \|^2_{\Om_1} -\frac12  <\phi, (- \De +   \bar  \mu_k  )\phi> \Big)   \ \ \   \textrm{  at   }  \ \ \     \phi_{\Om_1}    =   \phi_{k, \bom} 
     \\
 \end{split}
  \end{equation}
  Here     $ Z_{k, \bom} $   is a constant 
 and   $\phi_{k, \bom} :  \bbT^{-k}_{\sM +\sN -k} \cap  \Om_1  \to \bbR$   is defined by 
  \begin{equation}  \label{unknown}
\phi_{k, \bom}  
 =  \phi_{k, \bom}  ( \phi_{\Om^c_1},   \Phi_{k,\bom})
=  G_{k, \bom}\Big(  Q_{k, \bom} ^T  \ba \Phi_{k,\bom} +     [\De]_{\Om_1, \Om_1^c} \phi_{\Om^c_1} \Big) 
\end{equation}
where  
\begin{equation}  \label{sinsin}
G_{k, \bom}   =   \Big[ -\De  +   \bar  \mu_k   +  Q_{k, \bom}^T  \ba  Q_{k, \bom}  \Big]_{\Om_1}^{-1}  
\end{equation}
\end{thm}  
 \bigskip

\pr  Inserting   (\ref{tony2})  into  (\ref{mabel2} ) we have  
 \begin{equation}  \label{tony3}
 \begin{split}
&   \rho_{k,  \bom  }(\phi_{\Om^c_1},\Phi_{k,  \bom}  ) 
= \exp\Big( - \frac12<\phi_{\Om_1^c}, [- \De +   \bar  \mu_k]_{\Om^c_1}\phi_{\Om_1^c}>  \Big)  \cN_{k, \bom}^{-1}  \\
 &  \int         \exp  \Big(   - \frac12   \|  \ba^{\frac12}(  \Phi_{k,\bom}  - Q_{k, \bom}  \phi_{\Om_1} ) \|^2   
  -    <\phi_{\Om_1}, [- \De]_{\Om_1, \Om_1^c} \phi_{\Om^c_1}>  - \frac12 <\phi_{\Om_1}, [- \De +   \bar  \mu_0]_{\Om_1}\phi_{\Om_1}> 
   \Big)           d \phi_{\Om_1} \\
   \end{split}
  \end{equation}

We   do the integral   by  minimizing the exponent in  $ \phi_{\Om_1}$.
Taking the derivative in this variable and setting it equal to zero  gives the variational equation
\begin{equation}
[ - \De  +\bar   \mu_k  +   Q_{k, \bom} ^T \ba   Q_{k, \bom} ]_{\Om_1}  \phi_{\Om_1}  
=    Q_{k, \bom} ^T  \ba  \Phi_{k,\bom} +     [\De]_{\Om_1, \Om_1^c} 
\phi_{\Om_1^c}   
\end{equation}
and the solution is   $\phi_{\Om_1} =  \phi_{k, \bom}$ as given by  (\ref{unknown}).

Now  in   the exponent in  (\ref{tony3})  write   $\phi_{\Om_1}  =  \phi_{k, \bom}  + \cZ$   and integrate over $\cZ$ instead of $\phi_{\Om_1}$.   The 
term with no $\cZ$'s   comes outside the integral  and  gives the exponential in (\ref{bombshell}).   The term  linear in  $\cZ$  vanishes. 
The  term   quadratic  in $\cZ$  is    $-  \frac12   <\cZ, [- \De +   \bar  \mu_k +   Q^T_{k, \bom} \ba  Q_{k, \bom} ]_{\Om_1}\cZ  > $.
Thus we  have  the result  with 
\begin{equation}   
Z_{k, \bom}  =   \cN_{k, \bom}^{-1}    \int         \exp  \Big( 
    -  \frac12   <\cZ, [- \De +   \bar  \mu_k +   Q^T_{k, \bom} \ba  Q_{k, \bom}  ]_{\Om_1}\cZ  >   \Big)           d \cZ 
\end{equation}
\bigskip

\re   
The  result can also be written in the form
  \begin{equation}  \label{bombshell2}
  \begin{split}
 &  \rho_{k, \bom}(  \phi_{\Om^c_1},   \Phi_{k, \bom}) \\
  =  & Z_{k, \bom}  \exp \Big  (-\frac12  <\phi_{\Om^c_1}, [- \De +   \bar  \mu_k]_{\Om^c_1}\phi_{\Om^c_1}>  
 -  <\phi_{k, \bom}, [- \De]_{\Om_1,  \Om_1^c}     \phi_{\Om^c_1}>  -S_{k}(\Om_1,    \Phi_{k,\bom},  \phi_{k, \bom})\Big) \\
 \end{split}
  \end{equation}
where       
    \begin{equation}   \label{sun}
S_{k}(\Om_1,   \Phi_{k,\bom}, \phi)=
 \frac{1}{2}  \|\ba^{1/2}   (\Phi_{k,\bom}- Q_{k, \bom} \phi  )  \|^2_{\Om_1} 
  +  \frac12   <\phi , [- \De +   \bar  \mu_k]_{\Om_1}\phi  > 
   \end{equation}
This  action   can   also  be usefully  written in terms of the fundamental variables:   
\begin{lem}    
\begin{equation}  \label{rain}
S_{k}(\Om_1,    \Phi_{k,\bom},  \phi_{k, \bom})= \frac12  < \Phi_{k,\bom}, \De_{k, \bom}    \Phi_{k,\bom}>    +     \frac12  \Big<    \phi_{\Om_1^c}, 
  [\De]_{\Om^c_1, \Om_1}  G_{k, \bom}    [\De]_{\Om_1, \Om_1^c}\phi_{\Om_1^c}\Big>  
\end{equation} 
  where   
 \begin{equation} 
 \De_{k, \bom}   =   \ba   -   \ba    Q_{k, \bom}  G_{k, \bom} Q^T_{k, \bom}  \ba  
 \end{equation}
\end{lem}
\bigskip

\pr  From  (\ref{sun}) we have
  \begin{equation}   \label{sun2}
  \begin{split}
S_{k}(\Om_1,    \Phi_{k,\bom},  \phi_{k, \bom})=&
 \frac{1}{2}  \|\ba^{1/2}  \Phi_{k, \bom}   \|^2    -<Q^T_{k, \bom} \ba  \Phi_{k,\bom}, \phi_{k, \bom} >\\
  - & \frac12  \Big <\phi_{k, \bom} , \Big[- \De +   \bar  \mu_k  +  Q^T_{k, \bom} \ba  Q_{k, \bom}\Big ]_{\Om_1}\phi_{k, \bom}  \Big > \\
  \end{split}
   \end{equation}
 Insert   the  expression  for   $\phi_{k,\bom}$     and obtain    
\begin{equation}  \label{ping}
\begin{split}
S_{k}(\Om_1,    \Phi_{k,\bom},  \phi_{k, \bom})
=& \frac12   \| \ba^{1/2}\Phi_{k, \bom} \|^2 
  -  \Big<  Q_{k, \bom} ^T \ba \Phi_{k,\bom}, G_{k, \bom}\Big (Q^T_{k, \bom} \ba \Phi_{k, \bom}  +        [\De]_{\Om_1, \Om_1^c} \phi_{\Om^c_1}\Big)   \Big>  \\
  + & \frac12  \Big<  \Big(Q_{k, \bom} ^T\ba \Phi_{k,\bom} +     [\De]_{\Om_1, \Om_1^c} \phi_{\Om^c_1}\Big), 
  G_{k, \bom}\Big( Q_{k, \bom} ^T \ba\Phi_{k,\bom} +     [\De]_{\Om_1, \Om_1^c} \phi_{\Om^c_1}\Big)\Big> \\
 =&  \frac12   \| \ba^{1/2}\Phi_{k, \bom} \|^2 -  \frac 12  \Big< Q_{k, \bom} ^T \ba\Phi_{k,\bom}  , G_{k, \bom} Q^T_{k, \bom} \ba \Phi_{k, \bom}\Big>  \\
  + & \frac12  \Big<   \phi_{\Om^c_1}, 
  [\De]_{\Om^c_1, \Om_1}  G_{k, \bom}    [\De]_{\Om_1, \Om_1^c}   \phi_{\Om^c_1}\Big> \\  
\end{split}
\end{equation}
which is  (\ref{rain}).  \bigskip

\subsection{free flow - single step}  \label{single step}

Now  we  investigate  how to follow the free flow a step at a time.  This is in preparation for a similar step for the full model.
Assuming  the  representation   (\ref{bombshell})     in the next step we  would want to compute  (ignoring constants)
 \begin{equation} \label{thefirst}
      \int     \exp \Big(  -J_{\bom^+}\big( \Phi_{k+1}, \Phi_{k,\bom}, (\phi_{\Om^c_1},  \phi_{k, \bom}) \big)  \Big)   d\Phi_{k,\Om_{k+1}}  
\end{equation}
where   for     $\phi: \bbT^{-k}_{\sM + \sN -k}  \to \bbR$   and  $\Phi_{k+1}:  \Om_{k+1}^{(k+1)} \to \bbR$
\begin{equation}
\begin{split}
&J_{\bom^+}( \Phi_{k+1}, \Phi_{k,\bom}, \phi )  \\ 
=&   \frac12 \frac{a}{L^2} \|\Phi_{k+1}   -  Q \Phi_{k}\|^2_{  \Om_{k+1} }  +
 \frac{1}{2}  \|\ba^{1/2}   (\Phi_{k,\bom}- Q_{k, \bom} \phi )  \|^2_{\Om_1}  +  \frac12 \Big< \phi,   (-\De  +   \bar    \mu_k ) \phi \Big> 
\\
\end{split}
\end{equation}

 To evaluate this integral we need to    find the minimizer of $J_{\bom^+}\big( \Phi_{k+1}, \Phi_{k,\bom}, (\phi_{\Om^c_1},  \phi_{k, \bom}) \big) $
 in   $  \Phi_{k,  \Om_{k+1}}$.     Since   this function    is  the  minimum  
 of   
 $J_{\bom^+}( \Phi_{k+1}, \Phi_{k,\bom}, \phi ) $  in $\phi_{\Om_1}$,        we  can proceed by finding the minimum of   
  $J_{\bom^+}( \Phi_{k+1}, \Phi_{k,\bom}, \phi ) $    simultaneously  in    $\phi_{\Om_1},  \Phi_{k,  \Om_{k+1}}$.
\newpage

\begin{lem} {  \  }  \label{otto1}
\begin{enumerate}
\item   
The  unique  minimum  of  $J_{\bom^+}(\Phi_{k+1}, \Phi_{k,\bom}, \phi )$  
in   $ \phi_{\Om_1},  \Phi_{k,  \Om_{k+1}}$   comes 
at    $\phi_{\Om_1}=     \phi^0_{k+1, \bom^+}$  where  
\footnote{   $Q_{k+1, \bom^+ }$  is also written  $Q_{\bom^+, \bbT^{-k}}$.}
\begin{equation}    \label{eddie}
\begin{split}
   \phi^0_{k+1, \bom^+}( \phi_{  \Om_1^c}, \Phi_{k+1, \bom^+}  )
       =        G^0_{k+1,  \bom^+}  
\left( L^{-2} Q^T_{k+1, \bom^+ } \ba^{(k+1)}\Phi_{k+1, \bom^+}  +   [\De]_{\Om_1, \Om_1^c}   \phi_{ \Om_1^c}  \right)  \\
\end{split}
\end{equation}
with  
\begin{equation}  \label{eddie1}
G^0_{k+1,  \bom^+}  =    \Big [ - \De  +  \bar \mu_k  +  L^{-2} Q^T_{k+1, \bom^+ } \ba^{(k+1)}Q_{k+1, \bom^+ }   \Big]^{-1}_{\Om_1}
  \end{equation}
and at   $ \Phi_{k, \Om_{k+1} } =      \Psi_{k, \Om_{k+1}}(\bom^+ )$ where
\begin{equation}  \label{eddie2}
\begin{split}
 \Psi_{k, \Om_{k+1}}(\bom^+ ) \equiv   & \Psi_{k, \Om_{k+1}}( \Phi_{k+1},   \phi^0_{k+1, \bom^+})   \\
    =   &Q_{k}    \phi^0_{k+1, \bom^+}
- \frac{aL^{-2}}{a_k+ aL^{-2}}Q^T    Q_{k+1}  \phi^0_{k+1, \bom^+}  +  \frac{aL^{-2}}{a_k+ aL^{-2}}Q^T  \Phi_{k+1}  \\
  \end{split}
  \end{equation}
\item
  Let   $ \Psi_{k, \bom^+}$  be    $ \Phi_{k, \bom}$   with  $\Phi_{k, \Om_{k+1}}$   replaced
by   the minimizer   $  \Psi_{k, \Om_{k+1}}(\bom^+)$,   that is 
\begin{equation}
  \Psi_{k, \bom^+}    \equiv ( \Phi_{1, \de  \Om_1}  , \dots,   \Phi_{k, \de    \Om_k},    \Psi_{k, \Om_{k+1}}(\bom^+)    )
\end{equation}
Then  the    minimizer in  $\phi$  can also be  written  $ \phi_{k, \bom}(\phi_{\Om_1^c}, \Psi_{k, \bom^+})$  so we have the identity
\begin{equation}   \label{loopy}
 \phi^0_{k+1,\bom^+} = \phi_{k, \bom}(\phi_{\Om_1^c}, \Psi_{k, \bom^+})
 \end{equation}
\item
    The   value  of    $J_{\bom^+}( \Phi_{k+1}, \Phi_{k,\bom}, \phi ) $    at the minimizer is:
\begin{equation}   \label{nono}
\begin{split}
 &       \frac12 \Big< \phi_{\Om^c_1},   [-\De  +   \bar    \mu_k ]_{\Om^c_1} \phi_{\Om^c_1} \Big> 
   + \Big< \phi_{\Om^c_1}  ,   [-\De ]_{\Om_1, \Om_1^c}   \phi^0_{k+1, \bom^+} \Big> 
  +   S_{k+1}^0   ( \Om_1,  \Phi_{k+1,  \bom^+},     \phi^0_{k+1,\bom^+}      )  \\
 \end{split}
\end{equation}
where  
\begin{equation}
 \begin{split}
  S_{k+1}^0   ( \Om_1,   \Phi_{k+1,  \bom^+},  \phi   )=&   \frac{1}{2}  \sum_{j=1}^k  a^{(k)}_j \|  \Phi_{j}- Q_{j} \phi   \|^2_{ \de  \Om_j} +
\frac{ a_{k+1}}{2L^2} \| \Phi_{k+1}  - Q_{k+1}  \phi   \|^2_{\Om_{k+1}}
 \\
  +  & \frac12 \Big<   \phi,  \Big [-\De  +   \bar    \mu_k \Big ]_{\Om_1}   \phi \Big>  \\
\end{split}  
\end{equation}
\end{enumerate}
\end{lem}
\bigskip

\pr  
 Recalling  the expression  (\ref{29})  for   $  \| ( \ba^{(k)} )^{\frac12}(  \Phi_{k,\bom}  - Q_{k, \bom}  \phi ) \|^2 $
 we find that the variational equations for  $J$  in  $\phi =  \phi_{\Om_1}$  and  $\Phi_k =  \Phi_{k, \Om_{k+1}}$
 are   
\begin{equation}  \label{cherry}
 \begin{split}
 \Big(a_k  +  \frac{a}{L^2}    Q^TQ  \Big)\Phi_k   =  &  a_k Q_{k} \phi    +  Q^T  \Phi_{k+1}   \\
\Big( - \De  +\bar   \mu_k  +   Q_{k, \bom} ^T \ba^{(k)}   Q_{k, \bom} \Big)  \phi  
=&    Q_{k, \bom} ^T  \ba^{(k)}   \Phi_{k, \bom} +     [\De]_{\Om_1, \Om_1^c} 
\phi_{\Om_1^c}      \\
\end{split}
\end{equation}
Both of these we have seen before.
The  first equation is solved  by   $\Phi_k  =        \Psi_k (\Phi_{k+1}, \phi)$  defined in (\ref{psik}).
We  substitute this into the second   equation.   First note that   splitting  $\Phi_{k, \Om_k} =  \Phi_{k,  \de \Om_k}  +
 \Phi_{k,  \de \Om_{k+1}}$    we have
\begin{equation}
  Q_{k, \bom} ^T  \ba^{(k)}   \Phi_{k,\bom}  =   \sum_{j=1}^{k-1}   a^{(k)}_j   Q_j^T \Phi_{j, \de \Om_j}    + a_k   Q_k^T \Phi_{k,  \Om_{k+1}} 
\end{equation}
The  substitution  $\Phi_{k, \Om_{k+1} }= \Psi_{k}$  goes in the last  term  here,   and we  have  
 on $\Om_{k+1}$   
 \begin{equation}
a_kQ_k^T      \Psi_k  = a_k Q^T_k Q_{k}  \phi  
- a_{k+1}L^{-2}Q^T_{k+1} Q_{k+1} \phi  + a_{k+1}L^{-2}Q_{k+1}^T  \Phi_{k+1} 
\end{equation}
Then  the second  equation becomes 
\begin{equation}
\begin{split}
&\Big( - \De  +\bar   \mu_k  + \sum_{j=1}^k [Q_j^T a^{(k)}_j   Q_j   ]_{\de \Om_j}   +
 L^{-2}[Q^T_{k+1} a_{k+1} Q_{k+1}]_{\Om_{k+1}}  \Big )   \phi   \\
=& \sum_{j=1}^k   a^{(k)}_j   Q_j^T \Phi_{j, \de \Om_j}    + L^{-2} a_{k+1}Q_{k+1}^T   \Phi_{k+1}  +     [\De]_{\Om_1, \Om_1^c} 
\phi_{\Om_1^c}   \\
\end{split}
\end{equation}
This has the solution  $\phi  = \phi^0_{k+1,  \bom^+}$,   and with this choice the first equation is solved by  
 $\Phi_k   =        \Psi_k (\Phi_{k+1},  \phi^0_{k+1,  \bom^+}) \equiv  \Psi_{k,  \Om_{k+1}}( \bom^+)  $.  This establishes  (\ref{eddie}),  (\ref{eddie2}).

Replace   $\Phi_{k, \bom}$   by   $\Psi_{k, \bom^+} $  in  the second equation  in  (\ref{cherry}) and solve for  $\phi$.  We  find 
that    $\phi = \phi_{k, \bom}(\phi_{\Om_1^c}, \Psi_{k, \bom^+})$.   This gives the second representation for the minimizer 
  and establishes  (\ref{loopy}).

To  evaluate  $J$  at the minimum    split    the term   $  \frac{1}{2}  \|\ba^{1/2}   (\Phi_{k,\bom}- Q_{k, \bom} \phi )  \|^2_{\Om_1} 
$  into  a piece   in   $\Om_{k+1}$  and a piece  in   $\Om_1- \Om_{k+1}$.
Then      the value  at the minimum  is   
  \begin{equation}  \label{stung}
\begin{split}
&J_{\bom^+}( \Phi_{k+1},  \Psi_{k,\bom^+}, (\phi_{\Om_1^c},  \phi^0_{k+1, \bom^+} ))  
=    \frac{a}{2L^2} \|\Phi_{k+1}   -  Q   \Psi_{k}\|^2_{  \Om_{k+1} } +
 \frac{a_k}{2}  \|   \Psi_k- Q_{k}  \phi   \|^2_{ \Om_{k+1} } \\   
+  &  
  \frac{1}{2}  \sum_{j=1}^k  a^{(k)}_j \|  \Phi_{j}- Q_{j} \phi   \|^2_{ \de  \Om_j} 
  +  \frac12 \Big< \phi,   (-\De  +   \bar    \mu_k ) \phi \Big>  
  \ \ \  \textrm{ at  }   \ \ \    \phi_{\Om_1}  =   \phi^0_{k+1, \bom^+},  \Psi_k =   \Psi_{k,  \Om_{k+1}}( \bom^+) \\
\end{split}
\end{equation}
However  just as  in    (\ref{stringy2})  the first  two terms combine  to  give 
   $  \frac12  a_{k+1}L^{-2}  \|    \Phi_{k+1}- Q_{k+1}  \phi \|^2_{ \Om_{k+1} }  $
so this is the same as   
   \begin{equation}  \label{stung2}
\begin{split}
  &   \frac{1}{2}  \sum_{j=1}^k  a^{(k)}_j \|  \Phi_{j}- Q_{j} \phi   \|^2_{ \de  \Om_j} +
\frac{ a_{k+1}}{2L^2} \| \Phi_{k+1}  - Q_{k+1}  \phi   \|^2_{\Om_{k+1}}
  +   \frac12 \Big< \phi,   (-\De  +   \bar    \mu_k ) \phi \Big>   \ \ \  \textrm{ at  }   \ \ \    \phi_{\Om_1}  =   \phi^0_{k+1, \bom^+}
\end{split}
\end{equation}
This is the same as  (\ref{nono})   and  
   this completes the proof.
\bigskip

\res   We  develop some consequences of these  results.    
 Suppose  
 we expand around the minimizer  for    $J_{\bom^+}( \Phi_{k+1}, \Phi_{k,\bom},  (\phi_{\Om_1^c}, \phi_{k, \bom} )$ in  $\Phi_{k, \Om_{k+1}}$, namely
  $     \Psi_{k,\Om_{k+1}}( \bom^+)   $. 
Put   $\Phi_{k, \Om_{k+1}}=    \Psi_{k, \Om_{k+1}}(\bom^+)  +  Z$   and hence   
$\Phi_{k,\bom}   =    \Psi_{k, \bom^+}   +  (0, Z)  $.
Then   $ \phi_{k, \bom}  =  \phi_{k, \bom} ( \phi_{\Om_1^c},   \Phi_{k, \bom} ) $  becomes  
\begin{equation}
\begin{split}
 \phi_{k, \bom} \Big( \phi_{\Om_1^c},   \Psi_{k, \bom^+}    +(0,Z)    \Big)  = &   \phi_{k, \bom} ( \phi_{\Om_1^c},  \Psi_{k, \bom^+}    ) 
  +  a_k  G_{k, \bom}  Q_k ^T Z
=  \phi^0_{k+1, \bom^+} + \cZ_{k, \bom}  \\
\end{split}
\end{equation}
 Here   we have  used  (\ref{loopy})    and   defined   $ \cZ_{k, \bom}  = \phi_{k, \bom}(0,Z) = a_k  G_{k, \bom}  Q_k ^T Z$.
Now  we  claim  that   
\begin{equation}  \label{expand}
\begin{split}
&
J_{\bom^+}( \Phi_{k+1},  \Psi_{k, \bom^+}   +  (0, Z)  ,( \phi_{\Om_1^c} ,  \phi^0_{k+1, \bom^+ } + \cZ_{k, \bom} )  )\\
=&     \frac12 \Big< \phi_{\Om^c_1},   [-\De  +   \bar    \mu_k ]_{\Om^c_1} \phi_{\Om^c_1} \Big> 
   + \Big< \phi_{\Om^c_1}  ,   [-\De ]_{\Om_1, \Om_1^c}   \phi^0_{k+1, \bom^+} \Big>  +
     S_{k+1}^0   ( \Om_1,  \Phi_{k+1,  \bom^+},     \phi^0_{k+1,\bom^+}      )\\
      + & \frac12 \Big(Z,  \left[ \De_{k, \bom} +  \frac{a}{L^2} Q^TQ\right]_{\Om_{k+1}}  Z  \Big) \\
\end{split}
\end{equation}
That   the first three terms   are   the  value  at  $Z=0$  follows from the previous lemma.   The  linear  terms must vanish.      Thus we only have to look at the quadratic terms  in $Z$ which are
\begin{equation}     \label{worry}
\begin{split}
&   \frac{a}{2L^2}  \|  Q   Z \|^2_{  \Om_{k+1}}
+  S_{k}(\Om_1,   (0,Z), \cZ_{k, \bom}  )   
=      \frac{a}{2L^2}  \|    Q  Z \|^2_{  \Om_{k+1}}
 +  \frac12 \Big(Z,  \left[ \De_{k, \bom} \right]_{\Om_{k+1}}  Z  \Big) 
  \\
\end{split}
\end{equation}
The second form follows from  (\ref{ping}).
Hence  (\ref{expand})  is  established.

The  original integral    (\ref{thefirst})  with  $Z$ as the integration variable   would   now  be evaluated as  
\begin{equation}     \label{cloudy3}
\begin{split}
& \exp    \Big( - \frac12 \Big< \phi_{\Om^c_1},   [-\De  +   \bar    \mu_k ]_{\Om^c_1} \phi_{\Om^c_1} \Big> 
   - \Big< \phi_{\Om^c_1}  ,   [-\De ]_{\Om_1, \Om_1^c}   \phi^0_{k+1, \bom^+} \Big>  
     -   S_{k+1}^0   ( \Om_1,  \Phi_{k+1,  \bom^+},     \phi^0_{k+1,\bom^+}      ) \Big)\\
   &  \int    \exp \Big(     -  \frac12 \Big(Z,  \left[ \De_{k, \bom} +  \frac{a}{L^2} Q^TQ\right]_{\Om_{k+1}}  Z  \Big)   dZ   \\
\end{split}   
\end{equation}

Let  us also  check that this scales the way   we   expect.    As  in  section  \ref{averaging}  we   replace each  $\Om_j$  by  $L \Om_j$
and each  field  like    $\Phi_{j, \de \Om_j}$   by    $ [\Phi_{j,L} ]_{L\de \Om_j}  =[\Phi_{j, \de \Om_j}]_L $.       Since  $\bar \mu_k = L^{-2} \bar \mu_{k+1}$  and 
 $a^{(k)}_j  =  L^{-2} a^{(k+1)}_j$   and  $Q_j$ is scale invariant   we have    
\begin{equation}
\begin{split}
&
\Big(  - \De  +  \bar \mu_k  +  L^{-2} Q^T_{k+1, \bom^+ } \ba^{(k+1)}Q_{k+1, \bom^+ }             \Big )  f_L  \\
=  &   L^{-2}  \Big(  (-  \De  +  \bar \mu_{k+1}  + Q^T_{k+1, \bom^+ } \ba^{(k+1)}Q_{k+1, \bom^+ }    )  f  \Big)_L \\
\end{split}
\end{equation}
It  follows that 
\begin{equation}
G^0_{k, L\bom^+} f_L   =   L^2  [ G_{k, \bom^+}  f]_L
\end{equation}
and  hence   from  (\ref{eddie})  
\begin{equation}   \label{lumpy}
 \phi^0_{k+1, L \bom^+} \Big([\phi_{\Om_1^c}]_L,  [\Phi_{k+1, \bom^+}]_L\Big)      
 =  \Big[ \phi_{k+1,  \bom^+} (\phi_{\Om_1^c},  \Phi_{k+1, \bom^+})\Big]_L
\end{equation}
Then  
\begin{equation}   \label{oooo}
 S^0_{k+1} \Big( L\Om_1,  [\Phi_{k+1, \bom^+}]_L,     [\phi_{k+1,  \bom^+}]_L\Big)     
=    S_{k+1} \Big( \Om_1,  \Phi_{k+1, \bom^+},     \phi_{k+1,  \bom^+}\Big) 
    \end{equation}  
The  other terms  in the exponent in    (\ref{cloudy3})   scale   similarly,  and  thus  they  become     
\begin{equation}
\exp    \Big( - \frac12 \Big< \phi_{\Om^c_1},   [-\De  +   \bar    \mu_{k+1} ]_{\Om^c_1} \phi_{\Om^c_1} \Big> 
   - \Big< \phi_{\Om^c_1}  ,   [-\De ]_{\Om_1, \Om_1^c}   \phi_{k+1, \bom^+} \Big> 
      -  S_{k+1} \Big( \Om_1,  \Phi_{k+1, \bom^+},     \phi_{k+1,  \bom^+}\Big) 
 \Big)
\end{equation}
as expected.

\subsection{a variation}

We actually use a variation of  the previous section.  First  suppose  $\La$ is any   union of $M$-cubes  in $\tk$ and define
\be   \label{louie}
S^*_k( \La,\Phi_{k,\bom}, \phi)  
=
   \frac{ 1}{2}  \| \ba^{1/2}(\Phi_{k,  \bom}- Q_{k, \bom}   \phi)\|^2_{\La}
+   \frac12 \| \pa  \phi \|_{*, \La}^2   + \frac12  \bar  \mu_k  \|  \phi   \|^2_{\La}\\
\ee
 Here   $ \| \pa  \phi \|_{*, \La}^2$  contains 
half the bonds  that  cross the boundary of $\La$.     Precisely it is defined  for  $\phi:  \bbT^{-k}_{\sM + \sN -k} \to  \bbR $  by
\begin{equation}
\| \pa  \phi \|_{*, \La}^2  =  \sum_{<x,x'>  \in \La }L^{-3k}  | \pa \phi(x,x')|^2    +    \frac12
 \sum_{x \in \La,  x'  \in  \La^c }L^{-3k}  | \pa \phi(x,x')|^2   
\end{equation}
This has the advantage  that   if  $\La_1,  \La_2$   are  disjoint  (but possibly with a common boundary),  then 
\begin{equation}
\| \pa  \phi \|_{*, \La_1  \cup  \La_2 }^2 
=   \| \pa  \phi \|_{*, \La_1}^2   +  \| \pa  \phi \|_{*, \La_2}^2 
\end{equation}
A similar decomposition holds    for  $S^*_k( \La,\Phi_{k,\bom}, \phi)  $.

Also suppose the  free action is in a  set  $\La$  smaller  than  $\Om_1$.    In fact  suppose we
have  
 \begin{equation} \label{geometry}
      \Om_1 \supset  \Om_2  \supset  \cdots \supset \Om_k  \supset   \La  \supset  \Om_{k+1}
 \end{equation}
 with  separation between  $\Om_1^c$ and  $\La$.
In that  case    (\ref{louie}) becomes
 \be 
S^*_k( \La,\Phi_k, \phi)  
=
     \frac{ a_k}{2}  \| \Phi_k- Q_k   \phi\|^2_{\La}
+   \frac12 \| \pa  \phi \|_{*, \La}^2   + \frac12  \bar    \mu_k  \|  \phi   \|^2_{\La}
\ee
We     study what  happens   to  this  expression if  we  make   expansions  around
the minimizer for the original problem        $\phi  =\phi_{k, \bom}=  \phi_{k, \bom} (\phi_{\Om^c},\Phi_{k, \bom})$   (which satisfies useful identities).
Although this is not the minimizer for the current problem   we   will   obtain a similar result.

\begin{lem}  For    $Z:  \bbT^{0}_{\sM + \sN -k} \to  \bbR$ and    $\cZ :    \bbT^{-k}_{\sM +\sN-k}  \to   \bbR$
each defined on  a neighborhood of  $\La$  
\begin{equation}
\begin{split}
S^*_k( \La,\Phi_{k}  +Z     , \phi_{k,\bom}  + \cZ)  
=  &  S^*_k( \La,   \Phi_{k}, \phi_{k,\bom})      + S^*_k( \La,  Z, \cZ)    \\
+    &  a_k
< Z,   ( \Phi_k     - Q_k  \phi_{k,\bom})>_{\La} +  \sfb_{\La} (\pa \phi_{k, \bom},\cZ)       \\
\end{split}
\end{equation}
where  the boundary term is 
\begin{equation}
\sfb_{\La} (\pa \phi,\cZ)       \equiv 
    \frac12  \sum_{ x \in \La,  x'\in \La^c} L^{-2k}   \pa  \phi (x,x')  \Big( \cZ(x) + \cZ(x') \Big )  
\end{equation}
\end{lem}
\bigskip

\pr   Everything is quadratic so it suffices to identify cross terms.
These are
\be
\begin{split}
& a_k< Z,   ( \Phi_k     - Q_k  \phi_{k,\bom})>_{\La}  \\
-  & a_k< ( \Phi_k     - Q_k  \phi_{k,\bom}),  Q_{k}\cZ>_{\La} 
 +     <\pa  \phi_{k, \bom},\pa  \cZ>  _{*, \La}  + \bar   \mu_k <  \phi_{k, \bom}, \cZ>  _{ \La}\\
\end{split}
\end{equation}
In    appendix  \ref{zzz}  it is shown that    
  \begin{equation}  
  <\pa  \phi_{k, \bom},\pa  \cZ>  _{*, \La}   = <(-\De)   \phi_{k, \bom},  \cZ>  _{\La} 
+  \sfb_{\La} (\pa \phi_{k, \bom},\cZ)    
\end{equation}
Then our expression becomes   
\be
\begin{split}
& a_k< Z,   (  \Phi_k     - Q_k  \phi_{k,\bom})>_{\La}  \\
-  &  a_k < Q_{k}^T  \Phi_{k}, \cZ>_{\La}  +    
 \Big< (-\De +    \bar   \mu_k  + a_k Q_{k}^T   Q_{k} )\phi_{k, \bom}, \cZ\Big>  _{ \La}  +   \sfb_{\La} (\pa \phi_{k, \bom},\cZ)   
\\
\end{split}
\end{equation}
But the second and third terms  combine to zero    by the definition (\ref{unknown})   of  $\phi_{k, \bom}$.   There is no contribution from $\phi_{\Om_1}^c$ due to our separation assumption. This completes the proof.
\bigskip

Continuing with  our  assumption    on $\La$, we  now investigate how the single  step analysis  of section \ref{single step}
changes,  particularly  (\ref{expand}).     We     introduce   
\begin{equation}    \label{jstar}
J^*_{\La, \Om_{k+1}}( \Phi_{k+1}, \Phi_{k}, \phi )  
=  \frac{a}{2L^2} \|\Phi_{k+1}   -  Q \Phi_{k}\|^2_{  \Om_{k+1} }  + S^*_k( \La,\Phi_{k}, \phi)  
\end{equation}
and  expand in   $\Phi_{k, \Om_{k+1} }$   and  $\phi$  around   the minima  $ \Psi_{k, \Om_{k+1}}(\bom^+) $
and  $\phi^0_{k+1, \bom^+} =  \phi_{k,\bom}(\phi_{\Om_1^c}, \Psi_{k, \bom^+})   $  for the original problem  with $\bom^+$.
The    result is the following 

\begin{lem}  \label{otto2}
 For   $Z:  \Om_{k+1}^{(k)}  \to \bbR$  and     $\cZ_{k, \bom}  =  \phi_{k, \bom} (0,Z)$ 
\begin{equation}
\begin{split}
&J^*_{\La, \Om_{k+1}}\Big( \Phi_{k+1},   \Psi_{k, \bom^+}  +(0,  Z)  , \phi^0_{k+1, \bom^+} +  \cZ_{k, \bom} \Big)  \\
=  &   S^{*,0}_{k+1}  ( \La,   \Phi_{k+1,  \bom^+},\phi^0_{k+1, \bom^+})  +    \frac12 \Big<Z,  \left[ \De_{k, \bom} +     \frac{a}{L^2}   Q^T  Q  \right]_{\Om_{k+1}}  Z  \Big> +     R_{k, \bom, \La} +   \sfb_{\La} (\pa \phi_{k, \bom},\cZ)    \\
 \end{split}
\end{equation}
where    
\begin{equation}
\begin{split}
S^{*,0}_{k+1}  ( \La,   \Phi_{k+1,  \bom^+},\phi)  = & \frac{a_{k+1}}{2L^2}  \| \Phi_{k+1} - Q_{k+1} \phi \|^2_{\Om_{k+1} } 
+  \frac{a_k}{2}    \| \Phi_{k} - Q_{k} \phi \|^2_{\La  - \Om_{k+1}   }   \\
+  &   \frac12 \| \pa  \phi \|_{*, \La}^2   + \frac12  \bar    \mu_k  \|  \phi   \|^2_{\La} \\
\end{split}
\end{equation}
and    
\begin{equation}
 R_{k, \bom, \La} \equiv   - \frac{ 1}{2}  \|\ba^{1/2}Q_{k, \bom}   \cZ_{k, \bom} \|^2_{ \La^c} +
  \| \pa \cZ_{k, \bom}  \|_{*, \La}^2    +   \frac12  \bar    \mu_k  \|  \cZ_{k, \bom}   \|^2_{  \La^c} 
\end{equation}
\end{lem}
\bigskip

\re    $S^{*,0}_{k+1}  ( \La,   \Phi_{k+1,  \bom^+},\phi^0_{k+1, \bom^+})$  scales to   $S^{*}_{k+1}  ( \La,   \Phi_{k+1,  \bom^+},\phi_{k+1, \bom^+})$   as   in  (\ref{oooo}).
\bigskip

\pr   By the previous lemma,   and using again  (\ref{loopy}),   our expression is   
\begin{equation}
\begin{split}
  &    
   \frac{a}{2L^2} \|\Phi_{k+1}   -  Q  \Psi_{k, \Om_{k+1}}(\bom^+)   \|^2_{  \Om_{k+1} } +     \frac{a}{2L^2} \| Q Z\|^2_{  \Om_{k+1} } - \frac{a}{L^2} <Q Z, (\Phi_{k+1}   -  Q  \Psi_{k, \Om_{k+1}}(\bom^+))>     \\
   + & S^*_k( \La,    \Psi_{k, \Om_{k+1}}(\bom^+), \phi^0_{k+1, \bom^+})     +S^*_k( \La, (0, Z),  \cZ_{k, \bom} )   \\
+  &  a_k   < Z,   (  \Psi_{k, \Om_{k+1}}(\bom^+)     - Q_k \phi^0_{k+1, \bom^+} )>_{\Om_{k+1}} +   \sfb_{\La} (\pa \phi_{k, \bom},\cZ)      \\
 \end{split}
\end{equation}
However the   linear terms vanish since \begin{equation}
 \frac{a}{L^2} Q^T (\Phi_{k+1}   -  Q  \Psi_{k, \Om_{k+1}}(\bom^+) ) =  \frac{a_{k+1}}{L^2} Q^T(  \Phi_{k+1} - Q_{k+1} \phi^0_{k+1, \bom^+} )
 =  a_k  (  \Psi_{k, \Om_{k+1}}(\bom^+)     - Q_k \phi^0_{k+1, \bom^+})
\end{equation}
as one can check by    inserting the definition of  $ \Psi_{k, \Om_{k+1}}(\bom^+)   $ from   (\ref{eddie2}).    
  Also  as in  (\ref{stung}), (\ref{stung2})
\begin{equation}
 \frac{a}{2L^2} \|\Phi_{k+1}   -  Q  \Psi_{k, \Om_{k+1}}(\bom^+) \|^2_{  \Om_{k+1} } 
+    S^*_k( \La,   \Psi_{k, \Om_{k+1}}(\bom^+), \phi^0_{k+1, \bom^+})  =     S^{*,0}_{k+1}  ( \La,   \Phi_{k+1,  \bom^+},\phi^0_{k+1, \bom^+})
\end{equation}
Finally    
\begin{equation}
\begin{split}
   \frac{a}{2L^2} \|  QZ\|^2_{  \Om_{k+1} }
     +S^*_k( \La, (0, Z),  \cZ_{k, \bom} )  =  &  \frac{a}{2L^2} \|  QZ\|^2_{  \Om_{k+1} }
     +S_k( \Om_1, (0, Z) ,  \cZ_{k, \bom} )   + R_{k, \bom, \La}  \\
=  &      \frac12 \Big<Z,  \left[ \De_{k, \bom} +     \frac{a}{L^2}   Q^T  Q  \right]_{\Om_{k+1}}  Z  \Big> +     R_{k, \bom, \La} \\
\end{split}
\end{equation}
The last step   is from      (\ref{worry}).  This completes the proof.

 \subsection{random walk expansion}  \label{random2}

We  analyze the propagator   
\begin{equation}
G_{k, \bom}   = \Big[ -\De   + \bar  \mu_k  +  Q_{k, \bom} ^T \ba\   Q_{k, \bom} \Big]^{-1}_{ \Om_1}
\end{equation}
with Dirichlet boundary conditions,  
defined on functions on $\Om_1  \subset  \bbT^{-k}_{\sM+\sN-k}$.
We will  need     a random walk expansion for this operator   analagous to the global   expansion  explained   in  part I.

Recall  that  
$\bom  =    ( \Om_1, \Om_2,   \dots,  \Om_k)$  with   $\Om_j \supset  \Om_{j+1}$   and   $\Om_j$  a 
union   of  $L^{-(k-j)}M$  cubes.    For the random walk expansion  we   impose the separation  condition
\begin{equation}   \label{hundred}
d(\Om^c_j,  \Om_{j+1} )  \geq L^{-(k-j)}MR
\end{equation}
for  some    positive integer $R = \cO(1)$.
 Hence     $\de \Om_j  = \Om_j - \Om_{j+1}$  has  a minimum  width $L^{-(k-j)}MR$.
There is an associated     scaled distance   that  will  play a  role in what follows.  It is defined for  $(x,y)  \in  \cup_{j=1}^k  \de \Om_j^{(j)}$  by
\begin{equation}   \label{stingray}
d_{\bom}(x,y)   =  \inf_{\ga:    x \to  y}  \sum_{j=1}^k  L^{k-j}  \ell(\ga  \cap \de   \Om_j )
\end{equation}   
with  $\de \Om_k  =  \Om_k$.
Here  $\ga$ is a  path   joining   $x,y$    in the lattice  $\bbT^{-k}_{\sM+\sN-k}$  such that  in    $ \de  \Om_j$ the path
   $\ga$  consists of   $L^{-(k-j)}$ links   in       $\de \Om_j^{(j)}$.     The    factor  $L^{k-j}$  in $d_{\bom}(x,y)   $  means we count
these     links      as unit length.   For    $0 \leq  \de   \leq  1$  if    $MR$  is sufficiently  large 
this  satisfies  the bound  (Lemma 2.1  in  \cite{Bal84b})
\begin{equation}  \label{funnysum}
\sum_y   \exp \Big(   -  \de  d_{\bom}(x,y)   \Big)   \leq     \cO(1)\de^{-3}
\end{equation}

 Now  $\delta  \Om_j$    is partitioned into $L^{-(k-j)}M =   L^{-k+j+m}$  cubes   $\square_z$   centered on the points 
 $z   \in   \de  \Om^{(j+m)}_j $.       Correspondingly   there  is  a multiscale  partition 
 $\square_z$     of  $ \Om_1  =  \de   \Om_1  \cup  \de  \Om_2  \cup  \cdots   \cup \de \Om_{k-1}  \cup   \Om_k$.
   We   also   consider   enlargements  $\tilde \square_z$  which   are  centered on
 the same  points but have width  $3L^{-(k-j)}M$ in  $\de \Om_j$.    They provide a cover of $\Om_1$.

 The  random walk expansion  is  based on local   inverses  based on   the   cubes  $\tilde  \square  =  \tilde \square_z$
we   define 
\begin{equation}
G_{k, \bom}(\tilde  \square)   =\Big [ -\De   + \bar    \mu_k+    Q_{k, \bom} ^T \ba\  Q_{k, \bom} \Big]^{-1}_{\tilde  \square  \cap  \Om_1}
\end{equation}
Here  $[-\De]_{\tilde  \square  \cap  \Om_1}$  is taken with Neumann boundary conditions  on the part of the boundary 
of  $\tilde  \square  \cap  \Om_1$    in $ \Om_1$,  and Dirichlet boundary conditions of the part of the boundary  shared with  $\pa  \Om_1$.
Away from  $ \pa \Om_1$   it is just   $[-\De]_{\tilde  \square}$ with Neumann conditions,  as in  part I.
 \bigskip

Define 
\begin{equation}  \label{text}   
\De_y  =  L^{-(k-j)}  \textrm{   cubes     centered on   }   y   \in   \de \Om_j^{(j)}   
\end{equation}
These   give a finer  partition  of $\de  \Om_j$   and hence   also        a partition of  $\Om_1$.    A   basic   result is the following:

 \begin{lem}   \label{bonfire}
 Let   $\De_y  \subset  \tilde \square  \cap  \de  \Om_{j}$   and $    \De_{y'}  \subset   \tilde \square  \cap \de   \Om_{j'}$,  $|j-j'|\leq  1$.   Then  with   $\ga_0  = \cO(L^{-2})$ 
  \begin{equation}  \label{lefty}
\begin{split}
 |1_{\De_y}G_{k, \bom}(\tilde   \square) 1_{\De_{y'}}f|      \leq &  CL^{-2(k-j)}  
  e^{ - \frac12 \ga_0  d_{\bom}(y,y') }  \|f\|_{\infty}  \\
  |1_{\De_y} \pa  G_{k, \bom}(\tilde   \square) 1_{\De_{y'}}f|      \leq &  CL^{-(k-j)}  
  e^{ - \frac12 \ga_0  d_{\bom}(y,y') }  \|f\|_{\infty}  \\
   |1_{\De_y}\de_{\al}  \pa G_{k, \bom}(\tilde   \square) 1_{\De_{y'}}f|      \leq &  CL^{-(1- \al)(k-j)}  
  e^{ - \frac12 \ga_0  d_{\bom}(y,y') }  \|f\|_{\infty}  \\
  \end{split}
\end{equation}
\end{lem}
\bigskip

\re  
\begin{enumerate}
\item  Here   $\de_{\al}$ is the Holder derivative defined for $x \neq x'$ by  $
(\de_{\al} f) (x,x')  =  ( f(x)  -f(x'))d(x,x')^{-\al}$.
We  take  $\frac12 < \al < 1$.
\item  
 There is another way to state this bound which will be useful.
It  is    
\begin{equation}  \label{another}
\begin{split}
&\Big |1_{\De_{y}}G_{k, \bom}(\tilde   \square) 1_{\De_{y'}}f\Big|,  \  \     L^{-(k-j)}   \Big|1_{\De_{y}}\pa  G_{k, \bom}(\tilde   \square)1_{\De_{y'}} f\Big|,   \  \     L^{-(1 + \al)(k-j)}   \Big|1_{\De_{y}}\de_{\al}\pa  G_{k, \bom}(\tilde   \square)1_{\De_{y'}} f\Big|  \\
&\ \ \ \  \leq   CL^{-2(k-j')}  
  e^{  - \frac12 \ga_0  d_{\bom}(y,y') }  \|f\|_{\infty}  \\
\end{split}
\end{equation}
Since  $|j-j'| \leq 1$  this follows from  (\ref{lefty}).  
\end{enumerate}
\bigskip

\pr   First  suppose that  $\De_y, \De_{y'}  \subset   \tilde \square  \subset   \de \Om_j $
In    this circumstance  we  have  $\tilde  \square  \cap  \Om_1 =\tilde  \square $  and 
\begin{equation}
G_{k, \bom}(\tilde  \square)  
 = [ -\De   +  \bar \mu_k  + a_jL^{2(k-j)} Q_j^T  Q_j]^{-1}_{\tilde  \square}   
 \end{equation}
 We need to prove the bound with   $d_{\bom}(y,y')  = L^{k-j}d(y,y') $.    
   We  scale  up    the 
 the required estimate  from  $\bbT^{-k}_{\sM + \sN -k}$ to   $\bbT^{-j}_{\sM + \sN -j}$  and prove it there.    Replace  $f$ by     $f_{L^{-(k-j)}}$   
 where   $f:  \bbT^{-j}_{\sM + \sN -j} \to \bbR$,  and replace   $\tilde \square$  by  $L^{-(k-j)} \tilde \square$  where  now
    $\tilde \square$  is  now  a $3M$-cube in $ \bbT^{-j}_{\sM + \sN -j}$.   Then       $G_{k, \bom}(L^{-(k-j)} \tilde \square)  
 f_{L^{-(k-j)}}  = L^{-2(k-j)}  [ G_j(\tilde  \square) f  ]_{L^{-(k-j)}}$  where   
\begin{equation}  \label{gong0}
G_k(\tilde  \square)  =    [ -\De   +  \bar \mu_k  + a_k Q_k^T  Q_k]^{-1}_{\tilde  \square}   
 \end{equation}
 is the standard propagator on    $\tk$
 For  the   bounds  (\ref{lefty}) it now suffices to prove  that for    unit cubes  $\De_y,  \De_{y'}$
 and     $x,x' \in \De_y$,  $ \supp f  \subset \De_{y'}$:
 \begin{equation}   \label{gong1}
 \begin{split}
  |(G_k(\tilde   \square) f)(x)|, \ \    |(\pa G_k(\tilde   \square) f)(x)|,   \ \    |(\de_{\al}\pa G_k(\tilde   \square) f)(x,x')|     \      \leq &\  C  e^{  -  \ga_0  d(y,y' ) }  \|f\|_{\infty} \\
  \end{split}
 \end{equation}
 These are  already  established; see    \cite{Bal83b}  or  Appendix  D  in  part I.

  It    may happen  that  $\tilde  \square$  in  not in a single  $\de \Om_j$.     Suppose that   $ \tilde \square$  intersects  both 
  $\de \Om_j$  and $\de  \Om_{j+1}$.  Then 
  \begin{equation}
G_{k, \bom}(\tilde  \square)  
 =\Big [ -\De_{\tilde  \square}      +  \bar \mu_k  + a_jL^{2(k-j)} [Q_j^T  Q_j]_{\tilde \square \cap \delta \Om_j}
+ a_{j+1}L^{2(k-(j+1))} [Q_{j+1}^T  Q_{j+1}]_{\tilde \square \cap \delta \Om_{j+1} } \Big]^{-1}
 \end{equation}
 Since    $d_{\bom}(y,y')  \leq    L^{k-j}d(y,y') $  it   again  suffices to prove   (\ref{lefty}) with the $  L^{k-j}d(y,y')$ 
 in the exponential.
  Again  we  scale  up  to  $\bbT^{-j}_{\sM + \sN -j}$  where the propagator becomes   
   \begin{equation}
G'_{j}(\tilde  \square)  
 \equiv   \Big [ -\De_{\tilde  \square}   
   +  \bar \mu_j  + a_j [Q_j^T  Q_j]_{\tilde \square \cap \delta \Om_j}
+  \frac{a_{j+1}}{L^2} [Q_j^T  Q_j]_{\tilde \square \cap \delta \Om_{j+1} } \Big]^{-1}
 \end{equation}
We  must establish  that  $G'_{j}(\tilde  \square) $  satisfies      bounds  of the form     (\ref{gong1}), now with     $\De_y$ a unit cube,  $\De_{y'}$  an  $L$-cube,  and   $x,x' \in \De_y$,  $ \supp f  \subset \De_{y'}$:
In  this case  we use  the identity  (\cite{Bal84b},  p.  230)
\begin{equation}
G'_{j}(\tilde  \square)  =  G_j(\tilde \square)  +  a_j^2   G_j(\tilde \square) Q_j^T  C_j( \tilde \square \cap \de \Om_{j+1}) Q_j   G_j(\tilde \square) 
\end{equation}
Here all the pieces  have  pointwise bounds of the form we want  and this   yields the  the result.  For $G_j(\tilde \square) $   use   (\ref{gong1})     and   for $ C_j( \tilde \square \cap \de \Om_j)$ see  
 \cite{Bal83b}  or  Appendix  D  in  part I .
 Also note the following piece of the estimate.  If  the $L$-cube  $\De_{y'}$  is written  as a union of  unit cubes $\De_{y''}$  then  for
 $x \in \De_y$,  $ \supp f  \subset \De_y'$
\begin{equation}
\begin{split}
& |(G_j(\tilde \square) f)(x)| 
  \leq    \sum_{y''}    |(G_j(\tilde \square) 1_{\De_{y''}}f)(x)|  
 \leq   C    \sum_{y''}     e^{ - \ga_0  d(y,y'') } \|f\|_{\infty} 
\leq  C  e^{ -\frac12   \ga_0  d(y,y') } \|f\|_{\infty} \\
\end{split}
\end{equation}
Here    we  used  $d(y,y'') \geq   d(y,y')  -  L$

There is another special case that needs to be considered,   namely when  $\tilde \square$ touches or  intersects  $ \Om^c_1$.
In this case   $\tilde  \square  \cap  \Om_1$   may not be rectangular  and we may have mixed boundary conditions and so  the 
pointwise bounds   (\ref{gong1})  may  not hold.  However  we do still have
$L^2$ bounds even for  non-rectangular regions and mixed boundary conditions.    In the scaled version instead of  
(\ref{gong0})  we have  for   a $3M$ cube  $\tilde \square$  in  $\bbT^{-1}_{\sM + \sN  -1}$
\begin{equation} 
G_1(\tilde  \square)  =    [ -\De   +  \bar \mu_1  + a_1 Q^T  Q]^{-1}_{\tilde  \square  \cap  \Om_1}   
 \end{equation}
Instead of (\ref{gong1}) we have  for  unit cubes   $\De_y, \De_y'  \subset  \tilde  \square  \cap  \Om_1$ 
 \begin{equation}   \label{gong2}
  \|1_{\De_y} G_1(\tilde   \square)1_{\De_{y'}}   f\|_{2},  \  \        \leq   C
  e^{ - \ga_0  d(y,y' )}  \|f\|_{2} 
 \end{equation}
However   on this  $L^{-1}$ lattice  we have
$L^{-3}\|  f \| _{\infty, \De_y}  \leq      \| f\|_{2, \De_y}   \leq   \|f \|_{\infty, \De_y}$
  so  this    implies
 \begin{equation}   \label{gong3}
   |1_{\De_y} G_1(\tilde   \square)1_{\De_{y'}}   f|   \      \leq   C    e^{ - \ga_0  d(y,y' )}   \|f\|_{\infty} 
 \end{equation}
 Also on this  $L^{-1}$ lattice this implies bounds on the   derivatives.
 Thus   $G_1(\tilde \square)$   satisfies  the bounds (\ref{gong1}).   This completes the proof.
\bigskip

A   random walk or  path  is  a  sequence of points 
\begin{equation}
\om=    (  \om_0,  \om_1,  \dots,   \om_n  )
\end{equation}
in   $   \de \Om_1^{(1+m)} \cup \cdots  \cup     \de \Om_{k-1}^{(k-1+m)}   \cup   \Om_k^{(k+m)}$.
  These are the  centers of the cubes  in the multi scale partition  $\{\square_z\}$.  Successive points in the walk  are  required to be neighbors in the sense that  the larger cubes satisfy  $\tilde   \square_{\om_j}  \cap  \tilde \square_{\om_{j+1}} \neq \emptyset$.

\begin{thm}     \label{th}
The Green's function $G_{k, \bom}$ defined in  (\ref{sinsin})  has  a random walk expansion of 
 the form
\begin{equation}  \label{stinger}
G_{k, \bom}  =  \sum_{\om } G_{k,\bom,   \om}
\end{equation}
convergent  for    $M$   sufficiently  large.   It yields the bounds for      $ \De_y \subset     \de \Om_j $
and   $  \De_{y'}  \subset    \de \Om_{j'} $     as in  (\ref{text}):
  \begin{equation}  \label{lefty2}
  \begin{split}
 |1_{\De_{y}}G_{k, \bom} 1_{\De_{y'}} f|\  \leq   \  &    C L^{-2(k-j')}
   e^{ - \frac14 \ga_0  d_{\bom}(y,y') }   \|f\|_{\infty}  
    \\ 
 L^{-(k-j)} | 1_{\De_{y}} \pa G_{k, \bom} 1_{\De_{y'}} f|  \    \leq    &  \   C L^{-2(k-j')}
   e^{ - \frac14 \ga_0  d_{\bom}(y,y') }   \|f\|_{\infty}  \\
   L^{-(1+ \al)(k-j)} | 1_{\De_{y}} \de_{\al}\pa G_{k, \bom} 1_{\De_{y'}} f|  \    \leq    &  \   C L^{-2(k-j')}
  e^{ - \frac14 \ga_0  d_{\bom}(y,y') }   \|f\|_{\infty}  \\
\end{split}  
\end{equation} 
\end{thm}
\bigskip

\pr   We  sketch the proof   (see \cite{Bal83b},  \cite{Bal84b},  \cite{Bal96b} ).    Let  $0 \leq  h_z \leq  1$  be such  that     $h^2_z$  be a smooth   partition of 
unity subordinate to   the covering   $\{\tilde \square_z  \}$  of  $\Om_1$.  Thus  $\supp\  h_z  \subset  \tilde \square_z$
and   $\sum_z h_z^2  =1$   on a neighborhood of   $\Om_1$.  Taking advantage of the size of  $\tilde \square_z$   we   can arrange that
in   $\de \Om_j$
 \be
 |\pa  h_z|  \leq   \cO(1) (L^{-(k-j)}M)^{-1}   \hs| \pa  \pa  h_z|  \leq   \cO(1) (L^{-(k-j)}M)^{-2}
\ee
Define the parametrix  
\begin{equation}
G^*_{k, \bom}   =   \sum_z h_{z}  G_{k, \bom} (\tilde  \square_z)  h_{z}
\end{equation}
Then   
\begin{equation}
\begin{split}
 \Big[ -\De   + \bar \mu_k  +   Q_{k, \bom} ^T  \ba  Q_{k, \bom} \Big ]_{\Om_1}  G^*_{k, \bom}  
=  & \sum_{z} h_{z} \Big [ -\De   + \bar \mu_k  +   Q_{k, \bom} ^T  \ba  Q_{k, \bom} \Big ]_{\Om_1}   G_{k, \bom} (\tilde \square_z)  h_{z}\\
+ & \sum_{z} K_z   G_{k, \bom} (\tilde  \square_z)  h_{z}\\
\end{split}
\end{equation}
where
\begin{equation}
  K_z  =    \Big[     \Big [ -\De   + \bar \mu_k  +   Q_{k, \bom} ^T  \ba  Q_{k, \bom}\Big ]_{\Om_1},  h_z \Big]
\end{equation}  
In  the first  term,  since   $\supp\  h_z$ is well inside $\tilde \square_z$  we  impose   add  Neumann boundary conditions
on  $\tilde  \square_z  \cap  \Om_1$  with no change  and identify
  $ [ -\De   + \bar \mu_k  +   Q_{k, \bom} ^T  \ba  Q_{k, \bom}  ]_{\tilde \square_z  \cap  \Om_1} $.  This removes  the  operator 
  $  G_{k, \bom} (\tilde \square_z) $  and leaves us with   $\sum_z h_z^2  =1$
Thus we  have
\begin{equation}
\Big [ -\De   + \bar \mu_k  +   Q_{k, \bom} ^T  \ba  Q_{k, \bom} \Big ]_{\Om_1}  G^*_{k, \bom}  
=  I   -  \sum_{z} R_z    \equiv   I  - R
\end{equation}
where 
\begin{equation}
R_z   =  K_z G_{k, \bom}(\tilde \square_z) h_z
\end{equation}
Then   if the series converges
\begin{equation}
\begin{split}
G_{k, \bom} =  &   G^*_{k \bom}    (I -R)^{-1}  
=    G^*_{k \bom}   \sum_{n=0}^{\infty}   R^n  \\
=& \sum_{n=0}^{\infty}     \sum_{\om_0,   \om_1,  \dots,   \om_n}
\Big( h_{\om_0}  G( \tilde    \square_{\om_0})  h_{\om_0}  \Big)
  R_{\om_1}  \cdots   R_{\om_n}  \\
  \equiv &   \sum_{\om}  G_{k, \bom,  \om } \\
\end{split}
\end{equation}
The last line  defines  $G_{k, \bom,  \om } $.
We have used that  $R_zR_{z'}  =0$  unless  $\tilde \square_z  \cap  \tilde \square_{z'} \neq \emptyset$ to identify the
sum over walks.

We   estimate  $K_z f$.
First   for  $x  \in   \De_y \subset  \de \Om_j$   we have 
 \begin{equation}
|(  [  -\De, h_z] f)(x) |   \leq     \cO(1)   \Big (  (L^{-(k-j)}M)^{-2} \|1_{\De_y}f\|_{\infty}   + (L^{-(k-j)}M)^{-1} \|1_{\De_y}  \pa  f  \|_{\infty} \Big )
\end{equation}
Indeed  the    term  $ [  -\De, h_z]$  is local  and involves derivatives of  $h_z$,
  so   we  get  the  indicated factors.
    The   term  $\left[   Q_{k, \bom}^T \ba    Q_{k, \bom}, h_z\right]  f$       
can  also   be  expressed in term  of derivatives  of  $h_z$  since  it can be written  in $\de \Om_j$ as
\begin{equation}
 a_j^{(k)}\Big(  \Big[Q_j^T Q_j, h_z  \Big] f  \Big)  (x)   = a_j L^{2(k-j)}  L^{-3j}  \sum_{x' \in  B_j(x)} ( h_z(x') - h_z(x))f(x') 
\end{equation}
and so  is   estimated  by   
\begin{equation}
 a_j^{(k)} \Big|\Big( \Big[Q_j^T Q_j, h_z\Big] f  \Big)  (x) \Big|    \leq   \cO(1) L^{2(k-j)} M^{-1} \|1_{\De_y}f\|_{\infty}  
\end{equation}
Combining these  we  have  for   $x  \in   \De_y  \subset \de \Om_j$ 
 \begin{equation}
|(    K_z f)(x) |   \leq     \cO(1) M^{-1}  \Big (  L^{2(k-j)} \|1_{\De_y}f\|_{\infty}   + L^{(k-j)} \|1_{\De_y}  \pa  f  \|_{\infty} \Big )
\end{equation}
Combining   this bound      with the   bound   (\ref{another})    on  $G_{k, \bom}(\tilde  \square_z)$ 
yields  for $x  \in   \De_y \subset  \tilde \square_z  \cap \de \Om_j$  and   $\supp f   \subset  \De_{y'} \subset  \tilde \square_z  \cap  \de \Om_{j'}$: 
 \begin{equation}
 \begin{split}
&L^{-2(k-j)}|(K_z G_{k, \bom}(\tilde  \square_z)f)(x) |  \\
&   \leq       \cO(1) M^{-1}  \Big (   \|1_{\De_y} G_{k, \bom}(\tilde  \square_z)f\|_{\infty}   
+ L^{-(k-j)}  \|1_{\De_y}  \pa   G_{k, \bom}(\tilde  \square_z)f  \|_{\infty}  \Big)  \\
&  \leq    C M^{-1} L^{-2(k-j')} e^{ -  \frac12  \ga_0  d_{\bom}(y,y') } \|f\|_{\infty} \\
 \end{split}
\end{equation}
It  follows that  for the same  $x, f$
\begin{equation}     \label{style}
L^{-2(k-j)}|(R_zf)(x) |  \
 \leq  \  C M^{-1}  L^{-2(k-j')} e^{ - \frac12  \ga_0  d_{\bom}(y,y') } \|f\|_{\infty} 
 \end{equation}

Now consider  $G_{k, \bom,   \om}$  with  $|\om| = n$.    By   (\ref{another})  and  (\ref{style})   we have for  
 $x  \in   \De_y$  and   $\supp f   \subset  \De_{y'}$   with  $y_0 = y,  y_{n+1} = y'$
 \begin{equation}
\begin{split}
|( G_{k,\bom,   \om}f )(x)| 
 =  & \left|\Big(  \Big( h_{ \om_0} G_{k, \bom}(\tilde  \square_{\om_0})    h_{ \om_0} \Big)R_{\om_1}   \cdots    R_{\om_n}    f \Big)(x)  \right|\\   
 \leq    &   \sum_{y_1,\dots,  y_n}  \left|   \Big(  ( h_{\om_0} G_{k, \bom}(\tilde  \square_{\om_0})  h_{\om_0}\Big)1_{\De_{y_1}}  
R_{\om_1}    1_{\De_{y_2}} 
\cdots   1_{\De_{y_n}}   R_{\om_n}     f \Big)(x) \right|\\
 \leq   &C\Big (C M^{-1}\Big)^n  L^{-2(k-j')}  \sum_{y_1,\dots,  y_n}  \prod_{j=0}^n    e^{-\frac12  \ga_0 d_{\bom}(y_j,y_{j+1}) } \|f\|_{\infty}\\
\leq   & C\Big (C M^{-1}\Big)^n   L^{-2(k-j')}    e^{-\frac14  \ga_0 d_{\bom}(y_j,y_{j+1}) }  \|f\|_{\infty}\\
\end{split}
\end{equation}
In the last  step   we  use  
\begin{equation}
\sum_{j=0}^n   d_{\bom}(y_j,y_{j+1})   \geq     d_{\bom}(y,y') 
\end{equation}
to  extract a factor    $ e^{ - \frac14\ga_0  d_{\bom}(y,y') }$,   
and then   use (\ref{funnysum})  with  $\de   =   \frac12 \ga_0$  repeatedly.   

For convergence of the random walk expansion    we have     \begin{equation}  \label{sonic}
\begin{split}
|( G_{k, \bom}f) (x) |   \leq &    
\sum_{\om} |( G_{k,\bom, \om}f ) (x) | 
\leq  C    L^{-2(k-j')} e^{-\frac14  \ga_0 d_{\bom}(y_j,y_{j+1}) }\|f\|_{\infty}\sum_{n=0}^{\infty}  \sum_{\om:  |\om|  =n} \Big (C M^{-1}\Big)^n     \\
\end{split}
 \end{equation}
 But       for each  $\tilde \square_z$  there are  at most   $\cO(L^2)$  cubes $\tilde \square_{z'}$
 such that   $\tilde \square_z \cap  \tilde \square_{z'}  \neq  \emptyset$,  and so there are at most  $ \cO(L^{2n})$
 paths with  $| \om|  = n$.  Thus   for  $M$  sufficiently large
 the sum is bounded  by   
 $   \sum_{n=0}^{\infty}    (CL^2 M^{-1})^n    \leq  2
 $.
  This establishes  the bound  on   $G_{k, \bom}$.
  
      For the bound on $ \pa G_{k, \bom}$ there
  are terms with   $\pa G_{k, \bom}(\tilde  \square_{\om_0}) $   and we  need the extra factor  $L^{-(k-j)}$  to
  estimate it  by  (\ref{another}).  A similar remark applies to the Holder derivatives.  This completes the proof.
  \bigskip

As an  application we give an estimate  on  $\phi_{k, \bom} =\phi_{k, \bom} ( \phi,  \Phi_{k, \bom} )$ as defined in  (\ref{unknown}).
With  $\de \Om_j = \Om_j - \Om_{j+1}$  for  $j=1, \dots, k-1$ and      $\de \Om_k  = \Om_k$  define
\be 
\|  \Phi_{k, \bom}  \|_{\infty}   =   \sup_{1 \leq j \leq  k}   \|  \Phi_{j, \de  \Om_j}  \|_{\infty} 
\ee

\begin{lem}   \label{twoone} There is a constant  $C$ depending only on $L$ such that  on $\de \Om_j$
\be  \label{twoone1}
|\phi_{k, \bom} |,   \  \ \   L^{-(k-j)}  |\pa  \phi_{k, \bom} |,  \ \ \   L^{-(1 + \al)(k-j)} |\de_{\al} \pa  \phi_{k, \bom} |
\leq    C   \Big(   \|\phi\|_{\infty}  +   \|\Phi_{k,\bom} \|_{\infty} \Big) 
 \ee
 \end{lem}

\pr    $\phi_{k, \bom}$  is a sum of two terms.  The first  is 
\be 
\begin{split}
G_{k, \bom} Q_{k, \bom}^T \ba^{(k)} \Phi_{k, \bom}
=  & 
  \sum_{j'=1}^k  a^{(k)}_{j'}   G_{k, \bom} Q_{j'}^T  \Phi_{j', \de  \Om_{j'}} 
 \\
=  &
  \sum_{j'=1}^k   \sum_{y' \in \de \Om^{(j')}_{j'} } a_{j'}  L^{2(k-j')}G_{k, \bom}  1_{\De_{y'}}Q_{j'}^T  \Phi_{j', \de  \Om_{j'}} 
  \\
\end{split}
\ee
Then    on     $ \De_y \subset  \de  \Om_j$,   by   (\ref{lefty2})    and     (\ref{funnysum})   (note the cancellation of the factors  
$ L^{2(k-j')}$  by   (\ref{lefty2}) )
\be 
\begin{split}
\Big|G_{k, \bom} Q_{k, \bom}^T \ba^{(k)} \Phi_{k, \bom}\Big|
\leq     & 
 C  \sum_{j'=1}^k   \sum_{y'  \in  \de \Om^{(j')}_{j'} }   
e^{ - \frac14 \ga_0 d_{\bom}(y,y' )}  
\|Q_{j'}^T  \Phi_{j', \de  \Om_{j'}} \|_{\infty}
 \\
 =       & 
 C   \sum_{y' }   
e^{ - \frac14 \ga_0 d_{\bom}(y,y' )}  \| \Phi_{k, \bom} \|_{\infty} 
\leq           
 C \| \Phi_{k, \bom} \|_{\infty} \\
\end{split}
\ee
The second term  is   for  $\phi$ on $\Om_1^c$:
\be    G_{k, \bom}[ \De]_{\Om_1,  \Om^c_1}  \phi
=    \sum_{y' \in \de \Om_1^{(1)}}     G_{k, \bom}1_{\De_{y'}} [ \De]_{\Om_1,  \Om^c_1}  \phi
\ee
Then   on   $\De_y \subset  \de  \Om_j$,   by   (\ref{lefty2})   and     (\ref{funnysum})   )  
\be 
\begin{split}  \label{lunar}
  \Big|G_{k, \bom}[ \De]_{\Om_1,  \Om^c_1}  \phi    \Big| \
   \leq \  &
 C  \sum_{y'}  e^{ - \frac14 \ga_0 d_{\bom}(y,y' )}    L^{-2(k-1)} \|   [ \De]_{\Om_1,  \Om^c_1}  \phi  \|_{\infty}
  \leq  
   C \| \phi  \|_{\infty}\\
 \end{split}
\ee
In  the last   step we  used that   for      $ \phi:   \bbT^{-k}_{\sM + \sN -k} \to \bbR$   we have    $ \|  \De \phi \|_{\infty}  \leq  \cO(1)  L^{2k} \|\phi\|_{\infty}$      
Combining the two  bounds  gives the bound on   $\phi_{k, \bom}$.   The  bounds on the derivatives   are   similar. 
   \bigskip

\noindent  \textbf{Variations}:

\noindent(A.)   A  local version of   (\ref{twoone1})    will also be useful.     This   says   for  $L^{-(k-j)}$ cubes  $\De_y$ in  $\de  \Om_j$
\be   \label{twoone3} 
\begin{split}
& \Big| 1_{\De_y}  \phi_{k, \bom}\Big(  1_{\De_{y'} } ( \phi,   \Phi_{k, \bom})\Big)\Big|,\    L^{-(k-j)} \Big| 1_{\De_y} \pa  \phi_{k, \bom}\Big(  1_{\De_{y'} } ( \phi,   \Phi_{k, \bom})\Big)\Big|, \   \\
&
      L^{-(1 + \al)(k-j)} \Big| 1_{\De_y} \de_{\al} \pa  \phi_{k, \bom}\Big(  1_{\De_{y'} } ( \phi,   \Phi_{k, \bom})\Big)\Big|   
        \leq  \  C  e^{ - \frac14 \ga_0 d_{\bom}(y,y' )}    \Big(   \|\phi\|_{\infty}  +   \|\Phi_{k,\bom} \|_{\infty} \Big) \\
\end{split}
\ee
The follows  since only one term in the final sum over  $y'$ contributes. 
\bigskip

\noindent (B.)    We  can introduce a weakening  parameter
  $0 \leq  s_{\square}\leq   1$
for  each  $L^{-(k-j)}M$  square  $\square$   in  $  \de  \Om_j   $   and all  $1 \leq  j  \leq  k$.   Define
\begin{equation}
   s_{\om}  =  \prod_{\square \subset  X_{\om}}  s_{\square}  \hs    X_{\om}  = \bigcup_{j=1}^n  \tilde \square_{\om_j}
\end{equation}
and   define    $G_{k, \bom}(s)$  by  
\begin{equation}  \label{stinger2}
G_{k, \bom}(s)  =  
\sum_{\om} s_{\om}G_{k,\bom,  \om}  
\end{equation}   
In   the basic  convergence  proof of the lemma   we  do  not need all of the  $M^{-1}$,   a factor
$M^{-1/2}$   would  do.     Thus if  $s_{\square}$  is complex and   $|s_{\square}|  \leq   e^{\ka_1}$
we have  an extra  factor  $  \prod_{\square}    |s_{\square}| M^{-1/2}   $   to estimate.    This 
is less than  one provided   $e^{\ka_1}  \leq  M^{1/2}$ which we assume.

All the above 
results hold  with   $|s_{\square}|  \leq   e^{\ka_1}$.   In particular   theorem \ref{th}  holds  with  $G_{k, \bom}$  replaced by 
$G_{k, \bom}(s)$.  We  also  change    $\phi_{k, \bom}$ to  $\phi_{k, \bom}(s)$   by replacing     $G_{k, \bom}$ by 
$G_{k, \bom}(s)$.  Then    $\phi_{k, \bom}(s)$ satisfies the bounds of lemma \ref{twoone}  and  (\ref{twoone3})
\bigskip

\noindent  (C.)
Similar  results  hold  for  the  Green's function   $G^0_{k+1, \bom^+}$  defined in  (\ref{eddie1}).  This has a random
walk expansion which is a scaling up  of the expansion  (\ref{stinger}) for  $k+1$.  
In this case the statement of the theorem says that     for   $ \De_y \subset     \de \Om_j $
and   $  \De_{y'}  \subset    \de \Om_{j'} $:
\be  \label{lullaby}
\begin{split}
&\Big |1_{\De_{y}}G^0_{k+1, \bom^+} 1_{\De_{y'}} f \Big|,\   L^{-(k-j)}\Big | 1_{\De_{y}} \pa G^0_{k+1, \bom^+} 1_{\De_{y'}} f \Big|,\   L^{-(1+ \al)(k-j)} \Big| 1_{\De_{y}} \de_{\al}\pa G^0_{k+1, \bom^+} 1_{\De_{y'}} f  \Big| \\
&\hs    \leq     \   C L^{-2(k-j')}
   e^{ - \frac14 \ga_0  d_{\bom^+}(y,y') }   \|f\|_{\infty}  \\
 \end{split}
 \ee 
 Here       $j=1, \dots,  k+1$  with  $\de \Om_{k+1} = \Om_{k+1}$.   Now   $\De_y$ in  $\Om_{k+1}$ is  an  $L$-cube.
 Also    $d_{\bom^+}(y,y')  $  is  defined  as  in   (\ref{stingray}) but with the sum up to $k+1$,  so in $\Om_{k+1}$ paths are
    weighted by  $L^{-2}$.
   As  in   (\ref{twoone1})
the associated   fields  $\phi^0_{k+1, \bom^+}$  defined in   (\ref{eddie})  satisfy on  $\de  \Om_j$
\be   \label{dos} 
  |  \phi^0_{k+1, \bom^+}|, \   L^{-(k-j)} | \pa \phi^0_{k+1, \bom^+}|, \   L^{-(1 + \al)(k-j)} |\de_{\al} \pa \phi^0_{k+1, \bom^+}|      \leq   C \Big(  \| \phi\|_{\infty}  +  \|  \Phi_{k+1, \bom^+} \|_{\infty}   \Big)
   \ee
 This can be understood  as (\ref{twoone1})  for $k+1$ scaled up by $L$.   

Finally     one can introduce weakening parameters  $\{ s_{\square} \}$,  replacing   $G^0_{k+1, \bom^+}$ with  $G^0_{k+1, \bom^+}(s)$   and  
$\phi^0_{k+1, \bom^+}$  with 
$\phi^0_{k+1, \bom^+}(s)$,    and  obtain bounds of the same form.

\section{The full expansion}

\subsection{definitions and  notation}

In this section and the next we     introduce the concepts we need to state the main theorem.  A  basic  parameter  
is the scaled coupling constant  
\begin{equation}
\la_k   = \la_k^{\sN }  = L^{-(\sN- k)}  \la
\end{equation}
which satisfies   $\la_k  = L^k \la_0$.  This  is  our effective coupling constant after $k$ renormalization group steps.
We  always assume  $\la_k$ is sufficiently small depending on the parameter  $L,M$,  and in particular  $\log( - \la_k)  \geq  1$.

\subsubsection{small field  regions  } 
At   the $k^{th}$     stage of the  iteration we will introduce not one but two new small field regions
$\Om_k,  \La_k$   and the pair is denoted  $\Pi_k =  ( \Om_k, \La_k)$.   Each will be associated with the introduction of  characteristic functions   in a manner
yet to be explained.  
After  $k$  steps      there is  a sequence   of regions   with  $\Om_0 = \emptyset$  and   
\begin{equation}
\bpi  =    (\Pi_0,  \Pi_1, \dots  \Pi_n)  =   (\La_0,  \Om_1,  \La_1,    \Om_2,  \La_2,   \dots,   \Om_k, \La_k)
\end{equation}
These   are  decreasing: 
\begin{equation} 
\La_0   \supset  \Om_1  \supset  \La_1  \supset   \Om_2  \supset  \La_2  \supset  \cdots  \supset   \Om_k \supset  \La_k  
\end{equation}
All  these  regions  are subsets of      $\bbT^{-k}_{M+N-k}$  and      $\Om_j,  \La_j$   are  unions of     $L^{-(k-j)}M$  cubes.  
We  also    use the notation
\begin{equation}
\bom  =  (\Om_1,  \Om_2,  \dots,   \Om_k)   \hs   \bla  =  (\La_0,   \La_1,  \La_2,  \dots,   \La_k) 
 \end{equation}

 We   
require much stronger separation conditions  than those   defined  in  the previous section.  
Define  
\begin{equation}
r_k  =   r(\la_k)   =  ( -  \log \la_k ) ^r   = ((N-k) \log L   - \log   \la    )^r
 \end{equation}
 for  some postive integer  $r$.  
 We  assume  always  $\la_k$  is small  so  $- \log \la_k >0$ is  large and $r_k$ is large  and decreasing in $k$.
The separation requirement is that   
   \begin{equation}
d( (\bar  \La_{j-1})^c,  \Om_j )  \geq 5[r_j] L^{-(k-j)} M 
   \hs     d( \Om_j^c,  \La_j )  \geq 5[r_j] L^{-(k-j)}  M 
\end{equation} 
where  here  $ \bar   \La_{j-1}$ is all  $L^{-(k-j)}M$ cubes intersecting  $\La_{j-1}$.  (The  ''5'' is somewhat arbitrary.)

There are some special  cases.   It  may be     that  some  region  is the full torus    $\bbT^{-k}_{M+N -k}$.   In this case 
all   larger  regions are also the full torus  and  for these  there is no separation requirement. 
It  may also be  that  some  region is empty.   In this case all smaller regions are also empty and 
for these    there is no separation requirement.

 \subsubsection{polymers}  \label{polymersection}
  
 Recall   that a polymer      $ X \in  \cD_k$   is a   connected union of  $M$-cubes  in  $\bbT^{-k}_{\sM+\sN -k}$.
 
 A variation is a polymer with holes.   
   Given  a   final small field  region    $\Om_k$  (not necessarily connected) on the same torus,       suppose the large field region  $\Om_k^c$    has    connected components  $\Om^c_{k, \al}$   (the holes).
  We    define a subset  of  $\cD_k$   by
  \begin{equation}
  \cD_k( \bmod\  \Om^c_k   )  
  =\{  X  \subset  \cD_k:  \textrm{ for all }   \al   \textrm{  either  }  \Om^c_{k,\al} \subset  X \textrm{ or }
   \Om^c_{k,\al} ,    X  \textrm{ are disjoint } \}
   \end{equation}
 We  associate with  any   $X   \in     \cD_k( \bmod\  \Om^c_k   ) $    a  linear distance  $ d_M(X, \bmod\  \Om^c_k )$  on 
 scale  $M$  defined by  
 \begin{equation}
 M\  d_M(X, \bmod\  \Om^c_k  )  =   \inf_{  \tau  \textrm{  on  } X}    \ell(\tau)  
\end{equation}   
where the infimum is  over all continuuum   tree graphs $\tau$  contained in  $X$ and    intersecting   every $M$-cube  in
   $  X  \cap   \Om_k$,   and $ \ell(\tau)$ is the length of $\tau$.
   Thus  $d_M(X,  \textrm{  mod } \Om^c_k  )  $   measures the size  of  the components of   $X  \cap  \Om_k$  and  
   the distances between these components.   But   $ d_M(X,  \bmod\  \Om^c_k  )$  does not measure
   the  bulk of  $X  \cap  \Om_k^c$.   The idea is that decay in these regions (holes) will be taken  care of  elsewhere.     If     $X \subset   \Om_k$  then 
   $d_M(X,  \bmod\  \Om^c_k  )  = d_M(X)$  where   $d_M(X)$   is the infimum of the lengths of continuum  tree graphs intersecting  \textit{every} $M$-cube in  $X$.    In  general  
   $d_M(X, \bmod\  \Om^c_k  )  \leq    d_M(X)$. 
     For any $M$-cube     $\square  \in   \Om_k$  we  have  for a universal constants  $\ka_0,  K_0$
 \begin{equation}   \label{snow}
 \sum_{ X  \in    \cD_k( \textrm{  mod   }  \Om^c_k   ) ,  X \supset \square}
 e^{ -  \ka_0   d_M(X  \textrm{  mod } \Om^c_k    )}  \leq  K_0
 \end{equation}
 See  appendix \ref{tree} for  the proof.

Another variation is multiscale  polymers  $\cD_{k, \bom} $.   An element  $X$ of $\cD_{k, \bom}$  is  a connected
subset  of  $\tk$  with that   $X \cap  \de \Om_j$  is a union of  $L^{-(k-j)}$  cubes.     Let  
\be   |X|_{\bom}  =   \sum_{j=1}^k  |X \cap  \de \Om_j|_{L^{-(k-j)}M }
\ee
be  the total  number of  blocks  in  $X$.   Then  for  any elementary cube  $\square  \subset  \cD_{k, \bom}$ and $\ka_*$
large enough
\be   \label{lugnut8}
\sum_{X    \in   \cD_{k, \bom},     X  \supset  \square}  e^{-  \ka_*  |X|_{\bom} }   \leq    e^{- \frac12 \ka_*}
\ee
See  appendix   \ref{sanibel}  for the proof.

We  will also consider   $\cD^0_{k+1}$  which is connected unions of $LM$ cubes.  Also  for $\bom^+ = ( \Om_1, \dots  \Om_{k+1})$
with  $\Om_{k+1}$ a union of $LM$ cubes,    we  define  $\cD^0_{k+1}(  \bmod  \ \Om_{k+1})$  and  $\cD^0_{k+1, \bom^+}$  as above.
These are still in   $\tk$   but they will scale  to  $\cD_{k+1},  \cD_{k+1}(  \bmod \ \Om_{k+1})$  and   $ \cD_{k+1, \bom^+}$
on   $\bbT^{-k-1}_{\sM + \sN -k-1}$.

\subsubsection{localized functionals }

As   discussed   in   section \ref{averaging},     associated with  the   regions  $ (\Om_1,  \Om_2,  \dots,   \Om_k)$   are fundamental  fields   
\be  \label{sunshine3}
 \Phi_{k,\bom}  =  (  \Phi_{1, \de  \Om_1},  \dots,  \Phi_{k-1,  \de     \Om_{k-1}},   \Phi_{k,\Om_k})
\ee
where  $\de \Om_j  = \Om_j - \Om_{j+1}$    and    $ \Phi_{j, \de  \Om_j}$  is defined on the subset   $\de \Om_j^{(j)}$. We   want to consider  functions  of the fields of the form   $ H(X, \Phi_{k, \bom})$  with   $X \in   \cD_k( \bmod\  \Om^c_k   ) $  and 
 the property   that  they  only  depend on   $\Phi_{k, \bom}$  in  $X$.
 For bounded fields these  will satisfy  bounds like
\begin{equation}
|H(X, \Phi_{k, \bom})|  \leq  \const  \exp  \Big(  -  \ka_0   d_M(X,  \bmod\ \Om^c_k    )\Big)
\end{equation}
We   also  consider    sums  of these   denoted  for  any   union of  $M$-cubes $\La$   by   
\footnote{This is not a very good notation,  since if  $\La$ is connected this could refer  to the single 
function  $H(\La)$ rather  than the sum.  One should really  denote the sum  by a different symbol.   
But the notation is already overburdened,   so  we  instead we adopt the convention that if  the region
is given by a Greek letter like  $\La$ then we mean the sum}
\begin{equation}
H(\La)   =  \sum_{X \subset   \La}  H(X)
\end{equation}
For boundary  terms,  denoted by $B(\La)$ or some such,  we employ a  different convention summing   over  polymers  $X$ that cross
$\La$.   We  write
\begin{equation}
B(\La)  =  \sum_{X  \#  \La}  B(X)
\end{equation}
where $X  \#  \La$  means  $X$ crosses  $\La$ or 
\begin{equation}
X  \#  \La  \hspace{.3cm}   \Longleftrightarrow   \hspace{.3cm}    X \cap  \La \neq  \emptyset  \textrm{ and }  X \cap  \La^c \neq  \emptyset
\end{equation}

There  are also      the associated  smeared fields $ \phi_{k, \bom} $ on  $\bbT^{-k}_{\sM+\sN-k}$      as defined in  (\ref{unknown})
and we will also want to consider     functionals  of the fields of the form  $ H(X,  \phi_{k, \bom} ) $  or    $ H(X, \Phi_{k, \bom}, \phi_{k, \bom} )$.
These   are also required to depend on the indicated fields  in  $X$.    In this  case  the functional   depends on the fundamental fields  $ \Phi_{\bom}  $  outside of $X$, but only very weakly.    In fact it will be useful to have  a stricter  localization.  This will be accomplished by 
introducing  modifications  of  $\phi_{k, \bom} $  with   stricter localization, which we now explain in several steps.

\subsubsection{averaging operators again}

First  revisit the averaging    operator $Q_{k, \bom}$,   also denoted  $Q_{\bom,  \bbT^{-k}}$.  For $\phi$
on   $\bbT^{-k}_{\sM+\sN -k}$  we defined in  (\ref{sunshine0})
\begin{equation} 
Q_{k,\bom}  \phi  = Q_{\bom,  \bbT^{-k}}\phi =     (  [  Q_1   \phi]_{\de \Om_1},  \dots,   [ Q_{k-1}   \phi]_{ \de    \Om_{k-1}},  [Q_k \phi]_{\Om_k})
\end{equation}
As   a completion for this  averaging  we   define on  $Q_{k,\bom}  \phi $,   or more generally any  multiscale field 
  $\Phi_{k,\bom}$   of the form  (\ref{sunshine3}) the averaging operator
\begin{equation}
 Q_{  \bbT^{0}, \bom}    \Phi_{\bom} 
=   ( Q_{k-1}  \Phi_{1, \de \Om_1},  \dots,  Q_1 \Phi_{k-1,  \de  \Om_{k-1}},    \Phi_{k,    \Om_k}  )
\end{equation}
Then  we  have   
\begin{equation}      \label{concantate}
 Q_{  \bbT^{0}, \bom}   Q_{\bom,  \bbT^{-k}}\phi 
=       Q_{  \bbT^{0},   \bbT^{-k}}\phi 
=    Q_k  \phi
\end{equation}

\subsubsection{buffers}
\begin{figure}[t]  
\begin{picture}(250,200)(-100,0)
\thinlines
\put(10,10){\line(0,1){180} }
\put(10,10){\line(1,0){180} }
\put(20,10){\line(0,1){180} }
\put(10,20){\line(1,0){180} }
\put(10,190){\line(1,0){180}}
\put(10,180){\line(1,0){180}}
\put(180,10){\line(0,1){180}}
\put(190,10){\line(0,1){180}}

\multiput(10,40)(0,40){4}{\line(1,0){180}}
\multiput(40,10)(40,0){4}{\line(0,1){180}}

\multiput(60,10)(40,0){3}{\line(0,1){30}}
\multiput(60,160)(40,0){3}{\line(0,1){30}}
\multiput(10,60)(0,40){3}{\line(1,0){30}}
\multiput(160,60)(0,40){3}{\line(1,0){30}}

\multiput(10,30)(0,20){8}{\line(1,0){10}}
\multiput(180,30)(0,20){8}{\line(1,0){10}}
\multiput(30,10)(20,0){8}{\line(0,1){10}}
\multiput(30,180)(20,0){8}{\line(0,1){10}}

\linethickness{2pt}
\put(10,10){\framebox(180,180)}
\put(20,20){\framebox(160,160) }
\put(40,40){\framebox(120,120) }
\put(80,80){\framebox(40,40) }
\put(95,83){\text{$X$}}
\put(90,45){\text{$\Om_3(X)$}}
\put(90,25){\text{$\Om_2(X)$}}
\put(90,10){\text{$  \Om_1(X)$}}
\end{picture}
\caption{A minimal buffer  $X \subset  \Om_3(X) \subset \Om_2(X)  \subset  \Om_1(X)$  \label{minbuf}  }
\end{figure}
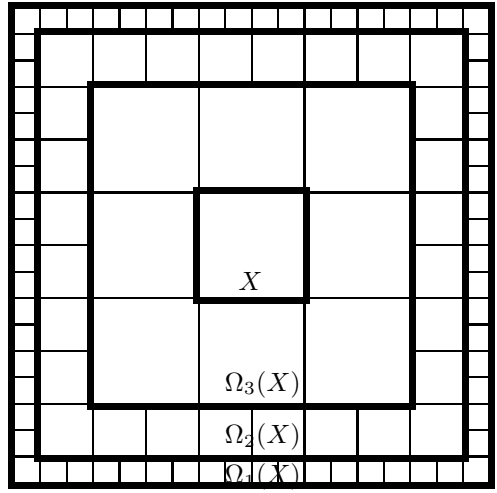

Let  $X$ be a union of  $M$ cubes in  $\bbT^{-k}_{\sM+ \sN  -k}$.   We  define    a minimal buffer 
\be
\bom(X)  =  (\Om_1( X ),  \Om_2( X ),  \dots,   \Om_k(X)) 
\ee
 around  $X$  as  follows.   The region     $\Om_k(X)$ is $X$ with  $R= \cO(1)$ layers of $M$-cubes added.  Then for 
 $j=k-1, \dots, 1$ we  successively define   $\Om_j$ to be  $\Om_{j+1}$  with  $R$ layers of  $L^{-(k-j)}M$ cubes added.
   Then  $X    \subset    \Om_k( X ) \subset  \Om_{k-1}( X ) \subset  \cdots  \subset  \Om_1( X ) $
 and the separation condition  (\ref{hundred}) is satisfied.
  The  regions  $\bom(X)$  are not generated
in the same way  as  the  $\bom$   in the main expansion,   and if $X$  is small a more typical picture of  $\bom (X)$  is 
shown in    figure  \ref{minbuf}.

We  also define $\square^{\sim n}$  to  be  $\square$  enlarged by $n $ layers of  $M$ blocks ,  and more generally    
\begin{equation}
 X^{\sim n}     =  \bigcup_{\square  \subset X}   \square^{\sim n}  =  \textrm{ $X$   enlarged by $n $ layers of  $M$ blocks }
\end{equation}  
Then    $\tilde X= X^{\sim}$  is the case  $n=1$.
Then we  have
\begin{equation}
\Om_1(X)  \subset  X^{\sim 2R}
\end{equation}

We  also   consider  larger buffers  
defining  \begin{equation}
  X^*   =  X    \textrm{ enlarged by $ [r_k] $  layers of  $M$ blocks }  =   X^{\sim [r_k]}
\end{equation}  
  We also define
      $X^{2*} =  X^{**}$,  $X^{3*} =  X^{***}$,  etc.
We  also define
\begin{equation}
  X^{\nat}  =  X_k  \textrm{ shrunk   by $ [r_k] $  layers of  $M$ blocks }  = (  (X^c)^*)^c
\end{equation}  
and       $X^{2  \nat} =  X^{\nat  \nat}$,  $X^{3 \nat} =  X^{\nat \nat \nat}$,  etc.

For a single  cube  $\square^{\nat *} = \square =  \square^{* \nat}$,   but in general
\be     X^{\nat *} \subset   X \subset   X^{* \nat}
\ee

 The definitions of  $X^{\sim},X^*, X^{\nat}$  vary with scale.   If   $X$ is specified as a union  of  $L^{-(k-j)}M$
 blocks,  then  the enlargements also  are taken  with   $L^{-(k-j)}M$ blocks.

\subsubsection{sharply  localized   fields}

\begin{enumerate}
\item
After  $k$  steps   the current small field region will be  $\La_k$   and  instead of  $\phi_{k, \bom}$   we  will want to consider a field  more sharply    localized  in  this  region;  but still not strictly localized which would be too singular.      These will be based on the buffer  $\bom(\La_k^*)$ which provides a gradual transition. 
We  would like to consider   the field  $ \phi_{k, \bom(\La^*)} $,    but  it has  to be expressed in terms  of the
fundamental  variables   $\Phi_{k,\bom}  $   and these  are   not  compatible  with  the  buffer.      The  buffer   $ \bom(\La^*)$
   is smaller  since  $\Om_1(\La_k^*) \subset  \Om_k$.
    Figure  \ref{sec}  might help in remembering the ordering of these regions.

      To  remedy this  consider the averaging operators  
 $Q_{ \bbT^0,  \bom(\La_k^*)}$   which take   functions    $ \Phi_{\bom(\La_k^*)}$
to  functions on the  unit  lattice.   Then  the adjoint  $Q^T_{ \bbT^0,   \bom(\La_k^*)}$   takes   functions  on the unit lattice       to  functions  $ \Phi_{\bom(\La_k^*)}$.
 It has the form
\be
  Q^T_{ \bbT^0,   \bom(\La_k^*)}    \Phi_k   =   \Big(  [Q_{k-1}^T \Phi_k]_{\de \Om_1(\La_k^*)},  \dots,
   [Q_1^T \Phi_k]_{\de \Om_{k-1}(\La_k^*)},    \Phi_{k, \Om_k(\La_k^*)} \Big)  
\ee
where   $ [Q_{k-j}^T \Phi_k]_{\de \Om_j(\La_k^*)}$ is defined  on  $\de \Om_j^{(j)}  \subset  \bbT^{-(k-j)}_{\sM +\sN -k}$.
 This    is   equal to  $\Phi_k$ everywhere,  but on an  increasingly fine lattice as one moves away  from $\La_k^*$.
In  $\Om_1(\La_k^*)^c  \subset  \tk$ we can take   $Q^T_k \Phi_k$.      The combination is denoted  
\be
 \tilde Q^T_{ \bbT^0,   \bom(\La_k^*)} \Phi_k=  \Big([Q^T_k \Phi_k]_{\Om_1(\La_k^*)^c}, Q^T_{ \bbT^0,   \bom(\La_k^*)}\Phi_k\Big)
\ee
and  we may consider
\be   \label{angie}
 \phi_{k, \bom(\La_k^*)} 
  \equiv   \phi_{k, \bom(\La_k^*)} \Big(\tilde   Q^T_{ \bbT^0,    \bom(\La_k^*)}   \Phi_{k}\Big) 
 \ee
 Note that this field  $ \phi_{k, \bom(\La^*)}  $ is
defined on  a neighborhood  of    $\Om_1(\La_k^*) \subset  \Om_k$.
and depends only on  $\Phi_k$.

\begin{figure} 
 \begin{picture}(250,100)(-100,0)
 \thicklines
\put(0,10){\line(0,1){50}}
\put(0,15){\vector(1,0){10}}
\put(10,20){$\Om_k$}
\put(63,10){\line(0,1){50}}
\put(65,10){\line(0,1){50}}
\put(70,10){\line(0,1){50}}
\put(55,70){$\bom(\La_k^*)$}
\put(75,10){\line(0,1){50}}
\put(75,15){\vector(1,0){10}}
\put(75,20){$\La_k^*$}
\put(100,10){\line(0,1){50}}
\put(100,15){\vector(1,0){10}}
\put(100,20){$\La_k$}
\put(200,10){\line(0,1){50}}
\put(200,15){\vector(1,0){10}}
\put(200,20){$\Om_{k+1}$}
\end{picture}
 \caption{ordering of   regions-I    \label{sec}  }
\end{figure}
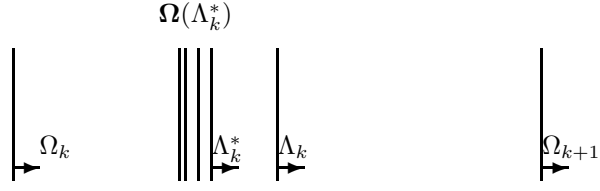

\item 
 Another  case  is   a field  defined and   localized   by  $\La_{k-1}, \Om_k,  \La_k$  as  above.  
  Suppose    $\bom  (  \La^*_{k-1} )  $ is  a buffering sequence of  length  $k-1$,   defined as  above,  but  finishing   with  unions of   $L^{-1}M$ cubes  .    We  adjoin   $\Om_k  \subset  \La_{k-1}$  and  define the     sequence    $  (\bom  (  \La^*_{k-1} ),  \Om_k )$   of length $k$ finishing with unions of   $M$ cubes.    Another  sequence
  of  length  $k$  is   $\bom  (  \La^{c,*}_{k} )$.    We combine them and form     
 \footnote{  In general    let  $\bom,  \tilde  \bom$  be  sequences   of the form  (\ref{sundry2}), and suppose
  that   $\Om^c_k$ and $\tilde \Om^c_k$  are  disjoint.   Then we  can define a new sequence  
  \[      \bom  \cap  \tilde \bom  =   \Big( \Om_1  \cap  \tilde  \Om_1,  \dots  ,   \Om_k  \cap  \tilde \Om_k \Big)
  \]
  If   $\bom, \tilde \bom$ 
 satisfy the separation condition   (\ref{hundred})  then so does  $  \bom  \cap  \tilde \bom $.   The  increments  for   $  \bom  \cap  \tilde \bom$    are  $   \de  \Om_j   \cup  \de   \tilde \Om_j$. }
  \be
  \bom( \La_{k-1},  \Om_k,  \La_k)  \equiv   \Big(\bom  (  \La^*_{k-1} ),  \Om_k  \Big)  \cap    \bom  (  \La^{c,*}_{k} )
 \ee

 Now define    a field   $ \phi_{k,  \bom( \La_{k-1},  \Om_k,  \La_k)}$   roughly localized  in  $\La_{k-1}^*  \cap  \La_k^{c,*}$    as   
  \be    \label{piggy}
        \phi_{k,  \bom( \La_{k-1},  \Om_k,  \La_k)}   \Big(  \Big[ \tilde  Q^T_{\bbT^{-1},  \bom(  \La^*_{k-1})} \Phi_{k-1}\Big ]_{ \Om^c_k},
        \  \Big[  \tilde  Q^T_{\bbT^{0},  \bom(  \La^{c,*}_{k})  }  \Phi_k  \Big] _{ \Om_k}    \Big)
        \ee
Note that    in          $\La_{k-1}^*  \cap  \La_k^{c,*}$   the field  in  parentheses  is just    $( \Phi_{k-1, \de  \Om_{k-1}},  \Phi_{k,  \Om_k} )$.      
 The ordering of the regions    is illustrated in figure \ref{sec2}.

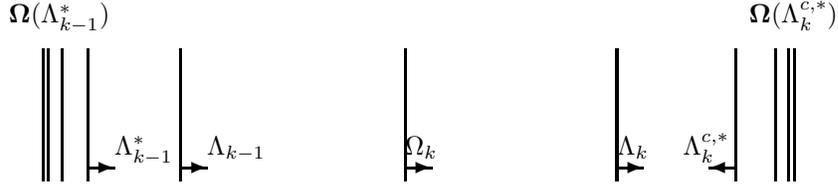
\begin{figure} 
 \begin{picture}(250,100)(-100,0)
 \thicklines
 \put(-17,10){\line(0,1){50}}
\put(-15,10){\line(0,1){50}}
\put(-10,10){\line(0,1){50}}
\put(0,10){\line(0,1){50}}
\put(0,15){\vector(1,0){10}}
\put(10,20){$\La^*_{k-1}$}
\put(35,10){\line(0,1){50}}
\put(35,15){\vector(1,0){10}}
\put(45,20){$\La_{k-1}$}
\put(120,10){\line(0,1){50}}
\put(120,15){\vector(1,0){10}}
\put(120,20){$\Om_k$}
\put(200,10){\line(0,1){50}}
\put(200,15){\vector(1,0){10}}
\put(200,20){$\La_{k}$}
\put(245,10){\line(0,1){50}}
\put(245,15){\vector(-1,0){10}}
\put(225,20){$ \La^{c,*}_k$}
\put(260, 10){\line(0,1){50}}
\put(265, 10){\line(0,1){50}}
\put(267, 10){\line(0,1){50}}
\put(-30,70){$ \bom(  \La^*_{k-1})$}
\put(250,70){$ \bom(  \La^{c,*}_k)$}
\end{picture}
 \caption{ordering of regions-II     \label{sec2}  }
\end{figure}

\item
We  also need still more  localized fields.  After  $k$ steps,  let     $\square$     be an $M$-cube  in  
$  \Om_k$.
 We  consider    the buffer 
 $\bom(    \square)$   and want to define   $ \phi_{k, \bom(  \square)} $.
   If   $\square$  is well inside   $\Om_k$    in the sense that  $\square^{\sim (2R+1)} \subset \Om_k $,   then  the buffer
     $\bom(   \square)$  is also in   $\Om_k$  and  
we  can  define       $ \phi_{k, \bom(   \square)} $   as  a function of $\Phi_k$  alone  by
\begin{equation}  \label{shortened2}
 \phi_{k, \bom(   \square)}  =  \phi_{k, \bom(   \square)} \Big( \tilde    Q^T_{\bbT^0,    \bom(   \square)}    \Phi_{k}\Big)
\end{equation}
If   however    $\square$  is near the boundary  of $\Om_k$     then       $ \phi_{k, \bom(  \square)} $  depends  on
$\Phi_{k-1}$ as well.   
 In this case  we  define
 \be   \label{shortened3}
 \phi_{k, \bom(  \square)} 
  =  \phi_{k, \bom(   \square)} \Big( \tilde    Q^T_{ \bbT^{0},    \bom(   \square)}   (Q\Phi_{k-1, \de \Om_{k-1}},  \Phi_{k, \Om_k})    \Big)
\ee

In   passing from $k$ to $k+1$  we will introduced  $\Phi_{k+1}$ in   $\Om_{k+1}$   and  $\Phi_k$ will be
 eliminated  on $\de  \Om_k = \Om_k- \Om_{k+1}$.    Then   for  $\square   \in  \de \Om_k$   we define  instead of    
 $ \phi_{k, \bom(  \square)}$
 \be  \label{sunscreen}
   \phi'_{k, \bom(  \square)} 
  =  \phi_{k, \bom(   \square)} \Big( \tilde    Q^T_{ \bbT^{0},    \bom(   \square)}   (Q\Phi_{k-1, \de \Om_{k-1}},  \Phi_{k, \de \Om_k},  Q^T\Phi_{k+1, \Om_{k+1}})  \Big)
\ee
This is the same as  $\phi_{k, \bom(   \square)} $  away from the boundary of $\de  \Om_k$

\end{enumerate}

\subsection{bounds on fields}

As a further preliminary we  define some small field regions.
A  basic parameter here  is      
\begin{equation}
p_k  =   p(\la_k)   =  ( -  \log \la_k ) ^p   = ((N-k) \log L   - \log   \la    )^p
 \end{equation}
for some positive integer  $p$   larger  than  $r$.  Also define 
\be 
  \al_k  =  \max \{  \bar \mu_k^{\frac12},  \la_k^{\frac14}  \}
\ee

We  start  with  a definition of  local small  field region:

  \begin{defn}  \label{s} For    $\square  \subset   \Om_k$,     $\cS_k(\square)$  is all  $(\Phi_{k-1,  \Om_{k-1}}, \Phi_{k, \Om_k})$  such that   
 \begin{equation}    \label{yass1}
\begin{split}
\left |   \Phi_{k}   -Q_{k} \phi_{k, \bom(\square)}  \right| 
  \leq &    p_k   \hs  \  \  \textrm{  on  } \  \tilde   \square \cap   \Om_k       \\
 | \pa  \phi_{ k,  \bom(\square)  }|
 \leq &      p_k  \hs  \  \  \textrm{  on  } \  \tilde   \square       \\ 
 |  \phi_{ k,   \bom(\square)  } |
   \leq &    p_k   \al_k^{-1}  \   \  \  \textrm{  on  } \  \tilde   \square        \\
\end{split}
\end{equation} 
If  $X$  is a union  of  $M$-cubes 
\be    \label{yass2}
\cS_k(X)   =  \bigcap_{\square  \subset   X}   \cS_k(\square)
\ee
\end{defn}

This  is similar to the  global   small field region $\cS_k$  defined in part  I.   The global field  $\phi_k$  has been replaced by 
the local  $ \phi_{ k,   \bom(\square)  } $,   and the  estimate    $ |  \phi_k   |
   \leq     p_k   \la_k^{-1/4} $    has been replaced  by   $ |  \phi_{ k,   \bom(\square)  } |
   \leq     p_k   \al_k^{-1}$.   Since  $\al_k^{-1}  =   \min \{  \bar \mu_k^{-\frac12},  \la_k^{-\frac14}  \}  \leq  \la_k^{-\frac14} $
this is sometimes   a tighter  estimate.     It is ultimately  the term  $\exp( - \bar \mu_k  \| \phi \|^2  )$ in the density  which allows this  refinement,
and we will find the stronger bound useful.
\bigskip

We  say that   $\square$  is   \textit{ well-inside }  $\Om_k$ if      $\square^{\sim (2R+1)} \subset \Om_k $.   In this case  $\cS_k(\square)$    is a condition on  $\Phi_k$ alone,
and the   bounds are more or less equivalent  to bounds on  $\Phi_k,  \pa \Phi_k$ for we have  the following result:
\footnote{   It   might seem  
more straightforward to  work  directly with    bounds on  $\Phi_k , \pa \Phi_k$.  The conditions    (\ref{yass1}) turn out to be more convenient since they  correspond directly  to pieces of the action,   and   behave better    under iteration. }

\begin{lem} {  \  }  \label{threeone}  Let     $\square$  be    well-inside  $\Om_k$  
\begin{enumerate}
\item
 If      $\Phi_k   \in  \cS_k(\square)$  then   on $\tilde \square$
\be     \label{scone2}
 |\Phi_k| \leq  2 p_k  \al_k^{-1}  \hs   |\pa \Phi_k|  \leq 3 p_k 
\ee
\item
Conversely   if   $\bar \mu \leq  1$ and      on 
$   \square^{ \sim (2R+1) } $    
   \be |\Phi_k| \leq    p_k  \al_k^{-1}  \hs   |\pa \Phi_k|  \leq     p_k   \ee
       then  $\Phi_k   \in  C  \cS_k(\square)$, i.e. the bounds  (\ref{yass1}) hold with a constant $C$ on the right.    
\end{enumerate}       
   \end{lem}
   \bigskip
   
   \pr  \cite{Bal82b},    \cite{Bal95}

 \noindent   
  (1.)  For   $x  \subset \tilde    \square$
   \begin{equation}  \label{base1}
   |\Phi_k(x) |  \leq   | \Phi_k(x)  -  Q_{k}    \phi_{k, \bom( \square)}(x)| 
   +      |   Q_{k}    \phi_{k, \bom(   \square)}(x))|  \leq    2 p_k \al_k^{-1} 
\end{equation}
Also for  $x,  x  + e_{\mu}  \in  \tilde  \square$ 
\begin{equation}    \label{base2}
\begin{split}
|(\pa_{\mu}  \Phi_k )(x) |  =    &  | ( \Phi_k )(x+ e_{\mu})- ( \Phi_k )(x)|  \\
\leq    & |Q_{k}    \phi_{k, \bom(  \square)}(x+ e_{\mu} ) 
-    Q_{k}    \phi_{k, \bom(   \square)}(x )|  + 2 p_k  \\
\leq  &  \sup_{x \in \tilde   \square}   | \pa   \phi_{k, \bom(    \square)}(x )|
+ 2p_k   \leq   3p_k \\
\end{split}
\end{equation} 
This proves the first part.
\bigskip

\noindent   (2.) We    need     $ |  \phi_{ k,   \bom(\square)  } |
   \leq       C   \al_k^{-1} p_k  $  on $\tilde \square$.   Since our assumptions imply   $(\square')^{\sim 2}  \subset   \Om_k(\square)$  for all
 $\square'  \subset  \tilde  \square$    we  can use
(\ref{twoone1})  to  estimate    on $\tilde \square$ that     $ |  \phi_{ k,   \bom(\square)  } |  \leq  C  \|  \Phi_k\|_{\infty}  $ with the supremum over 
$\square^{\sim (2R+1)}$.   Then the result follows from the bound on  $\Phi_k$.
 For the derivative we can get the same bound,  but we  need the better bound   $  |\pa    \phi_{ k,   \bom(\square)  } |
   \leq     C   p_k  $  on $\tilde \square$.

For this we use the following identity.   
   Let    $y$ be   unit lattice point  in   $\tilde \square$
and take    $x$   in a neighborhood of  $ \De_y$.  Then  the claim is that  
\be     \label{samsum}
\begin{split}
\Big[  \phi_{k, \bom(\square)} \Big(\tilde   Q^T_{ \bbT^0,    \bom(\square)}   \Phi_{k}\Big) \Big](x)   
 = & \Big[  \phi_{k, \bom(\square)} \Big(\tilde   Q^T_{ \bbT^0,    \bom(\square)}(\Phi_k -  \Phi_k(y)  ) \Big) \Big](x)   \\
   +  &  \Big[ 1  -  \bar \mu_k   G_{k, \bom(\square)}  \cdot  1 \Big](x)   \Phi_k(y)    \\
   \end{split}
\ee
Indeed    the identity holds if the second term on the right  is  
$ [  \phi_{k, \bom(\square)} (\tilde   Q^T_{ \bbT^0,    \bom(\square)}1  ) ](x)   \Phi_k(y) $ which is the same as 
$ [  \phi_{k, \bom(\square)} (1) ](x)   \Phi_k(y) $ .    However    since   $[-\De]_{\Om} \cdot  1 =  \De_{\Om, \Om^c} \cdot  1$  we  have
\be
\Big[- \De  + \bar   \mu_k +  Q^T_{k, \bom(\square)}\ba Q_{k, \bom(\square)} \Big ]_{ \Om_1(\square)}  \cdot  1 
= \Big( \De_{\Om_1(\square), \Om_1^c(\square)}  +\bar \mu_k  +  Q^T_{k, \bom(\square)}\ba \Big)  \cdot  1
\ee
   and   so
\be  G_{k,  \bom(\square)}    \Big(  \bar \mu_k  +  Q^T_{k, \bom(\square)}\ba  +   \De_{\Om_1(\square), \Om_1^c(\square)} \Big)  \cdot  1  =1
\ee
Therefore  
\be    \phi_{k, \bom(\square)} (   1  )  
=      G_{k, \bom(\square)}  \Big(   (Q^T_{k, \bom(\square)}\ba  +   \De_{\Om_1(\square), \Om_1^c(\square)}) \cdot  1 \Big)   
=      1  -  \bar \mu_k   G_{k, \bom(\square)}  \cdot  1
\ee
which  gives the result  (\ref{samsum}).

On  $\tilde \square  \subset  \Om_k( \square)  $   we have    $| \pa   G_{k, \bom(\square)}\cdot  1|   \leq  C$  by  (\ref{lefty2}).   Thus the   derivative of  the  second term   in  (\ref{samsum})
is bounded by    
 \be      \label{cindy}  \Big   |  \pa   (   1  -  \bar \mu_k   G_{k, \bom(\square)}  \cdot  1)    \Phi_k(y)  \Big |   \leq   
   C  \bar \mu_k  p_k \al_k^{-1}   \leq       C  \bar \mu_k^{\frac12}  p_k          \leq    C    p_k  
\ee
 By      (\ref{twoone1})
 for  derivatives     and the bound  on $\pa  \Phi_k$,   the derivative of the first term in  (\ref{samsum})   is bounded   on  $\tilde \square \subset  \Om_k( \square)  $    by  
  \be   \label{samsum3}
 \begin{split}
&  \sum_{y'}\big| \Big[  \pa \phi_{k, \bom(\square)} \Big(1_{\De_{y'}}  \tilde   Q^T_{ \bbT^0,    \bom(\square)}(\Phi_k -  \Phi_k(y)  ) \Big) \Big](x)\Big|\\
  \leq   &    C  \sum_{y'}   e^{ - \frac14 \ga_0 d_{\bom(\square)}(y,y' )}    \| 1_{\De_{y'}}  \tilde   Q^T_{ \bbT^0,    \bom(\square)}(\Phi_k -  \Phi_k(y)  )\|_{\infty}\\
 \leq   &   C   \sum_{y'}   e^{ - \frac14 \ga_0 d_{\bom(\square)}(y,y' )}    ( d(y,y')   + 1)p_k    \leq       C   p_k
 \end{split}
 \ee
 The last  holds  since   $d (y,y') \leq   d_{\bom(\square)} (y,y')$.
 Thus we  end with    $|\pa  \phi_{k, \bom(\square)}(x)|  \leq  C  p_k$.
   
 Finally  we  need    $| \Phi_{k}  -Q_{k} \phi_{k, \bom(\square)} |  \leq   C p_k$ on  $\tilde \square$. 
 Again    let  $y$  be a unit lattice point in  $\tilde \square$  and   consider   
$ \Phi_{k}(y)   -(Q_{k} \phi_{k, \bom(\square)})(y) $.  
  Insert  the  expression  (\ref{samsum}) for  $ \phi_{k, \bom(\square)}$. 
 Terms  arising from the first term in  (\ref{samsum})   are bounded by  $Cp_k$   as  in  (\ref{samsum3})
 (now for      $  \phi_{k, \bom(\square)} $  not  $ \pa \phi_{k, \bom(\square)} $).
The  remaining terms   are    
\be
\Phi_k(y)   -  \Big[ Q  (   1  -  \bar \mu_k   G_{k, \bom(\square)}  \cdot  1 )    \Big](y)   \Phi_{k}(y)=   \bar \mu_k  ( QG_{k, \bom(\square)}  \cdot  1 )(y)       \Phi_{k}(y)
\ee 
Since   $ |G_{k, \bom(\square)}  \cdot  1|  \leq  C$,  this   is bounded  by $C  p_k$   as  in    (\ref{cindy}).
This completes the proof.
\bigskip

  Here is a variation in which  $\square$ is allowed to approach the boundary of  $\Om_k$.   
  \begin{lem} {  \  }  \label{threetwo}  Let     $\square  \subset   \Om_k$.   
  \begin{enumerate}
  \item
 If      $ ( \Phi_{k-1, \Om_{k-1}},  \Phi_{k, \Om_k})  \in  \cS_k(\square)$  then   on $\tilde \square  \cap    \Om_k$
\be     
 |\Phi_k| \leq  2 p_k  \al_k^{-1}  \hs   |\pa \Phi_k|  \leq 3 p_k 
\ee
\item
Conversely   if  $\bar \mu  \leq  1$  and    on   
$   \square^{ \sim (2R+1) } $    the  field    $\Phi_k^{\#} \equiv  ( Q\Phi_{k-1, \Om_{k-1}},  \Phi_{k, \Om_k})$   satisfies
   \be
     |\Phi^\#_k| \leq  p_k  \al_k^{-1}  \hs   |\pa \Phi^\#_k|  \leq  p_k 
 \ee
 then    $ ( \Phi_{k-1, \Om_{k-1}},  \Phi_{k, \Om_k})    \in   C  \cS_k(\square)$.    
 \end{enumerate}
   \end{lem}
   
This is proved as in the previous lemma,    but now  $  \phi_{ k,   \bom(\square)  }$  is      $ \phi_{k, \bom(\square)} (\tilde   Q^T_{k, \bbT^0,    \bom(\square)}   \Phi^\#_{k})$  rather  than   $ \phi_{k, \bom(\square)}(\tilde   Q^T_{k, \bbT^0,    \bom(\square)}   \Phi_{k})$.
    \bigskip

  Next   we  introduce an analyticity domain    for our fundamental fields.
  
 \begin{defn}   Let  $\de$ be a fixed small positive number.
 Let   $\square$  be  an  $M$- cube  in       $  \Om_k \subset   \tk$.     Define     $ \cP_{k}(\square, \de )$   to be all complex-valued  
 $( \Phi_{k-1,  \de  \Om_{k-1}},   \Phi_{k,    \Om_k  })$  satisfying:
\begin{equation}  \label{lamp}
\begin{split}
 | \Phi_{k}  -  Q_{k}    \phi_{k, \bom(   \square )}|  \leq &  \la_k^{-\frac14 - \de} \  \  \textrm{  on  } \  \tilde   \square \cap   \Om_k       \\
 |\pa   \phi_{k, \bom(    \square )}|  \leq &  \la_k^{-\frac14 - \de}  \  \  \textrm{  on  } \  \tilde   \square   \\   
 |  \phi_{k, \bom(    \square )}|  \leq &    \la_k^{-\frac14 - \de}    \  \  \textrm{  on  } \  \tilde   \square    \\
\end{split}
\end{equation} 
For  $X \subset   \Om_k$  define  
\be    \cP_{k}(X, \de)   =   \bigcap_{\square \subset  X }    \cP_{k}(\square, \de)
\ee
\end{defn}

If  $\square$  is well inside   $\de \Om_k$      then  the bounds  (\ref{lamp}) are  
a  condition on   $\Phi_{k, \Om_k}$     alone.
As   in  lemma  \ref{threeone}   if the fields are in  $\cP_k(\square)$ then on  $\tilde \square \cap  \Om_k$
\be           \label{scone3}
 |\Phi_k| \leq  2  \la_k^{-\frac14 -\de}  \hs   |\pa \Phi_k|  \leq 3 \la_k^{-\frac14- \de} 
\ee
Note also that    $\cS_k(\square ) \subset  \cP_k(\square, \de )$.   

Once  we have introduced    $\Om_{k+1}, \Phi_{k+1}$  we  use a modified definition:

\begin{defn}
Let        $ \cP'_{k}(\square, \de )$   to be all complex-valued  
 $( \Phi_{k-1,  \de  \Om_{k-1}},   \Phi_{k,  \de  \Om_k  },    \Phi_{k+1,   \Om_{k+1}  })$  satisfying
the inequalities   (\ref{lamp})    but with   $ \phi_{k, \bom(    \square )}$  replaced by    $ \phi'_{k, \bom(    \square )}$
defined in  (\ref{sunscreen}).
For  $X \subset  \de \Om_k$  define   
\be     \cP'_{k}(X, \de)   =   \bigcap_{\square \subset  X }    \cP'_{k}(\square, \de)
\ee
\end{defn}
\bigskip

For  all fields at  once   a natural domain    might   be    
  $\cap_{j=1}^{k-1} [ \cP'_j (\de \Om_j,  \de)]_{L^{-(k-j)}}  \cap   \cP_j( \Om_k, \de)$.
Here  the    subscript   $[\dots ]_{L^{-(k-j)}}$  denotes that  the indicated function of fields on  $\bbT^{-j}_{\sM + \sN -j}$  is
to be scaled down to   a function of fields on  $\tk$.   However  we find   that in the final region we need     tighter restrictions near the boundaries relative to the bulk.    So instead we take the definition:

\begin{defn}   
\be 
    \cP_{k ,\bom}   = \bigcap_{j=1}^{k-1}\Big[  \cP'_j(\de  \Om_j, \de) \Big]_{L^{-(k-j)}}  \cap  \cP_k( \Om_k - \Om_k^{2\nat},  \de)  \cap   \cP_k(\Om_k^{ 2  \nat}, 2 \de)  
    \ee
\end{defn}
\bigskip

  We  also  define  a domain of analyticity   for   fields  on the fine lattice  $\tk$, as in part I.     
\begin{defn}
 Let  $\ep > 2\de $  be a fixed small positive number and   $X \subset  \tk$.
Then    $\cR_k =  \cR_k(X, \ep) $   is  all    functions  $  \phi: X  \to \bbC$    such  that : 
\begin{equation}  \label{rk}
\begin{split}
&  |\phi|  <  \la_k^{-1/4-3 \ep}  \\
& | \pa \phi|  <    \la_k^{-1/4- 2\ep}  \\
 & |\de_{\al}\pa \phi|    <    \la_k^{-1/4- \ep}  \\
\end{split}
\end{equation}
\end{defn}

\subsection{the main theorem}

We   repeatedly block  average   starting with  $\rho_0$   given  by  (\ref{den0}).     Given  $\rho_k( \Phi_k)$
we  define  first
  \begin{equation}   \label{lime}
\tilde  \rho_{k+1}(  \Phi_{k+1}   ) = 
  \cN_{ aL,   \bbT^1_{\sM+ \sN -k}}^{-1}     \int     \exp \left(
- \frac  12  aL    |\Phi_{k+1}- Q \Phi_{k}|^2   \right)  \rho_k(  \Phi_k)
   d\Phi_k  
\end{equation}
and then  scale by  
\begin{equation}
  \rho_{k+1}(  \Phi_{k+1}   )     =     \tilde  \rho_{k+1}(  \Phi_{k+1,L}   ) L^{-|\bbT^{1}_{\sM + \sN -k}|/2}
\end{equation}
The theorem will  assert that   after  $k$  steps  the 
density can be represented in the form
\begin{equation}     \label{representation2}
\begin{split}
  \rho_{k}(\Phi_k)  =     &  Z_{k}   \sum_{\bpi} \int d \Phi_{k,\bom^c}\  d W_{k,\bpi}\ 
  K_{k, \bpi}\  \cC_{k, \bpi}   \\
&   \chi_k ( \La_k)      \exp \Big( -S^+_k(\La_k ) +  E_k(\La_k ) +
 R_{k, \bpi}(\La_k)  +  B_{k,\bpi}( \La_k)         \Big) \\
\end{split}
\end{equation}
where 
 \begin{equation}  \label{orca}
\begin{split}
d \Phi_{k,\bom^c}  =  & \prod_{j=0}^{k-1}  
\exp \left(  - \frac{1}{2 } aL^{-(k-j-1)}|\Phi_{j+1}- Q \Phi_{j}|^2_{\Om_{j+1}^c}  \right)
 d\Phi^{(k-j)}_{j, \Om_{j+1}^c}  
\\
d W_{k,\bpi}   =  &  \prod_{j=0}^{k-1}      (2 \pi)^{-|  [ \Om_{j+1}  - \La_{j+1}]^{(j)}|/2}
  \exp  \Big(   - \frac{1}{2 } L^{-(k-j)} | W_j |^2_{\Om_{j+1}- \La_{j+1}}  \Big) 
    dW^{(k-j)}_{j,\Om_{j+1}- \La_{j+1}}
\\ 
  K_{k,\bpi}  =  &
  \prod_{j=0}^{k}  
      \exp\left( c_j|\Om_j^{c,(j-1)}|-S^{+,u}_{j,L^{-(k-j)}}(\La_{j-1}  -  \La_{j}     )+  \Big( \tilde   B_{j,L^{-(k-j)}}\Big) _{\bpi_j}(\La_{j-1}, \La_{j}) \right)     \\
 \cC_{k, \bpi}  =   &      \prod_{j=0}^k  \Big(  \cC_{j, L^{-(k-j)}}\Big)_{ \La_{j-1}, \Om_j, \La_j }  \\
\end{split}
\end{equation}
Here  for   $\Phi_{j+1, \Om_j^c}:  [ \Om^c_{j+1}]^{(j)} \to  \bbR$   we define
\be  \label{old97}
d \Phi^{(k-j)}_{j, \Om_{j+1}^c}  = [ L^{-(k-j)/2}]^{|(\Om_{j+1}^c)^{(j)}|}
\prod_{ x \in   [ \Om^c_{j+1}]^{(j)} }  d \Phi_j(x)
\ee
Besides our basic  variables  $\Phi_{k, \bom}$   we also employ      auxiliary variables      
\begin{equation}
W_{k, \bpi}  =  (W_{0, \Om_1-  \La_1},  \dots, W_{k-1,  \Om_k -  \La_k}) 
\end{equation}
  with  $W_{j, \Om_{j+1}  - \La_{j+1}}    :    [ \Om_{j+1}  - \La_{j+1}]^{(j)} \to \bbR$.  The measure   $  dW^{(k-j)}_{j,\Om_{j+1}- \La_{j+1}}
$  is defined as in   (\ref{old97}).

We  employ the convention that   $ \La_{-1}, \Om_0$  are  the full torus  $ \bbT^{-k}_{\sM + \sN -k}$.
\bigskip

To  state the main result it is convenient to assume  that  $\la,  \bar \mu$  are bounded,   and we  somewhat arbitrarily take  $\la  \leq  e^{-1}$ 
(so that  $- \log \la  \geq  1$) and  $\bar  \mu \leq  1$.    The dimensionless coupling constant  $\la/ \sqrt{\bar \mu}$ is still unrestricted.

\begin{thm}  \label{maintheorem}  Let   $0 <  \la  \leq  e^{-1}$ and  $0 < \bar \mu  \leq  1$.  
Let   $L$  be sufficiently large,  let  $M$ be sufficiently large (depending on $L$),  and  let  $\la_k$  be sufficiently small  (depending on $L,M$).
    Let     $\vep_k,    \mu_k$   be the coupling constants  selected   in part I. 
Then  the  representation    (\ref{representation2}), (\ref{orca})    holds with the following properties:
\begin{enumerate} 
\item   
$Z_k$  is the global normalization factor  of   part  I.   It  satisfies  $Z_0 =1$  and 
 \be
Z_{k+1}  =   Z_k \ \cN^{-1}_{a ,  \bbT^1_{\sM + \sN-k}}   (2 \pi)^{| \bbT^0_{\sM +\sN-k}|/2}( \det  C_k  )^{1/2}
\ee
\item
 The  characteristic function    $ \chi_k ( \La_k) $   forces the field  $\Phi_{k}$  to  be in the space $\cS_k(\La_k)$ defined in 
 (\ref{yass1}),(\ref{yass2}).  It has the form  
\begin{equation}   \label{cinnamon}
  \chi_k ( \La_k)  =   \prod_{\square \subset  \La_k}       \chi_k ( \square)  \hs     \chi_k ( \square  )  = \chi\Big( \Phi_k  \in \cS_k(\square)\Big)
  \end{equation}

\item  
   $\cC_{k,   \La_{k-1},  \Om_k,  \La_k}( \Phi_{k-1}, W_{k-1}, \Phi_k ) $   
 is a    collection of characteristic functions  forcing certain   fields to be large or small or both.    The  exact definition will be given in the course of the proof.
 It  does  not depend on  $\Phi_{k, \La_k^{\nat}}$   and  
  enforces   on  $\La_{k-1} - \Om_k$
 \be   \label{mexicanhat} | \Phi_{k-1}|  \leq  2 p_{k-1} \al_{k-1}^{-1}L^{\frac12}   \hs   |\pa   \Phi_{k-1}|  \leq  3 p_{k-1} L^{\frac32} 
 \ee
 and enforces   on    
  $\Om_k - \La_k$  for some constant $C_w$
    \begin{equation}   \label{sombrero}
      |\Phi_k|  \leq   \  3    p_{k-1}\al_{k-1}^{-1}L^{\frac12}   
        \hs \    |\pa \Phi_k|  \leq   \  4    p_{k-1}L^{\frac32}
        \hs       |W_{k-1}|  \leq   C_w p_{k-1}  L^{\frac12}    
 \end{equation}

     In the  expression   (\ref{orca}) this is scaled down to  
   \begin{equation}
\Big( \cC_{j,  L^{-(k-j)}}  \Big)_{ \La_{j-1},  \Om_j,  \La_j }  (  \Phi_{j-1},W_{j-1}, \Phi_{j} )
 \equiv    \cC_{j,  L^{k-j}( \La_{j-1},  \Om_j,  \La_j )}(  \Phi_{j-1, L^{k-j}},    W_{j, L^{k-j}} \Phi_{j, L^{k-j}} )
 \end{equation}

\item    The  bare      action   is  $S^{+}_k(\La_k ) =    S^{+}_k(\La_k, \Phi_{k},  \phi_{k, \bom(\La_k^*)} )$
where    $ \phi_{k, \bom(\La^*_k)} $  is defined in   (\ref{angie})   and 
\begin{equation}
\begin{split}
S^{+}_k(\La_k, \Phi_{k}, \phi ) =&   S^*_{k}(\La_k,  \Phi_{k}, \phi )
 +   V_k( \La_k,  \phi)    \\
S^*_k( \La_k,  \Phi_k, \phi)    
 =&\frac{a_k}{2}\| \Phi_k  -  Q_{k}   \phi  \|^2_{\La_k} 
+   \frac12 \| \pa  \phi  \|^2_{*,\La_{k}}   +  \frac 12  \bar \mu_k   \|  \phi\|^2_{\La_k} \\
 V_k( \La_k,  \phi)  = & \vep_k  \Vol (\La_k) +
\frac12  \mu_k  \|  \phi^2 \|_{\La_k}  + \frac14  \la_k  \int_{\La_k}   \phi^4  \\ 
\end{split}
\end{equation} 

\item  $  E_k(\La_k)   =   E_k(\La_k, \phi_{k,\bom(\La^*_k)}   ) $     are  the main corrections to the bare action and have the local structure 
 \begin{equation}
  E_k(\La_k)  =  \sum_{X \in \cD_k, X \subset \La_k}  E_k(X)    
\end{equation} 
The  functionals    $ E_k(X, \phi_{k,\bom(\La^*_k)}   )$  are  restrictions of  functionals  
  $   E_k(X, \phi )$  analytic  in  $\phi \in  \cR_k(X, \ep  )$   
and satisfying    there for  $\beta <  \frac14- 10 \ep$
\begin{equation}
|    E_k(X )|  \leq   \la_k^{\beta}   e^{ - \ka d_M(X)  }       
\end{equation} 
They  are  identical  with  the  global small field  functions    $E_k(X, \phi) $  of part I.

\item        $  R_{k, \bpi}(\La_k)=    R_{k, \bpi}(\La_k, \Phi_{k})   $  is a tiny remainder and   has   the    local   structure     
\begin{equation}   R_{k, \bpi}(\La_k)  =  \sum_{X \in \cD_k, X \subset \La_k}   R_{k, \bpi}(X)   
\end{equation} 
   The function  $ R_{k, \bpi}(X, \Phi_{k}) $  is analytic  in $\cP_k(X, 2 \de)$,   
  and satisfies   there  for  a fixed integer $n_0 \geq  4$:
\begin{equation}
|  R_{k, \bpi}(X ) |  \leq    \la_k^{n_0}    e^{ - \ka d_M(X)  } 
\end{equation}

\item   The  active  boundary term has the form          $    B_{k,\bpi} ( \La_k)  
 =  B_{k,\bpi} ( \La_k; \Phi_{k,\bom},  W_{k,\bpi})  $.
It has  the local  expansion   
\begin{equation}
   B_{k,\bpi}( \La_k)   =  \sum_{X \in \cD_k (\bmod \Om_k^c), X  \#  \La_k, X \subset   \Om_1 }  B_{k,\bpi}  (X)  
\end{equation}
The function   $  B_{k,\bpi}  (X, \Phi_{k,\bom},  W_{k,\bpi})  $    is analytic 
in   $\cP_{k, \bom}$    and     for some $B_w > C_w$ 
\be    \label{secretary}
 |W_j|  \leq     B_w \   p_j L^{\frac12(k-j)}    \  \textrm{ on } \   \Om_{j+1} -     \La_{j+1} 
 \ee
and  satisfies there  
\begin{equation}
| B_{k,\bpi}  (X) |  \leq  B_0  \la_k^{\beta}  e^{- \ka  d_M(X,  \bmod  \Om_k^c)}
\end{equation}
for some constant  $B_0$  depending   on $L,M$.

\item  
The  inactive   boundary term  has   the form      $\tilde   B_{k,\bpi}(\La_{k-1}, \La_k, \Phi_{k, \bom_k},  W_{k,\bpi})  $.   It   depends
on  the   variables  only   in  $\Om_1- \La_k$,   is analytic  in $\cP_{k, \bom}$   and   (\ref{secretary}) and   satisfies there  
   \begin{equation}
| \tilde   B_{k, \bpi}(\La_{k-1}, \La_k) |  \leq   B_0 \Big |  \La^{(k)}_{k-1} - \La^{(k)}_k  \Big|
\end{equation}
It is additive in the connected components of $\La_k^c$.
In the expression  (\ref{orca})  we have the  scaled version for $j \leq k$
\begin{equation}
\begin{split}
& \Big( \tilde  B_{j,L^{-(k-j)}}  \Big)_{\bpi_j}(\La_{j-1}, \La_j;  \Phi_{j,\bom_j},  W_{j,\bpi_j}) \\
 & \ \  \equiv      \tilde   B_{j,L^{k-j}\bpi_j}\Big(L^{k-j}\La_{j-1},L^{k-j} \La_j; ( \Phi_{j,\bom_j})_{L^{k-j}},  (W_{j,\bpi_j})_{L^{k-j}}\Big)\\
\end{split}
\end{equation}
where    $ \bom_j,  \bpi_j$   are    $\bom,  \bpi$ 
truncated at the $j^{th}$ level.

 \item  With  $\de \La_{k-1}  =  \La_{k-1} - \La_k$,  the  unrenormalized    action   is  
 $ S^{+,u}_k\Big(\de \La_{k-1},  \Phi_{k,\bom}, \phi_{k, \bom(\La_{k-1}, \Om_k, \La_k)}  \Big)$
where    $ \phi_{k,  \bom(\La_{k-1}, \Om_k, \La_k)  } $  is defined in   (\ref{piggy})   and 
\begin{equation}
\begin{split}
S^{+,u}_k(\de \La_{k-1} ,  \Phi_{k,\bom}, \phi) =&   S^*_k(\de \La_{k-1}  ,  \Phi_{k, \bom}, \phi )
 +   V^u_k( \de \La_{k-1} ,  \phi)   \\
 V^u_k( \La,  \phi)  =  &     L^3\vep_{k-1}  \Vol (\La) +
 \frac12   L^2     \mu_{k-1}  \| \phi^2\|_{\La}  + \frac 14 \la_{k}  \int_{\La}   \phi^4  \\ 
\end{split}
\end{equation} 
In the expression  (\ref{orca})  we have the  scaled version for  $j \leq k$
 \begin{equation}
\begin{split}
&S^{+,u}_{j,L^{-(k-j)}}(\de \La_{j-1} ,  \Phi_{j,\bom_j},\phi_{j, \bom(\de  \La^*_{j-1})}    ) \\
& \ \ =     S^{+,u}_j\Big( L^{k-j}(\de \La_{j-1} ),  \Phi_{j, \bom_j, L^{k-j}},  \
\phi_{j, \bom(\de  \La^*_{j-1}),  L^{k-j}} \Big) \\
\end{split}
\end{equation}
\bigskip
\end{enumerate}
\end{thm}

\res
\begin{enumerate}

\item  
The functions   $S^+_k(\La_k ),   E_k(\La_k ),  R_{k, \bpi}(\La_k)$,  and  
 $   B_{k, \bpi}(\La_k)$   depend on fields in the current   small field region  $\La_k$   and     
contribute to the next   RG   transformation.   The functions   $S^+_{j}(\La_{j-1}- \La_j)$  and
 $  \tilde   B_{j, \bpi_j}(\La_{j-1}, \La_j) $    do  not  depend on fields   in  $\La_k$  
     and  do  not contribute to  the  next RG transformation. 

\item  
The  statement   that  the coupling constants  $\vep_k,  \la_k,   \mu_k$  in  $V_k$     are chosen  as  in  part I    means
that  they satisfy  discrete dynamical equations   
 \begin{equation}  \label{recursive}
\begin{split}
\vep_{k+1}   =&  L^3 \vep_k  + \cL_1E_k   +  \vep_k^*(\la_k, \mu_k,  E_k) \\
\mu_{k+1}   =&   L^2 \mu_k  +  \cL_2E_k  + \mu_k^*(\la_k, \mu_k,  E_k)  \\
\la_{k+1}   =&  L\la_k   \\ 
E_{k+1}   =&    \cL_3 E_k  +  E^*_k(\la_k,    \mu_k,  E_k)  \\
 \end{split}
\end{equation}
and they are    tuned so that     $|\vep_{k}|  \leq  \cO(1) \la_k^{\beta}$  and  
 $|\mu_{k}|  \leq   \la_k^{\frac12 +  \beta}$.

The unrenormalized  potential   $V^u_j$  differs from the renormalized potential  $V_j$
only   in that the last corrections  to the coupling constants  are not  included.  That is we  have energy density 
 $ L^3\vep_{k-1} $  instead  of  $\vep_k  =   L^3\vep_{k-1}   + \dots$  and   mass $ L^2\mu_{k-1} $  instead of  $\mu_{k} =  L^2\mu_{k-1} + \cdots$.
  
\item   With more  work one   could probably identify the history dependent parts  of  $R_{k, \bpi}$   as
boundary terms  and   so  get  a new  $R_k$  independent of  $\bpi$.   We    note that in Balaban's models these
$R_k$   terms  get  additional contributions from a recycling operation that  converts some  large field contributions
back  to    small field contributions    (the "R - operation").  This  difficult step is not necessary for this model.
\end{enumerate}
  
To  check that our formula makes sense we need the following:

\begin{lem}    The bounds of the characteristic functions    $\cC_{k, \bpi}$ and  $\chi_k(\La_k)$
put the various fields in the analyticity domains for    $E_k(\La_k ),  R_{k, \bpi}(\La_k ),    B_{k, \bpi}(\La_k ),    \tilde   B_{j, \bpi_j}(\La_{j-1}, \La_j) $ 
\end{lem}
\bigskip
 
\pr  
First note that they  imply 
 \begin{equation}   \label{sombrero2}
 \begin{split} 
   | \Phi_k  |   \leq  &  3 p_{k-1}  \al_{k-1}^{-1  }L^{\frac12}   \hs    \hs   \textrm{    on    }   \Om_k  \\    
  |\Phi_j|  \leq   &  3   p_{j-1}\al_{j-1}^{-1}L^{\frac12(k-j)}  \ \ \ \ \    \textrm{    on    }   \Om_j -  \Om_{j+1}  \hs   j=1, \dots,  k-1\\ 
    |\Phi_0 |  \leq   &  2   p_{0}\al_{0}^{-1} L^{\frac12k}\  \hs  \textrm{    on    }   \La_0 -  \Om_1    \\
  \end{split}   
 \end{equation}
These  all are enforced   by   $\cC_{k, \bpi}$  except  that  it only gives the   first   on  $\Om_k-  \La_k$.   By  lemma \ref{threeone}
the   characteristic function  $\chi_k(\La_k)$  gives a stronger  bound   on  $\La_k$.

The bound on  $\Phi_k$ implies  $|\tilde   Q^T_{k, \bbT^0,    \bom(\La_k^*)}   \Phi_{k}|  \leq    3 p_{k-1}  \al_{k-1}^{-1  }L^{\frac12}$ on a neighborhood of    $ \Om_1(\La_k^*)$,  since the latter is contained in $\Om_k$.
Then     (\ref{twoone1}) gives     that  on $\La_k$
\be
|\phi_{k, \bom(\La_k^*)} |,   \  \ \   |\pa  \phi_{k, \bom(\La_k^*)} |,  \ \ \    |\de_{\al} \pa  \phi_{k, \bom(\La_k^*)} |
\leq    C  p_{k-1}\al_{k-1}^{-1}   \ee
But for  $\la_k$ sufficiently small  $ C  p_{k-1}\al_{k-1}^{-1}  \leq   \la_k^{-\frac14 - \ep}$    so  $ \phi_{k, \bom(\La_k^*)}  \in  \cR_k(X)$ for  $X \subset  \La_k$.
 Thus we are in the analyticity region for   $E_k(\La_k)$.   

 Similarly   for  $\la_k$ sufficiently small   $|\Phi_k|   \leq    3 p_{k-1}  \al_{k-1}^{-1 }$   implies for $\square \subset  \La_k$
 on  $\tilde \square$
 \be
|\Phi_k - Q_k\phi_{k, \bom(\square)} |,   \  \ \   |\pa  \phi_{k, \bom(\square)} |,  \ \ \     |\phi_{k, \bom(\square)} |
\leq    C  p_{k-1}\al_{k-1}^{-1}    \leq    \la_k^{-\frac14 - \de}   \ee
Thus we   are  in $\cP_k(\square)$  and hence in  $\cP_k(\La_k)$  which is  the domain for   $R_{k, \bpi}(\La_k)$.

Next   note that  (\ref{sombrero2}) implies that  $ |\Phi_{j,L^{k-j}}|  \leq     3   p_{j-1}\al_{j-1}^{-1}$     on   $L^{k-j}\de \Om_j$.  
Then   for an  $M$-cube   $\square$  well inside    $L^{k-j}  \de \Om_j$ we have on  $\tilde \square$
\be
\begin{split}
&|\Phi_{j,L^{k-j}}- Q_j\phi_{j, \bom(\square)}(\Phi_{j,L^{k-j}}) |,   \  \ \   |\pa  \phi_{j, \bom(\square)}(\Phi_{j,L^{k-j}}) |,
  \ \ \     |\phi_{j, \bom(\square)} (\Phi_{j,L^{k-j}})|  \\
&\leq    C  p_{j-1}\al_{j-1}^{-1}    \leq    \la_j^{-\frac14 - \de} \\
\end{split}
  \ee
The   bound says   for $j=k$   that   $\Phi_{k}    \in   \cP_k(\square, \de)$,  and   for  $j<k$    that    $\Phi_{j,L^{k-j}}    \in   \cP'_j(\square, \de)$  or     $\Phi_{j}    \in  [ \cP'_j(\square, \de)]_{L^{-(k-j)}}$. 
The same conclusion   holds for  any    $\square  \subset  L^{k-j}  \de \Om_j$,  but now involves fields from adjacent regions.
  Thus we  are in  $\cP_{k, \bom}$.
  Also   since  $C_w < B_w$ we have that 
$ |W_j|  \leq    C_w \   p_j L^{\frac12(k-j)}   $   implies   $ |W_j|  \leq    B_w \   p_j L^{\frac12(k-j)}   $.   Thus we  are in the analyticity domain for   $  B_{k, \bpi}(\La_k )$.
 
 The analysis for   $  \tilde   B_{j, \bpi_j}(\La_{j-1}, \La_j) $ 
is similar.

\subsection{initial  representation}

We   begin the proof of  theorem  \ref{maintheorem}  by  showing   that the representation holds  for  $k=0$.
We  have  initially  
\begin{equation}  
 \rho_0(\Phi_0)  =   \exp   \Big(  - S^+_0(\Phi_0)  \Big)
=\exp \left( - \frac12 \| \pa \Phi_0 \|^2  - \frac12   \bar  \mu_0  \|\Phi_0 \|^2 - V_0(\Phi_0)  \right) 
\end{equation}
We   break  into large and small field regions   as  follows.  For  each  $M$-cube  $\square  $  define characteristic functions by  
\begin{equation}  \label{knot}
\begin{split}
\chi_0  ( \square,   \Phi_0 )  =&
\prod_{x \in \tilde  \square}   \chi \Big (  | \pa  \Phi_0(x)|  \leq   p_0,     | \Phi_0(x)|  \leq  \al_0^{-1}p_0    \Big)
\end{split}
\end{equation} 
Then we write  with   $\zeta_0 (\square)  =   1-\chi_0  ( \square) $   and with   $Q_0$  a union
of  $M$ cubes
\begin{equation}  \label{knotty}
\begin{split}
1  =  &  \prod_{ \square  \subset   \bbT^0_{M+N}   } ( \zeta_0 (\square)  + \chi_0  ( \square)  )
=     \sum_{ Q_0}   
\prod_{ \square  \subset  Q_0  } \zeta_{0} (\square)
\prod_{ \square  \subset  Q_0^c  } \chi_{0} (\square)
\equiv      \sum_{Q_0}   
 \zeta_0( Q_0 )  \chi_0( Q_0^c  )\\
 \end{split}
 \end{equation}
   Then in  $Q_0^c$   the inequalities
$  | \pa  \Phi_0|  \leq   p_0,     | \Phi_0|  \leq  \al_0^{-1}p_0$  hold at every point, 
 whereas  in   every cube  in  $Q_0$   some  inequality is     violated   at   some  point.

Now  $Q_0^c$ is an adequate small field region, but for consistency with subsequent steps we  shrink it.
Let     $\La_0 =  (Q_0^c)^{5 \nat}$   or   $\La^c_0 =  Q^{5*}_0$    and rewrite (\ref{knotty}) as 
\be  1  =          \sum_{\La_0}    \cC_{0, \La_0} \    \chi_0( \La_0  )
\hs  \mathrm{where}    \hs   \cC_{0, \La_0} =    \sum_{Q_0:  Q_0^{4*}=  \La^c_0}    \zeta_0( Q_0 )  \chi_0( Q_0^c - \La_0)
\ee
 Juxtaposing this  with  $\rho_0$  and splitting the action on  $\La_0$  we  have
\begin{equation}  
 \rho_0 =  \sum_{\La_0}    \cC_{0, \La_0} \   \exp   \Big(  - S^+_0(\La^c_0)  \Big)     \chi_0( \La_0  ) \exp   \Big(  - S^+_0(\La_0)  \Big)
\end{equation}
This  is the  representation  (\ref{representation2})  if we make the following interpretations.     There are no integrals,  $Z_0=1$,
and   $\La_{-1} =  \Om_0  = \bbT^0_{\sM+ \sN}$.    The functions      $E_0, R_0,  B_0,   \tilde B_0$   are all zero.  
The as yet  undefined fields  $\phi_{0, \bom(\La_0^*)}$   and  $\phi_{0, \bom((\La^c_0)^*)}$   are just  $\Phi_0$   and  
$S^{+,u}_0(\La^c_0) =S^+_0(\La^c_0)$.       We  also interpret   $\phi_{0, \bom(\square)}$  as  $\Phi_0$    
and   $Q_0$  as the identity,  and then     $\chi_0(\square)$  defined in (\ref{knot})    is  the same  as (\ref{cinnamon}).
\bigskip

\subsection{new small field region}  \label{nsf}

 We have seen that  theorem  \ref{maintheorem}   is true for  $k=0$.   To complete the proof   we assume it is true for  $k$  and  generate   the representation for $k+1$.  This will occupy the rest of the paper.  The proof follows especially  \cite{Bal95}- \cite{Bal98c}.
 
 To  begin  insert  the expression   (\ref{representation2}) for  $\rho_k$  into the definition (\ref{lime})  of   $\rho_{k+1}$,   and
 bring the sums outside the integral.
 
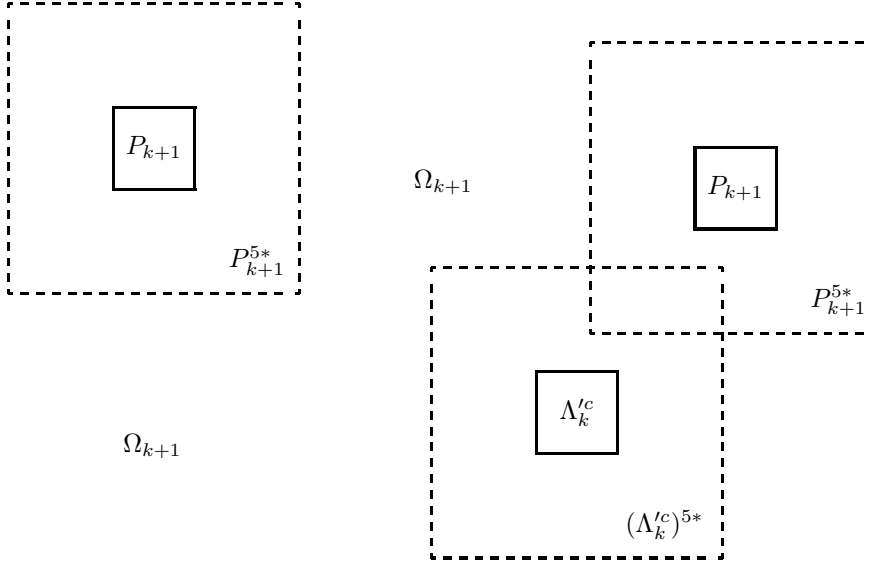
\begin{figure}[t] 
 \begin{picture}(350,220)
 \thicklines
 \put(80,150){\framebox(30,30){$P_{k+1}$}  }
 \put(40,110){\dashbox{3}(110,110)}
 \put (120,120){ \text{$P^{5*}_{k+1}$ }  }
  \put(240,50){\framebox(30,30){$\La'^c_{k}$}  }
 \put(200,10){\dashbox{3}(110,110)}
 \put (270,20){ \text{$(\La'^c_{k})^{5*}$ }  }
  \put(300,135){\framebox(30,30){$P_{k+1}$}  }
 \put(260,95){\dashbox{3}(110,110)}
 \put (340,105){ \text{$P^{5*}_{k+1}$ } }
 \put (80,50){ \text{$\Om_{k+1}$ } }
 \put(190,150){ \text{$\Om_{k+1}$ } }
 \end{picture}
 \caption{Illustrating   $\Om_{k+1}^c   = ( \bar \La_k )^{c,5*}  \cup  P_{k+1}^{5*}$.  Here   $\Om_{k+1}^c $  is the region inside the
 dotted lines.    \label{concomitant}        }
\end{figure}

In the current small field region $\La_k$ we  have  limits on the size  of  $\Phi_k$.   Because of the factor 
$  \exp \left(
- \frac  12  aL    |\Phi_{k+1}- Q \Phi_{k}|^2   \right)$
   large  $\Phi_{k+1}$  will   also be suppressed  in  $\La_k$.    To  take advantage of this  feature 
we proceed as follows.  
For any  $LM$-cube  $\square$  in  $\tk$   define the characteristic function
\begin{equation}  \label{nothing}
\begin{split}
\chi^q_k  ( \square,   \Phi_{k+1}, \Phi_k )  =&
\prod_{y \in  \square  \cap  \bbT^1_{\sM + \sN -k}}   \chi \Big (  |\Phi_{k+1} (y) - (Q  \Phi_k)(y) |  \leq   p_k \Big)
\end{split}
\end{equation} 
Also   let   
$  \bar \La_k $   be the union of all    $ LM$   cubes  intersecting     $  \La_k  $.
Then  with    $\zeta^q_k (\square)  =   1-\chi^q_k  ( \square) $  
  \begin{equation}
\begin{split}
1  =  &  \prod_{ \square  \subset   \bar \La_k}   \zeta^q_{k}  (\square)  + \chi^q_{k}   ( \square)  \\
=  &   \sum_{P_{k+1} \subset   \bar \La_k}   
\prod_{ \square  \subset  P_{k+1}  } \zeta^q_{k}  (\square)
\prod_{ \square  \subset   \bar \La_k -  P_{k+1}  } \chi^q_{k}  (\square)\\\
\equiv  &    \sum_{P_{k+1}   \subset   \bar \La_k}   
 \zeta^q_{k} ( P_{k+1} )  \chi^q_{k} ( \bar \La_k- P_{k+1} )\\
 \end{split}
 \end{equation}
 where     $P_{k+1}$  is a union of  $LM$-cubes.
 Now   given    $\bar \La_k$   and    $P_{k+1}$  define a     new small   field region $\Om_{k+1}$   by    (See figure  \ref{concomitant})
 \begin{equation}
 \Om_{k+1}  =  ( \bar \La_k)^{5 \nat} -    P^{5*}_{k+1} \hs  \textrm{ or }  \hs   \Om^c_{k+1}  =    (\bar \La_k)^{c,5*}  \cup   P_{k+1}^{5*} 
 \end{equation}
 Here the $*, \nat$ operations refer to adding or deleting layers of  $LM$-cubes.
 In generating  $\Om_{k+1}^c$ from  $(\bar \La_k)^c$ we add at least  $5[r_{k+1}]$ layers of $LM$-cubes  so  $d((\bar \La_k)^c,  \Om_{k+1})
 \geq 5 [r_{k+1}]LM$.  This is the required separation at this scale.

Now  classify   the terms in  the sum  by the union of $LM$-cubes    $\Om_{k+1}$  that they generate  and find 
 \begin{equation}   \label{snafu}
1  =      \sum_{\Om_{k+1}  \subset  \bar \La^{5 \nat }_k}  \cC^q_{k} (\La_k, \Om_{k+1} )  \chi^q_{k} ( \Om_{k+1} )
 \end{equation}
where 
\begin{equation}
\cC^q_{k}(\La_k,   \Om_{k+1} )  =  \sum_{P_{k+1}  \subset  \bar \La_k
:  \Om_{k+1}  =   (\bar \La_k)^{5 \nat  } -    P^{ 5*}_{k+1} }  
 \zeta^q_{k}( P_{k+1} )  \chi^q_{k}((\bar \La_k-P_{k+1})- \Om_{k+1} )
 \end{equation}
  We have
 \begin{equation}  \label{sub}
 |\Phi_{k+1} - Q \Phi_k|  \leq    p_k  \ \ \  \textrm{  on }  \ \ \    \Om_{k+1}
 \end{equation}

 Insert  (\ref{snafu})  under the integral sign in  our expression.
Split   the  integral   over   $\Phi_k$  into  an integral over   $\Phi_{k,  \Om_{k+1}^c }$
and an integral over   $\Phi_{k,  \Om_{k+1} }$.
We     define  
with  
 \begin{equation}    \bom^+  =  (\bom, \Om_{k+1})  =  (\Om_1,  \cdots,  \Om_{k+1})
   \end{equation}
   the measure
\begin{equation}
d \Phi^0_{k+1, \bom^{+,c} } =    \exp \left(
- \frac{aL}{2}  |\Phi_{k+1}- Q \Phi_{k}|^2_{\Om_{k+1}^c}  \right)  d \Phi_{k,  \Om^c_{k+1}} d \Phi_{k, \bom^c}
\end{equation}
We   also transfer the potential from  $S^+(\La_k)$  to  $E(\La_k)$  writing  $-S^+(\La_k) + E(\La_k)  = -S^*(\La_k) + E^+(\La_k)$ where
\begin{equation}
  E^+_k(\La_k)= E_k(\La_k ) -  V_k(\La_k) 
\end{equation}
Finally  moving the integral over    $ \Phi_{k,  \Om_{k+1}} $  inside and taking account that   $ K_{k, \bpi},  \cC_{k, \bpi}, 
  \cC^q_{k} (\La_k, \Om_{k+1} )  $ do not depend on   $ \Phi_{k,  \Om_{k+1}} $   we  have the expression:
  \begin{equation}     \label{representation3}
\begin{split}
\tilde    \rho_{k+1}( \Phi_{k+1}) 
 =     & Z_k      \cN^{-1}_{aL,  \bbT^1_{\sM+\sN -k   }  } 
 \sum_{\bpi, \Om_{k+1}}    \int   d \Phi^0_{k+1, \bom^{+,c} }d W_{k,\bpi}\  K_{k, \bpi} \  \cC_{k, \bpi} 
\  \cC^q_{k} (\La_k, \Om_{k+1} )   \\
    &   \int   d\Phi_{k, \Om_{k+1}}  
\exp \Big(- \frac12  a L  |\Phi_{k+1,L}- Q \Phi_{k}|^2_{ \Om_{k+1} }\Big) \\
& \chi_k ( \La_k)    \chi^q_{k} ( \Om_{k+1} ) \exp \Big(  -S^*_k(\La_k ) +  E^+_k(\La_k ) + R_{k, \bpi}(\La_k)  +  B_{k, \bpi}(\La_k)         \Big) \\
   \end{split}
\end{equation}
\bigskip

Let  us collect  the  bounds implied by the  characteristic  functions     $ \cC_{k, \bpi}  \cC^q_{k} (\La_k, \Om_{k+1} ) \chi_k ( \La_k)    \chi^q_{k} ( \Om_{k+1} )$. 

\begin{lem}  \label{suddsy}
 The characteristic  functions     enforce the following bounds:
  \begin{eqnarray}   \label{pine}
|\Phi_k|  \leq   3 p_{k-1} \al_{k-1}^{-1}L^{\frac12} \hs   & |  \pa   \Phi_k|  \leq   4 p_{k-1} L^{\frac32} \ \  & \textrm{  on  } \ \    \Om_k - \La_k
 \label{gb1}\\
|\Phi_k|  \leq      2 p_{k} \al_{k}^{-1} \hs   &  |\pa   \Phi_k|  \leq    3 p_{k}    \ \  \hs &  \textrm{  on  } \ \  \tilde   \La_k
 \label{gb1.5}  \\
  | \Phi_{k+1}|      \leq        3 p_k \al_k^{-1} \hs      & |\pa  \Phi_{k+1}|      \leq   4 p_k     \hs   &  \textrm{  on  } \  \Om_{k+1}  
  \label{gb2}  
  \end{eqnarray}
  In addition    $\Phi^\#_{k+1} = (Q \Phi_{k, \de \Om_k},  \Phi_{k+1,\Om_{k+1} } )$   satisfies  
 \be    \label{gb2.5}
   | \Phi^\#_{k+1}|      \leq      C p_k \al_k^{-1} \hs    |\pa  \Phi^\#_{k+1}|      \leq   C p_k      \hs     \textrm{  on  } \  \Om_{k}  
 \ee
 \end{lem}
 \bigskip

 \pr   The characteristic 
function  $\cC_{k, \bpi}$  enforces   the    bounds  (\ref{gb1})   on   $\Om_k- \La_k$ by assumption.
 The function $\chi_k(\La_k)$  says that for  $\square  \subset  \La_k$   we have      $\Phi_k \in \cS_k(\square)$.    Then by   lemma
 \ref{threeone}   we  have  on $\tilde \square$  the bounds   
   $|\Phi_k|  \leq  2 p_k\al_k^{-1}$   and     $|\pa \Phi_k|  \leq  3 p_k  $.   This gives   (\ref{gb1.5}).

 The  first bound   in   (\ref{gb1.5})  and  (\ref{sub})  give    $ |\Phi_{k+1}|  \leq   3 p_k \al_k^{-\frac14}  $  on  $\Om_{k+1}$.
 The second  bound follows  by  (\ref{sub})  as well   by the estimate  for  $y, y + L e_{\mu} \in  \Om_{k}^{(k+1)} $
 \be 
\begin{split} &  |\pa_{\mu}  \Phi_{k+1}(y)| =  L^{-1} | \Phi_{k+1}(y+ Le_{\mu})  -   \Phi_{k+1}(y)|  \\
&\ \ \ \leq     L^{-1} |(Q \Phi_k)(y+ Le_{\mu})  -  Q \Phi_k(y)|  +  L^{-1}2p_k  \leq   3p_k  +     L^{-1}2p_k  \leq  4 p_k  \\
\end{split} 
\ee 
Thus   (\ref{gb2}) is established.
 
 For    (\ref{gb2.5})  note that   (\ref{gb1}), (\ref{gb1.5})  imply that  $|\Phi_k|  \leq  C p_k \al_k^{-1}$  and  $|\pa \Phi_k|  \leq  C p_k$
 on  $\Om_k$.  Hence   $Q\Phi_k$ satisfies the same bounds on  $\de \Om_k$,  as does  $\Phi_{k+1}$ on  $\Om_{k+1}$.  
 The remaining issue is when   
a derivative crosses  the boundary of  $\Om_{k+1}$.    This is handled as follows.  Suppose   $y \in  \Om_{k}^{(k+1)} $
and   $y+ L e_{\mu} \in  \Om_{k+1}^{(k+1)} $.  Then 
\be
\begin{split}
(\pa_{\mu}   \Phi^\#_{k+1})(y)   =  &    L^{-1}   \Big(   \Phi_{k+1}  ( y + L e_{\mu})   -  (Q\Phi_k) (y)  \Big) \\
 =  &    L^{-1}   \Big(   \Phi_{k+1}  ( y + L e_{\mu})   -  (Q\Phi_k) (y+ L e _{ \mu})  \Big)
 +      L^{-1}   \Big(  (Q\Phi_k) (y+ L e _{ \mu})  -  (Q\Phi_k) (y)  \Big) \\
\end{split}
\ee
The first  term  is bounded  by  $C p_k$  by   (\ref{sub})    and the second term  is bounded  by  $C p_k$  by the bound on $\pa   \Phi_k$.
This completes the proof.

   \subsection{an  approximate    minimizer}
   
 The  two quadratic terms in the exponents in (\ref{representation3})   can be identified as  
 \begin{equation}   \label{studyJ1}
  \begin{split}
 J^*_{\La_k,\Om_{k+1}}(   \La_k,     \Phi_{k+1}, \Phi_{k},  \phi_{\bom(\La_k^*)})  
=&    \frac12 \frac{a}{L^2}  \|\Phi_{k+1}   -  Q \Phi_{k}\|^2_{  \Om_{k+1}}
+   S^*_k(\La_k,  \Phi_{k} ,   \phi_{k, \bom(\La_k^*)} )     \\
 \end{split}
\end{equation}
 defined   previously in  (\ref{jstar}).
 
 We  want to find the minimizer of this  functional  in the variable
  in   $\Phi_{k,  \Om_{k+1}}$.
This is not  the standard  setup because the action is restricted to $\La_k$,    but it is a problem we have anticipated.
Instead  we  use  an  approximate minimizer,  namely  the minimizer 
for   the full  problem on  
   \begin{equation}
   \bom'  \equiv   ( \bom(\La_k^*),  \Om_{k+1} )  
 \end{equation}
By lemma  \ref{otto1}      the minimum   in  $\Phi_{k, \Om_{k+1}}$   for that problem    comes  at  
 \begin{equation}  
   \Psi_{k, \Om_{k+1}}(\bom')=   Q_{k}   \phi^0_{k+1,\bom'}  
- \frac{aL^{-2}}{a_k+ aL^{-2}} Q^T  Q_{k+1} \phi^0_{k+1,\bom'}   + \frac{aL^{-2}}{a_k+ aL^{-2}} Q^T  \Phi_{k+1}   
\end{equation} 
where   $ \phi^0_{k+1,\bom'} $  is defined in   (\ref{eddie}).
Recalling  that     $ \phi_{k,\bom(\La_k^*)}=  \phi_{k,\bom(\La_k^*)}(  \tilde   Q^T_{\bbT^0,    \bom(\La_k^*)}    \Phi_k  ) $     the minimizer 
in  $\phi$     is   more precisely characterized as  
 $ \phi^0_{k+1,\bom'}   =   \phi^0_{k+1,\bom'}   (\hat   \Phi_{k+1,  \bom'} )$ 
where
\begin{equation}   \label{guessing}
 \hat   \Phi_{k+1,  \bom'}   = \Big(  [  \tilde   Q^T_{ \bbT^0,    \bom(\La_k^*)}    \Phi_k ]_{\Om^c_{k+1}},   \Phi_{k+1, \Om_{k+1}}  \Big)  
 \end{equation}  
 If  $k=0$  then   the minimizer   in $\Phi_{0,\Om_1}$   is  just   $ \Psi_{0, \Om_1}=     \phi^0_{1, \Om_1} $ as a separate calculation reveals.

Inserting  $ \Psi_{k, \Om_{k+1}}( \bom')$  into  $ \tilde   Q^T_{\bbT^0,    \bom(\La_k^*)}    \Phi_k  $         gives   
\begin{equation}    \label{55}
  \hat    \Psi_{k,\bom'}   \equiv    \Big(  [ \tilde   Q^T_{ \bbT^0,    \bom(\La_k^*)}    \Phi_k ]_{\Om^c_{k+1}},  \Psi_{k, \Om_{k+1}}(\bom') \Big)
\end{equation}
Note that  $    \hat   \Phi_{k+1,  \bom'}  $  and  $\hat    \Psi_{k,\bom'}  $ now include boundary fields  in $\Om_1(\La_k^*)^c$.  This is 
a different convention from chapter \ref{two}   where such fields  were treated separately.
The identity  \be    \label{such}
\phi^0_{k+1, \bom'} =     \phi_{k,\bom(\La_k^*)}( \hat \Psi_{k, \bom'} ) 
 \ee
  holds  by      lemma  \ref{otto1}.
(The fields  in    $\Om^c_{k+1}$ are  just  spectators in the proof of this identity.)

Now expand    $J^*_{\La_k,  \Om_{k+1}}(\La_k )  $
around  the minimizer  inserting    $\Phi_{k, \Om_{k+1}} =    \Psi_{k,\Om_{k+1}}( \bom' )  +  Z $.  Using the identity  (\ref{such})  this   entails 
\begin{equation}  \label{spinning}
\begin{split}
\hat  \Phi_{k, \bom( \La_k^*)}  = &  \hat  \Psi_{k,  \bom' }  + (0, Z)\\
\phi_{k, \bom(\La_k^*)}  = &            \phi^0_{k+1, \bom'} + \cZ_{k, \bom(\La_k^*)}  \\
\end{split}
\end{equation}
Here  $Z$  is a function on  the unit lattice  $\Om_{k+1}^{(k)}$   and   as before
\begin{equation}
 \cZ_{k, \bom(\La_k^*)}  =   \phi_{k, \bom(\La_k^*)}  (0, Z )  
   =   a_k  G_{k,   \bom(\La_k^*)}  Q_k^T  Z 
\end{equation}
Lemma  \ref{otto2}  is applicable  here  and so  
\footnote{ Because of the restriction to  $\La_k$ the substitution    $\hat  \Phi_{k, \bom( \La_k^*)}  =   \hat  \Psi_{k,  \bom' }  + (0, Z) $
is the same as  the expected   $\Phi_k  =  ( \Phi_{k, \La_k- \Om_{k+1}},  \Psi_{k,\Om_{k+1}}( \bom' )  +  Z)$.
The restriction  also  allows us to replace   $\Phi_{k+1,  \bom'}$  by  $\Phi_{k+1,  \bom^+}$  in   $  S^{*,0}_{k+1}(\La_k)$.
}   
\begin{equation}  \label{otto3}
\begin{split}
 & J^*_{\La_k,  \Om_{k+1}}(\La_k, \Phi_{k+1},   \hat  \Psi_{k, \bom'}  +(0,Z) ,  \phi^0_{k+1, \bom'} + \cZ_{k, \bom(\La_k^*)}  )     
=     S^{*,0}_{k+1}(\La_k, \Phi_{k+1,  \bom^+}, \phi^0_{k+1,\bom'}) \\ 
   +  &   
 \frac12 \Big <Z,  \left[ \De_{k, \bom( \La_k^*)} + \frac{a}{L^2} Q^TQ   \right]_{\Om_{k+1}}  Z  \Big>  
 +   R^{(1)}_{\bpi,  \Om_{k+1}} \\
\end{split}
\end{equation}
where  
\begin{equation}
\begin{split}
  R^{(1)}_{\bpi,  \Om_{k+1}}
  =&     
 \sfb_{\La_k}  \Big[  \pa \phi^0_{k+1,  \bom'},   \cZ_{k, \bom(\La_k^*)}  \Big] + \frac{ 1}{2}  \|\ba^{1/2}Q_{k, \bom(\La_k^*)}   \cZ_{k, \bom(\La_k^*)} \|^2_{  \La^c_k}
\\
  &
+\frac12   \| \pa \cZ_{k, \bom(\La_k^*)}  \|_{*, \La_k^c}^2    +  \frac12  \bar    \mu_k  \|  \cZ_{k, \bom(\La_k^*)}   \|^2_{ \La^c_k} 
\\
  \end{split}
 \end{equation} 
 
  The function $ R^{(1)}_{\bpi,  \Om_{k+1}}$  is tiny  
because   $Z$  is  localized  in $\Om_{k+1} $,  and  $ \cZ_{k, \bom(\La_k^*)} 
   $ is evaluated on      $\La_k^c$,   and the   operator  connecting these   distant 
   sets, namely  $ G_{k,   \bom(\La_k^*)} $,  has an  exponentially decaying kernel.   Furthermore  $ R^{(1)}_{\bpi,  \Om_{k+1}}$  
   has a local expansion.     However  we  postpone the demonstration
   of such facts       to  section \ref{localization}.

The idea is  now   to   make  the substitutions  (\ref{spinning})   in the integral (\ref{representation3}) and then    integrate over
 $Z$  instead of   $\Phi_{k, \Om_{k+1}} $,   taking advantage of  (\ref{otto3}).
This  substitution    is not  completely satisfactory  because when it appears in the characteristic function  it will  introduce non-local  non-analytic  dependence on  $\Phi_{k+1}$  everywhere inside   $\La_k$.    This we  want to avoid.      Instead we  replace it      by    a  more   local version.

\subsection{a  better   approximation}
\label{better}

To  develop the  more local  version    
we  introduce  some definitions.  In  $ \phi^0_{k+1,\bom'} $   we have the propagator   $G^0_{k+1, \bom'} $.  A weakened  propagator  $G^0_{k+1, \bom'}(s)$  is defined as in
section  \ref{random2}.   Here   $s =  \{  s_{\square} \} $ is a collection  of  variables indexed by  cubes   $\square$  associated with 
$\bom'$,   that  is    $\square  \subset   \de  \Om'_j$  is  an $L^{k-j}M$ cube, and in particular
 $\square$ is an $LM$ cube in  $\Om_{k+1}$.   This  gives   a  more local field
\begin{equation}     \label{unlikely}
\phi^0_{k+1,\bom'} (  s) 
=     G^0_{k+1,\bom'}(s) \Big(L^{-2} Q_{k+1, \bom'}^T  \ba^{(k+1)}    [\hat \Phi_{k+1,  \bom'} ]_{ \Om'_1}
+   \De_{  \Om'_1,  \Om_1^{' c} }Q_k^T \Phi_k\Big)
\end{equation}
with   $\hat  \Phi_{k+1,  \bom'}$ defined  in  (\ref{guessing}).
Then define  a more local minimizer   by 
\begin{equation}    \label{dingdong2}
  \Psi_{k,\Om_{k+1}}(\bom',  s)  =   Q_{k}  \phi^0_{k+1,\bom'}(s) 
-  \frac{aL^{-2}}{a_k+ aL^{-2}}   Q^T  Q_{k+1} \phi^0_{k+1,\bom'}(s) 
  +   \frac{aL^{-2}}{a_k+ aL^{-2}}  Q^T  \Phi_{k+1}   
\end{equation}

Next  let  $\square$ be  an  $LM$ cube    in  $\Om_{k+1}$, and let   $\square^*$  be the same with  $[r_{k+1}]$ layers of 
$LM$ cubes added.   We define  
\begin{equation}
G^0_{k+1, \bom' } (\square^* )   =    G^0_{k+1, \bom'}\Big(s_{\square^*} =1,  s_{\square^{*,c}}=0  \Big)
\end{equation}
which has  no    coupling outside  of   $\square^*$. This does not depend on the full extent of $\bom'$; we could as well take $\bom^+$ here. Similarly  define  $\phi^0_{k+1,\bom'} ( \square^*) $  
and  $ \Psi_{k,\Om_{k+1}}(\bom',   \square^*)$.  
 Then we  define a  more localized field  $ \Psi^{\loc}_{k,\Om_{k+1}}(\bom',x)$  to be  equal  to   $ \Psi_{k,\Om_{k+1}}(\bom',  \square^*,x) $  for    $x \in  \Om_{k+1}^{(k)} \cap   \square$.  This can also be written
\begin{equation}
 \Psi^{\loc}_{k,\Om_{k+1}}(\bom') =   \sum_{\square \subset  \Om_{k+1}}  1_{\square}   \Psi_{k,\Om_{k+1}}(\bom',  \square^*)       \end{equation}
with the spectator fields present this gives  
\begin{equation}   
  \hat   \Psi^{\loc}_{k,\bom'}   \equiv    \Big(  [ \tilde   Q^T_{ \bbT^0,    \bom(\La_k^*)}    \Phi_k ]_{\Om^c_{k+1}},  \Psi^{\loc}_{k, \Om_{k+1}}(\bom') \Big)
\end{equation}

Now  make the  change of variables    
\be   \label{a1}
\Phi_{k, \Om_{k+1} } =  \Psi^{\loc}_{k,\Om_{k+1}}(\bom')  +Z
\ee
With the spectator fields present this says that  
\begin{equation}  \label{a2}
\begin{split}
 \hat \Phi_{k, \bom'}   =&  \hat \Psi^{\loc}_{k, \bom'}  + (0,Z)  
 = \hat  \Psi_{k,   \bom'}  + (0,Z)  +    (0,   \de   \Psi_{k,\Om_{k+1}}(\bom')) \\
    \de   \Psi_{k,\Om_{k+1}}(\bom')    \equiv   &     \Psi^{\loc}_{k,\Om_{k+1}}(\bom')-   \Psi_{k,\Om_{k+1}}(\bom') \\
\end{split}
 \end{equation}
 Then we have  as  a modification of (\ref{spinning})  
  \begin{equation}  \label{a3}
 \begin{split}
 \phi_{k, \bom(\La_k^*)}
 =&  
\phi^0_{k+1,\bom'} +  \cZ _{k,   \bom(\La_k^*)}  +   \de   \phi_{k,\bom'}\\
\de   \phi_{k,\bom'} =  &  \phi_{k, \bom(\La_k^*)}(  0,   \de   \Psi_{k,\Om_{k+1}}(\bom')  ) \\
\end{split}
\end{equation}
  We will see eventually     that  $\de    \Psi_{k,\Om_{k+1}}(\bom') $  and  
$\de   \phi_{k,\bom'}$   are  tiny.

Now  we  make the substitutions   (\ref{a1}),  (\ref{a2}),  (\ref{a3}) in   (\ref{representation3}).
and  integrate over   $Z$ instead of     $\Phi_{k, \Om_{k+1}}$
Instead  of   and taking advantage of  (\ref{otto3})    
we have 
\begin{equation}  \label{otto4}
\begin{split}
 & J^*_{\La_k, \Om_{k+1}}( \Phi_{k+1}, \hat  \Psi_{k,\bom'} + (0,Z)   +  (0,  \de   \Psi_{k,\Om_{k+1}}(\bom') )  ,  \phi^0_{k+1,\bom'} 
 +  \cZ _{k,   \bom(\La_k^*)}  +   \de   \phi_{k,\bom'} )    \\
 =  &      J^*_{\La_k, \Om_{k+1}}( \Phi_{k+1},  \hat  \Psi_{k, \bom'}  + (0,Z)   ,  \phi^0_{k+1,\bom'} +  \cZ _{k,   \bom(\La_k^*)}  ) +
   R^{(2)}_{\bpi,  \Om_{k+1}}  \\
=  &   S^{*,0}_{k+1}(\La_k, \Phi_{k+1,\bom^+}, \phi^0_{k+1,\bom'})
 + \frac12 \Big<Z,  \left[ \De_{k, \bom( \La_k^*)}
 + \frac{a}{L^2} Q^TQ   \right]_{\Om_{k+1}}  Z  \Big> +       R^{(1)}_{\bpi,  \Om_{k+1}}   +     R^{(2)}_{\bpi,  \Om_{k+1}}  \\
\end{split}
\end{equation} 
Here the   first    equality  defines  $  R^{(2)}_{\bpi,  \Om_{k+1}}  $.
 We   also  have
\begin{equation}
\begin{split}
  E_k^+ ( \La_k,  \phi^0_{k+1,\bom'} +   \cZ _{k,   \bom(\La_k^*)}  +   \de   \phi_{k,\bom'})
=  &    E_k^+ ( \La_k,  \phi^0_{k+1,\bom'} +   \cZ _{k,   \bom(\La_k^*)}  ) 
 +  R^{(3)}_{\bpi,  \Om_{k+1}} \\
\end{split}
\end{equation}
which defines  $ R^{(3)}_{\bpi,  \Om_{k+1}} $.      The  characteristic functions in  $\Om_{k+1}$   now have the  form  
\begin{equation}  \label{omega}
\begin{split}
  \chi_k(  \La_k)         =&  \prod_{\square \in    \La_k}
   \chi_k\Big(  (  \Phi_{k,  \de \Om_{k}},   \Psi^{\loc}_{k, \Om_{k+1}}(\bom')  +Z)  \in \cS_k(\square)  \Big)\\
   \chi^q_{k}(  \Om_{k+1} ) 
   = &\prod_{\square \subset   \Om_{k+1}}  \chi^q_{k}\Big(\square,   \Phi_{k+1},  \Psi^{\loc}_{k, \Om_{k+1}}(\bom')  +Z  \Big)\\
 \end{split}  
\end{equation}
We introduce the abbreviations   
\begin{equation}  \label{stutz}
\begin{split}
 S^{*,0}_{k+1}(\La_k)  =&  S^{*,0}_{k+1}(\La_k, \Phi_{k+1,\bom^+}, \phi^0_{k+1,\bom'}) \\
  E^+_k (\La_k) =&  E^+_k \big(\La_k,    \phi^0_{k+1,\bom'}   +   \cZ _{k,   \bom(\La_k^*)}   \big) \\
 R_{k,\bpi} (\La_k)  =& R_{k, \bpi}(\La_k, \Phi_{k, \La_k - \Om_{k+1}},  \Psi^{\loc}_{k, \Om_{k+1}}( \bom')   +Z)   \\
 B_{k,\bpi} (\La_k)  =   & B_{k, \bpi}(\La_k, \Phi_{k, \bom \cap  \Om^c_{k+1}}  ,  \Psi^{\loc}_{k, \Om_{k+1}}( \bom')   +Z,  W_{k, \bpi})   \\ 
 \end{split}
\end{equation}
We  also write $ R^{(0)}_{\bpi,  \Om_{k+1}}   =R_{k,\bpi} (\La_k)$,  and  then  tiny terms are collected in   
\begin{equation}
R^{(\leq 3))}_{\bpi,  \Om_{k+1}}  = R^{(0)}_{\bpi,  \Om_{k+1}}  +   R^{(1)}_{\bpi,  \Om_{k+1}}   +     R^{(2)}_{\bpi,  \Om_{k+1}} +     R^{(3)}_{\bpi,  \Om_{k+1}}   
\end{equation}

 Making all these changes  (\ref{representation3}) becomes
  \begin{equation}     \label{representation4}
\begin{split}
&\tilde    \rho_{k+1}( \Phi_{k+1}) 
 =      Z_k     \cN^{-1}_{aL,  \bbT^1_{\sM+\sN -k   }  } 
  \sum_{\bpi, \Om_{k+1}}    \int   d \Phi^0_{k+1, \bom^{+,c} }d W_{k,\bpi}\  K_{k, \bpi} \  \cC_{k, \bpi} 
 \  \cC^q_{k} (\La_k, \Om_{k+1} )   
      \\
&      
      \exp \Big(-  S^{*,0}_{k+1}(\La_k) \Big) 
  \int   d Z\
  \chi_k( \La_k) \   \chi^q_{k}(  \Om_{k+1}  )   
  \exp    \Big( -
 \frac12 \Big<Z,  \left[ \De_{k, \bom(\La_k^*)} + \frac{a}{L^2} Q^TQ   \right]_{\Om_{k+1}}  Z     \Big>  \Big)   \\
 &
     \exp\Big( E^+_k(\La_k)    + R^{(\leq 3))}_{\bpi,  \Om_{k+1}}  +  B_{k, \bpi}(\La_k)  \Big) \\
   \end{split}
\end{equation}

\subsection{fluctuation integral}  \label{bingo}

In the last expression we have  the fluctuation  integral with the  measure $ \exp (  - \frac12 <Z, C_{k,\bom'} ^{-1} Z>)  dZ$  where
\begin{equation}
C_{k,\bom'}   =  \big[\De_{k,  \bom(  \La^*_k)}  +\frac{a}{L^2} Q^TQ \Big]_{\Om_{k+1}}^{-1}
\end{equation}
(If  $k=0$ it is  $C_{0, \Om_1} =   [ - \De  + a L^{-2} Q^TQ]_{\Om_1}^{-1}  $).
This is an   operator on functions on the unit lattice   $\Om_{k+1}^{(k)}$.
We  would like  to make  the   change  of  variables    
$Z =C_{k,  \bom'}^{1/2}W_k$   
which would yield   the   ultra-local     measure    $( \det  C^{1/2} _{k,  \bom'}) \exp(  - \frac12   \|W_k\|^2  )dW_k$.  This would  move the non-locality into
other terms   where it  is easier to handle.
However    non-locality inherent in   $C_{k,   \bom'}^{1/2}W_k$   is awkward  in the characteristic functions   since it occurs in a non-analytic setting.
The problem is similar to the problem  of the   non-local  minimizer   and the solution is the same.

We  introduce a  more     local  approximation  to  $C_{k,    \bom'}^{1/2}$  defined as  follows \cite{Bal96b}.
Start with the representation
\begin{equation}
\begin{split}
C^{1/2}_{k, \bom'  }  \label{z1}
  =&     \frac{1}{\pi}  \int_0^{\infty}  \frac{dr}{\sqrt{r}}  C_{k, \bom', r} \\
C_{k, \bom', r}  =  &  \Big[   \De_{k,  \bom(\La^*_k)}+ \frac{a}{L^2}Q^TQ    +r \Big]_{\Om_{k+1}}  \\
\end{split}
\end{equation}
In  appendix  \ref{moonshine}   we establish that  
\begin{equation}  \label{z2}
 C_{k, \bom', r} =  \Big[ A_{k,r}  +   a_k^2  A_{k,r} Q_k  G_{k, \bom',  r} Q_k^T A_{k,r}\Big]_{\Om_{k+1}}
\end{equation}
where 
\begin{equation}  \label{z3}
\begin{split}
A_{k,r}   =&   \frac{1}{a_k+r}  (I - Q^TQ)   +   \frac{1}{ a_k + aL^{-2}  +r}   Q^T Q \\
B_{k,r}  = &   \frac{r}{a_k+r}  (I - Q^TQ)   +   \frac{aL^{-2}  +r}{ a_k + aL^{-2}  +r}   Q^T Q \\
G_{k, \bom', r}  = &  \Big[ -\De  + \bar \mu_k  
+ \left[ Q_{k, \bom(\La_k^*)}^T  \ba  Q_{k, \bom(\La_k^*)} \right]_{\Om_{k+1}^c}    +a_k  
\left[   Q_k^T   B_{k,r} Q_k \right]_{\Om_{k+1}}  \Big]_{\Om_1(\La_k^*)}^{-1} \\
\end{split}
\end{equation}

 Now  for  $0 \leq r \leq \infty$ the Green's function   $G_{k, \bom', r}$   has  random walk expansions  just  as its extreme values  $G^0_{k+1,  \bom'}$ and  $G_{k, \bom(\La_k^*)}$.  We  choose a version   based on   multiscale   cubes  $\square$ 
    in   $\bom'$,  just as for  $G^0_{k+1,  \bom'}$.  (For more details   see lemma \ref{th2} to follow).  Then there is a weakened  
 Greens function  $G_{k, \bom', r}(s)$  defined for     $s = \{  s_{\square} \}$.
 Correspondingly we defne
 \begin{equation}  \label{needed2}
 \begin{split}
 C^{1/2}_{k, \bom'  }(s)
  =&     \frac{1}{\pi}  \int_0^{\infty}  \frac{dr}{\sqrt{r}}  C_{k, \bom', r}(s) \\
 C_{k, \bom', r}(s) =&  \Big[ A_{k,r}  +   a_k^2  A_{k,r} Q_k  G_{k, \bom',  r}(s) Q_k^T A_{k,r}\Big]_{\Om_{k+1}}\\
  \end{split}
\end{equation}
For  $\square$ an  $LM$ cube in   $\Om_{k+1}$    
 we   decouple   the $[r_{k+1}]$ enlargement 
  $\square^*$  from the complement by  considering  
 \begin{equation}
 G_{k, \bom', r}(\square^*)  
 \equiv     G_{k, \bom', r}(    s_{\square^*} =1 ,    s_{(\square^*)^c} =0  )
 \end{equation}
 and   we have the   associated  $ C_{k, \bom', r}(\square^*), C^{1/2}_{k,  \bom'}( \square^*)$.

 The  local   approximation on  $\Om_{k+1}^{(k)}$    is 
 \begin{equation}
 ( C^{1/2}_{k,  \bom'} )^{\loc}
  =   \sum_{\square \subset  \Om_{k+1}}  1_{\square}   C^{1/2}_{k,  \bom'}( \square^*)  
  \end{equation}
The difference  
 \begin{equation}
\de  C^{1/2}_{k,  \bom'}=C^{1/2}_{k,  \bom'} -( C^{1/2}_{k,  \bom'} )^{\loc}
\end{equation}
is tiny  as we will see.

We  will also  see that  $ ( C^{1/2}_{k,  \bom'} )^{\loc}$  is invertible,   and  we   make  the change of variables  $  Z   =  ( C^{1/2}_{k,  \bom'} )^{\loc}W_k$
where    $W_k:  \Om_{k+1}^{(k)} \to  \bbR$.
The   quadratic form    $\frac12  <Z, C_{k,\bom'} ^{-1} Z>$   becomes
\begin{equation}
\frac12   <    ( C^{1/2}_{k,  \bom'} )^{\loc}W_k,      C^{-1}_{k,  \bom'}  ( C^{1/2}_{k,  \bom'} )^{\loc}W_k>=
\frac 12   \|W_k  \|^2_{\Om_{k+1}}  -     R^{(4)}_{\bpi, \Om_{k+1}}
  \end{equation}
 where 
 \begin{equation}
  R^{(4)}_{\bpi, \Om_{k+1}}
  =     <  C^{-1/2}_{k,  \bom'}W_k, \de C^{1/2}_{k,  \bom'}W_k>  
   - \frac12     <  \de C^{1/2}_{k,  \bom'}W_k, C^{-1}_{k,  \bom'}  \de C^{1/2}_{k,  \bom'}W_k>   
\end{equation}
is tiny.
The   change  of   variables   also  
introduces  
 \begin{equation}
\det \Big( \Big(      C^{1/2}_{k,  \bom'} \Big) ^{\loc}\Big)  =  \det     C^{1/2}_{k,  \bom'}\   \exp  (    R^{(5)}_{\bpi, \Om_{k+1}})
 \end{equation}
 where    
  \begin{equation}   \label{onionstew}
 \begin{split}
     R^{(5)}_{\bpi, \Om_{k+1}}
   =&  \tr \log      (  ( C^{1/2}_{k,  \bom'} )^{\loc})  -  \tr    \log     (   C^{1/2}_{k,  \bom'} )  
  = - \sum_{n=1}^{\infty} \frac{1}{n} \tr \Big( (    C^{-1/2}_{k,  \bom'}   \de C^{1/2}_{k,  \bom'} )^n\Big) \\
   \end{split}
   \end{equation}
For  the last identity see   (2.41)   in  \cite{Bal98a}.  This is also  tiny.

 We  also want to replace   $ \det   (  C^{1/2}_{k,  \bom'} )$  by the global determinant  $ \det   (  C^{1/2}_{k} )  $.   We  have  
  \begin{equation}  \label{stay1}
 \begin{split}
&  \det   (  C^{1/2}_{k,  \bom'} ) \\
=  &   \det   (  C^{1/2}_{k} ) \exp \Big(   - \frac 12  \tr  \log  \Big(  \   \Big[\De_{k,  \bom(  \La^*_k)}  +\frac{a}{L^2} Q^TQ \Big]_{\Om_{k+1}}\Big) +     \frac 12  \tr  \log  \Big(  \  \De_{k}  +\frac{a}{L^2} Q^TQ   \Big)\Big) \\
\end{split}
  \end{equation}
  A  variation  of our formula for   $C^{1/2}_{k, \bom'  } $  is  
   \begin{equation}   \label{z5}
 \begin{split}
 & \log \Big(  [\De_{k,  \bom(  \La^*_k)}  +\frac{a}{L^2} Q^TQ]_{\Om_{k+1}} \Big) \\ = &   \log a_k   [I - Q^TQ]_{ \Om_{k+1}}    +  \log(a_k + aL^{-2}) [Q^TQ]_{ \Om_{k+1}}   -a_k^2  \int_0^{\infty}
 [ A_{k,r} Q_k G_{k,\bom',  r} Q_k^T A_{k,r}]_{ \Om_{k+1}}   dr    \\
\end{split}
\end{equation}
provided the integral converges.  
See     (3.22) in \cite{Bal96b}   for  the derivation.    A similar formula holds for   $ \log \Big(  \De_{k}  +\frac{a}{L^2} Q^TQ \Big)$.
Thus the exponent in  (\ref{stay1})  can be written
\be 
\begin{split}
   \label{stay2}
& \frac12   \log a_k \tr  [I - Q^TQ]_{ \Om^c_{k+1}}    + \frac12  \log(a_k + aL^{-2})  \tr [Q^TQ]_{ \Om^c_{k+1}} \\
  & -  \frac12  a_k^2  \int_0^{\infty}
\tr [ A_{k,r} Q_k G_{k,\bom',  r} Q_k^T A_{k,r}]_{ \Om_{k+1}}   dr 
  + \frac12   a_k^2  \int_0^{\infty}  \tr[ A_{k,r} Q_k G_{k,  r} Q_k^T A_{k,r}]  dr   \\
\end{split} 
\ee
But   for  $\Om  \subset  \tk$,   $\tr   [Q^TQ]_{ \Om}$   means
\footnote{   Keep in mind that for    $\Om  \subset  \tk$  we have    $|\Om^{(k)}|  =  \Vol( \Om )$.   We prefer to write it  in the first 
form  which is scale invariant.  }  
\begin{equation}
\tr   [Q^TQ]_{ \Om^{(k)}} =\tr   [QQ^T]_{ \Om^{(k+1)}}  = \tr   [I]_{ \Om^{(k+1)}}
  =  |\Om^{(k+1)}|  =  L^{-3}  |\Om^{(k)}|
\end{equation}
and similarly  $  \tr  [I - Q^TQ]_{ \Om} = (1 - L^{-3})   |\Om^{(k)}|  $.
So the first  two  terms   in   (\ref{stay2})  are  
\be  \frac12  \Big((1 - L^{-3})     \log a_k    +   L^{-3}   \log(a_k + aL^{-2})  \Big) |\Om^{c,(k)}_{k+1}|
\equiv  \frac 12   b_k |\Om^{c,(k)}_{k+1}|
\ee
The second two   terms    in  (\ref{stay2})  can be  written 
\begin{equation}  \label{stay3}
\begin{split}
    -  & \frac 12   a_k^2  \int_0^{\infty}  \Big(
\tr   [ A_{k,r} Q_k (G_{k,\bom',  r}-  G_{k,r}) Q_k^T A_{k,r}]_{ \Om_{k+1}} - \tr   [ A_{k,r} Q_k  G_{k,r} Q_k^T A_{k,r}]_{ \Om^c_{k+1}}\Big)  dr      \\
   \end{split}
\end{equation}
The  first   term here is   defined to be   $ R^{(6)}_{\bpi, \Om_{k+1}}$   and the second    term  is   $\frac12  a_k^2 b'_k  |\Om^{c, (k)}_{k+1}|$    where   
  \be b'_k   =    \int \Big(A_{k,r} Q_k  G_{k,r} Q_k^T A_{k,r}  \Big)(x,x)  dr   \ee
 is  independent of  $x$.  We will see  that $b_k'$ is bounded in  $k$.
 Therefore  we  have  
 \be
  \det   (  C^{1/2}_{k,  \bom'} )  =  \det   (  C^{1/2}_{k} )\exp\Big(    \frac 12   b''_k |\Om^{c,(k)}_{k}|        +          R^{(6)}_{\bpi, \Om_{k+1}}                    \Big)
 \ee
  where  $  b''_k   = b_k  +    a_k^2   b'_k $.

 We  also   introduce the Gaussian measure with identity covariance
 \be 
 d \mu_{ \Om_{k+1} }(W_k)  =  ( 2\pi  )^{- \frac 12   |\Om^{(k)}_{k+1}| }  \exp    \Big(- \frac12  \| W_k\|^2   \Big) \ d W_k
 \ee
 Normalization factors are  rearranged as  
\begin{equation}  \label{sugar}
 (2 \pi)^{\frac 12| \Om_{k+1}^{(k)}|}   =    (2 \pi)^{\frac 12|  \bbT^0_{\sM+\sN -k} |}  (2 \pi)^{-\frac12 |\Om^{c, (k)}_{k+1}|  }   
\end{equation}
The first term contributes to  
\begin{equation}
Z^0_{k+1}  \equiv   \cN_{aL,  \bbT^1_{\sM+\sN -k} }^{-1} (2 \pi)^{\frac12 |  \bbT^0_{\sM+\sN -k} |}  ( \det   C_k )^{1/2}   Z_k  
\end{equation}
The second term combines    with    $ \exp\Big(    \frac 12   b''_k |\Om^{c,(k)}_{k}|   \Big) $     to  give    
$ \exp\Big(    \frac 12   c_{k+1} |\Om^{c,(k)}_{k}|    \Big)$
where  $   c_{k+1}=  b_k''    - \log 2 \pi     $  is bounded in $k$.

  There  are  still  more changes.
The    field  
 $   \cZ_{k, \bom(\La_k^*)} 
   =   a_k  G_{k,   \bom(\La_k^*)}  Q_k^T  Z  $
   now  becomes  
\begin{equation}
\begin{split}
 \cW^{\loc}_{k,\bom'}     \equiv   
 &    a_k  G_{k,   \bom(\La_k^*)}  Q_k^T      ( C^{1/2}_{k,  \bom'} )^{\loc}  W_k  \\
=  &    a_k  G_{k,   \bom(\La_k^*)}  Q_k^T \      C^{1/2}_{k,  \bom'} W_k  + 
     a_k  G_{k,   \bom(\La_k^*)}  Q_k^T \  \de  C^{1/2}_{k,  \bom'}   W_k    \\
  \equiv  & \cW_{k,\bom'}   +  \de   \cW_{k,\bom'}   \\
\end{split}
\end{equation}
and   
   $\de     \cW_{k,\bom'}$  will be  tiny.
   Therefore we can write  
 \begin{equation}
 \begin{split}
    E_k^+ ( \La_k,  \phi^0_{k+1,\bom'} +\cW^{\loc}_{k,\bom'}    )
    =&   E_k^+ ( \La_k,  \phi^0_{k+1,\bom'} +   \cW_{k,\bom'}    )  
    +       R^{(7)}_{\bpi, \Om_{k+1}} \\
    \end{split}
    \end{equation}  
 which defines a tiny term     $  R^{(7)}_{\bpi, \Om_{k+1}} $.

 Now we  rewrite  (\ref{representation4}).
 The  characteristic  functions    $ \chi_k(\Om_{k+1}) ,    \chi^q_{k}( \Om_{k+1} ) $ 
 are the same  as   (\ref{omega}), except   that    $Z$   is   replaced by   $( C^{1/2}_{k,  \bom'} )^{\loc} W_k $.
 Now  $  E_k^+ ( \La_k  )  $ stands for   $  E_k^+ ( \La_k,  \phi^0_{k+1,\bom'} +   \cW_{k,\bom'}    )  $
Also     $ R^{(\leq 3))}_{\bpi,  \Om_{k+1}},    B_{k, \bpi}(\La_k) $  are the same as before, but  with 
 $Z$   replaced by   $( C^{1/2}_{k,  \bom'} )^{\loc} W_k $  and   $  \cZ_{k, \bom(\La_k^*)} $
  replaced by  $ \cW^{\loc}_{k,\bom'}  $.
All tiny terms are collected in    $R^{(\leq 7)}_ { \bpi, \Om_{k+1}}  =    R^{(0)}_{\bpi, \Om_{k+1}} + \dots+  R^{(7)}_{\bpi, \Om_{k+1}} $.  (Actually parts
of   $R^{(6)}_{\bpi, \Om_{k+1}}$  are not tiny,  but they are bounded and  will eventually end up as boundary terms.)
Making all these changes   (\ref{representation4}) becomes
  \begin{equation}     \label{representation5}
\begin{split}
\tilde    \rho_{k+1}( \Phi_{k+1}) 
 =  &    Z^0_{k+1}   \sum_{\bpi, \Om_{k+1}}   
  \int   d \Phi^0_{k+1, \bom^{+,c} }d W_{k,\bpi}\  K_{k, \bpi} \  \cC_{k, \bpi}  \cC^q_{k} (\La_k, \Om_{k+1} )  
 \\
    &    
\         \exp \Big(-  S^{*,0}_{k+1}(\La_k) \Big)   \exp  \Big(  c_{k+1}  |\Om^{c,(k)}_{k+1}|    \Big)   \\
&   \int    d \mu_{ \Om_{k+1} }(W_k)  
  \chi_k( \La_k) \   \chi^q_{k}(  \Om_{k+1}  ) 
     \exp\Big( E^+_k(\La_k)    + R^{(\leq 7)}_{\bpi,  \Om_{k+1}}  +  B_{k, \bpi}(\La_k)  \Big) \\
   \end{split}
\end{equation}

\subsection{estimates}

We  collect some estimates on  these operators.  First we  elaborate on the statement that  $G_{k, \bom',  r}$
has   a  random  walk expansion.   This means repeating the  analysis of section  \ref{random2}  with some modifications.   Actually
things are a little easier here  since      we  will not need  derivatives and  $L^2$ bounds will   suffice.     We  will be  a little  more general  and consider  $G_{k,  \bom^+,  r}$
where  $\bom^+ = (\bom, \Om_{k+1}) =   ( \Om_1, \dots,  \Om_{k+1})$  satisfies the minimal separation conditions   (\ref{hundred})  and   where
\begin{equation}  \label{general}
G_{k,  \bom^+, r}  =   \Big[ -\De  + \bar \mu_k  
+ \left[ Q_{k, \bom}^T  \ba  Q_{k, \bom} \right]_{\Om_{k+1}^c}    +a_k  
\left[   Q_k^T   B_{k,r} Q_k \right]_{\Om_{k+1}}  \Big]_{\Om_1}^{-1} \
\end{equation}
The inverse taken with Dirichlet  boundary conditions.
Instead of theorem  \ref{th}  we
have:

\begin{lem}  \label{th2}
The Green's function $G_{k, \bom^+,r}$       has  a random walk expansion 
\begin{equation} 
G_{k, \bom^+,r}  =  \sum_{\om } G_{k,\bom^+,r,    \om}
\end{equation}
based  on multiscale cubes  $\square$    for  $\bom^+$.
It converges in $L^2$ norm   for      $M$   sufficiently  large  and    yields the following   bounds.  There are  constants  $C$ (depending on $L$) and
$\ga = \cO(L^{-2})$  so   
for    $  \De_y \subset     \de \Om_j $
and   $  \De_{y'}  \subset    \de \Om_{j'} $  and  all  $r \geq  0$
\be    \label{security}
 \|1_{\De_y} G_{k, \bom^+,r}  1_{\De_y'}  f\|_{2}  \  \leq   \      C L^{-2(k-j')}
  e^{   -    \ga  d_{\bom^+}(y,y' ) }  \|f\|_{2}  
\ee
\end{lem}
\bigskip

\pr   The proof follows the analysis in section \ref{random2}, see  also   \cite{Bal96b}.  First  note that   
\be   \label{summer}
  a_k   Q_k^T   B_{k,r} Q_k  =
\frac{ a_k r}{a_k  +r}  Q_k^TQ_k    +   \frac{a_k^2aL^{-2}}{(a_k +r)   (a_k + aL^{-2} +r)}Q_{k+1}^T Q_{k+1}
\ee
This is bounded  below by  $\cO(1)L^{-2} Q_{k+1}^T Q_{k+1}$  for  $0 \leq  r \leq  1$  and  by    $\cO(1) Q_{k}^T Q_{k}$
for   $r \geq  1$.   
It  follows that   for an  $LM$ cube    $\square$  with $\tilde  \square \subset  \Om_{k+1}$,  and  Neumann boundary conditions on $\tilde \square$,
\be
\label{teeny}
  \Big[ -\De  + \bar \mu_k  
+ \left[ Q_{k, \bom}^T  \ba  Q_{k, \bom} \right]_{\Om_{k+1}^c}    +a_k  
\left[   Q_k^T   B_{k,r} Q_k \right]_{\Om_{k+1}}  \Big]_{\tilde  \square  }  \geq  \one [   - \De   +  \cO(1)  L^{-2}]_{\tilde  \square  } 
\ee
If      $G_{k, \bom^+,r}(\tilde \square)$  is    the    the inverse  of the operator on the left,   then
\be
  \| G_{k, \bom^+,r}(\tilde \square)  f\|_2,      \    \| \pa  G_{k, \bom^+,r}(\tilde \square)   f\|_2    \  \leq   \  C  \|f\|_{2}
\ee
This can be extended   to   all $\square  \subset    \Om_{k+1}$.     
More  generally    for an    $L^{-(k-j)}M$ cube  in     $\square \subset    \de \Om_j$  we have
\be
  \| G_{k, \bom^+,r}(\tilde \square)  f\|_2,      \   L^{-(k-j) }   \|\pa G_{k, \bom^+,r}(\tilde \square)   f\|_2    \  \leq   \      C L^{-2(k-j)}   \|f\|_{2}
\ee
This improves to the local bound  for  scaled cubes  $\De_y \subset \de \Om_j$ and $\De_{y'} \subset  \de \Om'_j$  and  $|j-j'|\leq 1$
  \begin{equation}  \label{righty2}
  \begin{split}
 \|1_{\De_y} G_{k, \bom^+,r}(\tilde \square)  1_{\De_y'}  f\|_{2}          \  \leq   \  &    C L^{-2(k-j')}
  e^{  -  \ \ga  d_{\bom^+}(y,y' )  } \|f\|_{2}  \\
  L^{-(k-j) }  \|1_{\De_y} \pa   G_{k, \bom^+,r}(\tilde \square)  1_{\De_y'}  f\|_{2}      \  \leq   \  &     C L^{-2(k-j')}
  e^{  -   \ga  d_{\bom^+}(y,y' ) }  \|f\|_{2}  \\
\end{split}
\end{equation}
To see this for   $\square  \subset  \Om_{k+1}$    one shows that
the bound   (\ref{teeny})   still   holds when   the left side is replaced by   $e^{-q\cdot x}[ \cdots  ]  e^{q\cdot x}$  for
   $q  =     \cO( L^{-2})       $.   This  yields a bound  on $ \|1_{\De_y} G_{k, \bom^+,r}(\tilde \square)  1_{\De_y'}  f\|_{2}  $ with a factor   $e^{q\cdot (y-y')}$ and one chooses
$q$ in the direction  $-(y-y')$.    See  the appendix E  of    part I  for more details on this type of argument.

The random walk expansion  is  based  on the  estimates   (\ref{righty2})   as in  theorem  \ref{th}.   Now  estimates are 
all   in  $L^2$ norms     and the bound   
(\ref{security})  follows.  This completes the proof.

\begin{lem}   \label{stem}       {   \  }   
\begin{enumerate}
\item   For  $f:  \Om_{k+1}^{(k)}  \to   \bbR$  
\begin{equation} 
\begin{split} \label{around}
 \Big|  C^{1/2}_{k,  \bom'}f\Big|, \ \   \Big |( C^{1/2}_{k,  \bom'})^{\loc} f\Big|    \leq &   C \|f\|_{\infty}   \\  
  | \de C^{1/2}_{k,  \bom'} f|    \leq &   C    e^{-r_{k+1}}  \|f\|_{\infty}   \\  
   \end{split} 
\end{equation}
  \item      $ C^{1/2}_{k,  \bom'}$   and    $( C^{1/2}_{k,  \bom'})^{\loc}$
are invertible  and  
\be  \label{steaming}
  \Big|  C^{-1/2}_{k,  \bom'}f\Big|, \ \   \Big |\Big[ ( C^{1/2}_{k,  \bom'})^{\loc} \Big]^{-1}f\Big|    \leq   C \|f\|_{\infty}  
 \ee
\end{enumerate}
\end{lem}
\bigskip

\pr
First  define   $ D_{k, \bom',  r}   =  [ Q_k  G_{k, \bom',  r} Q_k^T]_{\Om_{k+1}}
$.  This has the kernel  for  $y,y' \in  \Om_{k+1}^{(k)}$
\be    D_{k, \bom',  r} (y,y') =< Q_k^T \de_y , G_{k, \bom',  r}  Q_k^T \de_{y'} > =  < 1_{\De_y}, G_{k, \bom',  r}  1_{\De_{y'}}>
\ee
where  $\De_y,  \De_{y'}$ are unit cubes.  
 By the lemma  with   $j=j'=k+1$   this    satisfies
\be 
 | D_{k, \bom',  r} (y,y')  |  \leq  C    \|1_{\De_y}\|_2   \|1_{\De_y'}\|_2         e^{  -  \ga d_{\bom'} (y,y' )  }
 =  C        e^{  -  \ga d (y,y' )  }
\ee
and so  $
 | D_{k, \bom',  r}  f |  \leq     C  \|f\|_{\infty}
$.
Combining this with  
$   |A_{k,r}f|  \leq  \cO(1)(1+r)^{-1}   \|f\|_{\infty}   $
we  can estimate
$C_{k, \bom', r} =  A_{k,r}  +   a_k^2  A_{k,r}  D_{k, \bom',  r} A_{k,r}$.  by
$  \label{around3}
 |   C_{k,  \bom', r } f |    \leq     C (1+r)^{-1}   \|f\|_{\infty}   
$.  This gives   $|C^{1/2}_{k, \bom'  }f|
 \leq    C  \|f\|_{\infty} $.

The   same estimates hold for the weakened versions based on    $G_{k, \bom',  r}(s) $.  These are denoted      $D_{k, \bom',  r}(s) $,   $ C_{k,  \bom', r }(s)$,  $C^{1/2}_{k, \bom'  }(s)$   and   satisfy the same bounds.    Specializing  to    
 $ s_{\square^*} =1 ,    s_{(\square^*)^c} =0 $  we  get the same bounds for   $G_{k, \bom',  r}(\square^*) $  and hence for   $D_{k, \bom',  r}(\square^*) $,   $ C_{k,  \bom', r }(\square^*)$     and   $C^{1/2}_{k, \bom'  }(\square^*)$.   The bound   on  $ (C^{1/2}_{k,  \bom'})^{\loc} $   follows 
 from the bound on    $C^{1/2}_{k,  \bom'}(\square^*)$.

 The bound  on  $\de   C^{1/2}_{k,  \bom'}=   C^{1/2}_{k,  \bom'}-( C^{1/2}_{k,  \bom'})^{\loc}$
 follows from    a modification of   lemma  \ref{th2} which says   for  $y,y'   \in \square  \subset  \Om_{k+1}$
\begin{equation}
\|1_{\De_y}\Big( G_{k,  \bom',  r}(\square^*)    - G_{k, \bom',  r}\Big)1_{\De_{y'} } f  \|_2  \leq   Ce^{-r_{k+1}}  e^{   -    \ga  d_{\bom'}(y,y' ) } \|f\|_{2}  
\end{equation}
This is true since in the random    walk expansion for the difference, any path   must  start in  
$\square$,  exit $\square^*$,  and then return to $\square$.    Thus the  minimum   number of steps  is at  approximately   $2[r_{k+1}]$
and  this enables  us to extract a factor     $e^{-r_{k+1}}$.  Running through the above argument   gives a bound 
$ | (C^{1/2}_{k,  \bom'}- C^{1/2}_{k,  \bom'}(\square^*))f|  \leq    Ce^{-r_{k+1}}\|f\|_{\infty}$   which gives the
bound  on    $ \de C^{1/2}_{k,  \bom'}$.   This completes the proof of part  1.

For part  2,
   $ C^{1/2}_{k,  \bom'}$  is invertible since    $ C_{k,  \bom'}$  is invertible  and 
we  can    write 
\begin{equation}
  C^{-1/2}_{k,  \bom'}  =     C^{-1}_{k,  \bom'}   C^{1/2}_{k,  \bom'} 
    =\Big [ \De_{k, \bom(\La_k^*) }  + \frac{a}{L^2}  Q^TQ\Big]_{\Om_{k+1}}   C^{1/2}_{k,  \bom'} 
  \end{equation}
 Restricted to     $\Om_{k+1}$  we have   $ \De_{k, \bom(\La_k^*) }=  a_k + a_k^2Q_k  G_{k,  \bom(\La_k^*)}Q_k^T$.
  Since   $|G_{k,  \bom(\La_k^*)} f |  \leq   C \|f\|_{\infty}$ follows  from  theorem  \ref{th}      we  have   
\begin{equation}
\Big| \Big[ \De_{k, \bom}  + \frac{a}{L^2}  Q^TQ\Big]_{\Om_{k+1}} f \Big|   \leq  C  \|f\|_{\infty}
\end{equation}
Combined     with the bound   on $ C^{1/2}_{k,  \bom'} $   this   yields 
   $ |  C^{-1/2}_{k,  \bom'}f|    \leq   C \|f\|_{\infty}  $

Finally we  write    $(C^{1/2}_{k,  \bom'} )^{\loc}  =   C^{1/2}_{k,  \bom'}   -   \de C^{1/2}_{k,  \bom'}$
and then  the    inverse is realized  as 
\begin{equation}
\Big[ (C^{1/2}_{k,  \bom'} )^{\loc} \Big]^{-1}
 =   C^{-1/2}_{k,  \bom'} \sum_{n=0}^{\infty}      (   \de C^{1/2}_{k,  \bom'}  C^{-1/2}_{k,  \bom'}   )^n
 \end{equation}
The convergence and the bound then follow  from    $|C^{-1/2}_{k,  \bom'}  f|  \leq  C \|f\|_{\infty}$
and    $| \de C^{1/2}_{k,  \bom'}  f|  \leq  e^{-r_{k+1}} \|f\|_{\infty}$,  since we  can assume  $C e^{-r_{k+1}} < \frac12$.
\bigskip

\begin{lem}  \label{r5}
$R^{(5)}_ { \bpi, \Om_{k+1}}$ has a   local expansion  in $LM$ cubes
\begin{equation}
R^{(5)}_ { \bpi, \Om_{k+1}}  = \sum_{\square  \subset  \Om_{k+1}}  R^{(5)}_ { \bpi, \Om_{k+1}}(\square)
 \hs
| R^{(5)}_ { \bpi, \Om_{k+1}}(\square)|  \leq    C (LM)^3 e^{ -\frac12 r_{k+1}  }  
 \end{equation}  
\end{lem}
\bigskip

\pr  $R^{(5)}_ { \bpi, \Om_{k+1}}$   given by  (\ref{onionstew})    has  the local expansion 
\begin{equation}   \label{spiffy}
   R^{(5)}_{\bpi, \Om_{k+1}}(\square)
    =-  \sum_{n=1}^{\infty} \frac{1}{n}\tr \Big( 1_{\square}(    C^{-1/2}_{k,  \bom'}   \de C^{1/2}_{k,  \bom'} )^n\Big)
  =  -  \sum_{n=1}^{\infty} \frac{1}{n}  \sum_{y  \in \square}\Big(   \Big(  C^{-1/2}_{k,  \bom'}   \de C^{1/2}_{k,  \bom'} )^n \Big)  \de_y \Big)(y) 
\end{equation}
From the bounds (\ref{around})  and  (\ref{steaming})   we have
\begin{equation}
\Big|\Big(   \Big(  C^{-1/2}_{k,  \bom'}   \de C^{1/2}_{k,  \bom'} )^n \Big)  \de_y \Big)(y) \Big|
\leq     ( Ce^{-r_{k+1}})^n \|  \de_y  \|_{\infty}  \leq     ( Ce^{-r_{k+1}})^n  
\end{equation}
Summing over  $y \in \square$ gives the factor   $(LM)^3$  and summing over $n$  gives the result.
\bigskip

\begin{lem}  \label{r6}
$R^{(6)}_ { \bpi, \Om_{k+1}}$
has  a    local expansion   in $LM$ cubes
\begin{equation}
R^{(6)}_ { \bpi, \Om_{k+1}}  = \sum_{\square  \subset  \Om_{k+1}}  R^{(6)}_ {k, \bpi, \Om_{k+1}}(\square)
 \hs
| R^{(6)}_ { \bpi, \Om_{k+1}}(\square)|  \leq   C (LM)^3 
 \end{equation}  
 If   $\square  \subset \Om_{k+1}-  \Om_{k+1}^{\nat}$  then    
 the bound improves  to  
\be  \label{improved}
 | R^{(6)}_ { \bpi, \Om_{k+1}}(\square)|  \leq   C (LM)^3 e^{ - r_{k+1}  }  
\ee 
\end{lem}
\bigskip

\pr    $R^{(6)}_ { \bpi, \Om_{k+1}}$ is  given in  (\ref{stay3}).  With    $ D_{k, \bom',  r}   =   [Q_k  G_{k, \bom',  r} Q_k^T]_{\Om_{k+1}}$  
and    $ D_{k,  r}  =  [ Q_k  G_{k,   r} Q_k^T]_{\Om_{k+1}}$     the  expansion holds with   
\begin{equation}
R^{(6)}_ { \bpi, \Om_{k+1}}(\square)= -  \frac 12   a_k^2  \int_0^{\infty}
  \sum_{y \in \square}   \Big(A_{k,r}  (D_{k,\bom',  r}-  D_{k,r}) A_{k,r}  \Big)(y,y)  dr  
\end{equation}
Since  $\|A_{k,r}  \de_y\|_2   \leq    \cO(1)(1+r)^{-1}$   and     $| D_{k, \bom',  r} (y,y') - D_{k, r} (y,y')|
 \leq     C        e^{  -  \ga d (y,y' )  }$ we have 
 \be
   \Big|\Big<A_{k,r} \de_y, \Big(D_{k,\bom',  r}-  D_{k,r}\Big) A_{k,r}\de_{y}  \Big> \Big|
 \leq C  \| A_{k,r} \de_y\|^2_2         \leq  C         \frac{1}{(1+r)^2} 
 \ee
 This gives   the announced
 \begin{equation}
R^{(6)}_ { \bpi, \Om_{k+1}}(\square)\leq   C(LM)^3       \int_0^{\infty}  \frac{1}{(1+r)^2}   \leq    C (LM)^3  
\end{equation}

If    $\square  \subset  \Om_{k+1}^{\nat}$  then   for  $y,y'   \in \square$
 \begin{equation}
\|1_{\De_y} \Big(  G_{k, \bom',  r}- G_{k,  r}   \Big)1_{\De_{y'}}   f  \|_2 
 \leq   Ce^{-r_{k+1}} e^{ - \ga d_{\bom}(y,y' ) }   \|f\|_2 
\end{equation}
This holds  since  in the random walk expansion  for   $1_{\De_y} \Big(  G_{k, \bom',  r}- G_{k,  r}   \Big)1_{\De_{y'}} $  any  path  must start in  $\Om_{k+1}^{\nat}$, pass through   $\Om_{k+1}^c$,   and then  return to $\Om_{k+1}^{\nat}$.   But  $\Om_{k+1}^{\nat}$ and
$\Om^c_{k+1}$  are separated by  $[r_{k+1}]$ layers of  $LM$ cubes.      Thus the minimum  number of 
steps is approximately $2[r_{k+1}]$,  which allows us to extract  a factor   $e^{-r_{k+1}}$.
 This gives an extra factor  of  $e^{-r_{k+1}}$  in      the estimate on   $| D_{k, \bom',  r} (y,y') - D_{k, r} (y,y')|$  and hence in the result.

Similar estimates show that  $b_k'$  is bounded.

\subsection{new  characteristic  functions}

In $\Om_{k+1}$  the    characteristic functions   are  limitations on the field  
  $  \Psi^{\loc}_{k,\Om_{k+1}}(\bom') + ( C^{1/2}_{k,  \bom'} )^{\loc}W_k $   and hence  on  the background field    $\phi^0_{k+1, \bom'}(\square^*)$  and the fluctuation field $W_k$.  This structure is awkward and we  replace it  by  separate cleaner  characteristic functions    for  the background and  fluctuation.

Anticipating that the term       $\exp(  -S^{*,0}_{k+1}(\La_k)) $   in (\ref{representation5})    suppresses  large fields,  
we   introduce    for  each    $LM$  cube  $\square$  well inside  $\Om_{k+1}$, the characteristic function    $ \chi^0_{k+1}( \square) $
enforcing   the  inequalities  in  $\tilde \square$:  
\begin{equation}  \label{not}
\begin{split}
 | \Phi_{k+1}  -  Q_{k+1}    \phi^0_{k+1, \bom^+(     \square )}|  \leq &  p_{k+1}L^{-\frac12} \\
 |\pa   \phi^0_{k+1, \bom^+(   \square )}|  \leq &    p_{k+1}  L^{-\frac32} \\   
 |  \phi^0_{k+1, \bom^+(   \square )}|  \leq &   p_{k+1} \al_{k+1}^{-1}  L^{-\frac12} \\
\end{split}
\end{equation} 
Here $\bom^+(   \square )$ has  $k+1$ layers   and 
 $ \phi^0_{k+1, \bom^+(   \square )}= \phi^0_{k+1, \bom^+(   \square )}(\tilde Q^T_{\bbT^1, \bom^+(\square)} \Phi_{k+1})$ as defined in (\ref{eddie}).  Call functions  satisfying   (\ref{not})   $\cS^0_{k+1}(  \square)$,   since when  we  scale   later on  these will   become the   inequalities 
defining  
 $\cS_{k+1}(  \square)$,   and    $ \chi^0_{k+1}( \square) $ will scale to    $ \chi_{k+1}( \square) $.

Now  we  write  with   $ \zeta^0_{k+1}( \square) = 1-   \chi^0_{k+1}( \square) $ 
\begin{equation}
\begin{split}
1 =&   \prod_{\square  \subset   \Om^{\nat}_{k+1}  } \zeta^0_{k+1}( \square) +  \chi^0_{k+1}( \square) \\
= &   \sum_{Q_{k+1}  \subset  \Om^{\nat}_{k+1} }\prod_{\square  \subset  Q_{k+1}} \zeta^0_{k+1}( \square) 
\prod_{\square  \subset  \Om_{k+1}^{\nat} - Q_{k+1} }  \chi^0_{k+1} (\square)  \\
\equiv &   \sum_{Q_{k+1}  \subset  \Om^{\nat}_{k+1} } \zeta^0_{k+1}( Q_{k+1})    \chi^0_{k+1} (\Om^{\nat}_{k+1} - Q_{k+1})  \\
\end{split}
\end{equation}

Anticipating that the factor       $\exp(  - 1/2  \|W_k\|^2_{\Om_{k+1} } )$ in  (\ref{representation5})  suppresses  large  $W_k$    we  define 
  \begin{equation} 
   \chi^w_{k}  ( \square,   W_k )  =
\prod_{x \in  \square  \cap  \bbT^0_{\sM + \sN -k}}  
\chi \Big ( |W_k(x)|  \leq  p_{0,k} \Big)     
\end{equation} 
for  $p_{0,k}  \leq  p_k$.
Then we write  with   $ \zeta^w_{k} (\square)  =   1-   \chi^w_{k}  ( \square) $
\begin{equation}
\begin{split}
1  =  &  \prod_{ \square  \subset  \Om_{k+1}}  \zeta^w_{k} (\square)  + \chi^w_{k}  ( \square)  \\
=  &   \sum_{R_{k+1}  \subset    \Om_{k+1}  }   
\prod_{ \square  \subset   R_{k+1}  }  \zeta^w_{k} (\square)
\prod_{ \square  \subset   \Om_{k+1}  -R_{k+1}   }    \chi^w_{k} (\square)\\
\equiv  &    \sum_{R_{k+1}  \subset    \Om_{k+1} }    
  \zeta^w_{k}( R_{k+1} )   \chi^w_{k}(    \Om_{k+1} -R_{k+1} )\\
 \end{split}
 \end{equation}

  Now  we   have 
   \begin{equation}   \label{breakup}
1=   \sum_{Q_{k+1},R_{k+1} }  
\zeta^0_{k+1}( Q_{k+1} ) 
 \zeta^w_{k}(  R_{k+1}  )   \chi^0_{k+1} ( \Om^{\nat}_{k+1} - Q_{k+1}  )   \chi^w_{k}(    \Om_{k+1}-R_{k+1}  )  
\end{equation}
The new large field regions   $Q_{k+1}, R_{k+1}$   generate  a new small field region   $\La_{k+1}$, also  a union of  $LM$  cubes,   defined by  
\begin{equation}
 \La_{k+1}  =  \Om_{k+1}^{5\nat }  -(  Q^{5*}_{k+1}  \cup  R^{5*}_{k+1})  \hs \textrm{ or  }  \hs
 \La^c_{k+1}  =  (\Om_{k+1}^c)^{5*}  \cup   Q^{5*}_{k+1}  \cup  R^{5*}_{k+1} 
\end{equation}
We  write   $Q_{k+1}, R_{k+1}  \to   \La_{k+1}$   and classify the terms  in  (\ref{breakup})  by  the  $\La_{k+1}$ that they  generate.
Thus we have  
  \begin{equation}  \label{breakup2}
1=   \sum_{\La_{k+1}   \subset  \Om^{5\nat }_{k+1} } \cC_{k+1}(\Om_{k+1},\La_{k+1})
\end{equation}
where
  \begin{equation}  \label{breakup3}
 \cC_{k+1}(\Om_{k+1},\La_{k+1})= 
   \sum_{         Q_{k+1},R_{k+1} \to \La_{k+1}   }
\zeta^0_{k+1}( Q_{k+1} ) 
 \zeta^w_{k}(  R_{k+1}  )   \chi^0_{k+1} ( \Om^{\nat}_{k+1} - Q_{k+1}  )   \chi^w_{k}(    \Om_{k+1}-R_{k+1}  )  
\end{equation}
The sum is still restricted by  $P_{k+1} \subset  \Om_{k+1}^{\nat}$  and   $Q_{k+1} \subset  \Om_{k+1}$.  

Note  that   $ \cC_{k+1}(\Om_{k+1},\La_{k+1})$  enforces that   the bounds (\ref{not})  and   $ |W_k|  \leq  p_{0,k} $
hold  on  $\La_{k+1}^{4*}$.    To see this  it suffices  to show that every term in the sum  (\ref{breakup3})  has  this 
property.   But  the term with    $ Q_{k+1},R_{k+1}$  enforces  the bounds  (\ref{not})   on   $\Om^{\nat}_{k+1}-Q_{k+1}$
and the  bound    $ |W_k|  \leq  p_{0,k} $  on   $\Om_{k+1}-R_{k+1}$.   Since both these sets contain  $\La_{k+1}^{4*}$
we have the result.

We  insert  (\ref{breakup2})  under the integral sign   in  (\ref{representation5}).
The sum is now over          
 \begin{equation}
\bpi^+=  (\bpi, \Om_{k+1},  \La_{k+1} )=  (  \Om_1, \La_1,  \dots,   \Om_{k+1}, \La_{k+1})
\end{equation}
and
   \begin{equation}     \label{representation6}
\begin{split}
\tilde    \rho_{k+1}( \Phi_{k+1}) 
 =  &    Z^0_{k+1}   \sum_{\bpi^+}    \int   d \Phi^0_{k+1, \bom^{+,c} }d W_{k,\bpi}\  K_{k, \bpi} \  \cC_{k, \bpi}  \cC^q_{k} (\La_k, \Om_{k+1} )  
 \cC_{k+1}(\Om_{k+1},\La_{k+1}) \\
    &    
\         \exp \Big(-  S^{*,0}_{k+1}(\La_k) \Big)   \exp  \Big(  c_{k+1}  |\Om^{c,(k)}_{k+1}|    \Big)   \\
&   \int    d \mu_{ \Om_{k+1} }(W_k)  
  \chi_k( \La_k) \   \chi^q_{k}(  \Om_{k+1}  ) 
     \exp\Big( E^+_k(\La_k)    + R^{(\leq 7)}_{\bpi,  \Om_{k+1}}  +  B_{k, \bpi}(\La_k)  \Big) \\
   \end{split}
\end{equation}

  \subsection{more   estimates }  
We collect some estimates for future use.

\subsubsection{bounds on the fluctuation field}
The characteristic  functions can now be written   
\begin{equation}  \label{pinky0}
\Big[ \cC_{k, \bpi}\   \cC^q_{k} (\La_k, \Om_{k+1} )   \chi_k ( \La_k  )\    \chi^q_{k}(  \Om_{k+1}  )  \ \Big] \cC_{k+1}(\Om_{k+1},\La_{k+1})
\end{equation}
The  bracketed   characteristic functions  still  enforce the bounds of  lemma  \ref{suddsy},   except that  the bound  (\ref{gb1.5})   on  $\Phi_k$
now only  holds  on  $\tilde  \La_k - \Om_{k+1}$  since  $\Phi_k$ on  $\Om_{k+1}$  was relabeled.    The relabeling does not affect
any of the remaining bounds.    In a slightly weaker form we  can summarize them  as
  \begin{eqnarray}   
|\Phi_k|  \leq      C p_{k} \al_{k}^{-1} \hs   &  |\pa   \Phi_k|  \leq    C p_{k}    \ \  \hs &  \textrm{  on  } \ \    \Om_k- \Om_{k+1}
 \label{go1}  \\
  | \Phi_{k+1}|      \leq        3 p_k \al_k^{-1} \hs      & |\pa  \Phi_{k+1}|      \leq   4 p_k     \hs   &  \textrm{  on  } \  \Om_{k+1}  
  \label{go2}   
 \end{eqnarray}
 as  well  as   
 \be    
  | \Phi^\#_{k+1}|      \leq      C p_k \al_k^{-1} \hs     |\pa  \Phi^\#_{k+1}|      \leq   C p_k      \hs      \textrm{  on  } \  \Om_{k}  
 \ee

Our immediate goal is to get a bound on $W_k$  on its whole domain $\Om_{k+1}$.  We start with:
\newpage

\begin{lem}  {  \  }    \label{singsong2}
The bracketed characteristic functions  in (\ref{pinky0})
enforce the following inequalities:    
\begin{enumerate}
\item   For an $LM$-cube  $\square \subset  \Om_{k+1}$
\begin{equation}  \label{not1}
\begin{split}
 | \Phi_{k+1}  -  Q_{k+1}    \phi^0_{k+1, \bom'}|  \leq & C p_k    \hs       \  \ \   \textrm{  on  } \  \tilde \square  \cap  \Om_{k+1}  \\
 |\pa   \phi^0_{k+1, \bom'}|  \leq &  Cp_k   \hs   \    \  \ \   \textrm{  on  } \   \tilde \square  \\   
 |  \phi^0_{k+1, \bom'}|  \leq &   C p_k \al_k^{-1 }  \hs  \textrm{  on  } \   \tilde \square   \\
\end{split}
\end{equation} 
The same bounds hold for    $  \phi^0_{k, \bom'}(s),   \phi^0_{k, \bom'}(\square^*)$   and  
\begin{equation}  \label{skunky2}
\Big| \phi^0_{k+1,\bom'} ( \square^* )   -   \phi^0_{k+1,\bom'} \Big|
\leq        e^{-r_{k+1}} \   
\end{equation}
\item 
 For  $\square \subset  \Om_{k+1}$ we have  on  $\tilde \square \cap  \Om_{k+1}$
\begin{equation}   \label{swish}
 | \Psi_{k,\Om_{k+1}}(\bom')|  \leq     C p_k  \al_k^{-1}  \hs     |\pa   \Psi_{k,\Om_{k+1}}(\bom')|  \leq     C p_k 
\end{equation}
The same holds  for   $ \Psi_{k,\Om_{k+1}}(\bom',s), \Psi_{k,\Om_{k+1}}(\bom', \square^*)$
   and    $ \Psi^{\loc}_{k,\Om_{k+1}}(\bom')$    and     
\be   \label{salt}
| \de \Psi_{k, \Om_{k+1}}(\bom')|  = |  \Psi^{\loc}_{k,\Om_{k+1}}(\bom')-   \Psi_{k,\Om_{k+1}}(\bom') |   \leq       e^{-r_{k+1}} 
 \ee
 \end{enumerate}
\end{lem}
\bigskip

\pr
Instead of considering   $\phi^0_{k+1, \bom'} ( \hat  \Phi_{k+1, \bom'} ) $   with 
  $  \hat  \Phi_{k+1, \bom' }  = (\tilde   Q_{ \bbT^0,  \bom(\La_k^*)} ^T    \Phi_{k,\de \Om_k},  \Phi_{k+1,\Om_{k+1} } )  $   we  start  with 
    $\phi^0_{k+1, \bom'} (   \Phi^{\star}_{k+1,  \bom' } ) $  
where 
\be      \Phi^{\star}_{k+1,  \bom' }      = \tilde   Q_{ \bbT^1,  \bom'} ^T   (Q \Phi_{k, \de \Om_k},  \Phi_{k+1,\Om_{k+1} } ) =  \tilde    Q_{ \bbT^1,  \bom'} ^T    \Phi^\#_{k+1}           \ee
We  claim that the bounds    (\ref{not1})   hold  for     $\phi^0_{k+1, \bom'} (   \Phi^{\star}_{k+1  \bom' } ) $.    This  follows from the bounds on
$\Phi_{k+1}^\#$  by an  argument  very  similar to   lemma \ref{threetwo}.    The differences are that we are considering   the pre-scaled field 
$\phi^0_{k+1, \bom'} $   and  we  need the bounds  on the basic  fields on the larger  set   $\Om'_1 =  \Om_1(\La_k^*)  \subset  \Om_k$.

 Next     note that  since   $ Q_{ \bbT^1,  \bom'}  $ is the identity  on
$\Om_{k+1}$   and   since  it  is  $ Q Q_{ \bbT^0, \bom(\La_k^*)} $  on  $\de  \Om_k$   we  have  
\be    \hat  \Phi_{k+1,   \bom' }   -    \Phi^{\star}_{k+1,  \bom' } 
 =     \Big( \tilde  Q^T_{ \bbT^0, \bom(\La_k^*)}   ( I - Q^TQ ) \Phi_{k,\de \Om_k},  0 \Big) \ee
 The expression  $ ( I - Q^TQ ) \Phi_{k,\de \Om_k} $    can be written as a function of   $\pa   \Phi_{k,\de \Om_k}$
 Using the bound on  this  function it    is  bounded  by   $C p_k$. Therefore we  have    by   (\ref{dos})
\be
\begin{split}
|\phi^0_{k+1, \bom'} (  \hat \Phi_{k+1,  \bom' } ) -   \phi^0_{k+1, \bom'} (   \Phi^{\star}_{k+1,   \bom' } )  |  \leq    &
C p_k  \\
|\pa \phi^0_{k+1, \bom'} (  \hat \Phi_{k+1,  \bom' } ) - \pa   \phi^0_{k+1, \bom'} (   \Phi^{\star}_{k+1, \bom' } )  |  \leq    &
C p_k  \\
\end{split}
\ee
Hence the bounds    (\ref{not1})  also hold  for  $ \phi^0_{k+1, \bom'} ( \hat  \Phi_{k+1,  \bom' } )$  as  claimed.

This argument holds equally well for  the weaked version     $  \phi^0_{k,   \bom'}(s)$  and  $  \phi^0_{k,    \bom'}(\square^*)$   is a special case..
To bound the difference   $\phi^0_{k+1,\bom'} ( \square^* )   -   \phi^0_{k+1,\bom'}$  we  claim that    for  $y,y' \in   \tilde  \square$
\begin{equation}
|1_{\De_y} \Big(  G^0_{k+1, \bom'}- G^0_{k+1, \bom'}(\square^*)   \Big)1_{\De_{y'}}   f  | 
 \leq   Ce^{-r_{k+1}} e^{ - \ga d_{\bom'}(y,y' ) }   \|f\|_{\infty} 
\end{equation}
This is a variation of   (\ref{lullaby})  with the following observation.  In the random walk expansion for this expression we  start and end   in  $\tilde \square$,  and   only terms  which   
which  exit  $\square^*$  contribute.    These  paths   must  have  at least  $[r_{k+1}]$ steps.  Hence we can extract a factor 
 $M^{-\frac 14[r_{k+1}]}  \leq  e^{-2r_{k+1}}$
when estimating the expansion.  With this  modification the bound (\ref{skunky2})  follows  as  in  (\ref{dos});   here we use  
 $ e^{-r_{k+1}}Cp_k \al_k^{-1}  \leq  1$.

The minimizer can be written
 \begin{equation}    \label{ding}
  \Psi_{k,\Om_{k+1}}(\bom')  =   Q_{k}  \phi^0_{k+1,\bom'}
+ \frac{aL^{-2}}{a_k+ aL^{-2}}   Q^T \Big(  \Phi_{k+1}  -   Q_{k+1} \phi^0_{k+1,\bom'} \Big)
\end{equation} 
We  claim that  on $\tilde \square \cap  \Om_{k+1}$
\begin{equation}
 | \Psi_{k,\Om_{k+1}}(\bom')|  \leq     C p_k  \al_k^{-1}  \hs     |\pa   \Psi_{k,\Om_{k+1}}(\bom')|  \leq     C p_k 
\end{equation}
These follow more or less directly from the bounds   (\ref{not1}).   The bound   $| Q^T \Big(  \Phi_{k+1}  -   Q_{k+1} \phi^0_{k+1,\bom'} \Big)|
\leq    C p_k$,  implies a bound  of the same form on the derivative,   since  it is defined  on a unit lattice.
The same argument gives bounds on      $\Psi_{k,\Om_{k+1}}(\bom',s)$  and   $ \Psi_{k,\Om_{k+1}}(\bom',  \square^* )$.
The bound    $|\Psi^{\loc}_{k,\Om_{k+1}}(\bom')|  \leq   C p_k  \al_k^{-1}  $  follows as well.

We  also   have  by (\ref{skunky2})    on $\tilde \square \cap  \Om_{k+1}$
 \begin{equation}  \label{tiny0}
 \begin{split}
&   |   \Psi_{k,\Om_{k+1}}(\bom',  \square^*)-\Psi_{k,\Om_{k+1}}(\bom')| \\
  =& \Big | Q_{k} \Big( \phi^0_{k+1,\bom'}(\square^*) - \phi^0_{k+1,\bom'}\Big) 
-  \frac{aL^{-2}}{a_k+ aL^{-2}}   Q^T  Q_{k+1}  \Big( \phi^0_{k+1,\bom'}(\square^*) - \phi^0_{k+1,\bom'}\Big)\Big |
\leq      e^{-r_{k+1}}  \\
\end{split}
\end{equation}
This implies the bound on   $ \Psi^{\loc}_{k,\Om_{k+1}}(\bom')-   \Psi_{k,\Om_{k+1}}(\bom') $

Finally we  need the bound $|\pa  \Psi^{\loc}_{k,\Om_{k+1}}(\bom')|  \leq   C p_k  $.    There is a potential problem here when the 
derivative crosses the boundary of a cube  $\square$.    Suppose $\square_1, \square_2$ are adjacent cubes and $x \in \square_1  \cap  \Om_{k+1}^{(k)}$ and  $x + e_{\mu}  
 \in \square_2  \cap  \Om_{k+1}^{(k)}$.  We  need  to show     $   | \Psi_{k,\Om_{k+1}}(\bom',  \square^*_2,x+ e_{\mu})  -
 \Psi_{k,\Om_{k+1}}(\bom',  \square^*_1,x)|  \leq  Cp_k$.     However just as  in   (\ref{tiny0})  one can show  
$| \Psi_{k,\Om_{k+1}}(\bom',  \square^*_1,x)- \Psi_{k,\Om_{k+1}}(\bom',  \square^*_2,x)|  \leq   e^{-r_{k+1}}  $.  This  
reduces the estimate  to     $   |\pa  \Psi_{k,\Om_{k+1}}(\bom',  \square^*_2,x)  |  \leq  Cp_k$,  which we know since we  control 
this derivative on $\tilde  \square_2$.

\bigskip

 \begin{lem}  {  \  }    \label{singsong1}
  The   bracketed characteristic functions in  (\ref{pinky0})
enforce the  inequality     on  $\Om_{k+1}$
  \begin{equation}  \label{gb3}
    |W_k |  \leq    C p_{k}
     \end{equation}
\end{lem}
\bigskip

\re  A bound  with  $ C p_{k}\al_k^{-1} $ is easy.  The issue is to eliminate the  $\al_k^{-1}$.
\bigskip

\pr    
  First    note  that      $ \chi^q_{k}(  \Om_{k+1}  )$   is  now  saying that  on $\Om_{k+1}$
  \be
 \Big | \Phi_{k+1} -   Q \Big ( \Psi^{\loc}_{k, \Om_{k+1}} (\bom')   +( C^{1/2}_{k,  \bom'} )^{\loc} W_k \Big) \Big |  \leq    p_k
  \ee
 By   (\ref{salt}) this implies
   \be
  \Big  | \Phi_{k+1} -   Q \Big ( \Psi_{k, \Om_{k+1}} (\bom')   +( C^{1/2}_{k,  \bom'} )^{\loc} W_k \Big) \Big |  \leq   2 p_k
  \ee
  However  (\ref{ding}) can be rearranged to say 
  \be    \label{ding2}
      \Phi_{k+1} -   Q  \Psi_{k, \Om_{k+1}} (\bom')   =  \frac{a_k}{a_k + a L^{-2}} (\Phi_{k+1} - Q_{k+1}\phi^0_{k+1, \bom'})
    \ee
 This    is bounded  by  $Cp_k$  by  (\ref{not1}).  
  Thus we have    on $ \Om_{k+1} $
  \be     |Q (C^{1/2}_{k,  \bom'} )^{\loc} W_k|  \leq   C p_k  \label{mars}
  \ee

 We   supplement this with a bound on  $  \pa (C^{1/2}_{k,  \bom'} )^{\loc} W_k $.
 For this we look to the bounds  of   $\chi_k(\Om_{k+1})$  which say that   
$ (\Phi_{k, \de \Om_k},  \Psi^{\loc}_{k,\Om_{k+1}}(\bom') + ( C^{1/2}_{k,  \bom'} )^{\loc}W) $
 is  in    $\cS_k( \square)$  for any  $\square \subset  \Om_{k+1}$.   By lemma  \ref{threetwo}   this  implies that 
 on  $\tilde  \square \cap  \Om_{k+1}$
 \be
  |\Psi^{\loc}_{k,\Om_{k+1}}(\bom') + ( C^{1/2}_{k,  \bom'} )^{\loc}W | \leq   2 p_k \al_k^{-1}
 \hs       |\pa\Big(\Psi^{\loc}_{k,\Om_{k+1}}(\bom') + ( C^{1/2}_{k,  \bom'} )^{\loc}W \Big) | \leq  3p_k 
\ee
By the previous    lemma  $\Psi^{\loc}_{k,\Om_{k+1}}(\bom') $  alone satisfies these bounds  with  constants   $C p_k \al_k^{-1},
 Cp_k$.   Therefore   on  $\Om_{k+1}$
 \begin{equation}
\label{snickers}
   |(C^{1/2}_{k,  \bom'} )^{\loc} W_k|  \leq  Cp_k \al_k^{-1}   \hs   |\pa (C^{1/2}_{k,  \bom'} )^{\loc} W_k|  \leq  Cp_k
\end{equation}

Now  by   (\ref{mars}) and the second bound in  (\ref{snickers})  we  have
\be
 |(C^{1/2}_{k,  \bom'} )^{\loc} W_k|   \leq  
      |  (C^{1/2}_{k,  \bom'} )^{\loc} W_k  - Q(C^{1/2}_{k,  \bom'} )^{\loc} W_k| 
+  | Q (C^{1/2}_{k,  \bom'} )^{\loc} W_k|  
\leq     C  p_k
\ee
Finally     the bound on  $W_k$   follows     from  this result and (\ref{steaming}) which gives
\begin{equation}
|W_k|     = \Big| \Big[ (C^{1/2}_{k,  \bom'} )^{\loc} \Big]^{-1}(C^{1/2}_{k,  \bom'} )^{\loc}   W_k\Big|
\leq   C  \|   (C^{1/2}_{k,  \bom'} )^{\loc}  W_k \|_{\infty}  \leq       C p_k
\end{equation}
This completes the proof.
\bigskip

\subsubsection{redundant characteristic functions}
Now write the characteristic functions as   
 \begin{equation}  \label{pinky1}
\Big[ \cC_{k, \bpi}\   \cC^q_{k} (\La_k, \Om_{k+1} )   \chi_k ( \La_k-  \La^{**}_{k+1}  )\    \chi^q_{k}(  \Om_{k+1} - \La^{**}_{k+1} ) 
 \cC_{k+1}(\Om_{k+1},\La_{k+1}) \Big]  \chi_k (  \La^{**}_{k+1}  )\    \chi^q_{k}(  \La^{**}_{k+1} ) 
\end{equation}
The  bracketed characteristic functions here   still enforce the bounds   (\ref{go1}), (\ref{go2}).    Indeed the  bounds on  $\Phi_k$
on  $\de \Om_k$
and  $\Phi_{k+1}$  on    $\Om_{k+1} -  \La^{**}_{k+1}$   are unaffected by the loss of    $\chi_k (  \La^{**}_{k+1}  )\    \chi^q_{k}(  \La^{**}_{k+1} ) $,     and    the new   $ \cC_{k+1}(\Om_{k+1},\La_{k+1})$  supplies  the even stronger  bounds  on $\La_{k+1}^{4*}$:
\be  \label{slumber}
  | \Phi_{k+1} |  \leq  2  p_{k+1} \al_{k+1}^{-1}L^{-\frac12}
\hs      |\pa   \Phi_{k+1} |  \leq  3  p_{k+1} L^{-\frac32}
\ee

 We  want to show that the     last  two characteristic functions in (\ref{pinky1})  are     redundant.  
 But first we show that the remaining  terms  in the    bracket   have no dependence  on $W_k$    in  $\La_{k+1}$    This is  an  important   simplification for our fluctuation integral.

   \begin{lem} 
$ \chi_k ( \La_k-  \La^{**}_{k+1}  )\    \chi^q_{k}(  \Om_{k+1} - \La^{**}_{k+1} )  $
does not depend on  $W_k$ in  $\La_{k+1}$.
\end{lem}
\bigskip

\pr   The Green's function  $G_{k, \bom',r}(\square^*)$  only connects points in $\square^*$,   thus   $G_{k, \bom',r}(\square^*)f$  on  $\square$
only depends on $f$  on  $\square^*$.   Hence   $C^{\frac12}_{k, \bom'}(\square^*) f$  on  $\square$ only depends on  $f$ in  $\square^*$.
It follows that   $(C^{1/2}_{k,  \bom'} )^{\loc}f   $  on   a set  $X$  only depends  on   $f$  in  $X^*$.

Now   
 $ \chi^q_{k}(  \Om_{k+1} - \La^{**}_{k+1} ) $   depends on  $(C^{1/2}_{k,  \bom'} )^{\loc}W_k$  in    $( \La^{**}_{k+1})^c $
 and so  on $W_k$  in    $(( \La^{**}_{k+1})^c)^*  \subset  ( \La_{k+1}^*)^c  \subset    \La_{k+1}^c$.
The function     $\chi_k ( \La_k-  \La^{**}_{k+1}  )$   also   depends  on  $(C^{1/2}_{k,  \bom'} )^{\loc}W_k$  in    $( \La^{**}_{k+1})^c $,
but  also  on  $\phi_{k,  \bom(\square)} (  (C^{1/2}_{k,  \bom'} )^{\loc}W_k )$  on $\tilde \square$     for  $\square  \subset     \La_k-  \La^{**}_{k+1} $.
By  a similar argument this depends  on  $W_k$  on  $(( \La^{**}_{k+1})^c)^{**}  \subset  \La_{k+1}^c $.  This completes the proof.
\bigskip

Next  a preliminary  result:

\begin{lem}  \label{replace} The bracketed   characteristic functions  in  (\ref{pinky1})  enforce the following inequalities
\begin{enumerate}
\item   For    an $LM$-cube   $\square$ in   $\Om^{\nat}_{k+1}$ we have   on $\tilde \square$  
 \be  \label{note7}
|  \phi^0_{k+1,\bom^+( \square)}- \phi^0_{k+1, \bom'}  |,    \ 
|\pa   \phi^0_{k+1,\bom^+( \square)}-  \pa  \phi^0_{k+1, \bom'} |   \leq   CM^{-\frac12}  p_{k+1}   
 \ee  
\item  For an  $M$ cubes  $\square$  in  $\Om^{\nat}_{k+1}$   let    $ \phi^{\min}_{k,\bom(\square)} =   \phi_{k,  \bom(\square)}(\tilde Q^T_{\bbT_0, \bom(\square)}\Psi_{k, \Om_{k+1}}(\bom')  )$ 
(i.e. we replace  $\Phi_{k, \Om_{k+1}}$ with the minimizer). 
 Then   on $\tilde \square$  
\be  \label{note6}
|  \phi^{\min}_{k,\bom(\square)}- \phi_{k, \bom(\La_k^*)}(\hat   \Psi_{k, \bom'} ) |,   \   
|\pa  \phi^{\min}_{k,\bom(\square)}-  \pa  \phi_{k,  \bom(\La_k^*)}(\hat   \Psi_{k, \bom'} )  |   \leq   CM^{-\frac12}  p_{k}   
 \ee  
 \end{enumerate}
\end{lem}
\bigskip

\pr        Straightforward estimates  would give an  unwanted factor  $\al_{k+1}^{-1}$
so  we  must   take another path. 
  First  we  note that  we can replace   $ \phi^0_{k+1, \bom'} =    \phi^0_{k+1, \bom'}(  \hat  \Phi_{k+1, \bom'})$  by
   $ \phi^0_{k+1, \bom'} =    \phi^0_{k+1, \bom'}( 0,  \Phi_{k+1})$.   This  is so since the difference  
$ \hat  \Phi_{k+1, \bom'}-    ( 0,  \Phi_{k+1, \Om_{k+1}})$  is localized  in   $\Om^c_{k+1}$ and so   is  $[r_{k+1}]$ $LM$-cubes away from
  $\Om^{\nat}_{k+1}$.   Thus a straightforward estimate  generates   a factor  $e^{-r_{k+1}}$  which is enough to dominate   
   $\al_{k+1}^{-1}  \leq   \la_{k+1}^{- \frac14}$  and   $M^{\frac12}$.

 Next we   use the representation developed in  lemma  \ref{threeone},
 now in the prescaled version.     Let   $\De_y$ be an  $L$-cube in  $\tilde \square$.   Then   for   $x$ in a neighborhood of    $ \De_y $ 
 as in  (\ref{samsum}): 
 \be     \label{samsum5}
\begin{split}
\Big[  \phi^0_{k+1, \bom^+(\square)} \Big(\tilde   Q^T_{ \bbT^1,    \bom^+(\square)}   \Phi_{k+1}\Big) \Big](x)   
 = & \Big[  \phi^0_{k+1, \bom^+(\square)} \Big(\tilde   Q^T_{ \bbT^1,    \bom^+(\square)}(\Phi_{k+1} -  \Phi_{k+1}(y)  ) \Big) \Big](x)   \\
  +  &   \Big[    1  -  \bar \mu_k   G^0_{k+1, \bom^+(\square)}  \cdot  1 \Big](x)   \Phi_{k+1}(y) \\
\Big[  \phi^0_{k+1, \bom'} \Big(  0,  \Phi_{k+1}\Big) \Big](x)   
 = & \Big[  \phi^0_{k+1, \bom'} \Big(0,  \Phi_{k+1} -  \Phi_{k+1}(y)   \Big) \Big](x)   \\
   +  &   \Big[    1  -  \bar \mu_k   G^0_{k+1, \bom'}  \cdot (0, 1_{\Om_{k+1}}) \Big](x)   \Phi_{k+1}(y) \\   
   \end{split}
\ee
As in lemma \ref{threeone}     we can use the bound (\ref{slumber})   on $\pa \Phi_{k+1}$  to estimate the   the first term  in (\ref{samsum5})
 by    
\be 
\begin{split}
  & \sum_{y'}    \Big| \phi^0_{k+1, \bom^+(\square)} \Big( 1_{\De_{y'}}\tilde   Q^T_{ \bbT^1,    \bom^+(\square)}(\Phi_{k+1} -  \Phi_{k+1}(y)   \Big)(x) \Big|
\\  
 \leq  & \sum_{y'}   C     e^{-\frac14 \ga_0d_{\bom^+(\square)}(y,y') } \|1_{\De_{y'}} Q^T_{ \bbT^1,    \bom^+(\square)}(\Phi_{k+1} -  \Phi_{k+1}(y)  )  \|_{\infty}  \\
    \leq  & \sum_{y'}   C     e^{-\frac 14   \ga_0d_{\bom^+(\square)}(y,y') }(d(y,y') +1)   p_{k+1}       
  \leq       C   p_{k+1} \\
\end{split}
\ee
We  need a smaller constant here.
This occurs    in terms  coming  from  $\De_{y'}$  not in  $ \square^{\sim 2}$.
This is because in the    random walk expansion  the  Green's functions  must have at least one step to get  from $\tilde \square$ to  $\square^{\sim 2}$.   This enables us to extract  a factor  $M^{-\frac12} $  without spoiling the convergence of the expansion.    Thus we get the bounds $ CM^{-\frac12}  p_{k+1}  $.
The  same works for an estimate on  these distant    terms   in  $\phi^0_{k+1, \bom'}$.

Now consider  terms with   $\De_y  \subset   \square^{\sim 2}$.    
In this case  the difference of the two  first    terms in  (\ref{samsum5})   is
\be  \sum_{y' \subset  \square^{\sim 2}}\Big[ \Big(  G^0_{k+1, \bom^+(\square)} - G^0_{k+1, \bom'} \Big)
 Q^T_{k+1} L^{-2}a_{k+1}    \Big(  1_{\De_{y'}}(\Phi_{k+1} -  \Phi_{k+1}(y)  ) \Big)\Big](x)
\ee
 This   reduces  the problem to an estimate on the difference of the Green's functions.  
 We  claim that  for  $y,y'  \in   \square^{ \sim  2}$
\be    \label{spin0}
|1_{\De_y} \Big(  G^0_{k+1, \bom^+(\square)}- G^0_{k+1, \bom'}   \Big)1_{\De_{y'}}  f  | 
 \leq   CM^{-\frac12}  e^{ - \one  L^{-2} d(y,y' ) }   \|f\|_{\infty}
\ee
This is so since     $ [G^0_{k+1, \bom^+(\square)}]^{-1}$   and  $ [G^0_{k+1, \bom'} ]^{-1}$ agree  on a set containing  $\square^{\sim 2}$.
Thus in a random walk expansion connecting points  in  $\square^{\sim 2}$  they have the same leading term which cancels.   The remaining 
terms  have more than one step and supply the factor  $M^{-\frac12} $.
Again  we have  a bound    $CM^{-\frac12}  p_{k+1}$,

Finally  consider the difference of the last two terms  in   (\ref{samsum5})   which is 
\be   \bar \mu_k     \Big[  ( G^0_{k+1, \bom^+(\square)}\cdot 1- G^0_{k+1, \bom'}  \cdot (0, 1_{\Om_{k+1}}) \Big](x)   \Phi_{k+1}(y) 
\ee
We   can replace the  $(0, 1_{\Om_{k+1}})$  here  by   $1$  since the difference is  $\cO(e^{-r_{k+1}})$ and is completely negligible.
Then  we  must estimate  
\be   \bar \mu_k     \Big[  \Big( G^0_{k+1, \bom^+(\square)}- G^0_{k+1, \bom'} \Big) \cdot  1 \Big](x)   \Phi_{k+1}(y) 
\ee
Again we  first localize  in  $\square^{\sim 2}$  and then use the estimate   (\ref{spin0}).
Then this term is bounded  by  $\bar \mu_k ( CM^{-\frac12}  p_{k+1})\al_{k+1}^{-1}$.   Since   $\bar \mu_k  \al_{k+1}^{-1}
\leq       \bar \mu_k  \al_{k}^{-1}   \leq  \bar \mu_k^{\frac12}   \leq 1$
we  again have the  bound  $CM^{-\frac12}  p_{k+1}$.

This  completes the estimate on   $  \phi^0_{k+1,\bom^+( \square)}- \phi^0_{k+1, \bom'} $.   The estimate on the derivatives
follows in just the same way, now using estimates on the derivatives of the Green's functions. 

The  proof  of the estimates  on  $ \phi^{\min}_{k,\bom(\square)}- \phi_{k, \bom(\La_k^*)}(\hat   \Psi_{k, \bom'} ) $  also  follows in the same 
way.   Here  the relevant input is the bounds  (\ref{swish})  on    $\Psi_{k, \Om_{k+1}}(\bom')$. 
 This completes the proof.

 \begin{lem}      \label{vanishing}
 The bracketed characteristic functions in   (\ref{pinky1}) enforce
\begin{equation}
 \chi_{k}(  \La^{**}_{k+1} )   \chi^q_{k}(  \La^{**}_{k+1}   ) =1
\end{equation} 
  \end{lem}
 \bigskip

\pr   
 To show    $ \chi_{k}(  \La^{**}_{k+1} )=1 $  we  must show  that  
  $ \Psi^{\loc}_{k,\Om_{k+1}}(\bom') + ( C^{1/2}_{k,  \bom'} )^{\loc}W_k  \in \cS_k( \square) $  for any $M$-cube  $\square \subset 
  \La_{k+1}^{**}$.  We  argue separately that   $( C^{1/2}_{k,  \bom'} )^{\loc}W_k   \in  \frac  12  \cS_k(\square)$
and that   $ \Psi^{\loc}_{k,\Om_{k+1}}(\bom')   \in  \frac  12  \cS_k(\square)$.

For  the $( C^{1/2}_{k,  \bom'} )^{\loc}W_k$ bounds    start with the  fact  that         $|W_k|  \leq  p_{0,k}$ on $\La_k^{4*}$.  
 Since  
     $(C^{1/2}_{k,  \bom'} )^{\loc}W_k$  on   $\La_k^{3*}$   depends  on  $W_k$  on  $\La_k^{4*}$   we have   on   $\La_k^{3*}$ by  (\ref{around}) 
\be  \label{summers}
  |(C^{1/2}_{k,  \bom'} )^{\loc}W_k|  \leq  Cp_{0,k}  
   \hs         |\pa (C^{1/2}_{k,  \bom'} )^{\loc}W_k|  \leq  Cp_{0,k}            
   \ee
The  second  bound follows from the first  since we  are  on a unit lattice.  
Then    $(Cp_{0,k})^{-1}p_k(C^{1/2}_{k,  \bom'} )^{\loc}W_k $  satisfies the same bounds with  only   a $p_k$ on the right side.
By  lemma   \ref{threeone}   for     $\square  \subset   \La_k^{**}$ we  have  
  $(Cp_{0,k})^{-1}p_k(C^{1/2}_{k,  \bom'} )^{\loc}W_k   \in C   \cS_k( \square)$  or   
   $(C^{1/2}_{k,  \bom'} )^{\loc}W_k   \in Cp_{0,k}p_k ^{-1}  \cS_k(\tilde \square)$.
   But  for  $p_0<p$  and   $\la_k$ sufficiently small    $Cp_{0,k}p_k ^{-1}    =   C(- \log \la_k)^{p_0-p}  < \frac12 $
 so we have the result.

For the   $ \Psi^{\loc}_{k,\Om_{k+1}}(\bom') $ bounds    recall     that  for  an $LM$  cube    $\square   \subset    \La_{k+1}^{4*}$,       $\Phi_{k+1}$   satisfies  the bounds   (\ref{not})      on  $\tilde \square$.
For  $M$  sufficiently large   $CM^{-\frac 12}p_{k+1} $  is smaller than anything on  the right side of these equations.   Thus 
by  (\ref{note7})     we  may replace  $  \phi^0_{k+1, \bom(   \square )}$  by   $ \phi^0_{k+1, \bom'}  $.  Then 
 for   $\square  \subset    \La_{k+1}^{4*}$   on $\tilde \square$:
   \begin{equation}  \label{not4}
\begin{split}
 | \Phi_{k+1}  -  Q_{k+1}   \phi^0_{k+1, \bom'} |  \leq &2  p_{k+1}L^{-\frac12} \\
 |\pa   \phi^0_{k+1, \bom'}  |  \leq &   2 p_{k+1}  L^{-\frac32} \\   
 |  \phi^0_{k+1, \bom'} |  \leq & 2 p_{k+1} \al_{k+1}^{-1}  L^{-\frac12} \\
\end{split}
\end{equation}

 Next recall  that  
   \begin{equation}    \label{ding3}
  \Psi_{k,\Om_{k+1}}(\bom')  =   Q_{k}  \phi^0_{k+1,\bom'}
+ \frac{aL^{-2}}{a_k+ aL^{-2}}   Q^T (  \Phi_{k+1}  -   Q_{k+1} \phi^0_{k+1,\bom'})
\end{equation} 
This lets us  replace    $\Phi_{k+1}  -  Q_{k+1}   \phi^0_{k+1, \bom'}$   by   $  \Psi_{k,\Om_{k+1}}(\bom')  -   Q_{k}  \phi^0_{k+1,\bom'}$  in the above  inequality.   Furthermore  in  (\ref{such})    we have   already noted the identity 
$  \phi^0_{k+1, \bom'}  =  \phi_{k,  \bom(\La_k^*)}(  \hat \Psi_{k, \bom'}  )$.   Thus the above  inequalities  (\ref{not4})  become 
on  $(\La_{k+1}^{4*}) ^{\sim}$ 
   \begin{equation}  \label{not5}
\begin{split}
 |  \Psi_{k,\Om_{k+1}}(\bom')  -   Q_{k}   \phi_{k,  \bom(\La_k^*)}(  \hat \Psi_{k, \bom'}  ) |  \leq &2  p_{k+1}L^{-\frac12} \\
 |\pa    \phi_{k,  \bom(\La_k^*)}(  \hat \Psi_{k, \bom'}  ) |  \leq &   2 p_{k+1}  L^{-\frac32} \\   
 |  \phi_{k,  \bom(\La_k^*)}(  \hat \Psi_{k, \bom'}  )|  \leq & 2 p_{k+1} \al_{k+1}^{-1}  L^{-\frac12} \\
\end{split}
\end{equation}
   Now for  $\square \subset  \La^{**}_{k+1}$  (\ref{note6})       lets  us replace   $ \phi_{k,  \bom(\La_k^*)}(  \hat \Psi_{k, \bom'}  )$ 
       by   $  \phi^{\min}_{k,\bom(\square)} =   \phi_{k,  \bom(\square)}( \tilde   Q^T_{\bbT_0, \bom(\square)}\Psi_{k, \Om_{k+1}}(\bom')    )$  for  $M$ sufficiently large,      and  so  on  $\tilde \square$
   \begin{equation}  \label{snappy}
\begin{split}
 |  \Psi_{k,\Om_{k+1}}(\bom')  -   Q_{k}    \phi^{\min}_{k,\bom(\square)} |  \leq &3  p_{k+1}L^{-\frac12} \\
 | \pa    \phi^{\min}_{k,\bom(\square)}  |  \leq &   3 p_{k+1}  L^{-\frac32} \\   
 |   \phi^{\min}_{k,\bom(\square)} |  \leq & 3 p_{k+1} \al_{k+1}^{-1}  L^{-\frac12} \\
\end{split}
\end{equation} 
Since  $ \al_{k+1}^{-1} \leq   L^{-\frac14}   \al_{k}^{-1} $  and   $p_{k+1}  \leq   (1 + \log L)^p  p_k$    this   says that 
    $ \Psi_{k,\Om_{k+1}}(\bom')  \in \frac14  \cS_k(\square)$  for  $L$  sufficiently large.
But   $ |  \Psi^{\loc}_{k,\Om_{k+1}}(\bom')-   \Psi_{k,\Om_{k+1}}(\bom') |   \leq       e^{-r_{k+1}} $  from   (\ref{salt}). 
 Hence    $ \Psi^{\loc}_{k,\Om_{k+1}}(\bom')  \in      \frac12  \cS_k(\square)  $.   This completes the proof  that  
 $ \chi_{k}(  \La^{**}_{k+1})=1 $.
\bigskip

  Now   we  show      $ \chi^q_{k}(  \La^{**}_{k+1} ) =1 $,  that is we  show   that  on  $\square  \subset  \La_{k+1}^{**}$     
  \begin{equation}
  | \Phi_{k+1}  -    Q\Big (   \Psi^{\loc}_{k,\Om_{k+1}}(\bom') + ( C^{1/2}_{k, \bom})^{\loc}W \Big)|  \leq   p_k  \end{equation}
  By  (\ref{salt})  and      (\ref{summers})  this reduces to  showing that 
   $  | \Phi_{k+1}  -     Q   \Psi_{k,\Om_{k+1}}(\bom')|  \leq \frac12  p_k$.
  This follows by      (\ref{ding2})   and then     (\ref{not4}):
   \begin{equation}
  | \Phi_{k+1}  -     Q   \Psi_{k,\Om_{k+1}}(\bom')|  \leq     |\Phi_{k+1} -  Q_{k+1}   \phi^0_{k+1, \bom'} |  \leq  2p_{k+1}L^{-\frac12}
  \leq   \frac12  p_k  
  \ee

\subsubsection{bounds for analyticity domains}

   In   an expression   like    $ R_{k, \bpi}(\La_k, \Phi_{k,\La_k- \Om_{k+1}} \Psi^{\loc}_{k,\bom'}  +  ( C^{1/2}_{k,  \bom'} )^{\loc}W_k)$,  we   need  to know how  large we  can allow the fundamental fields
$\Phi_k, \Phi_{k+1}$  to be,    and still stay in the region of analyticity  $\cP_k(\La_k, 2 \de)$ for  $ R_{k, \bpi}(\La_k)$.  The following is a result  in that direction.  The proof is similar to  lemma  \ref{vanishing}.

Define for an  $LM$-cube $ \square  \subset  \Om_{k+1}$  the domain    $\cP^0_{k+1}(\square, \de)$
to be  all  fields   satisfying 
\begin{equation}  \label{startrek}
\begin{split}
 | \Phi_{k+1}  -  Q_{k+1}    \phi_{k+1, \bom^+(   \square )}|  \leq &  \la_{k+1}^{-\frac14 - \de}L^{-\frac12} \  \  \textrm{  on  } \  \tilde   \square \cap   \Om_{k+1}       \\
 |\pa   \phi_{k+1, \bom^+(   \square )}|  \leq &  \la_{k+1}^{-\frac14 - \de}L^{-\frac32}  \  \  \textrm{  on  } \  \tilde   \square   \\   
 |  \phi_{k+1, \bom^+(   \square )}|  \leq &    \la_{k+1}^{-\frac14 - \de} L^{-\frac12}   \  \  \textrm{  on  } \  \tilde   \square    \\
\end{split}
\end{equation} 
For  $X \subset   \Om_{k+1}$  define  
\be    \cP^0_{k+1}(X, \de)   =   \bigcap_{\square \subset  X }    \cP^0_{k+1}(\square, \de)
\ee

\begin{lem}   \label{technical}  
If  
 \be    \label{sandy}
 (\Phi_{k, \de \Om_k}, \Phi_{k+1, \Om_{k+1}} ) 
\   \in \  \cP'_k(\Om^{\nat}_k-   \Om_{k+1}, \de)  \cap   \cP^0_{k+1}( \Om_{k+1}- \Om_{k+1}^{2\nat},  \de)
 \cap       \cP^0_{k+1}( \Om^{2\nat}_{k+1}, 2 \de)
 \ee
  then   
\be   \Big(\Phi_{k, \de \Om_k},   \Psi_{k, \Om_{k+1}}^{\loc}  (\bom')\Big) \   \in    \  \frac12   \cP_k(\Om^{\nat}_k, 2 \de) 
 \ee
\end{lem}
\bigskip

\res
\begin{enumerate}
\item   On the domain (\ref{sandy}) we have
\be   \label{somewhat}
\begin{split}
|\Phi_{k}| \leq   & 2     \la_{k}^{-\frac14 -  \de}   \textrm{   on  } \Om^{\nat}_k -\Om_{k+1}     \\
|\Phi_{k+1}| \leq & 2  \la_{k+1}^{-\frac14 -  \de}L^{-\frac12}   \textrm{   on  } \Om_{k+1} -  \Om^{2 \nat}_{k+1} \\
 |\Phi_{k+1}| \leq   &  2  \la_{k+1}^{-\frac14 - 2 \de}L^{-\frac12}  \textrm{   on  } \Om^{2 \nat}_{k+1}  \\
\end{split}
\ee
We  choose    (\ref{sandy})    because it   contains  the domain  
 \be   \cP^0_{k+1, \bom^+}  \equiv \bigcap_{j=1}^k  [ \cP'_j( \de \Om_j,  \de)]_{L^{-(k-j)}}   \cap    \cP^0_{k+1}( \Om_{k+1}- \Om_{k+1}^{2\nat},  \de)
 \cap       \cP^0_{k+1}( \Om^{2\nat}_{k+1}, 2 \de)
\ee
which    scales to   $\cP_{k+1, \bom'}$.   Note also that    $  \cP_k(\Om^{\nat}_k, 2 \de) $  is contained in  
 $ \cP_k(\La_k, 2 \de) $.

\item 
The domain suffers some shrinkage in  $\Om^{\nat}_k-  \Om_{k+1}^{2\nat}$  (we  need $  \de$ to get $2 \de$), 
   but not in  $\Om_{k+1}^{2\nat}$ .
The shrinkage   in   $\Om^{\nat}_k -  \Om_{k+1}^{2\nat}$   is tolerable since it is not repeated.  Shrinkage in  $\Om_{k+1}^{2\nat}$  might be repeated and eventually would lead to problems.
\end{enumerate}
\bigskip

\pr   For an   $M$-cube  $\square \subset   \Om^{\nat}_k$,we  must  show that   
 $(\Phi_{k, \de \Om_k},   \Psi_{k, \Om_{k+1}}^{\loc}  (\bom')\Big) \   \in    \  \frac12   \cP_k(\square, 2 \de)$.
 We  distinguish several cases.  
 \bigskip
 
 \noindent    (A.)      $\square \subset  \Om^{\nat}_{k+1}$.  
 Start by considering  a $LM$-cube   $\square \subset    \Om^{\nat}_{k+1}$.      Our assumptions imply that  on
  $  \tilde   \square  $   ($LM$-cube enlargement)
\begin{equation}  \label{lamp1}
 | \Phi_{k+1}  -  Q_{k+1}    \phi^0_{k+1, \bom^+(   \square )}|,  \ 
 |\pa   \phi^0_{k+1, \bom^+(    \square )}|,   \      
 |  \phi^0_{k+1, \bom^+(    \square )}|  \     \leq    \  \la_{k+1}^{-\frac14 - 2\de}  L^{-\frac12}
\end{equation} 
where    $ \phi^0_{k+1, \bom^+(    \square )} $   only depends on  $\Phi_{k+1}$.
We claim that  on $\tilde \square$
\be  \label{note9}
|  \phi^0_{k+1,\bom(\square)}- \phi^0_{k+1, \bom'} |,   \   
|\pa   \phi^0_{k+1,\bom( \square)}-  \pa  \phi^0_{k+1, \bom'} |   \leq   CM^{-\frac12}   \la_{k+1}^{-\frac14 - 2\de}
 \ee  
 This follows by a variation of  the argument leading to    (\ref{note7}),  now with  the larger bounds   (\ref{somewhat})  on the fields. Since
  fields  and derivatives have the same weight the identities   (\ref{samsum5})   are    not required here. 
  The main idea  is  to split the contribution of  $\Phi_{k+1}$    into a piece  not in  $\square^{\sim 2}$
  where the fields individually have the claimed bound,   and  a piece in  $\square^{\sim 2}$
where the difference of the Green's functions supplies the factor  $M^{-\frac12}$.

 Given  (\ref{note9}),     then   for $M$ large      we  can replace   $\phi^0_{k+1, \bom^+(    \square )}$  by    $\phi^0_{k+1, \bom'} $ in  (\ref{lamp1}) 
and  have    on  $(\Om^{\nat}_{k+1})^{\sim}$ \begin{equation}  \label{lamp2}
 | \Phi_{k+1}  -  Q_{k+1}  \phi^0_{k+1, \bom'}|,  \ 
 |\pa    \phi^0_{k+1, \bom'}|,   \      
 |  \phi^0_{k+1, \bom'}|  \     \leq    \   2 \la_{k+1}^{-\frac14 - 2\de}  L^{-\frac12}
\end{equation}

Now       replace    $|\Phi_{k+1}  -  Q_{k+1}   \phi^0_{k+1, \bom'}|$   by  the smaller    $ | \Psi_{k,\Om_{k+1}}(\bom')  -   Q_{k}  \phi^0_{k+1,\bom'}|$ and  use
the identity   
$  \phi^0_{k+1, \bom'}  =  \phi_{k,  \bom(\La_k^*)}(  \hat \Psi_{k, \bom'}  ) $ which still holds with complex fields. 
This yields    on  $(\Om^{\nat}_{k+1})^{\sim}$
   \begin{equation}  \label{lamp3}
 |  \Psi_{k,\Om_{k+1}}(\bom')  -   Q_{k}   \phi_{k,  \bom(\La_k^*)}(  \hat \Psi_{k, \bom'}  ) |,
   \    |\pa    \phi_{k,  \bom(\La_k^*)}(  \hat \Psi_{k, \bom'}  ) |,    \ 
 |  \phi_{k,  \bom(\La_k^*)}(  \hat \Psi_{k, \bom'}  )|\   \leq \   2\la_{k+1}^{-\frac14 -2 \de} L^{-\frac12}
\end{equation} 

The bounds   (\ref{somewhat}) imply that   $    |\phi^0_{k+1, \bom'}|  \leq  C  \la_{k+1}^{-\frac14 -2\de} $
and it follows that    $ |\Psi_{k,\Om_{k+1}}(\bom') |  \leq  C  \la_{k+1}^{-\frac14 -2\de} $ on its full domain.
Then if    $\square$ is    an $M$-cube in   $\Om^{\nat}_{k+1}$,      
by a  variation  of     (\ref{note6})  we  have   (again    no special identity required)  on $\tilde \square$  ($M$-cube  enlargement):
\be  \label{note8}
|  \phi^{\min}_{k,\bom(\square)}- \phi_{k, \bom(\La_k^*)}(\hat   \Psi_{k, \bom'} ) |,   \   
|\pa   \phi^{\min}_{k,\bom( \square)}-  \pa  \phi_{k,  \bom(\La_k^*)}(\hat   \Psi_{k, \bom'} )  |   \leq     CM^{-\frac12}   \la_{k+1}^{-\frac14 - 2\de}   
 \ee  
 Hence    for $M$ large   we  can replace   $ \phi_{k, \bom(\La_k^*)}(\hat   \Psi_{k, \bom'} )$  by  $ \phi^{\min}_{k,\bom(\square)}$
    in         (\ref{lamp3}) and obtain
    on $\tilde \square$ 
       \begin{equation}  \label{lamp4}
 |  \Psi_{k,\Om_{k+1}}(\bom')  -   Q_{k}   \phi^{\min}_{k,\bom(\square)}|, \ 
  |\pa     \phi^{\min}_{k,\bom(\square)} |, \    |   \phi^{\min}_{k,\bom(\square)}|\  \leq
    \ 3\la_{k+1}^{-\frac14 -2 \de} L^{-\frac12}
  \end{equation} 
The same holds  with   $ \hat \Psi_{k, \bom'} $  replaced by  $ \hat \Psi^{\loc}_{k, \bom'} $   and   $ 4\la_{k+1}^{-\frac14 -2 \de} L^{-\frac12}$
on  the right  side.   But  for $L$ large,   $4\la_{k+1}^{-\frac14 - 2\de} L^{-\frac12} =   4\la_{k}^{-\frac14 -2 \de} L^{-\frac34-2 \de}  \leq    \frac12\la_{k}^{-\frac14 -2 \de}$.  Thus we  have 
 $ \Psi_{k, \Om_{k+1}}^{\loc}  (\bom')   ) \   \in    \  \frac12   \cP_k(\square, 2 \de)$  as required.
\bigskip

 \noindent    (B.)    $\square  \subset   \Om_{k+1}  - \Om^{\nat}_{k+1}$.  This time we take a more direct approach.
The   bounds  (\ref{somewhat})
imply   that  on  $ [(\Om_{k+1}^{\nat})^c]^{\sim}$ ($LM$-cube enlargement)  
we have   $   |\phi^0_{k+1, \bom'}|  \leq  C  \la_{k+1}^{-\frac14 -\de} $.
The  point here  is  that the weaker  bound  $\cO( \la_{k+1}^{-\frac14 - 2 \de})$    in  $\Om^{2 \nat}_{k+1}$  is offset
by a factor  $e^{-r_{k+1}} $  due to the  distance between     $ [(\Om_{k+1}^{\nat})^c]^{\sim}$  and    $\Om^{2 \nat}_{k+1}$.
It follows that   on $ [(\Om_{k+1}^{\nat})^c]^{\sim} \cap   \Om_{k+1}$ we have  
\be   \label{substandard}     | \Psi_{k,\Om_{k+1}}(\bom')|  \leq   C  \la_{k+1}^{-\frac14 -  \de }
\ee

Let  $\square$ be an $M$-cube in   $\Om_{k+1}  -  \Om^{\nat}_{k+1}$,   and  now
    $ \phi^{\min}_{k,\bom( \square)}
    =  \phi_{k,  \bom(\square)}\Big( \tilde Q^T_{\bbT_0, \bom(\square)}(\Phi_{k, \de \Om_k},\Psi_{k, \Om_{k+1}}(\bom') ) \Big)$.  We   claim that  on $\tilde \square$   ($M$-cube enlargement)
  \begin{equation}  \label{lamp5}
 |  \Psi_{k,\Om_{k+1}}(\bom')  -   Q_{k}  \phi^{\min}_{k,\bom( \square)} |, \ 
  |\pa   \phi^{\min}_{k,\bom( \square)} |, \    |  \phi^{\min}_{k,\bom( \square)}|\  \leq
      C\la_{k+1}^{-\frac14 - \de} 
  \end{equation}     
 For example to bound   $  |  \phi^{\min}_{k,\bom( \square)}|$    we  need  a bound   on   $\Psi_{k,\Om_{k+1}}(\bom')  $
    on  $\square^{\sim((2R+1)}\cap  \Om_{k+1}$.   But this is included in the $LM$ enlargement   in (\ref{substandard})  , assuming  $L > 2R +1$.  A bound of the same form holds with   $\Psi_{k, \Om_{k+1}}(\bom') $  replaced by   $\Psi^{\loc}_{k, \Om_{k+1}}(\bom')$. 
    This  gives the result  since for   $\la_k$ sufficiently small    $ C\la_{k+1}^{-\frac14 - \de}  \leq   \frac12   \la_{k}^{-\frac14 -2 \de} $
   \bigskip 
    
 \noindent    (C.)   $\square \subset   \Om^{\nat}_{k} -  \Om_{k+1} $.   Argue as  in the previous case.   Note that now we are trying to prove
on  $\tilde \square$
 \begin{equation}  \label{lamp6}
 |  \Phi_{k}  -   Q_{k}  \phi^{\min}_{k,\bom( \square)} |, \ 
  |\pa   \phi^{\min}_{k,\bom( \square)} |, \    |  \phi^{\min}_{k,\bom( \square)}|\  \leq
     \frac12    \la_{k+1}^{-\frac14 -2 \de}        \end{equation}

\newpage

\subsection{adjustments}

Consider  the   definition  (\ref{breakup3})  of  $\cC_{k+1}(\Om_{k+1},\La_{k+1})$.
Since     $ \La_{k+1} \subset    \Om^{\nat}_{k+1} - Q_{k+1}$
 and  $ \La_{k+1} \subset  \Om_{k+1} - R_{k+1}$,
 we can extract  $ \chi^0_{k+1}(\La_{k+1})$   and
$\chi^w_{k}( \La_{k+1})$   from each term in this sum   and write 
\begin{equation}
 \cC_{k+1}(\Om_{k+1},\La_{k+1}) =  \chi^0_{k+1}(\La_{k+1})\chi^w_{k}( \La_{k+1})  \cC'_{k+1}(\Om_{k+1},\La_{k+1})
  \end{equation}
where 
   \begin{equation}   \label{lonely}
 \begin{split}
& \cC'_{k+1}(\Om_{k+1},\La_{k+1})    
 \\
=  &   \sum_{     Q_{k+1},R_{k+1} \to \La_{k+1}    }
 \zeta^0_{k+1}( Q_{k+1} ) 
 \zeta^w_{k}(  R_{k+1}  ) \chi^0_{k+1}(   \Om^{\nat}_{k+1} - (Q_{k+1} \cup \La_{k+1}) )
    \chi^w_{k}(  \Om_{k+1} - (R_{k+1} \cup  \La_{k+1})  ) \\
    \end{split}
 \end{equation}
 Using  lemma \ref{vanishing}   the characteristic functions now have the form
\begin{equation}
 \cC^0_{k+1, \bpi^+}\    \chi^0_{k+1}(  \La_{k+1} )   \chi^w_{k}(  \La_{k+1})
\end{equation}
  where  $ \cC^0_{k+1, \bpi^+} = \cC_{k, \bpi}\ \cC^0_{k+1, \La_k, \Om_{k+1}, \La_{k+1}}$  and   
\begin{equation}
 \cC^0_{k+1, \La_k, \Om_{k+1}, \La_{k+1}} =   \cC_k^q(\La_k, \Om_{k+1})  \chi_k ( \La_k-  \La^{**}_{k+1})  \chi^q_k(   \Om_{k+1}- \La^{**}_{k+1})  
   \cC'_{k+1}(\Om_{k+1},\La_{k+1})    
\end{equation}
Then    $ \cC^0_{k+1, \La_k, \Om_{k+1}, \La_{k+1}} $  does not depend on $\Phi_{k+1,  \La^{\nat}_{k+1}}$.
 The function    $ \cC^0_{k+1, \bpi^+}$  enforces  on  $\La_{k} - \Om_{k+1}$   (due to    $ \chi_k ( \La_k-  \La^{**}_{k+1})$)
 \be \label{mexicanhat1}
   | \Phi_{k}|  \leq  2 p_{k} \al_{k}^{-1}   \hs   |\pa   \Phi_{k}|  \leq  3 p_{k}
 \ee
 and on    
  $\Om_{k+1} - \La_{k+1}$    by   (\ref{go2}) and (\ref{gb3})  
    \begin{equation}   \label{sombrero1}
      |\Phi_{k+1}|  \leq   \  3    p_{k}\al_{k}^{-1}   
        \hs \    |\pa \Phi_{k+1}|  \leq   \  4    p_{k}
        \hs       |W_{k}|  \leq   C p_{k}      
 \end{equation} 
 The function   $\cC_{k+1,  \bpi^+}$  will be  $ \cC^0_{k+1, \bpi^+}$  scaled down   and the  bounds   (\ref{mexicanhat1}),(\ref{sombrero1}) 
 scale  to the required bounds   (\ref{mexicanhat}),(\ref{sombrero})  for  $k+1$, provided   $ C_w  >C$.

In  the expression  (\ref{representation6}) for  $ \tilde    \rho_{k+1}( \Phi_{k+1})   $  we  split the fluctuation  measure by  
\be
d  \mu_{\Om_{k+1} }(W_k)   =    d  \mu_{\Om_{k+1}-  \La_{k+1}} (W_k) \     d  \mu_{  \La_{k+1}} (W_k)   
\ee
With the first factor we  form   $d W^0_{k+1,\bpi^+}   \equiv   d W_{k,\bpi} \   d  \mu_{\Om_{k+1}-  \La_{k+1}} (W_k) $.
 For the second  factor    we  introduce  the probability measure
 \begin{equation}
d\mu_{\La_{k+1}}^*(W_k)  =   (\cN^w_{ k,  \La_{k+1}})^{-1}  \chi^w_{k}(  \La_{k+1})   d  \mu_{  \La_{k+1}} (W_k)   
\end{equation}     
Since the measure is a product over sites in  $\La_{k+1}^{(k)}$,  the    normalization factor can be  written
\begin{equation}
  \cN^w_{ k,  \La_{k+1}} =\int    \chi^w_{k}(  \La_{k+1})   d  \mu_{  \La_{k+1}} (W_k)   
 =   \exp  ( -  \vep_k^0 | \La^{(k)}_{k+1} |  )   =    \exp  ( -  \vep_k^0\Vol(  \La_{k+1} )  )  
 \end{equation}
This  defines   $  \vep_k^0 $  which is the same  as in part  I.

Now   (\ref{representation6})  becomes   
   \begin{equation}     \label{representation7}
\begin{split}
\tilde    \rho_{k+1}( \Phi_{k+1}) 
 =    &  Z^0_{k+1}   \sum_{\bpi^+}    \int   d \Phi^0_{k+1, \bom^{+,c} }d W^0_{k+1,\bpi^+}\  K_{k, \bpi} \  \cC^0_{k+1, \bpi^+}         
\        \exp  \Big(  c_{k+1}  |\Om^{c,(k)}_{k+1}|    \Big)   \\
&  \chi^0_{k+1}(  \La_{k+1} )   \exp \Big(-  S^{*,0}_{k+1}(\La_k)  \Big)
\Xi_{k,  \bpi^+}  \\
&   \\
   \end{split}
\end{equation}
Here     we  have   defined the fluctuation integral 
\begin{equation}
\Xi_{k,  \bpi^+}   =  \exp  ( -  \vep_k^0\Vol(  \La_{k+1} )  )
 \int    d \mu^*_{ \La_{k+1} }(W_k)  
     \exp\Big( E^+_k(\La_k)    + R^{(\leq 7)}_{\bpi,  \Om_{k+1}}  +  B_{k, \bpi}(\La_k)  \Big)
       \end{equation}

We  further define  
  $\de E^+_k(X, \phi,  \cW)$  by
    \begin{equation}
  E^+_k \big(X,   \phi +  \cW  )
 = E^+_k \big(X,   \phi    ) 
  +   \de  E^+_k \big(X,   \phi, \cW    )
\end{equation}
This is  the same as the definition in part I,   but  here we have  different fields  $\phi =   \phi^0_{k+1,\bom'}$
and   $\cW =    \cW_{k,  \bom'} $.
 The   factor  $ E^+_k \big(\La_k,   \phi^0_{k+1,\bom'}    ) $ does not depend on  $W_{k}$  and    can be  moved outside the integral.   Inside the integral we  have   $ \de    E^+_k(\La_k) =  \de    E^+_k(\La_k, \phi^0_{k+1,\bom'}, \cW_{k,  \bom'}  ) $
 and  
 with  these    adjustments   
\begin{equation}   \label{prime}
\Xi_{k,  \bpi^+}  =     
 \exp  \Big(-  \vep_k^0 \Vol(  \La_{k+1} )  +  E^+_k \big(\La_k,   \phi^0_{k+1,\bom'} \big)   \Big)
  \Xi'_{k,  \bpi^+}  
\end{equation}
where
\begin{equation}
\Xi'_{k,  \bpi^+}   =  \int    d \mu^*_{ \La_{k+1} }(W_k)  
     \exp\Big( \de   E^+_k(\La_k)    + R^{(\leq 7)}_{\bpi,  \Om_{k+1}}  +  B_{k, \bpi}(\La_k)  \Big)
\end{equation}

 \subsection{localization}  \label{localization}

We  will be giving   a local structure to  the fluctuation integral by a cluster expansion.  As input to this  we give localization expansions
for the integrand.

  We  start with $\de E^{+}_k \big(\La_k)   =   \sum_{X \subset  \La_k} \de E^{+}_k \big(X)$.  The function   $\de E^{+}_k \big(X,   \phi^0_{k+1,\bom'} ,  \cW_{k,\bom'}  )$ depends  on   $
 \cW_{k,\bom'}= a_k  G_{k,   \bom(\La_k^*)}  Q_k^T \  (  C^{1/2}_{k,  \bom'}   W_k )  
$
 only  in  $X$,  but depends  on  $W_k$   in  all of   $\Om'_1 =  \Om_1(\La_k^*)$.   We  need a sharper  localization  in  $W_k$.

 \begin{lem}    \label{first}   For $(\Phi_{k, \de \Om_k},  \Phi_{k+1,  \Om_{k+1}})$  in   (\ref{sandy})  ( hence 
 satisfying  (\ref{somewhat})),  and     $  |W_k| \leq  B_wp_k$:
 \begin{equation}   \label{city}
 \begin{split}
&\de E^{+}_k \big(\La_k,   \phi^0_{k+1,\bom'},  \cW_{k,\bom'}  )
 =   \sum_{ Y \in \cD^0_{k+1}: Y \subset   \La_{k+1}}   (\de    E^+_k)^{\loc}( Y,  \phi^0_{k+1,\bom'},W_k  ) \\
 +& 
 \sum_{Y  \in  \cD^0_{k+1}(\bmod \Om^c_{k+1}), Y \#   \La_{k+1}}  B^{(E)}_{k, \bpi^+}(Y)  \   +  \  \tilde B_{k+1, \bpi^+}   \textrm{ terms } \\
\end{split}
\end{equation}
where
\begin{enumerate}
\item     For    $Y \in \cD^0_{k+1}$ the leading term     $ (\de    E^+_k)^{\loc}( Y,   \phi,W_k)$ is exactly the global small field expression from part I. 
  It   depends on  $\phi, W_k$ only in $Y$,   is  analytic  in  $\phi \in \frac12 \cR_k$  and    $  |W_k| \leq  B_wp_k$   and satisfies  there
\begin{equation}
 |   (\de  E^+_k)^{\loc}( Y, \phi,W_k  )|   \leq      \cO(1)L^3  \la_k^{\frac14- 10 \ep}  e^{-L (\ka- 2\ka_0-2) d_{LM}(Y)  }
\end{equation}
\item   For  $Y  \in  \cD^0_{k+1}( \bmod \  \Om^c_{k+1})$ the boundary term   $  B^{(E)}_{k, \bpi^+}(Y,\Phi_{k, \de \Om_k},   \Phi_{k+1, \Om_{k+1}},   W_k )  $  depends on the fields only in $Y$.  It     is 
analytic   the stated domain    
 and satisfies there
  \begin{equation} 
   \label{boundary}
|B^{(E)}_{k, \bpi^+}(Y )|   \leq    \cO(1)L^3 \la_k^{\frac14 - 10 \ep}  e^{ - L  ( \ka-2\ka_0-3)  d_{LM}(Y,  \bmod \Om^c_{k+1})  }
\end{equation}
\end{enumerate}
\end{lem}
\newpage

\res
\begin{enumerate}
\item   
  The  expression     ``$ \tilde B_{k+1, \bpi^+}$   terms''  will be used repeatedly.   It  refers to functions    
 localized in  $\La^c_{k+1}$ which are  bounded  by 
$C|\La^{(k)}_k - \La^{(k)}_{k+1}| = C\  \Vol(\La_k - \La_{k+1})$.   Local    structure is no longer important for these terms.
\item  The bounds   (\ref{dos})  and  (\ref{somewhat})  and  $2 \de < \ep$    yield   $   | \phi^0_{k+1,\bom'}  |  \leq    C   \la_k^{-\frac14 - 2  \de}  \leq  \frac12  \la_k^{- \frac14 - 3\ep} $
with similar bounds on derivatives.   Thus   $ \phi^0_{k+1,\bom'} $  is in  $ \frac12  \cR_k$   as  required. 

Note also that   (\ref{twoone1})  and  (\ref{around})   and   (\ref{somewhat}) show that   $ \cW_{k,\bom'}= \phi_{k,   \bom(\La_k^*)} (   C^{1/2}_{k,  \bom'}   W_k )$  satisfies   
 $| \cW_{k,\bom'}|  \leq   C    \|  C^{1/2}_{k,  \bom'}   W_k \|_{\infty}  \leq      CB_w p_k$.
\end{enumerate}

\pr 
\noindent  (A.)
We  study  $\de  E^{+}_k (X,   \phi, \cW_{k,\bom'}  ) $  for  $\phi \in  \frac12 \cR_k$.    We argue as in lemma 17  of Part I.   
There  are two parts    $\de  E^{+}_k   =  \de   E_k   -  \de    V_k$.        The potential  is supported on cubes  $\square$  and 
has the form  $ \de   V_k (\square,   \phi, \cW_{k,\bom'} )= \de   V_k (\square,   \phi+ \cW_{k,\bom'})  -   \de   V_k (\square,   \phi) $.
 The bounds   $|\phi|  \leq  \la_k^{-\frac14 -3 \ep}$  and     $| \cW_{k,\bom'}|  \leq  CB_w p_k$   imply as in part I that
         $ | \de   V_k (\square,   \phi, \cW_{k,\bom'}  )|  \leq   \la_k^{\frac14- 10 \ep} $.  (We  had
 a sharper bound on $W_k$ there, but the argument stills holds.)

Furthermore      if  $|t|  \leq  \la_k^{-\frac14}$
we  have     $|t\cW_{k, \bom'}| \leq   CB_wp_k\la_k^{-\frac14}  \leq   \frac12  \la_k^{-\frac14- 3\ep} $.  There are similar bounds on derivatives
  and so  $t\cW_{k, \bom'} \in  \frac12 \cR_k$ 
Thus     $t \to    E_k \big(X,   \phi^0_{k+1,\bom'}  + t\cW_{k,\bom'}  )$  is analytic  in
$|t| \leq     \la_k^{-1/4} $     we have the  representation 
\begin{equation}
\de  E_k \big(X,   \phi, \cW_{k,\bom'}  )
=   \frac{1}{2 \pi i}  \int_{|t|  =     \la_k^{-1/4} }   \frac{dt}{t(t-1)}   E_k \big(X,   \phi  + t\cW_{k,\bom'}  )
\end{equation}  
Now     $ |E_k \big(X,   \phi )|  \leq  \la_k^{\beta} e^{-\ka d_M(X)}$ for  $\phi \in \cR_k$   is our basic assumption,
 and  hence   $|\de  E_k \big(X,   \phi,   \cW_{k,\bom'}  )|  \leq  \one  \la_k^{\frac14 + \beta} e^{-\ka d_M(X)}$.
Altogether then      
\begin{equation}
  | \de  E^{+}_k  (X,   \phi, \cW_{k,\bom'}  )|  \leq  \one  \la_k^{\frac14- 10 \ep}  e^{- \ka d_M(X)  }
\end{equation}

 We  also reblock  to  get  an element of   $\cD^0_{k+1, \bom'}$;   since  $X \subset  \La_k  \subset  \Om_k$  this 
 means  preserving  the $M$ cubes in  $\de \Om_k$  and replacing  $M$ cubes  by  $LM$  cubes  in  $\Om_{k+1}$.  
  We   define  
 for   $Y  \in  \cD^0_{k+1, \bom'}$    
\be   \label{twinkie}
 (\de  E^+_k)' ( Y)  =  \sum_{X:  \bar X  =Y,  X   \subset   \La_k }  (\de  E^+_k)' (X)
  \ee
  where for $X \in \cD_k$,   $\bar X$  is the smallest element of $\cD^0_{k+1, \bom'}$ containing $X$.   We  postpone the estimate on this quantity.
   Then  we have     $\de  E_k^+ (\La_k)  =  \sum_{Y \cap  \La_k \neq  \emptyset }   (\de  E^+_k)' ( Y )$.  

\bigskip
\noindent   (B.) 
In    $ \cW_{k,\bom'} $  we have the propagator  $ G_{k,   \bom(\La_k^*)} $.  This has a random walk expansion 
based on the cubes of  $\bom(\La_k^*)$.  It is convenient  to use a modification  in which we use the cubes of 
$\bom'  =  (\bom(\La_k^*),  \Om_{k+1})$ instead,   i.e. we take $LM$-cubes in  $\Om_{k+1}$  rather  than $M$ cubes.
This gives a new  random walk expansion,  but it leads to exactly the same bounds as the expansion of  theorem  \ref{th}.
This  is true  since  the basic  estimate  on  $G_{k, \bom(\La_k^*}(\tilde \square)$  of lemma  \ref{bonfire}   holds  for  $LM$-cubes
just as well as    $M$-cubes,    and since  the fact  that  a cube can have  $\cO(L^2)$ neighbors  is already built into
the proof of theorem \ref{th}.

Then we  can     introduce   the propagator  $ G_{k,   \bom(\La_k^*)} (s)$ with parameters  $s =  \{ s_{\square} \}$    which weaken  the coupling  through  multiscale   cubes  $\square$   in       $\bom'$.   We  also  use  again  the weakened operator
  $ C^{1/2}_{k,  \bom'}(s) $   defined in section  \ref{bingo}.      Replace  $ \cW_{k,\bom'}$
    by  $ \cW_{k,\bom'}(s) = a_k  G_{k,   \bom(\La_k^*)}(s)  Q_k^T \  (  C^{1/2}_{k,  \bom'}(s)   W_k )$.  
   This   does not spoil  any of our estimates,  even    if  we  allow  $s_{\square}$  to be complex and   satisfy  $|s_{\square}|  \leq   M^{\frac12}  \equiv  e^{\ka_1}$.
Then  we have a decoupling expansion   
\begin{equation} \label{gong}
\begin{split}
&  (   \de  E^+_k)'(Y,  \phi, \cW_{k,  \bom'}   )  =
  \sum_{Y_1  \supset Y}     (\de E_k^{+})'(Y, Y_1,  \phi,  W_k  )\\
\end{split}
\end{equation} 
where $Y_1  \in \cD^0_{k+1,     \bom'}$ is   a multiscale  polymer  
and   
\begin{equation}    \label{stingy2}
\begin{split}
& (\de E_k^{+})'(Y,   Y_1,   \phi , W_k  )
=    \int   ds_{Y_1-Y}  \frac{\  \ \pa}{ \pa s_{Y_1-Y}}  \left[ 
(\de E^{+}_k)' (Y,    \phi,  \cW_{k,\bom'}(s)  )
\right]_{s_{Y_1^c} = 0,   s_Y  =1}\\
\end{split}
\end{equation} 
The function   $(\de E_k^{+})'(Y, Y_1,  \phi ,  W_k  )$ depends on   $\bom', \La_k$,      has fields   strictly localized in  $Y_1$, and vanishes unless  $Y_1 \subset  \Om_1(\La_k^*)  \subset    \Om_k$.
  As  in  part  I    we  use  Cauchy bounds to estimate the derivatives  
$ \pa/\pa s_{Y_1-Y}$ for  $|s_{\square}| \leq  1$.  This gains  a  factor  $e^{-(\ka_1-1)}$ for each $\square  \subset  Y_1-Y$   and  gives  an overall improvement of our bounds,  still postponed,  by a factor of     $\exp(  - (\ka_1-1) |Y_1-Y|_{\bom'}    )$.

Now  for   $Y_1  \in \cD^0_{k+1,     \bom' }$   define  the function   $ (\de  E^+_k)''( Y_1 ) = (\de  E^+_k)'' ( Y_1, \phi, W_k ) $  
by
\begin{equation}   \label{arson2}
  (\de  E^+_k)'' ( Y_1 ) 
  =   \sum_{Y \subset Y_1}  ( \de E^+_k)'(Y,   Y_1  )
\end{equation}
This depends  on  $\bom', \La_k$  and we have   
\begin{equation}  \label{arson1}
\de  E^+_k (\La_k  )  =  \sum_{Y_1 \cap  \La_k  \neq  \emptyset }      (\de  E^+_k)''( Y_1 ) 
\end{equation}

\bigskip
\noindent (C.)   Consider terms in  (\ref{arson1})  with   $Y_1  \subset  \La_{k+1}$  and hence $Y_1 \in    \cD^0_{k+1}$. 
 In this case     in the expression  (\ref{stingy2})    we  are   evaluating  
  $\cW_{k,\bom'}(s)$  with  $s_{\La_{k+1}^c}= 0$.   In the random walk expansions defining this object only
  paths in $\La_{k+1}$ contribute,   and these  are the same  for  $ G_{k,   \bom(\La_k^*)}(s),  C^{1/2}_{k,  \bom'}(s) $
  and the global   $ G_{k}(s),  C^{1/2}_{k} $ with $LM$ cubes.  Hence in this circumstance    $\cW_{k,\bom'}(s)$   is the same as
  the global   $\cW_{k}(s)$   and hence   $ \de E_k^{+}(Y,   Y_1 )$  is the  same as the  global function of part I.
  Then   $ (\de  E^+_k)''( Y_1) $  is independent of  $\bom',  \La_k$   and is equal to  
to the global function  $ (\de  E^+_k)^{\loc}( Y_1 )$ of  part  I.

For  $Y_1 \subset  \La_{k+1}$ we have   $ |Y_1-Y|_{\bom'}  =   |Y_1-Y|_{LM}$  and as in part I this leads to the estimate
\be
| (\de  E^+_k)^{loc} ( Y_1 )|  \leq     \cO(1) L^3 \la_k^{\frac14 - 10 \ep} e^{- L (\ka- 2\ka_0-2)  d_{LM}(Y_1)}
\ee
   The      sum  of these terms in   (\ref{arson1}) is the  desired   expression
\begin{equation}  \label{arson4}
  \sum_{Y_1  \in  \cD^0_{k+1},  Y_1 \subset  \La_{k+1}}      (\de  E^+_k)^{loc}(Y_1) 
\end{equation}

\bigskip
\noindent  (D.)   Now consider terms in   (\ref{arson1})  with   $Y_1 \cap  \La^c_{k+1}
\neq  \emptyset$.  Weaken the coupling  in  $( \de E^+_k)''( Y_1;  \phi^0_{k+1,\bom'} , W_k  ) $
by    replacing $\phi^0_{k+1, \bom'} $ by  $\phi^0_{k+1, \bom'}(s) $ where again  $  s =  s_{\square}$  is 
indexed by  elementary cubes  in   $\bom'$.
Then  we have a second   decoupling expansion   
\begin{equation} 
   ( \de E^+_k)''( Y_1;  \phi^0_{k+1,\bom'} , W_k  )  =
  \sum_{Y_2 \supset Y_1}    ( \de E^+_k)''( Y_1,  Y_2;   \Phi_{k, \de \Om_k},   \Phi_{k+1, \Om_{k+1}} , W_k  )
\end{equation} 
where  for   $Y_2  \in \cD^0_{k+1,     \bom'}$
 \begin{equation}    \label{stingy3}
\begin{split}
&   ( \de E^+_k)''( Y_1,   Y_2;   \Phi_{k, \de \Om_k},   \Phi_{k+1, \Om_{k+1}} , W_k  )\\
=  &  \int   ds_{Y_2-Y_1}  \frac{\  \ \pa}{ \pa s_{Y_2-Y_1}}  \left[ 
(\de E^{+}_k)'' \big(Y_1, \phi^0_{k+1, \bom'}(s) ,  W_k  )
\right]_{s_{Y_2^c} = 0,   s_{ Y_1}  =1}\\
\end{split}
\end{equation} 
The function   $ ( \de E^+_k)''( Y_1, Y_2 )$
  depends  on the fields only     in $Y_2$.   The derivatives
  improve our estimates by  a factor   $  \exp(  -  (\ka_1-1) |Y_2-Y_1|_{\bom'}  )$.
  If we  now define  
  \be   
   ( \de E^+_k)'''(     Y_2)  =   \sum_{Y_1 \subset  Y_2,Y_1 \cap  \La^c_{k+1}
\neq  \emptyset}     ( \de E^+_k)''( Y_1, Y_2  )
  \ee
 then our     expression    becomes   $     \sum_{Y_2 }  ( \de E^+_k)'''(     Y_2 )  $

\bigskip
\noindent  (E.)  
Next   we pass from   polymers   $Y_2  \in  \cD^0_{k+1, \bom'}$        to polymers    $Z  \in  \cD^0_{k+1}$.    We   define   $  ( \de E^+_k)^{(iv)}(Z  )  =    ( \de E^+_k)^{(iv)}(Z, \Phi_{k, \de \Om_k},   \Phi_{k+1, \Om_{k+1}} , W_k   ) $  by  
\be   
  ( \de E^+_k)^{(iv)}(Z )  =    \sum_{Y_2:   \bar  Y_2    = Z }   ( \de E^+_k)'''(Y_2  )
\ee   
where now  $\bar  Y_2$ is the smallest   element of  $\cD^0_{k+1}$   containing  $Y_2$.
 The function  $  ( \de E^+_k)^{(iv)}(Z )  $  vanishes unless   $ Z   \cap  \La^c_{k+1}   \neq  \emptyset,  Z \cap \La_k \neq  \emptyset $
 so  we  can write our expression as  
\be    \label{foreign}  
 \sum_{Z  \in  \cD^0_{k+1}, Z   \cap  \La^c_{k+1}   \neq  \emptyset,  Z \cap \La_k \neq  \emptyset }      ( \de E^+_k)^{(iv)}(Z  ) 
  \ee

\bigskip
\noindent  (F.)  We  estimate $  ( \de E^+_k)^{(iv)}(Z )  $. 
Collecting all the contributions,  and dropping conditions   $X \subset  \La_k$   and    $Y_1 \cap  \La^c_{k+1} \neq  \emptyset$ and   we have  
\be
\begin{split}
& | ( \de E^+_k)^{(iv)}(Z )  |
 \leq   \    \cO(1)  \la_k^{\frac14 - 10 \ep} \  \sum_{Y_2:  \bar Y_2 =Z} \  \sum_{Y_1 \subset  Y_2}  \sum_{Y  \subset  Y_1} 
 \sum_{X: \bar  X  =  Y  }   \\
&    \   \
   \exp \Big( -  (\ka_1-1) |Y_2-Y_1|_{\bom'}   - (\ka_1-1) |Y_1-Y|_{\bom'} -  \ka   d_{M}(X)    \Big)\\
   \end{split}
 \ee
 Now   $ Md_M(X)  \geq   LM d_{LM}(\bar   Y)$  since  a tree joining the $M$ cubes in $X$ will also join the $LM$ cubes in $\bar  Y  \in  \cD^0_{k+1}$.
 Thus we  can extract  a factor   $\exp( -L(\ka -\ka_0) d_{LM}  (\bar Y))$ leaving  $\exp  ( -  \ka_0 d_M(X) )$.
 Now    
\be    |Y_2-Y_1|_{\bom'}  +   |Y_1   -Y|_{\bom'}  =  |Y_2- Y|_{\bom'}  \geq      |Y_2- \bar Y|_{\bom'}      \geq   |Z - \bar Y|_{LM}
\ee
The  last    step follows since in passing  from  $Y_2-  \bar  Y$ to $Z-\bar  Y$  
 we  replace each elementary  $\cD^0_{k+1, \bom'}$
cube  in  $Y_2- \bar  Y $    by  the   $LM$ cube  containing it,  and  this cannot increase the number of elementary cubes.
Then we can extract a factor   $ \exp ( -  (\ka_1/2-1) |Z- \bar Y|_{LM})$  which for $M$ sufficiently large is less than 
$ \exp(- L(\ka - \ka_0)  |Z- \bar  Y|_{LM})$.  
Now   use the inequality  from  part I :
 \be
   |Z   -  \bar Y|_{LM} +    d_{LM} (\bar   Y)    \geq   d_{LM} ( Z)
\ee
to dominate the extracted factors  by   
  $ \exp(  - L( \ka - \ka_0)   d_{LM}(Z)    )$.
Thus we have
 \be   \label{stovepipe}
\begin{split}
 | ( \de E^+_k)^{(iv)}(Z )  | \ 
 \leq  & \    \cO(1)  \la_k^{\frac14 - 10 \ep}   e^{  - L( \ka - \ka_0)   d_{LM}(Z)    }  \sum_{Y_2:  \bar Y_2 =Z} \  \sum_{Y_1 \subset  Y_2}  \sum_{Y  \subset  Y_1}   \sum_{X: \bar X  =Y}\\
&   \exp \Big( - \frac12 \ka_1 |Y_2-Y|_{\bom'}  -      \ka_0    d_{M}(X)    \Big)\\
   \end{split}
 \ee 
 For the sum  over  $Y_1$ we drop connectedness conditions  and take
 \be   \sum_{Y \subset Y_1 \subset  Y_2} 1 \ \leq  \ \textrm{  number of subsets of  cubes in  } Y_2-Y_1\   \leq   2^{|Y_2-Y|_{\bom'}}
 \ee
 This is absorbed by replacing  $\frac12\ka_1 $ by  $\frac14 \ka_1$ in  (\ref{stovepipe}).
  Now use   $  \sum_{ Y_2 \subset Z  } \  \sum_{Y \subset  Y_2}  \leq       \sum_{Y \subset Z} \sum_{Y_2 \supset  Y}$.  Then 
the sum over  $Y_2$  is estimated by  lemma \ref{donut2}   in the appendix   (here $Y,Y_2$ are connected and we use $LM$ cubes)  
\be   
 \sum_{Y_2  \supset Y}   e^{ - \frac14 \ka_1 |Y_2-Y|_{\bom'} }  \leq    \exp \Big( C e^{-\frac18 \ka_1}  |Y|_{\bom'}  \Big) 
 \leq     \exp \Big(   C e^{-\frac18 \ka_1}   |Z|_{LM}  \Big)   \leq  \one  e^{ d_{LM}(Z)  }
\ee
The  second inequality   holds  since  $ |Y|_{\bom'}  \leq  L^3 |Z|_{LM}$.     In the last step we have used the inequality   $|Z|_{LM} \leq   \one   (1 + d_{LM}(Z)  )  $  and suppressed the constants by  taking
$M$ and hence $\ka_1$ large enough.
In the final  sum   $\sum_{Y \subset Z} \sum_{X: \bar X =Y}  = \sum_{X \subset Z}$   and  
\be   \sum_{X \subset  Z}   \  e^{ -  \ka_0      d_M(X)    } \leq   \one  |Z|_{M}  =   \one L^3  |Z|_{LM}   \leq  \one L^3  (d_{LM} (Z) +1) 
\leq  \one L^3  e^{d_{LM}(Z)} 
\ee
Then  taking   $ L( \ka  - \ka_0)-2  \geq    L( \ka  - \ka_0-2) $ yields  
\be  \label{snuffit}
 | ( \de E^+_k)^{(iv)}(Z )  |
 \leq    \    \cO(1) L^3  \la_k^{\frac14 - 10 \ep}  e^{  -   L ( \ka - \ka_0-2)  d_{LM} ( Z) } 
 \ee

\bigskip
\noindent (G.)  
Terms in  (\ref{foreign})    with   $
Z \subset    \La^c_{k+1}$   are  the     $  \tilde B_{k+1, \bpi^+}$  terms  in  (\ref{city}).    These are estimated by 
\begin{equation} 
\begin{split}
& \cO(1)L^3   \la_k^{\frac14 - 10 \ep}  \sum_{Z:      Z \subset    \La^c_{k+1}, Z  \cap \La_k  \neq  \emptyset} e^{- L(\ka- \ka_0-2)d_{LM}(Z) } \\
 \leq  \    &  \cO(1)  L^3   \la_k^{\frac14 - 10 \ep}  \sum_{\square  \subset  \bar \La_k - \La_{k+1}}   \sum_{Z  \supset  \square}   
  e^{- L(\ka-\ka_0-2)d_{LM}(Z) }     \leq      \cO(1)L^3   \la_k^{\frac14 - 10 \ep}  |  \bar  \La_k - \La_{k+1}|_{LM}\\
 \end{split}
\end{equation}
where the sum is over  $LM$ cubes $\square$.   
Since   $|\bar   \La_k - \La_{k+1}|_{LM}  =  M^{-3}       | \bar  \La^{(k+1)}_k - \La^{(k+1)}_{k+1}|$    this is a bound of the required form.

\bigskip
\noindent  (H.)  
The remaining terms in    (\ref{foreign})  satisfy  $Z  \#   \La_{k+1}$     and   yield the  active boundary terms  
$ B^{(E)}_{k, \bpi^+}  $ terms  in  (\ref{city}).
First note that    
each   such      $Z $     determines    a  $Z^+   \in   \cD^0_{k+1}( \bmod \Om^c_{k+1})$    by 
taking the union with   all  connected  components of  $\Om_{k+1}^c$   connected to    $Z$,  written  
$Z\to Z^+$.    We  define  
 \begin{equation}
 B^{(E)}_{k, \bpi^+}( Y)   = \sum_{Z   \#   \La_{k+1},   Z ^+ =Y  }    (\de  E^+_k)^{(iv)}( Z) 
\end{equation}
Then  we  have    
\begin{equation}
 \sum_{ Z  \#  \La_{k+1}}    (\de  E^+_k)^{(iv)}( Z) 
 =  \sum_{Y  \in   \cD^0_{k+1}( \bmod   \Om^c_{k+1}),Y  \#  \La_{k+1} }
  B^{(E)}_{k, \bpi^+}( Y)  
 \end{equation}
 To  estimate   $ B^{(E)}_{k, \bpi^+}( Y) $   first  note that
 \begin{equation}  \label{siouZ}
 d_{LM}(Z)   \geq   d_{LM}(Z^+, \bmod\  \Om_{k+1}^c)  
\end{equation}
Indeed  let   $\tau$ be a minimal tree joining the cubes in    $Z $  of length  $\ell(\tau) = LMd_{LM}(Z)$.   Then 
$\tau$  is also a tree joining the cubes in    $Z^+  \cap  \Om_{k+1}$  since   $Z^+ \cap  \Om_{k+1}=   Z \cap  \Om_{k+1}
\subset  Z $.   Hence   $\ell(\tau)  \geq   LMd_{LM}(Z^+, \bmod  \Om_{k+1}^c)$ and hence the result.  Using this  and  (\ref{snuffit})
gives    
\be 
\begin{split}  \label{drum1}
|  B^{(E)}_{k, \bpi^+}( Y) |  \leq   &   \cO(1) L^3  \la_k^{\frac14 - 10 \ep}  e^{-(L (\ka- \ka_0-2)-\ka_0)d_{LM}(Y, \bmod  \Om_{k+1}^c) }   
    \sum_{ Z   \subset  Y,  Z   \#   \La_{k+1} }  e^{- \ka_0 d_{LM}(Z) }   \\
\end{split}
\end{equation}    
But the sum  is bounded by  
\be  \label{drum2}
\begin{split}
   \one |  Y \cap \La_{k+1}|_{LM}    \leq &  |  Y \cap \Om_{k+1}|_{LM}   \leq  \one ( d_{LM}(   Y \cap \Om_{k+1} )  +1  )   \\
 = & \one (     d_{LM}(Y, \bmod\  \Om_{k+1}^c)  +1    )     \leq  \one   e^{d_{LM}(Y, \bmod  \Om_{k+1}^c) }    \\
 \end{split}
\ee
The coefficient of   $ d_{LM}(   Y \cap \Om_{k+1} )$  is then   $L (\ka- \ka_0-2)-\ka_0-1  \geq  L (\ka- 2\ka_0-3)$ and we have the result.
 \bigskip

 \begin{lem}    \label{second}   The   function   $ R^{(\leq   7)}_{ \bpi, \Om_{k+1}} $
can be written     
  \begin{equation}
  R^{(\leq   7)}_{ \bpi, \Om_{k+1}}   = \sum_{Y \in \cD^0_{k+1}: Y  \subset  \La_{k+1} }    R^{\loc}_{k,\bpi^+}( Y) 
  +    \sum_{  Y \in \cD^0_{k+1}( \bmod  \Om^c_{k+1}), Y  \# \La_{k+1}}     B^{(R)}_{k, \bpi^+}( Y)  
  + \ \  \tilde B_{k+1, \bpi^+}   \textrm{  terms  }  
 \end{equation}
 Here   $  R^{\loc}_{k,\bpi^+}( Y,    \Phi_{k+1}, W_k) $  and $  B^{(R)}_{k, \bpi^+}( Y,   \Phi_{k,\de  \Om_k},  \Phi_{k+1, \Om_{k+1}},  W_{k})  $ 
 are  strictly localized in the fields.    They are  analytic for  $ \Phi_k,  \Phi_{k+1}$  in    the domain (\ref{sandy})  and  $|W_k|  \leq   B_wp_k$.    
 On      this   domain they  satisfy     
\begin{equation} \label{lulu0}
\begin{split}
|   R^{\loc}_{k,\bpi^+}( Y) |  \leq  &   \cO(1)L^3  \la_k^{n_0}   e^{-L(\ka - \ka_0-2) d_{LM}(Y)}  \\
| B^{(R)}_{k, \bpi^+}( Y) | \leq  &      \cO(1) L^3 \la_k^{n_0}   e^{-L (\ka- 2 \ka_0 - 3) d_{LM} (Y,  \bmod  \Om^c_{k+1})}  \\
 \end{split}
\end{equation}
  \end{lem}
  \bigskip

\pr  The function   $  R^{(\leq  7)}_{ \bpi^+, \Om_{k+1}}(\La_k)  $  has  many parts,  which we  consider one by one.  

\bigskip

\noindent  \textbf{ The term $R^{(0)}_{\bpi,  \Om_{k+1}}=R_{k,\bpi}(\La_k)$}.     This   original  term   
  after the change of  variables has the form     
\begin{equation}
R_{k, \bpi}(\La_k)  =   \sum_{X \in \cD_k,  X  \subset  \La_k}  
   R_{k, \bpi  }(X, \Phi_{k, \de \Om_k}, \Psi^{\loc}_{k,\Om_{k+1}}(\bom')  +   ( C_{k,\bom'}^{1/2})^{\loc}W_k)  \\ 
\end{equation}
Our  hypotheses on the  fields     and lemma \ref{technical}    imply that   $(\Phi_{k, \de \Om_k},\Psi^{\loc}_{k,\Om_{k+1}}(\bom') )
 \in  \frac12\cP_k(\La_k, 2 \de)$.
   We  argue below that   $|W_k|  \leq   B_wp_k$  implies that  
  $(\Phi_{k, \de \Om_k}, (C_{k,\bom'}^{1/2})^{\loc}W_k)
 \in  \frac12\cP_k(\La_k, 2 \de)$.
 Then   the sum  is  in   $\cP_k(\La_k, 2 \de)$,    hence we 
are in the analyticity domain for 
$ R_{k, \bpi  }(X)$,   and  hence
\begin{equation}  \label{olligarch}
    |R_{k, \bpi}(X)| \leq     \la_k^{n_0}e^{-\ka d_M(X) }
\end{equation}
For the missing piece let  $W'_k  =  (C_{k,\bom'}^{1/2})^{\loc}W_k$.
By the bound  (\ref{around})   
 $| W'_k |  \leq   C p_k$.   Then  for  $\square  \subset  \Om_{k+1}^{\nat}$  we  have  $|W'_k - \phi_{k , \bom}(\square)(W'_k)|,
|\pa   \phi_{k , \bom(\square)}(W'_k)|$,  and   $ | \phi_{k , \bom(\square)}(W'_k)|$   all  bounded by  $Cp_k$. 
 Since  $Cp_k \leq  \frac12 \la_k^{-\frac14 - 2\de}$   this  gives   $W'_k \in    \frac12\cP_k(\square , 2 \de)$.
If    $\square  \subset   \La_k  -  \Om_{k+1}^{\nat}$   argue as in parts  (B.), (C.) of lemma \ref{technical}  to obtain  
$ ( \Phi_{k, \de  \Om_k}, W'_k) \in    \frac12\cP_k(\square , 2 \de)$.  Altogether  then  we have   $ ( \Phi_{k, \de  \Om_k}, W'_k) \in    \frac12\cP_k(\La_k , 2 \de)$.

Now   reblock  as in (\ref{twinkie})   defining    for  $Y \in  \cD^0_{k+1, \bom'}$  
\be   \label{twinkie2}
  R'_{k, \bpi  } ( Y)  =  \sum_{X:  \bar X  =Y, X \subset  \La_k }   R_{k, \bpi  }  (X)
  \ee
 Then   $R_{k, \bpi}(\La_k)  =   \sum_{Y }    R'_{k, \bpi  } ( Y) $.

We  localize   further  as  follows.    Again introduce   parameters  $s = \{ s_{\square} \}$  indexed by  the cubes of $\bom'$.  Referring to the definitions in section \ref{better},  for an  $LM$ cube  $\square$ in  $\Om_{k+1}$   we  replace    $G^0_{k+1 , \bom'}(\square^*)=G^0_{k+1 , \bom'}(s_{\square^*}=1,    s_{\square^{*,c}} = 0) $ 
with  normal coupling  in  $\square^*$ 
   by   $G^0_{k+1 , \bom'}(\square^*,s)=G^0_{k+1 , \bom'}(s_{\square^*},   s_{\square^{*,c}} = 0) $  with weakened coupling inside  $\square^*$.
Correspondingly we  replace  $\phi^0_{k+1,  \bom'}(\square^*)$  by  $\phi^0_{k+1,  \bom'}(\square^*,s)$,   we  replace   
 $\Psi_{k,\Om_{k+1}}(\bom',  \square^*) $   by    $\Psi_{k,\Om_{k+1}}(\bom',  \square^*,s) $,  
and  we  replace     $ \Psi^{\loc}_{k,\Om_{k+1}}(\bom')$  by    $\Psi^{\loc}_{k,\Om_{k+1}}(\bom',   s)$.
Also  referring  to  the  definitions in section \ref{bingo},  
we  replace   $ G_{k, \bom',  r}(\square^*) $
by    $ G_{k, \bom',  r}(\square^*,s) $ with weakened coupling inside  $\square^*$,  we  replace  $ C_{k, \bom', r}(\square^*) $ by  $ C_{k, \bom', r}(\square^*,s) $,
we  replace  $C^{1/2}_{k,  \bom'}( \square^*)$    by  $C^{1/2}_{k,  \bom'}( \square^*,s) $,   and we  replace
$ ( C^{1/2}_{k,  \bom'})^{\loc}$  by   $ ( C^{1/2}_{k,  \bom'})^{\loc}( s)$.   None of these   changes affect  the bounds on the fields,
 even for  $|s_{\square}|  \leq  e^{\ka_1}$.  
Finally instead of      
$ R'_{k, \bpi  }(Y,\Phi_{k, \de \Om_k},  \Psi^{\loc}_{k,\Om_{k+1}}(\bom')  +   ( C_{k,\bom'}^{1/2})^{\loc}W_k) $  we  introduce
\begin{equation}   
\label{sudden}
  R'_{k, \bpi  }(Y,s) = R_{k, \bpi  }\Big(Y, \Phi_{k, \de \Om_k}, \Psi^{\loc}_{k,\Om_{k+1}}(\bom',s)  +   ( C_{k,\bom'}^{1/2})^{\loc}(s)W_k \Big)
\end{equation}

Now    make a decoupling expansion   as   in the previous lemma.     
We  have  
\begin{equation}   \label{x1}
 R'_{k, \bpi}(Y, \Phi_{k, \de \Om_k},  \Psi^{\loc}_{k,\Om_{k+1}}(\bom')  +   ( C_{k, \bom'}^{1/2})^{\loc}W_k)
=  \sum_{Y_1  \supset  Y } ( R_{k, \bpi^+})'(Y, Y_1,  \Phi_{k, \de \Om_k},  \Phi_{k+1, \Om_{k+1}},  W_k  )
\end{equation}
where  for  $Y_1  \in  \cD^0_{k+1, \bom'}$
\begin{equation}
\begin{split}
&  ( R_{k, \bpi^+})'(Y, Y_1 )
=  \int   ds_{Y_1-Y}  \frac{\  \ \pa}{ \pa s_{Y_1-Y}}  \left[ 
  R'_{k, \bpi}(Y, s)
\right]_{s_{Y_1^c} = 0,   s_Y =1}\\
\end{split}
\end{equation} 
depends on  the indicated fields  only  in  $Y_1$.   
Note that in this case  only  terms  with  $Y_1 \subset  \Om^*_{k+1}$ contribute  because  of  the sharper localization. 
Again    using   Cauchy bounds  on the derivatives,  we   improve our estimates by a factor  $ \exp (   - (\ka_1-1) |Y_1-Y|_{\bom'} )$.

Next   define  
\begin{equation}
( R_{k, \bpi^+})''( Y_1 )  = \sum_{Y \subset  Y_1} ( R_{k, \bpi^+})'(Y, Y_1 ) 
 \end{equation}
and then    $  R_{k, \bpi}(\La_k) 
=    \sum_{Y_1}      R''_{k, \bpi^+}( Y_1) $. 
Further  we  reblock to  $Z  \in \cD^0_{k+1}$   defining  
\be
( R_{k, \bpi^+})'''( Z  )  = \sum_{Y_1:   \bar  Y_1  = Z   }    ( R_{k, \bpi^+})''( Y_1 ) 
\ee
and  then     
\be    \label{sunstroke}
 R_{k, \bpi}(\La_k) 
=    \sum_{Z  \cap \La_k  \neq  \emptyset}      R'''_{k, \bpi^+}(Z)   
\ee
\bigskip

Collecting  our  estimates we have 
\be
\begin{split}
& |  R'''_{k, \bpi^+}(Z)    |
 \leq   \    \cO(1)  \la_k^{n_0} \  \sum_{Y_1:  \bar Y_1 =Z} \   \sum_{Y  \subset  Y_1} 
 \sum_{X: \bar  X  =  Y  }       \   \
   \exp \Big(   - (\ka_1-1) |Y_1-Y|_{\bom'} -  \ka   d_{M}(X)    \Big)\\
   \end{split}
 \ee
This is estimated just as   in part  (F.) of the previous lemma,  except that  now there is just a sum over  $Y_1$ instead of $Y_1,Y_2$.
The result is  
\be  \label{x2}
 |  R'''_{k, \bpi^+}(Z)    |  \leq     \cO(1)L^3  \la_k^{n_0}e^{  -   L ( \ka - \ka_0-2)  d_{LM} ( Z) } 
\ee

Finally     in    (\ref{sunstroke})     divide the terms  into three classes
as  in the previous lemma.  Terms   with   $Y \subset   \La_{k+1}$
contribute  to   $  R^{loc}_{k,\bpi^+}$.
Terms with   $Y \subset   \La_{k+1}^c$   are  the     $\tilde  B_{k+1,\bpi^+}$  terms,   and have  the correct bounds
as   before.       Terms   with   $ Y  \#   \La_{k+1}$
   the   boundary  terms.
 We  adjoin connected components of  $\Om_{k+1}^c$ as  before,   and get    a contribution to    $ B^{(R)}_{k, \bpi^+}( Y)$.
\bigskip

\noindent    \textbf{ The term $R^{(3)}_{\bpi, \Om_{k+1}}$}.
    We  have  $
  R^{(3)}_{\bpi,  \Om_{k+1}}  =   \sum_{X \subset  \La_k}  R^{(3)}_{\bpi,  \Om_{k+1}} (X) 
 $
   where   for $X \in  \cD_k$
    \begin{equation}
  R^{(3)}_{\bpi,  \Om_{k+1}} (X) =    E_k^+ ( X,  \phi^0_{k+1,\bom'} +   \cW^{\loc} _{k,   \bom'}  +   \de   \phi_{k,\bom'})
-      E_k^+ ( X,  \phi^0_{k+1,\bom'} +   \cW^{\loc} _{k,   \bom'}  ) 
 \end{equation}
 Our  field  bounds  imply   $ \phi^0_{k+1,\bom'} +   \cW^{\loc} _{k,   \bom'}   \in \frac12 \cR_k$.     They  also   imply,  arguing 
 as  in   lemma \ref{singsong2},   that    $|\de \Psi_{k, \Om_{k+1}}(\bom')|  \leq   e^{-r_{k+1}}$   which   yields the bound on $\La_k$:
 \be 
     |   \de   \phi_{k,\bom'}   |   =     | \phi_{k, \bom(\La_k^*)}  ( 0,  \de \Psi_{k, \Om_{k+1}}(\bom')  )|
 \leq   C   e^{-r_{k+1}}   \ee
 with    similar bounds on the derivatives.    Hence  for  complex    $|t| \leq   e^{r_{k+1}}$  we have  
     $t  \de   \phi_{k,\bom'}  \in   \frac12  \cR_k$.   Now  
  $ t \to   E_k^+ (X,  \phi^0_{k+1,\bom'} +  \cW^{\loc} _{k,   \bom'}  + t  \de   \phi_{k,\bom'})$
 is  analytic  in    $|t| \leq   e^{r_{k+1}}$   and   we  can  write  
\be
  R^{(3)}_{\bpi,  \Om_{k+1}} (X) = \frac{1}{2\pi i}    \int_{ |t| =   e^{r_{k+1} }  } \frac{1}{t(t-1)}  E_k^+ ( X,  \phi^0_{k+1,\bom'} +    \cW^{\loc} _{k,   \bom'} +  t \de   \phi_{k,\bom'})
 \ee
Using  the bound  $|E_k^+(X)|  \leq   \one  \la_k^{-12 \ep}  e^{-\ka d_M(X)}$  on  $\cR_k$,    we have    (since  $e^{-r_{k+1}}  =  \cO( \la_k^n)$ for any  $n$)
\begin{equation}  \label{unlikely2}
 | R^{(3)}_{\bpi,  \Om_{k+1}} (X) |  \leq    \cO(1)    e^{-r_{k+1}}   \la_k^{-12 \ep} e^{- \ka d_M(X)}  \leq   \la_k^{n_0}  e^{- \ka d_M(X)}
   \end{equation}

Now    reblock  as in (\ref{twinkie})   defining   $ ( R^{(3)})'_{\bpi,  \Om_{k+1} } ( Y)$    for  $Y \in  \cD^0_{k+1, \bom'}$.  
Next  
   replace   $ \phi^0_{k+1,\bom'}$  by    $ \phi^0_{k+1,\bom'}(s)$  and     $ \cW^{\loc} _{k,   \bom'}$ by  
$  \cW^{\loc} _{k,   \bom'}(s)  \equiv     a_k  G_{k,   \bom(\La_k^*)}(s)  Q_k^T    (  ( C^{1/2}_{k,  \bom'} )^{\loc} (s) W_k )  $.
 Furthermore we  replace  $\de   \phi_{k,\bom'}$
by   $\de   \phi_{k,\bom'}(s)  \equiv  a_k  G_{k,   \bom(\La_k^*)}(s)  Q_k^T  \de    \Psi_{k,\Om_{k+1}}(\bom',s) $
   where    $\de    \Psi_{k,\Om_{k+1}}(\bom',s) =   \Psi^{\loc}_{k, \Om_{k+1}} (\bom',s) - \Psi_{k, \Om_{k+1}} (\bom',s)$.   Bounds are
   unaffected and   we  get  a  weakened form  $ ( R^{(3)})'_{  \bpi,  \Om_{k+1} } ( Y,s)$   analagous  to  (\ref{sudden}).
   Finally    proceed with the decoupling and reblocking   as in   (\ref{x1}) - (\ref{x2})  and obtain  
 contributions  to  $ R^{loc}_{k,\bpi^+}$,
$\tilde  B_{k+1,\bpi^+}$,   and      $ B^{(R)}_{k, \bpi^+}( Y)$  of the required form.
\bigskip

 \noindent  \textbf{ The term $R^{(7)}_{\bpi, \Om_{k+1}}$}.
 This is entirely similar  to   $R^{(3)}_{\bpi, \Om_{k+1}}$.  We  have
 $
  R^{(7)}_{\bpi,  \Om_{k+1}}  =   \sum_{X \subset  \La_k}  R^{(7)}_{\bpi,  \Om_{k+1}} (X) 
 $
  where
  \be
    R^{(7)}_{\bpi, \Om_{k+1}} (X) = E_k^+ (X,  \phi^0_{k+1,\bom'} +\cW_{k,\bom'}   + \de   \cW_{k,\bom'}    )    -   E_k^+ (X,  \phi^0_{k+1,\bom'} +\cW_{k,\bom'}    )
 \ee
   By the bound  (\ref{around}) on  $\de C^{1/2}_{k, \bom'}  $   we  have  on  $\La_k$:
 \begin{equation}
 |\de   \cW_{k,\bom'} |  =  a_k | G_{k, \bom(\La_k^*)}Q_k^T (\de C^{1/2}_{k, \bom'}  W_{k})|
 \leq   C e^{-r_{k+1}}  p_k 
 \end{equation}
So    we  can replace  $\de   \cW_{k,\bom'}$  by  $t\de   \cW_{k,\bom'}$  with  $|t|  \leq    e^{r_{k+1}}$  and
 still stay in the region of analyticity.    Hence 
 we  have the representation
 \begin{equation}
  R^{(7)}_{\bpi,  \Om_{k+1}} (X) = \frac{1}{2\pi i}    \int_{ |t|   =   e^{r_{k+1}}}  \frac{dt}{t(t-1)} 
  E_k^+ (X,  \phi^0_{k+1,\bom'} +  \cW_{k,\bom'}  +t  \de   \cW_{k,\bom'}  ) 
  \end{equation} 
  and the bound
   \begin{equation}
  |R^{(7)}_{\bpi,  \Om_{k+1}} (X ) |  \leq   \cO(1)  e^{-r_{k+1}}  \la_k^{-12 \ep} e^{- \ka  d_M(X)}  \leq   \la_k^{n_0}  e^{- \ka  d_M(X)}
  \end{equation}
 Now reblock, decouple, and reblock again   exactly as in the previous case,  with the same result.
  \bigskip

  \noindent  \textbf{ The term $R^{(1)}_{\bpi, \Om_{k+1}}$.  }
  After the change of variables  and a localization we have   
 \begin{equation}  \label{b2}
\begin{split}
  R^{(1)}_{\bpi,  \Om_{k+1}}
  =&  \sum_{\square \textrm{ on } \pa  \La_k}  \sfb_{\La_k}   \Big [ \pa  \phi^0_{k+1,\bom'} ,1_{ \square}  \cW^{\loc}_{k,\bom( \La_k^*)} \Big]   \\
-& \frac 12  \sum_{\square \subset   \La^c_k}  \Big(   a_k   \| Q_{k,\bom( \La_k^*)}  \cW^{\loc}_{k,\bom'}\|^2_{\square}  +  \|  \pa   \cW^{\loc}_{k,\bom'} \|^2_{*,\square}
  + \bar \mu_k   \|  \cW^{\loc}_{k,\bom'}\|^2_{\square}    \Big)   \\
 \end{split}
 \end{equation} 
 where the sum  is over  $M$-cubes $\square$.
 Then     $
 R^{(1)}_{\bpi,  \Om_{k+1}}=  \sum_{X  \subset   (\La_k^c)^{\sim}}  
   R^{(1)}_{\bpi,  \Om_{k+1}}(X)
$
if we   say   $ R^{(1)}_{\bpi,  \Om_{k+1}}(X)$   vanishes  for  $|X|_M  \geq  2$.

Now      $(   C^{1/2}_{k,  \bom'})^{\loc}  W_{k}$  is localized in  $\Om_{k+1}$  which is  separated from  $\La^c_k$  by  at least  $5[r_{k+1}]$
layers  of  $LM$-cubes.   Thus from the random walk expansion  for  $G_{k,   \bom(\La_k^*)} $  joining points in  $\La^c_k$  and $\Om_{k+1}$   we  can extract a factor  $e^{-r_{k+1}}$.  Hence by a variation of
(\ref{twoone1})
and  have    have    on     $(\La^c_k)^{\sim}$:
\begin{equation}     \label{sound}
|\cW^{\loc}_{k,\bom'}|
=  \Big |  \phi_{k, \bom(\La_k^*) } \Big( ( C^{1/2}_{k,  \bom'} )^{\loc}  W_k \Big)\Big |  \leq  Ce^{-r_{k+1}}  \|  ( C^{1/2}_{k,  \bom'} )^{\loc}  W_k \|_{\infty}  
 \leq    C e^{-r_{k+1}}  p_{k}
\end{equation}
 Near   $\pa \La_k$ we are in $  \Om_k(\La_k^*)$  and  as  noted previously   $|\pa  \phi^0_{k+1,\bom'}|   \leq   C \la_k^{-\frac14- 2\de}$.  Also  $|\sfb_{\La_k} [ \pa \phi, 1_{\square}  \cW]|  \leq  M^2\|  \pa \phi  \|_{\infty} \| \cW \|_{\infty}$.  These lead to the bound
\begin{equation}
 \sfb_{\La_k}   \Big [ \pa  \phi^0_{k+1,\bom'} ,1_{ \square}  \cW^{\loc}_{k,\bom( \La_k^*)} \Big]   \leq    M^2 ( C \la_k^{-\frac14- 2\de})(C e^{-r_{k+1}}  p_k )
 \leq      \la_k^{n_0}  
 \end{equation}
 
Now  consider  the   the term  $ \|  \cW^{\loc}_{k,\bom'}\|^2_{\square}  $    for  $\square  \subset   \La_k^c$.
We  have  
\be  \label{sound2}
 \|  \cW^{\loc}_{k,\bom'}\|^2_{\square}  \leq    M^3    \|\cW^{\loc}_{k,\bom'} \|_{\infty} ^2
\leq  M^3 (C e^{-r_{k+1}}  p_k )^2  \leq      \la_k^{n_0}     
\ee
The   term  $ \|  Q_{k,\bom( \La_k^*)}  \cW^{\loc}_{k,\bom'}\|^2_{ \square}  $  is treated similarly.
The  derivative   term    needs more attention.
For  an  $L^{-(k-j)}$ cube    $\De_y \subset  \de \Om_j(\La_k^*) \cap  \square$,  by a variation of  (\ref{twoone3})
\be
\begin{split}
|1_{\De_y} \pa   \cW^{\loc}_{k,\bom'}|  
=&  \Big| \sum_{y' \in  \Om^{(k)}_{k+1}}  1_{\De_y}   \pa   \phi_{k, \bom(\La_k^*) } \Big(1_{\De_{y'}}(  C^{1/2}_{k,  \bom'} )^{\loc}  W_k \Big) \Big| \\
\leq  &  \sum_{y' \in  \Om^{(k)}_{k+1}}CL^{k-j} e^{-r_{k+1}}  e^{- \frac14 \ga_0 d_{\bom(\La_k^*)}(y,y')  } \|(  C^{1/2}_{k,  \bom'} )^{\loc}  W_k\|_{\infty }  \\
\leq   &  C e^{-r_{k+1}}  \|(  C^{1/2}_{k,  \bom'} )^{\loc}  W_k\|_{\infty } 
\leq     C e^{-r_{k+1}}  p_k  \\
\end{split}
\ee
Here we  have used the fact that  because      of our separation conditions  we  have   for  $y \in \de \Om_j^{(j)}(\La_k^*)$
  and  $y' \in \Om^{(k)}_{k+1} \subset \Om^{(k)}_k(\La_k^*)$
\be 
d_{\bom(\La_k^*)}(  y,y'  )   \geq   RM \max \{ |k-j|-1,  0  \}
\ee
Then  for $M$ large enough  the  factor  $ \exp(- \frac14 \ga_0 d_{\bom(\La_k^*)}(y,y') )$  is      enough to dominate the   $L^{k-j}$  and  give convergence of the sum over  $y'$.
Now we have as before  
\be
\|  \pa   \cW^{\loc}_{k,\bom'} \|^2_{*,\square}  \leq  M^3  (   C e^{-r_{k+1}}  p_k )^2   \leq  \la_k^{n_0}
\ee
Altogether  then   
\be    \label{summertime}  |R^{(1)}_{\bpi,  \Om_{k+1}}(\square)|  \leq   \one   \la_k^{n_0}
\ee
Now reblock, decouple, and reblock again  as before,  with the same result.
  \bigskip

\noindent  \textbf{ The term $R^{(2)}_{\bpi, \Om_{k+1}}$}.
First we  localize    $J^*_{\La_k,   \Om_{k+1}} $  defined in   (\ref{jstar}).      
\be 
J^*_{\La_k, \Om_{k+1}} ( \Phi_{k+1}, \Phi_k, \phi)  = \sum_{\square \subset  \Om_{k+1}}   \frac{a}{2L^2}  \|\Phi_{k+1}   -  Q \Phi_{k}\|^2_{  \square}   
+   \sum_{\square \subset   \La_k} S^*_k(\square,  \Phi_{k} ,   \phi )  \equiv    \sum_{\square}  J^*_{\La_k, \Om_{k+1}}(\square) 
\ee
where the sum is over  $M$-cubes $\square$.
For   $|\Phi_k|,  | \Phi_{k+1}|,  |\phi|,  | \pa \phi|     \leq C  \la_k^{-\frac14 - 2 \de}$   we  have the  bound   
\be     |J^*_{\La_k, \Om_{k+1}}(\square)|   \leq   C   M^3    \la_k^{-\frac12 - 4 \de}
\ee
Then   $ R^{(2)}_{\bpi,  \Om_{k+1}}   = \sum_\square    R^{(2)}_{\bpi,  \Om_{k+1}} (\square)$  where  
 \be 
\begin{split}
& R^{(2)}_{\bpi,  \Om_{k+1}}(\square)\\
 =
   &     J^*_{\La_k, \Om_{k+1}}\Big(\square,  \Phi_{k+1}, \hat  \Psi_{k,\bom'} + (C_{k, \bom'}^{1/2}W)^{\loc} 
     +    \de   \Psi_{k,\Om_{k+1}}(\bom'),    \ \   \phi^0_{k+1,\bom'} +  \cW^{\loc}_{k,   \bom(\La_k^*)}  +   \de   \phi_{k,\bom'} \Big )    \\
 -  &     J^*_{\La_k, \Om_{k+1}}\Big(\square,  \Phi_{k+1},  \hat  \Psi_{k, \bom'}  + ( C_{k, \bom'}^{1/2}W )^{\loc},
   \phi^0_{k+1,\bom'} +  \cW^{\loc}_{k,   \bom(\La_k^*)}  \Big) \\
\end{split}
\ee 
The hypotheses of the lemma give    the fields  the    $ C  \la_k^{-\frac14 - 2 \de}$  bound  as indicated in earlier  steps,   and so
$   | R^{(2)}_{\bpi,  \Om_{k+1}}(\square) |     \leq   C   M^3    \la_k^{-\frac12 - 4 \de}$.
However we  also know that      $ \de   \Psi_{k,\Om_{k+1}}(\bom')$  and   $ \de   \phi_{k,\bom'} 
 $  are 
$\cO(e^{-r_{k+1}})$    so  we  can multiply these  factors by  complex  $|t|  \leq   e^{r_{k+1}}$  and
still have the same  bound.   Therefore  
\begin{equation}  \label{sam2}
\begin{split}
   R^{(2)}_{\bpi,  \Om_{k+1}}(\square)
 =  &   \frac{1}{2\pi i}   \int_{ |t|   =   e^{r_{k+1}}}  \frac{dt}{t(t-1)} 
    J^*_{\La_k, \Om_{k+1}}\Big(\square,  \Phi_{k+1},  \hat \Psi_{k,\bom'} + (C_{k, \bom'}^{1/2}W)^{\loc} 
     +   t \de   \Psi_{k,\Om_{k+1}}(\bom'),    \\   & \hspace{2in}   \phi^0_{k+1,\bom'}+  \cW^{\loc}_{k,   \bom(\La_k^*)}  +  t \de   \phi_{k,\bom'}  \Big) dt   \\
\end{split}
\end{equation} 
which   leads  to the estimate 
\begin{equation}  \label{conch}
| R^{(2)}_{\bpi,  \Om_{k+1}}(\square)|   \leq    C   M^3    \la_k^{-\frac12 - 4 \de} e^{-r_{k+1}}
 \leq  \la_k^{n_0} 
\end{equation}
 Now reblock, decouple, and reblock again  as before,  with the same result.
  \bigskip

\noindent  \textbf{ The term $R^{(4)}_{\bpi, \Om_{k+1}}$}.
First     write  $R^{(4)}_{\bpi, \Om_{k+1}}  =  \sum_{\square  \subset  \Om_{k+1}}  R^{(4)}_{\bpi, \Om_{k+1}}(\square)$  where 
the sum is over  $LM$-cubes $\square$  and   
\begin{equation}
R^{(4)}_{\bpi, \Om_{k+1}}(\square)    
  =     <  C^{-1/2}_{k,  \bom'}W_k,1_{\square} \de C^{1/2}_{k,  \bom'}W_k>  
   - \frac12     <  \de C^{1/2}_{k,  \bom'}W_k, 1_{\square}C^{-1}_{k,  \bom'}  \de C^{1/2}_{k,  \bom'}W_k>   
\end{equation}
 Then  taking bounds    from  lemma  \ref{stem}  and using  $|W_k|  \leq  B_w p_k$    we  have  
 \begin{equation}
 \begin{split}
  |C^{-1/2}_{k,  \bom'}W_k|  \leq  & Cp_k  \\
 |\de C^{1/2}_{k,  \bom'}W_k|  \leq  & Cp_k  e^{-r_{k+1}}  \\
 |C^{-1}_{k,  \bom'}  \de C^{1/2}_{k,  \bom'}W_k|    \leq  &  Cp_k e^{-r_{k+1}}   \\
 \end{split}
 \end{equation}
Therefore 
\begin{equation}  \label{muddy}
|R^{(4)}_{\bpi, \Om_{k+1}}(\square)    |  \leq    M^3 Cp_k e^{-r_{k+1}}     \leq  \la_k^{n_0}
\end{equation}
 Now reblock, decouple, and reblock again  as before,  with the same result.
\bigskip
 
 \noindent  \textbf{ The term  $  R^{(5)}_{ \bpi,  \Om_{k+1}}   $  }     The bounds  of lemma  \ref{r5} suffice.  These  terms contribute
 to   $R_{k, \bpi^+}^{\loc}$   and  $\tilde  B_{k+1, \bpi^+}$.
\bigskip

\noindent
 \textbf{ The term      $  R^{(6)}_{ \bpi,  \Om_{k+1}}  $  }  Take  the result       $  R^{(6)}_{ \bpi,  \Om_{k+1}} = \sum_{\square \subset  \Om_{k+1} } R^{(6)}_{ \bpi,  \Om_{k+1}} (\square) $ of   lemma \ref{r6},
 and  split into  $\square \subset  \La_{k+1}$  and   $\square \subset    \La^c_{k+1}$.    In the former  case   $ |R^{(6)}_{ \bpi,  \Om_{k+1}} (\square)|  \leq  \la_k^{n_0}$
 by  (\ref{improved})  and the terms contribute to  $R_{k, \bpi^+}^{\loc}$.   In the  latter  case   $ |R^{(6)}_{ \bpi,  \Om_{k+1}} (\square)|  \leq  CM^3   =  C\   \Vol(\square)$
  and the terms contribute to   $\tilde  B_{k+1, \bpi^+}$.   This completes the proof.
\bigskip

 \begin{lem}    \label{third} 
   \begin{equation}
 \begin{split}
 B_{k, \bpi}(\La_k)  
  =  &  \sum_{       Y  \in \cD^0_{k+1}(\bmod  \Om^c_{k+1}), Y \#   \La_{k+1}     }   
    B^{(B)}_{k, \bpi^+}(Y) 
   +   \tilde  B_{k+1, \bpi^+}  \textrm{  terms }
\end{split}
\end{equation}
where   $ B^{(B)}_{k, \bpi^+}(Y,  \Phi_{k+1, \bom^+}, W_{k+1,\bpi^+}, W_{k, \La_{k+1}})   $  is  strictly local in the fields.  It    is analytic   in  $\Phi_{k+1, \bom^+}$  in  $ \cP^0_{k+1, \bom^+} $   and   $|W_j|  \leq     B_w \   p_j L^{\frac12(k-j)}$  for    $j=0,1, \dots,  k$,
and satisfies there
   \begin{equation}
|  B^{(B)}_{k, \bpi^+}(Y) |  \leq   
  \la_k^{n_0}   e^{ -L( \ka  - \ka_0  -3)   d_{LM}(Y, \bmod  \Om^c_{k+1}) }
\end{equation}
 \end{lem}
\bigskip

\re  The proof is similar to the proof   for   the   first  term in   lemma  \ref{second},  except  that now   there are holes which localize 
 spectator variables
left over from  the early stages of the expansion.
 \bigskip

\pr   We  are studying 
\begin{equation}  \label{spitoon}
 B_{k, \bpi}(\La_k)    =
 \sum_{     X \in \cD_k(\bmod  \Om^c_{k}), X \#   \La_k   }           B_{k, \bpi}  \Big(X, \Phi_{k, \bom \cap  \Om^c_{k+1}},    \Psi^{\loc}_{k, \Om_{k+1}}(\bom') 
 +   ( C_{k, \bom'}^{1/2})^{\loc}W_{k} , W_{k,\bpi}\Big)  
\end{equation}
Our assumptions on the fields and   lemma  (\ref{technical})      imply   that  $ (\Phi_{k, \bom \cap  \Om^c_{k+1}},    \Psi^{\loc}_{k, \Om_{k+1}}(\bom') 
 +   ( C_{k, \bom'}^{1/2})^{\loc}W_{k} )    \in   \cP_{k, \bom}$.   
Together with the bounds on $W_j$  this shows that we are in the domain  of  analyticity for   $  B_{k, \bpi} (X)$
and have  
\begin{equation}  \label{snoopy}
|  B_{k, \bpi} (X)  |  \leq  B_0      \la_k^{\beta}  e^{ - \ka     d_M(X, \bmod  \Om^c_{k} ) }
\end{equation}
If   it happens that  $X \subset  \Om_{k+1}^c$,  then there is no dependence on  the nonlocal fields   $\Psi^{\loc}_{k, \Om_{k+1}}(\bom') 
 +   ( C_{k, \bom'}^{1/2})^{\loc}W_{k}$.  Such terms sum up to a contribution to  $\tilde B_{k+1, \bpi^+}$.

 For the remaining terms we     reblock  to  polymers   $Y \in \cD^0_{k+1}$ by  
 \be  \label{able}
   B'_{k, \bpi} (Y ) =     \sum_{ \bar X  = Y:  X \cap \Om_{k+1}  \neq  \emptyset ,  X \#   \La_k,  }   B_{k, \bpi} (X)   
    \ee
 which  depends on the same variables.   
 
 The only non-locality  in $ B'_{k, \bpi} (Y )$ comes from  the  $\Psi^{\loc}_{k, \Om_{k+1}}(\bom') 
 +   ( C_{k, \bom'}^{1/2})^{\loc}W_{k}$    in  $Y \cap  \Om_{k+1}$.  Hence we  temporarily  treat   $  B'_{k, \bpi} (Y )$ 
 as localized in  $Y \cap  \Om_{k+1}$.    We  introduce  weakening parameters  $\{ s_{\square}  \}$  for 
 elementary cubes  $\square$ in  $\bom'$,   replace   $ \Psi^{\loc}_{k, \Om_{k+1}}(\bom') 
 +   ( C_{k, \bom'}^{1/2})^{\loc} W_{k} $  by  $ \Psi^{\loc}_{k, \Om_{k+1}}(\bom',s) 
 +   ( C_{k, \bom'}^{1/2})^{\loc}(s) W_{k} $   and call the result   $  B'_{k, \bpi} (Y, s ) $. As in  (\ref{able})  this  is  a sum over  certain  $ B_{k, \bpi} (X,s) $ which    satisfy   (\ref{snoopy}).

Now  we  make the decoupling expansion  based  on  $Y \cap  \Om_{k+1}$.   It is a little different from the previous
expansions since  $Y  \cap  \Om_{k+1}$ is not necessarily connected.
We  have  
\begin{equation}  
\begin{split}
& B'_{k, \bpi}(Y, \Phi_{k, \bom \cap  \Om^c_{k+1}} ,  \Psi^{\loc}_{k,\Om_{k+1}}(\bom')  +   ( C_{k, \bom'}^{1/2})^{\loc}W_k)  \\
&   \hs    = \sum_{Y_1  \supset  (Y \cap  \Om_{k+1}) }  B_{k, \bpi^+}(Y, Y_1,   \Phi_{k+1, \bom^+}, W_{k+1, \bpi^+} ,  W_{k, \La_{k+1} } )\\
\end{split}
\end{equation}
Here    $Y_1$  is a multiscale object   for $\bom'$   which   may not be connected, but has  the property that each connected component  of  $Y_1$  
contains at  least one  connected component  of  $Y$.   We  define   
\begin{equation}
 (  B_{k, \bpi^+})'(Y, Y_1 )
=  \int   ds_{Y_1-(Y\cap \Om_{k+1})}  \frac{\  \ \pa}{ \pa s_{Y_1-(Y\cap \Om_{k+1})}}  \left[ 
  B'_{k, \bpi}(Y, s)
\right]_{s_{Y_1^c} = 0,   s_{Y \cap  \Om_{k+1}} =1}
\end{equation} 
which     depends on  the indicated fields  only  in  $Y \cup  Y_1$. 
We  have made the identifications  $ \Phi_{k+1, \bom^+}  =   ( \Phi_{k, \bom \cap  \Om^c_{k+1}},   \Phi_{k+1, \Om_{k+1}})$ 
  and  $W_{k+1, \bpi^+}   = ( W_{k, \bpi},  W_{k, \Om_{k+1} - \La_{k+1}})$.  
Once  again
   only  terms  with  $Y_1   \subset  \Om^*_{k+1}$ contribute here.  
Using   Cauchy bounds   we   improve our estimate  on  $  B'_{k, \bpi} (Y, s ) $ by a factor  
$ \exp (   - (\ka_1-1) |Y_1-(Y \cap  \Om_{k+1})|_{\bom'} )$.

Let   $\bar Y_1$  be  all $LM$ cubes intersecting $Y_1$  and     let   $      Z_0   =   Y  \cup  \bar    Y_1 $.
 This is connected and hence an element of  $\cD^0_{k+1}$.
Then  let  $Z=Z_0^+$  be the union  of  $Z_0$  with any connected components  of  $\Om_{k+1}^c$ connected to    $Z_0$. 
 Then    $Z  \in \cD^0_{k+1} (  \bmod  \Om_{k+1}^c)$  and the composite process is denoted  $Y, Y_1 \to Z$.  We  define for such $Z$
\be   
  (B_{k, \bpi^+})''( Z )   =    \sum_{Y, Y_1  \to Z, \  Y_1 \supset  (Y \cap  \Om_{k+1}) }      ( B_{k, \bpi^+})'(Y, Y_1 )
\ee
and  then  
\be   \label{spitoon2}
 B_{k, \bpi}(\La_k)    =     
 \sum_{     Z \in \cD^0_{k+1}(\bmod  \Om^c_{k+1} ), Z \#  \La_k,   Z  \cap  \Om_{k+1} \neq  \emptyset  }        (B_{k, \bpi^+})''( Z )    
  +   \tilde  B_{k+1, \bpi^+}  \     terms 
 \ee

Collecting our estimates  we  have   
\be
\begin{split}
|(B_{k, \bpi^+})''( Z )  |\
\leq  & \   \cO(1)  B_0 \la_k^{\beta}\\  
  \sum_{Y, Y_1  \to Z,   Y_1  \supset  Y \cap  \Om_{k+1} }  &e^{   - (\ka_1-1) |Y_1-(Y \cap  \Om_{k+1})|_{\bom'} }   
    \sum_{\bar X  =  Y,  X \cap  \Om_{k+1} \neq  \emptyset, X \# \La_k}     e^{- \ka    d_M(X, \bmod  \Om^c_{k}) }   \\
\end{split}
\ee

To  extract a decay factor first note that  for $\bar X  =Y$
\be   d_M(X, \bmod \Om_k^c)   \geq    Ld_{LM}(Y, \bmod \Om_{k+1}^c)  
   \ee
This follows since if $\tau$ is a minimal tree on the $M$-cubes  in $X \cap  \Om_k$  with $\ell(\tau)  =  Md_M(X, \bmod \Om_k^c) $,
then it is also a tree on the  $M$-cubes in  $X \cap  \Om_{k+1}$,   and hence on the $LM$ cubes   in  $Y \cap  \Om_{k+1}$.   So 
$\ell(\tau)  \geq    LMd_{LM}(X, \bmod \Om_{k+1}^c)$.   We also note that  
\be   |Y_1-(Y \cap  \Om_{k+1})|_{\bom'}    \geq  |\bar  Y_1  -  (Y \cap  \Om_{k+1})|_{LM} 
\ee
and  we  can take this factor with  a coefficient  $L$  by borrowing from  the $\ka_1-1$.
   Now
we claim that  
\be    \label{shonuff}
LM |\bar  Y_1  -  (  Y \cap  \Om_{k+1})|_{LM} +    L Md_{LM}(  Y, \bmod\  \Om^c_{k+1})\
\geq  \  LM d_{LM}(Z   , \bmod\  \Om^c_{k+1})
\ee
To  see this  let   $\tau$ be a minimal  tree  on the  $LM$ cubes in    $Y \cap   \Om_{k+1}$   with
  $\ell(\tau) =  LM d_{M}(X, \bmod  \Om_k) $.   Let   $\{ \tau_{\al} \} $   be  trees  on  the $LM$ cubes   on the    connected components  of  
   $\bar  Y_1  -  (Y \cap  \Om_{k+1})$.  Then
$\tau$    joined to the   $\{ \tau_{\al} \} $   gives   a tree  $\tau'$   with  $\ell(\tau')$ equal to the right side of   (\ref{shonuff}).   See lemma 20 
in part I for more details.
The  tree   $\tau'$ is  constructed to  connect the    $LM$ cubes  in
\be
\begin{split}
 & ( Y  \cap   \Om_{k+1} )   \cup   (\bar  Y_1  -  (  Y \cap  \Om_{k+1})) 
  =     ( Y  \cap   \Om_{k+1} )   \cup   \bar  Y_1     \\
    &    \supset     (  Y  \cup   \bar  Y_1  )  \cap    \Om_{k+1}   
 =   Z_0  \cap  \Om_{k+1}   =  Z \cap  \Om_{k+1}     \\
  \end{split} 
\ee
This shows that   $\ell(\tau')  \geq   LM d_{LM}(Z, \bmod  \Om_{k+1})$,  and hence (\ref{shonuff}) is established.

By the above remarks our estimate becomes  
\be
\begin{split}
|(B_{k, \bpi^+})''( Z )  |
\leq  \ &  \cO(1)   \la_k^{\beta}B_0\   e^{-L(\ka - \ka_0 )d_{LM}(Z   , \bmod\  \Om^c_{k+1}) }
   \\
&  \sum_{Y, Y_1  \to Z,   Y_1  \supset  Y \cap  \Om_{k+1} }  e^{   - \ka_1/2 |Y_1-(Y \cap  \Om_{k+1})|_{\bom'} }   
    \sum_{\bar X  =  Y,  X \cap  \Om_{k+1} \neq  \emptyset}     e^{- \ka_0    d_M(X, \bmod  \Om^c_{k}) }   \\
 \end{split}
\ee
Relax the sum over  $Y,Y_1$  to just   $Y \subset  Z,  Y_1 \supset  Y \cap  \Om_{k+1}$.    The sum over  $Y_1$  is   estimated  by lemma
\ref{donut2}  by 
\be    \sum_{  Y_1  \supset  Y \cap  \Om_{k+1} }  e^{   - \ka_1/2 |Y_1-(Y \cap  \Om_{k+1})|_{\bom'} }  
\leq  e^{C e^{- \ka_1/4}|Y \cap  \Om_{k+1}|_{LM}  } \leq   \one  e^ { d_{LM}(Z   , \bmod\  \Om^c_{k+1}) }
\ee
The second  step follows  by   $  |Y \cap  \Om_{k+1}|_{LM}    \leq   |Z \cap  \Om_{k+1}|_{LM}  \leq      \one(d_{LM}(Z,  \bmod\  \Om^c_{k+1}) +1)  $ as in       (\ref{drum2}).      The constants are suppressed by   taking $M$ and hence $\ka_1$ sufficiently large.
Identifying  $\sum_{Y \subset Z}  \sum_{\bar X  = Y} $  as  $\sum_{X \subset  Z}$  and using   (\ref{snow})
the remaining sum is   dominated by 
\be   
\begin{split}
&  \sum_{X  \subset Z,  X \cap  \Om_{k+1} \neq  \emptyset}     e^{- \ka_0    d_M(X, \bmod  \Om^c_{k}) }  
 \leq  \one   |Z  \cap  \Om_{k+1}|_M  \\   & =     \one  L^3 |Z  \cap  \Om_{k+1}|_{LM} \leq   \one L^3 e^ { d_{LM}(Z   , \bmod\  \Om^c_{k+1}) }\
\end{split}
\ee
Thus   we obtain  for  $Z  \#  \La_k$  and  $Z \cap  \Om_{k+1} \neq  \emptyset$
\be
|(B_{k, \bpi^+})''( Z )  |\leq    \cO(1) L^3 B_0 \la_k^{\beta}\   e^{-L(\ka - \ka_0 -2)d_{LM}(Z   , \bmod\  \Om^c_{k+1})}
\ee

In the sum   (\ref{spitoon2})   consider  terms  with  $Z  \subset   \La^c_{k+1}$.    These  are  $\tilde B_{k+1, \bpi^+}$  terms  for they have the proper localization,  and the sum of these terms   can be estimated
by $  \cO(1) L^3 B_0 \la_k^{\beta}  $  times
\be   
\begin{split}
& \sum_{   Z   \in   \cD^0_{k+1} ( \bmod  \Om_{k+1}^c),Z \# \La_k, Z \subset  \La^c_{k+1}       }
 e^{-L(\ka - \ka_0 -2)d_{LM}(Z   , \bmod\  \Om^c_{k+1})}
 \\
  &  \hs \leq     \one|\bar  \La_k  -  \La_{k+1}|_{LM}  \leq    \one|  \La^{(k+1)}_k  -  \La^{(k+1)}_{k+1}|       \\
 \end{split}
\ee
The  last step follows  since the number of $LM$ cubes in  $\bar  \La_k  -  \La_{k+1}$ is less than the number of  $M$ cubes  in 
 $  \La_k  -  \La_{k+1}$,   which is less than the number of $L$ cubes in   $  \La_k  -  \La_{k+1}$.

Now consider  terms in  (\ref{spitoon2})  with  $Z  \#    \La_{k+1}$.  
  These are the terms  $(B^{(B)}_{k, \bpi^+})( Z )=(B_{k, \bpi^+})''( Z )$.   Since  also   $Z  \#   \La_k$  we  have  $Z \# \Om_{k+1}$
  and so  $Z$ must have cubes in $\Om_{k+1}$  on the boundary     and  in  $\La_{k+1}$,  and these are necessarily    a distance  at least  $r_{k+1}LM$  apart.
  Then  any tree  joining the $LM$ cubes in   $Z \cap  \Om_{k+1}$  must  have length at least  $r_{k+1}LM$. Hence  
  $LMd_{LM}(Z   , \bmod\  \Om^c_{k+1})  \geq  LM r_{k+1}$    and therefore     $d_{LM}(Z   , \bmod\  \Om^c_{k+1})  \geq  r_{k+1}$.
We  use    this to extract a tiny   factor  $e^{-r_{k+1}}$
leaving  say   $e^{     -    L(\ka- \ka_0 -3)   d_M(Z, \bmod  \Om^c_{k})  }$. 

Then  for  $\la_k$ sufficiently small   take   $ \cO(1) L^3 B_0  \la_k^{\beta} e^{-r_{k+1}}   \leq  \la^{n_0}_k$.
This   is the basic mechanism which keeps the boundary terms from growing. Terms  which survive many steps  must  have  a  large extent     and  hence  a  tiny value.     Altogether then  we have the announced bound  
\be    
|  B^{(B)}_{k, \bpi^+}(Z) |  \leq   
  \la_k^{n_0}   e^{ -L( \ka  - \ka_0  -3)   d_{LM}(Z, \bmod  \Om^c_{k+1}) }
\ee
This completes the proof.
\bigskip

\noindent
\textbf{Summary:}
We  collect the boundary    terms  by 
\begin{equation}
B^{\loc}_{k, \bpi^+}(Y)    =    B^{(E)}_{k, \bpi^+}(Y ) +    B^{(R)}_{k, \bpi^+}( Y)  
+   B^{(B)}_{k, \bpi^+}(Y) 
\end{equation}
Inserting the results of the last   three  lemmas   we
have  for the fluctuation integral:
\begin{equation}  \label{primeprime}
\begin{split}
&\Xi'_{k,  \bpi^+}   = \exp\Big(   \tilde B_{k+1} \textrm{   terms   }    \Big)\Xi''_{k,  \bpi^+}  \\
& \Xi''_{k,  \bpi^+}   
    =   \int d\mu^*_{\La_{k+1}}(W_k)
    \exp    \Big(   \sum_{ Y  \subset  \La_{k+1}}  ( \de   E^+_k)^{\loc} (Y )       +  R^{\loc}_{k, \bpi^+}(Y))+  
     \sum_{ Y  \#  \La_{k+1}  }   B^{loc}_{k, \bpi^+}(Y)    \Big) \\
      \end{split}        
\end{equation}
Here $Y  \in   \cD^0_{k+1}(  \bmod  \Om^c_{k+1})$  and  $ ( \de   E^+_k)^{\loc} (Y)= ( \de   E^+_k)^{\loc}(Y, \phi,  W_{k}  )$   at  $\phi = \phi^0_{k+1,\bom'}$.
It   is analytic in  
$\phi \in \frac 12  \cR_k$   and  $|W_k|  \leq  B_w p_k$   and  (relaxing the bounds a bit)   satisfies  there
\begin{equation}   
| ( \de   E^+_k)^{\loc}(Y)|  \leq     \cO(1)L^3  \la_k^{\frac14 - 10 \ep}  e^{   -L( \ka  - 3\ka_0  -3) d_{LM}(Y)  }
\end{equation}
Also    $ R^{\loc}_{k, \bpi^+}(Y,  \Phi_{k+1}, W_k)$   is    analytic      in    the domain (\ref{sandy})  and  $|W_k|  \leq   B_wp_k$ and satisfies there  
\begin{equation}
| R^{\loc}_{k, \bpi^+}(Y) | \leq   \cO(1)L^3 \la_k^{n_0}  e^{  -L( \ka  -3 \ka_0  -3) d_{LM}(Y)  } 
\end{equation}
Also 
$ B^{\loc}_{k, \bpi^+}(Y,  \Phi_{k+1, \bom^+}, W_{k+1, \bpi^+},  W_{k, \La_{k+1}}  )$
is     analytic in the domain  $\Phi_{k+1, \bom^+}  \in  \cP^0_{k+1}, \bom^+$  and   $|W_j|  \leq     B_w \   p_j L^{\frac12(k-j)}$            and satisfies there
\begin{equation}
| B^{\loc}_{k, \bpi^+}(Y)|
 \leq   \cO(1)L^3   \la_k^{\frac14 -10 \ep}  e^{  -L ( \ka  - 3\ka_0  -3) d_{LM}(Y,\bmod  \Om^c_{k+1} ) }
 \end{equation}

 We   can  also  write   
 \be     \Xi''_{k,  \bpi^+}   
    =   \int      d\mu^*_{\La_{k+1}}(W_k)
    \exp    \Big(   \sum_Y  H_{k,  \bpi^+} (Y)  \Big)  \ee
where    the sum is over  $Y  \in   \cD^0_{k+1}(  \bmod  \Om^c_{k+1})$ and  
 \be   
  H_{k,  \bpi^+} (Y)  =  \Big(  ( \de   E^+_k)^{\loc} (Y )       +  R^{\loc}_{k, \bpi^+}(Y)\Big)1_{Y \subset  \La_{k+1}}+  
        B^{loc}_{k, \bpi^+}(Y)  1_{Y \#  \La_{k+1}}
 \ee       
This     is analytic in the smallest of the three domains  which is   
\be   \label{500}
\Phi_{k+1, \bom^+}  \in   \cP^0_{k+1, \bom^+}    \hs   |W_j|  \leq     B_w \   p_j L^{\frac12(k-j)}
\ee
  Furthermore  if  $Y \subset  \La_{k+1}    \subset  \Om_{k+1}$    then  $d_{LM}(Y)  =  d_{LM}(Y, \bmod\  \Om_{k+1})$ and so 
    \begin{equation}
| H_{k,  \bpi^+} (Y) |
 \leq   \cO(1) L^3\la_k^{\frac14 -10 \ep}  e^{  -L ( \ka  - 3\ka_0  -3) d_{LM}(Y,\bmod  \Om^c_{k+1} ) }
 \end{equation}

\subsection{cluster  expansion}    
\label{cluster2}

We   want  to  perform  a cluster expansion on the  fluctuation integral  $ \Xi''_{k,  \bpi^+} $,    and  also  isolate  the most  important 
terms which come  from  $ ( \de E_k^+ )^{\loc} (Y  )$.     Accordingly   we    introduce   
some   variables   $  t,u  $   which  parametrize   the contribution of the other  terms  and    define    
\begin{equation}
 \Xi''_{k, \bpi^+} (t,u) 
=        \int   d\mu^*_{\La_{k+1}}(W_k)   \exp    \Big(   \sum_Y    H_{k, \bpi^+}(t,u,Y)     \Big) 
\end{equation}
where 
\begin{equation}
H_{k, \bpi^+}  (t,u,   Y) = \Big( ( \de E_k^+)^{\loc}( Y) 
  +  t \   R^{\loc}_{k, \bpi^+}(Y) \Big)1_{Y \subset  \La_{k+1}} +
u\  B^{\loc}_{k, \bpi^+}(Y)   1_{Y \#  \La_{k+1}}
\end{equation}
We   are interested  in     $\Xi''_{k,   \bpi^+} (1,1) = \Xi''_{k, \bpi^+}  $, but start with a
more general result:

\begin{lem}  (cluster  expansion)  \label{cluster}  For  $r_0 = \cO(1)$ sufficiently small  and     
$|t|  \leq   r_0  L^{-3} \la_k^{-n_0}$      and        $| u |  \leq  r_0 L^{-3} \la^{-\frac14+ 10 \ep  }$  
\begin{equation}
\label{output}
\Xi''_{k,  \bpi^+}(t,u)  =  \exp  \left(  \sum_{  Y \cap   \La_{k+1}  \neq  \emptyset}
  H^\#_{k, \bpi^+}(t,u,Y   )   \right)   
\end{equation}
where $Y   \in \cD^0_{k+1} (\bmod \   \Om^c_{k+1})$    and   $  H^\#_{k, \bpi^+}(t,u,Y   ) =   H^\#_{k, \bpi^+}(t,u,Y,\phi,   \Phi_{k+1,\bom^+},  W_{k+1\bpi^+},  )   $  is   evaluated at
$\phi =   \phi^0_{k+1,  \bom'}$.
The function is  analytic  in $t,u$,    and  in    $\phi \in \frac12  \cR_k$  and  $ \Phi_{k+1,\bom^+},  W_{k+1\bpi^+}$ in    (\ref{500})  and  
   on this   domain  it   satisfies 
\begin{equation}  \label{key}  
 |   H^\#_{k, \bpi^+}(t,u,Y  )    |  \leq     \cO(1)   e^{ - L  ( \ka- 6 \ka_0-6)  d_{LM}(Y,  \bmod \Om^c_{k+1})  }
\end{equation}
\end{lem}
\bigskip

\pr     Since   integrand is well-localized   and the  measure is ultralocal,    we  can use  the   cluster expansion  with holes which can be found
in  appendix   \ref{fancycluster},  now with  $LM$ cubes.     In the domain  (\ref{500}),  which puts  $W_k$ in the support  of  $\mu^*_{\La_{k+1}}$,  
we  have     
\begin{equation}  
\begin{split}
|u|\  |B^{\loc}_{k, \bpi^+}(Y)|   \leq  & \cO(1)  r_0    e^{ -   L ( \ka- 3\ka_0-3)  d_{LM}(Y,  \bmod \Om^c_{k+1}) }
 \leq   \frac 13  c_0  e^{ -  L ( \ka- 3\ka_0-3)  d_{LM}(Y,  \bmod \Om^c_{k+1}) }\\
 | t |\ |R^{\loc}_{k, \bpi^+}(Y)|   \leq     
 & \cO(1) r_0   e^{ -  L ( \ka- 3\ka_0-3) d_{LM}(Y)  }
 \leq   \frac 13  c_0  e^{ - L ( \ka- 3\ka_0-3)  d_{LM}(Y,  \bmod \Om^c_{k+1}) }\\
\end{split}
\end{equation}
and      $|( \de E_k^+)^{\loc}( Y)| $  satisfies the same bound    for  $\la_k$  sufficiently small.    Thus altogether
\begin{equation}   \label{keykey}  
|H_{k,  \bpi^+}  (t,u,   Y)|      
 \leq    c_0   e^{ -  L ( \ka- 3\ka_0-3)  d_{LM}(Y,  \bmod \Om^c_{k+1})  }
\end{equation}
This is the input for the cluster expansion.   The output  is  the  representation  (\ref{output})  with a bound   which implies  (\ref{key}).
\bigskip

 \begin{lem}  (removal of boundary terms)  \label{thirdcl}
 \begin{equation}  
  \Xi''_{k, \bpi^+} (1, 1) 
=   \exp \Big  (  \sum_{Y \#  \La_{k+1}}
 \ B^\#_{k, \bpi^+}(Y)   \Big)
 \Xi_{k, \bpi^+} (1,0)
  \end{equation}
 with $ B^\#_{k, \bpi^+}(Y) =   B^\#_{k, \bpi^+}(Y,\phi,     \Phi_{k+1,\bom^+},  W_{k+1,\bpi^+}) $ evaluated at   $\phi =   \phi^0_{k+1,  \bom'}$.
It  is  analytic   in  $\phi \in \frac12  \cR_k$  and    (\ref{500})    and satisfies there
 \begin{equation}  \label{whale}
|B^\#_{k, \bpi^+}(Y) |  \leq    \cO(1) L^3  \la^{\frac14- 10 \ep  }  e^{ -   L( \ka- 6 \ka_0-6)  d_{LM}(Y,  \bmod \Om^c_{k+1})  }
\end{equation}
 \end{lem}
\bigskip

\pr We  have   
\begin{equation}   \label{hotdog}
 \Xi''_{k,  \bpi^+} (1, 1)
 =    \exp  \left(    \sum_{Y  \cap  \La_{k+1} \neq  \emptyset} 
 (  H^\#_{k, \bpi^+}(1,1,   Y) - H^\#_{k, \bpi^+}(1,0,   Y) )  \right)  \Xi_{k,\bpi^+} (1,0) 
\end{equation}
so  the identity  holds   with   $ B^\#_{k, \bpi^+}(Y)   =   H^\#_{k, \bpi^+}(1,1,   Y) - H^\#_{k, \bpi^+}(1,0,   Y)$ or  
\begin{equation}
 B^\#_{k, \bpi^+}(Y)  
   =   \frac{1}{2\pi i}   \int_{|u|  =   r_0 L^{-3} \la_k^{-\frac14 + 10 \ep}  }  \frac{du}{u(u-1)}      H^\#_{k, \bpi^+}(1,u,   Y)  \\
\end{equation}
Note that if     $Y  \subset   \La_{k+1}$  then       no  boundary term  $B^{\\loc}_{k, \bpi^+}(Y)$    can contribute.  This is a consequence of the  local influence property of the   cluster expansion.    Hence    
$ H^\#_{k,  }(1,u,   Y)$  is  independent of  $u$ and   so   $ B^\#_{k, \bpi^+}(Y) =0$.    Therefore  the sum in  (\ref{hotdog})    is actually over
   $Y \#  \La_{k+1}$ as  claimed. 
  The  bound  (\ref{key})   on    $ H^\#_{k, \bpi^+}(1, u,  Y) $     for    $|u|  =  r_0 L^{-3} \la_k^{-\frac14   + 10 \ep} $  now   implies  that
  $ B^\#_{k, \bpi^+}(Y) $   satisfies    (\ref{whale}).  This completes the proof.  \bigskip

 \begin{lem}  (removal of tiny terms)  \label{fourth}
 \begin{equation}  
  \Xi''_{k,  \bpi^+} (1, 0) 
=   \exp \Big  (  \sum_{Y   \subset    \La_{k+1}} \ R^\#_{k, \bpi^+}(Y)   \Big)
 \Xi''_{k,  \bpi^+} (0,0) 
 \end{equation}
 with     $ R^\#_{k, \bpi^+}(Y)=   R^ \#_{k, \bpi^+}(Y , \phi,  \Phi_{k+1} )$    evaluated at     $\phi =   \phi^0_{k+1,  \bom'}$.
 It  is  analytic  in  $\phi \in \frac12  \cR_k$   and  $\Phi_{k+1}    \in \cP^0_{k+1}(\La_{k+1}, 2 \de)$ and satisfies there
 \begin{equation}     \label{whale2}
|R^\#_{k, \bpi^+}(Y) | 
 \leq    \cO(1)L^3  \la_k^{n_0}  e^{- L ( \ka- 6\ka_0-6) d_{LM}(Y)}
\end{equation}
\end{lem}
\bigskip

\pr  Now  we  are     back to the standard cluster expansion with no holes   and   all polymers  contained in $\La_{k+1}$.   
$H_{k, \bpi^+}(t,0,Y)$  is    analytic  in $|t|  \leq    r_0 L^{-3} \la_k^{-n_0} $
with the bound   $|H_{k, \bpi^+}(t,0,Y)| \leq    c_0  e^{- L ( \ka-3 \ka_0-3) d_{LM}(Y)}$.     Therefore   $  H^\#_{k, \bpi^+}(t,0,   Y) $  is  analytic 
in the same domain with bound  
\be    \label{key4} |H^\#_{k, \bpi^+}(t,0,   Y)|   \leq     \one  e^{- L ( \ka- 6\ka_0-6) d_{LM}(Y)}
\ee
Now  we  have   
\begin{equation}   \label{hotdog2}
 \Xi''_{k,  \bpi^+} (1, 0)
 =    \exp  \left(    \sum_{Y  \subset  \La_{k+1}} 
 (  H^\#_{k, \bpi^+}(1,0,   Y) - H^\#_{k, \bpi^+}(0,0,   Y) )  \right)  \Xi''_{k,\bpi^+} (0,0) 
\end{equation}
so  the identity  holds   with   
\begin{equation}
\begin{split}
 R^\#_{k, \bpi^+}(Y)   = &  H^\#_{k, \bpi^+}(1,0,   Y) - H^\#_{k, \bpi^+}(0,0,   Y)   
   =   \frac{1}{2\pi i}   \int_{|t|  =   r_0 L^{-3} \la_k^{-n_0}  }  \frac{dt}{t(t-1)}      H^\#_{k, \bpi^+}(t,0,   Y)  \\
\end{split}   
\end{equation}
The bound   (\ref{key4})  now implies that   $ R^\#_{k, \bpi^+}(Y) $  satisfies    (\ref{whale2}).  This completes the proof.
\bigskip

  Now we  are  reduced to   $ \Xi''_{k,  \bpi^+} (0,0) 
$    which  is
 \begin{equation}
 \Xi''_{k,  \bpi^+} (0,0) 
=        \int   d\mu_{\La_{k+1}}^*(W_k)   \exp    \Big(   \sum_{Y  \subset  \La_{k+1}}    ( \de E_k^+)^{loc} (Y, \phi,W_k  )  \Big) 
\end{equation}
at  $\phi =   \phi^0_{k+1,  \bom'}$.
Before the evaluation  in $\phi$  
 this is  just the quantity considered in part I,  except  that the sum over $Y$  is restricted to $\La_{k+1}$  
 and the measure is restricted to  $\La_{k+1}$.

 \begin{lem}  (leading terms)
 \begin{equation}
 \Xi''_{k,  \bpi^+}(0,0 )  =  \exp  \Big(  \sum_{Y  \subset      \La_{k+1} }
 E_k^\#(Y, \phi)   \Big)  \hs  \textrm{  at  }  \ \ \   \phi =   \phi^0_{k+1,  \bom'}
  \end{equation}
  where    $ E_k^\#(Y, \phi)  $  is analytic in  $\frac12 \cR_k$  and    satisfies  there     
     \begin{equation}  
| E_k^\#(Y, \phi)   |   \leq   \cO(1)L^3 \la_k^{\frac14 - 10 \ep }  
   e^{- L(\ka-  6\ka_0 - 6) d_{LM}(Y)}
   \end{equation}
  $E_{k}^\#(Y, \phi)  $  is identical  with the  function  constructed in  
   the global small field analysis in part  I.  
  \end{lem}
\bigskip

\pr  This is  again  the standard cluster expansion,   taking account that  $ ( \de E_k^+)^{\loc} (Y  )$   in analytic in $\frac12 \cR_k$  and 
has the bound   $  \cO(1) L^3 \la_k^{\frac14 - 10 \ep}  e^{   -L( \ka  -3 \ka_0  -3) d_{LM}(Y)  }  $.  
The  function  $ E_{k}^\#(Y, \phi)  $   defined here is the same as the global definition in part I,  even though here we are only summing over
polymers in $\La_{k+1}$.    This is so since    by the local influence property and the fact that  the 
 $ (\de E_k^+)^{\loc}(Y)$  are   the same.
This completes the proof.
\bigskip

Combining the  above results  yields      
\begin{equation}  \label{fifth}
 \Xi''_{k, \bpi^+}   = 
      \exp  \Big(    E_k^\#(\La_{k+1})   
 +    R^\#_{k, \bpi^+}(\La_{k+1})  
+ B^{\#}_{k, \bpi^+}( \La_{k+1})   \Big)
 \end{equation}
 where  $ E_k^\#(\La_{k+1})  =   E_k^\#(\La_{k+1},  \phi^0_{k+1, \bom'}   )$.  
Inserting this back into   (\ref{primeprime}) and   (\ref{prime})  yields  
\begin{equation}
\begin{split}
\Xi_{k,  \bpi^+}  =     &
 \exp  \Big(-  \vep_k^0 \Vol(  \La_{k+1} )  +  E^+_k \big(\La_k \big)    \\
 &   +   E^\#_{k}( \La_{k+1}  )    
 +    R^\#_{k, \bpi^+}(\La_{k+1})  
+ B^{\#}_{k, \bpi^+}( \La_{k+1})  + \tilde  B_{k+1, \bpi^+}  \textrm{ terms }  \Big)  \\
\end{split}
\end{equation}
Finally inserting  this  back  in  (\ref{representation7})
   \begin{equation}     \label{representation7.5}
\begin{split}
\tilde    \rho_{k+1}( \Phi_{k+1}) 
 =    &  Z^0_{k+1}   \sum_{\bpi^+}    \int   d \Phi^0_{k+1, \bom^{+,c} }d W^0_{k+1,\bpi^+}\  K_{k, \bpi} \  \cC^0_{k+1, \bpi^+}         
\        \exp  \Big(  c_{k+1} |\Om^{c,(k)}_{k+1}|    \Big)   \\
&  \chi^0_{k+1}(  \La_{k+1} )   \exp \Big(-  S^{*,0}_{k+1}(\La_k) -  \vep_k^0 \Vol(  \La_{k+1} )  +  E^+_k \big(\La_k \big)\\  &   +   E^\#_{k}( \La_{k+1}  )    
 +    R^\#_{k, \bpi^+}(\La_{k+1})  
+ B^{\#}_{k, \bpi^+}( \La_{k+1})  + \tilde  B_{k+1, \bpi^+}  \textrm{ terms }  \Big)  \\
   \end{split}
\end{equation}

\subsection{scaling}

We  scale  and   evaluate  
$\rho_{k+1} ( \Phi_{k+1} )   =  \tilde    \rho  (  \Phi_{k+1,L} ) L^{-|\bbT^1_{\sM + \sN -k}|/2}$  where now
 $\Phi_{k+1}$ is   defined on  $\bbT^{0}_{\sM+ \sN -(k+1)}$.  
We  make the following changes  in  this expression.    This follows the discussion 
in section  \ref{freeflow}.  

\begin{itemize}

\item  Identify   $   Z_{k+1}   =Z^0_{k+1}  L^{-|\bbT^1_{\sM + \sN -k}|/2}  $

\item   The sum over    regions     $\bom^+ = ( \Om_1,  \dots,  \Om_{k+1})$  with   $\Om_j$  a  union of  $L^{-(k-j)}M$ blocks 
in   $\bbT^{-k}_{\sM+\sN -k}$   is    relabeled    as     $L \bom^+ =   ( L\Om_1,  \dots, L \Om_{k+1})$  
where  now   $\bom^+ = ( \Om_1,  \dots,  \Om_{k+1})$  with   $\Om_j$  a  union of  $L^{-(k+1-j)}M$ blocks 
in   $\bbT^{-k-1}_{\sM+\sN -k-1}$.   Similarly the sum over   $\bla^+$ is replaced by a sum over $L \bla^+$
and the sum  over  $\bpi^+$ is replaced by a sum  over $L \bpi^+$.

\item  The fields    $\Phi_{ \bom^+} =  ( \Phi_{1, \de  \Om_1},  \dots,  \Phi_{k+1, \de   \Om_{k+1}})$  defined
on subsets of   $\bbT^{-k}_{\sM+\sN-k}$  which has  become     $\Phi_{L \bom^+} =  ( \Phi_{1,L\de   \Om_1},  \dots,  \Phi_{k+1,L \de  \Om_{k+1}})$.   Now   make a change of variables  replacing    $\Phi_{j,L\de \Om_j}$    by   $[\Phi_{j,L}  ]_{L\de \Om_j}  =    [\Phi_{j,\de \Om_j}]_L$.  Then      $\Phi_{L \bom^+} $  becomes
 $\Phi_{ \bom^+,L} =  ( \Phi_{1,\de   \Om_1,L},  \dots,  \Phi_{k+1, \de   \Om_{k+1},L})$.     Furthermore   the measure
  $d \Phi^0_{k+1 , \bom^{+,c}} $  becomes  $d \Phi_{k+1, \bom^{+,c}}$.

\item   Similarly  we  make a change of variables in  $W$ replacing   $   W_{j, L \Om_j -L \La_j} $   by      $[W_{j,  \Om_j - \La_j}]_L $.    
Then  $d W^0_{k+1,\bpi^+}   $  becomes  $d W_{k+1, \bpi^+}$.

\item   Under these changes         $\chi^0_{k+1}(\La_{k+1})$  becomes  $\chi_{k+1}(\La_{k+1})$.    Also if we  define
\be  \cC_{k+1, \La_k,  \Om_{k+1},  \La_{k+1}} (\Phi_k, W_k,  \Phi_{k+1}) =
 \cC^0_{k+1, L\La_k, L \Om_{k+1}, L \La_{k+1}} (\Phi_{k,L}, W_{k,L},  \Phi_{k+1,L})
\ee
then     $ \cC^0_{k+1,  \bpi^+}  $  becomes    $ \cC_{k+1,  \bpi^+}   $   as defined in (\ref{orca}).

  \item  We  have  already noted  in  (\ref{lumpy}) that   $\phi^0_{k+1,  \bom^+}$  becomes   $\phi_{k+1, \bom^+, L}$. 
    More to the point here   $\phi^0_{k+1, \bom'} =  \phi^0_{k+1,  \bom'} ( \hat \Phi_{k+1, \bom'} )$
becomes    
\be     \phi^0_{k+1, L \bom'} (  [\hat   \Phi_{k+1, \bom'}]_L )  =     \phi_{k+1, \bom',L} 
\ee
where  now    $\phi_{k+1, \bom'} =  \phi_{k+1,  \bom'} ( \hat  \Phi_{k+1, \bom'} )$
with  $\hat    \Phi_{k+1, \bom'} =  ( [\tilde  Q^T_{ \bbT^{-1},  \bom(\La_k^*)} \Phi_k]_{\Om_{k+1}^c},  \Phi_{k+1, \Om_{k+1}})$.

\item 
  As  noted  earlier the action        $S^{*,0}_{k+1}(\La_{k}, \Phi_{k+1, \bom^+}, \phi^0_{k+1,\bom'} )$
becomes   
\begin{equation}
S^{*,0}_{k+1}(L\La_{k}, \Phi_{k+1, \bom^+, L}, \phi_{k+1,\bom',L}  )
=    S^*_{k+1}(\La_{k}, \Phi_{k+1, \bom^+}, \phi_{k+1,\bom'})   
\end{equation}
We  split this as   $
S^*_{k+1}(\La_{k})   =   S^*_{k+1}(\La_{k}- \La_{k+1})  +  S^*_{k+1}(\La_{k+1}) $

\item   In    $E^+_k(\La_{k})  =  E_k(\La_{k})  -V_k(\La_{k}) $  we have   the potential  $V_{k} (\La_{k}, \phi^0_{k+1, \bom'}      )$.  This      becomes 
\begin{equation}
V_{k} (L\La_{k}, \phi_{k+1, \bom',L}      )
=  V^u_{k+1}(\La_{k}, \phi_{k+1, \bom'} )  
\end{equation}
and we  split this  as    $ V^u_{k+1}(\La_{k})   =    V^u_{k+1}(\La_{k}  - \La_{k+1})  +   V^u_{k+1}(\La_{k+1})$.

\item  Before  scaling         $E_{k}( \La_{k},\phi^0_{k+1,\bom'}) =   \sum_{X \in \cD_k, X  \subset  \La_{k}}E_{k}( X, \phi^0_{k+1,\bom'})  $
we  apply a   reblocking operation.  For    $Y  \in \cD_{k+1}^0$     define
\begin{equation}
(\cB E)(Y )   = \sum_{X \in \cD_k,  \bar X  = Y }E_{k}( X)  
\end{equation}
where  $\bar  X$  is  the union of all $LM$ cubes intersecting  $X$.
Then  $E_{k}( \La_{k})   =  \sum_{ Y \subset  \La_{k}}  (\cB E_k)(Y)$.
Upon scaling this becomes  (since $L  \cD_{k+1}  =  \cD^0_{k+1}$)
\begin{equation}
\begin{split}
E_k(L \La_{k}, \phi_{k+1,\bom',L})  & =  \sum_{Y \in \cD^0_{k+1}, Y \subset  L\La_{k}}  (\cB E_k)(Y, \phi_{k+1,\bom',L})\\
= & \sum_{X  \in   \cD_{k+1}, X \subset  \La_{k}} (\cB E_k)(LX, \phi_{k+1,\bom',L})\\
= & \sum_{X  \in   \cD_{k+1}, X \subset  \La_{k}} (\cB E_k)_{L^{-1}}(X, \phi_{k+1,\bom'})\\
\equiv    &  (\cB E_k)_{L^{-1}}(\La_{k}, \phi_{k+1,\bom'})\\
\end{split}
\end{equation}

\item  The function   $E^\#_{k}( \La_{k+1}, \phi^0_{k+1,\bom'})$  is already reblocked, but otherwise is treated the same way
Under  scaling it  becomes
\begin{equation}
E^\#_k(L \La_{k+1}, \phi_{k+1,\bom',L}) 
\equiv  ( E^\#_k)_{L^{-1}}(\La_{k+1}, \phi_{k+1,\bom'}) 
\end{equation}
Similarly     $R^\#_{k,  \bpi^+}( \La_{k+1}, \phi^0_{k+1,\bom'},  \Phi_{k+1})$  scales to  
\begin{equation}
R^\#_{k, L \bpi^+}(L \La_{k+1}, \phi_{k+1,\bom',L}, \Phi_{k+1, L}) 
\equiv  [( R^\#_k)_{L^{-1}}]_{ \bpi^+}(\La_{k+1}, \phi_{k+1,\bom'}, \Phi_{k+1}) 
\end{equation}

\item  Now consider  $ B^\# _{k, \bpi^+}( \La_{k+1},\phi^0_{k+1,\bom'},  \Phi_{k+1,\bom^+},  W_{k+1,\bpi^+}) $
  which  is also reblocked,  and    has the local decomposition
\begin{equation}  
 B^{\#} _{k, \bpi^+}( \La_{k+1}) 
=   \sum_{Y \in  \cD^0_{k+1}( \bmod   \Om_{k+1}^c),Y \#  \La_{k+1}}    B^{\#} _{k, \bpi^+}(Y) 
\end{equation}
Upon  scaling this becomes \begin{equation}
\begin{split}
& B^\# _{k, L \bpi^+}( L\La_{k+1},\phi_{k+1,\bom',L},  \Phi_{k+1, \bom^+,L},  W_{k+1,\bpi^+,L}) \\
= &  \sum_{Y \in  \cD^0_{k+1}( \bmod   L\Om_{k+1}^c),Y \# L \La_{k+1}}  
   (  B^{\#} _{k,L \bpi^+})(Y,\phi_{k+1,\bom',L},  \Phi_{k+1,\bom^+,L},  W_{k+1,\bpi^+,L}) \\
= &  \sum_{X \in  \cD_{k+1}( \bmod   \Om_{k+1}^c),X \#  \La_{k+1}}  
   (  B^{\#} _{k,L \bpi^+})(LX,\phi_{k+1,\bom',L},  \Phi_{k+1, \bom^+,L},  W_{k+1,\bpi^+,L}) \\
= &  \sum_{X \in  \cD_{k+1}( \bmod   \Om_{k+1}^c),X \#  \La_{k+1}}    
  [ (B^{\#}_k)_{L^{-1}}]_{ \bpi^+}(X,\phi_{k+1,\bom'},  \Phi_{k+1, \bom^+},  W_{k+1,\bpi^+}) \\
 \equiv &   [ (  B^{\#}_k)_{L^{-1}}]_{ \bpi^+}( \La_{k+1}, \phi_{k+1,\bom'},  \Phi_{k+1, \bom^+},  W_{k+1,\bpi^+})\\
\end{split}
\end{equation}
Here  we have  used that  $  L \cD_{k+1}( \bmod\   \Om_{k+1}^c)  =   \cD^0_{k+1}( \bmod \  L\Om_{k+1}^c) $.

\item  Finally  consider   $K_{k, \bpi}$  
         which scales to 
\begin{equation}
\begin{split}
&[ K_{k, L^{-1}}]_{\bpi}  \\
&  \equiv       \prod_{j=0}^k  
      \exp\Big(c_j  | \Om_j^{c,(j-1)}| -S^{+,u}_{j,  L^{-(k+1-j)} }(\La_{j-1} -  \La_j     )
       + \big(\tilde   B_{j,L^{-(k+1-j)} }\big ) _{ \bpi_j}(\La_{j-1},  \La_j)  \Big) \\
     \end{split}
\end{equation}
\item   Collect   all the scaled  ''$ \tilde  B_{k+1, \bpi^+}$    terms'' 
   into a single  term    $\tilde  B_{k+1, \bpi^+} (\La_k, \La_{k+1}) $ 
\end{itemize}

  With all  these changes
   \begin{equation}     \label{representation8}
\begin{split}
   \rho_{k+1}( \Phi_{k+1})     =     &Z_{k+1}  \sum_{\bpi^+}  \int d \Phi_{ \bom^{+,c}}  d W_{\bpi^+}   \  [ K_{k, L^{-1}}]_{\bpi} \ 
  \cC_{k+1,\bpi^+}   \\
&  \exp  \Big(   c_{k+1}  |\Om^{c,(k)}_{k+1}|    -  S^*_{k+1}(\La_k- \La_{k+1})  -     V^u_{k+1}(\La_k- \La_{k+1})  +  \tilde  B_{k+1, \bpi^+}(\La_k,  \La_{k+1})    \Big)  \\
&  \chi_{k+1}(  \La_{k+1} )   \exp \Big(-  S^*_{k+1}(\La_{k+1}) -  \vep_k^0L^3 \Vol(  \La_{k+1} )  -   V^u_{k+1} ( \La_{k+1}) 
+   (\cB E_k)_{L^{-1}}(\La_{k})  \\
 &   +    ( E^\#_k)_{L^{-1}}(\La_{k+1})  
 +   [( R^\#_k)_{L^{-1}}]_{ \bpi^+}(\La_{k+1}) +   [ (  B^{\#}_k)_{L^{-1}}]_{ \bpi^+}( \La_{k+1}) \Big) \\
    \end{split}
\end{equation}

\subsection{the RG flow}
 
Now  we  show  that  the coupling constant flow  follows     the  global   analysis of   part  I,   even though 
the effective action is localized.    To  do this we  need  to  process the   terms  $(\cB E_k)_{L^{-1}}(\La_{k+1})$ and
$    ( E^\#_k)_{L^{-1}}(\La_{k+1})  $.

First consider a more general  case.  Let    $\La  \subset    \bbT^{-k-1}_{\sM + \sN -k-1} $,   and  
$\phi: \La   \to  \bbR$,   and   suppose  $E( \La) =  \sum_{X \subset  \La}  E(X)$
  with   $E(X, \phi)$  translation invariant.     Following the analysis in part I   
 we  make the following definitions. 
  If   $X$ is small  ($X  \in \cS$) then  $\cR E(X)$  is 
defined by 
\begin{equation}  \label{renorm}
\begin{split}
& E( X, \phi)  
=   \al_0(E,X)  \Vol( X )
+  \al_{2}(E,X)   \int_X  \phi^2        + \sum_{\mu}  \al_{2, \mu}(E,X)       
   \int_X  \phi \  \pa_{\mu} \phi   \    +   \cR E( X, \phi)    \\
\end{split}
\end{equation} 
where
\be 
\begin{split}
\al_0(E, X)   =& \frac{1}{ \Vol (X)}  E(X,0)  \hs   \al_{2}(E,X) =      \frac{1}{2 \ \Vol(X)}   
 E_2 ( X, 0; 1,1 )\\
   \al_{2, \mu}(E,X)    =&\frac{1}{\Vol(X)} \left( E_2 ( X, 0; 1,x_{\mu} - x^0_{ \mu}) -   \frac{1 }{ \Vol(X)}  E_2 ( X, 0; 1,1)\int_X  x_{\mu} - x^0_{ \mu}\right)\\
\end{split}
\ee
The last   is independent of  the base point  $x^0$,  which we take to be  in $X$.  With these choices  $\cR E(X)$  is  normalized
for  small sets,  that is the function  and certain derivatives  vanish at zero.
 If   $X$ is large then  $\cR E(X)   =  E(X)$.

  Summing this over  $X  \subset  \La$  we  find
 \begin{equation}  \label{renorm2}
 \begin{split}
  E(\La)
 =   &  -  \sum_{\square \subset  \La}   \vep_{\La} (E,\square) \Vol(  \square ) -   \frac12  
 \sum_{\square \subset  \La}    \mu_{\La}(E, \square)   \|  \phi \|^2_{\square} \\
  - & \sum_{\mu}  \sum_{\square \subset  \La}    \nu_{\La, \mu }(E, \square)  \int_{\square}  \phi \pa_{\mu}  \phi
+    \sum_{X \subset \La} ( \cR E)(X) \\
\end{split}
\end{equation} 
where
\begin{equation}
\begin{split}
\vep_{\La}(E,\square)  =  &    -   \sum_{ \square \subset    X  \subset  \La,  X \in \cS}   \al(E,X)  \hs
\frac12\  \mu_{\La}(E,\square)  = - \sum_{ \square \subset    X  \subset  \La,  X \in \cS}  \al_{2}(E,X)  \\
\nu_{\La, \mu }(E,\square)  =& - \sum_{ \square \subset    X  \subset  \La,  X \in \cS}    \al_{2, \mu}(E,X)      \\
\end{split}
\end{equation}

Now  if   $\square$ is well inside   $\La$ then    $X \in \cS$  and  $X  \supset  \square$  imply  $X \subset  \La$  so we can drop 
the latter condition from the sums.
 Then    $\vep_{\La}(E, \square),
  \mu_{\La}(E,  \square), \nu_{\La, \mu }(E, \square)$
 are    independent of  $\square$
and  $\La$    and  agree with the global  quantities  which are denoted   $\vep(E),  \mu(E), \nu_{ \mu }(E)$.
Furthermore   the lattice symmetries imply  that  $\nu_{ \mu }(E)=0$.
Then we   write   
 \begin{equation}  \label{renorm3}
 \begin{split}
  E(\La) =
 & 
  -   \vep ( E) \Vol(  \La ) -  \frac12    \mu(  E)    \|  \phi \|^2_{\La}
 + \sum_{X \subset \La} ( \cR E)(X) 
    -  \sum_{\square \subset  \La} (  \vep_{\La} (E,\square) - \vep ( E) )\Vol(  \square ) \\  - & \frac12  
 \sum_{\square \subset  \La}   (\mu_{\La}(E, \square)-    \mu(  E) )   \|  \phi \|^2_{\square}
 +     \sum_{\mu}  \sum_{\square \subset  \La} (   \nu_{\La, \mu }(E, \square)-  \nu_{ \mu }(   E) ) \int_{\square}  \phi \pa_{\mu}  \phi  \\
  \end{split}
\end{equation} 
The  last  three terms    can be treated as boundary terms.   Indeed we have   
\begin{equation}
\begin{split}
 \sum_{\square \subset  \La} (  \vep_{\La} (E,\square) - \vep ( E) )\Vol(  \square )
=  & \sum_{\square \subset  \La}  \Big(  \sum_{X \in \cS,   X  \supset  \square,    X \#  \La}  \al_0(E,X)\Big)\Vol(  \square )
\\
=  & \sum_{X \in \cS,     X \#  \La}  \al_0(E,X) \Big(   \sum_{\square \subset  \La\cap X}  \Vol(  \square ) \Big)
\\
=  & \sum_{X \in \cS,     X \#  \La}   \al_0(E,X)   \Vol( \La \cap  X )
\\
\end{split}
\end{equation}
Similarly   
\begin{equation}
\frac12 \sum_{\square \subset  \La}   (\mu_{\La}(E, \square)-    \mu(  E) )   \|  \phi \|^2_{\square}
   =   \sum_{X \in \cS,      X \#  \La}   \al_{2,0}(X)    \|  \phi \|^2_{X \cap  \La}
\end{equation}
and 
\be 
   \sum_{\mu}  \sum_{\square \subset  \La} (   \nu_{\La, \mu }(E,  \square)-  \nu_{ \mu }(   E) ) \int_{\square}  \phi \pa_{\mu}  \phi  
=   \sum_{X \in \cS,      X \#  \La}     \sum_{\mu}      \al_{2, \mu}(E,X)  
    \int_{X  \cap  \La}  \phi \  \pa_{\mu} \phi     
\ee
we  combine the last three  terms defining  for  $X  \in \cS$ only  
\be   \cT_{\La} E  ( X,  \phi) 
=      \al_0(E,X)  \Vol( \La \cap  X ) +      \al_{2}(E,X)  \|  \phi \|^2_{X \cap  \La} 
 +  \sum_{\mu}      \al_{2, \mu}(E,X)  
    \int_{X  \cap  \La}  \phi \  \pa_{\mu} \phi     \ee
Now   (\ref{renorm3})  becomes  
 \begin{equation}  \label{renorm5}
 \begin{split}
  E(\La) =     -   \vep ( E) \Vol(  \La ) -  \frac12    \mu(  E)    \|  \phi \|^2_{\La}
 + \sum_{X \subset \La} ( \cR E)(X)  +  \sum_{X   \# \La, X \in \cS}   \cT_{\La} E  ( X  ) 
 \end{split}
\end{equation}

Now return  to our specific problem.   In the exponential in   (\ref{representation8})   we  pick out the terms  
\begin{equation}  \label{poof}
\begin{split}
&  -  \vep_k^0L^3 \Vol(  \La )  - V^u_{k+1} ( \La)   +    (\cB E_k)_{L^{-1}}(\La)  
  +    ( E^\#_k)_{L^{-1}}(\La)  \\
  = &    -  (\vep_k^0 + \vep_k)   L^3 \Vol(  \La )   -   \frac12   L^2\mu_k \| \phi \|^2_{\La} 
 -  \frac14  L \la_k  \int_{\La}  \phi^4   +  (\cB E_k)_{L^{-1}}(\La)  +   ( E^\#_k)_{L^{-1}}(\La)\\
 \end{split}
 \end{equation}
  evaluated at  $\La  = \La_{k+1}$  and   $\phi=   \phi_{k+1, \bom'}$.  
Applying   (\ref{renorm5})  to the last two terms  we have  
\begin{equation}
\begin{split}
 (\cB E_k)_{L^{-1}}(\La)   = &    -   \vep \Big(  (\cB E_k)_{L^{-1}}\Big) \Vol(  \La  )
 -  \frac12    \mu \Big(   (\cB E_k)_{L^{-1}}\Big)    \|  \phi_k \|^2_{\La}\\
 +& \sum_{X \subset \La} \Big( \cR  (\cB E_k)_{L^{-1}}\Big)(X) 
 +   \sum_{ X  \# \La}  \cT_{\La}  (\cB E_k)_{L^{-1}}  ( X  ) \\
 \equiv    &  -  \cL_1(E_k) \Vol(  \La ) -  \frac12   \cL_2(E_k)   \|  \phi \|^2_{\La}
 +\sum_{X  \subset  \La} (\cL_3 E_k)(X )        + \sum_{  X  \# \La}   \Big(   \cT_{\La}  (\cB E_k)_{L^{-1}} \Big) ( X  )  \\
  \end{split}
  \end{equation}
\begin{equation}
\begin{split}
 ( E^\#_k)_{L^{-1}}(\La)   = &    -   \vep \Big(  ( E^\#_k)_{L^{-1}}\Big) \Vol(  \La  )
 -  \frac12    \mu \Big(   ( E^\#_k)_{L^{-1}}\Big)    \|  \phi_k \|^2_{\La}\\
 +& \sum_{X \subset \La} \Big( \cR  ( E^\#_k)_{L^{-1}}\Big)(X) 
 +   \sum_{ X  \# \La}  \cT_{\La}  ( E^\#_k)_{L^{-1}}  ( X  ) \\
 \equiv    &    -  ( \vep_k^* -L^3 \vep_k^0) \Vol(  \La ) -  \frac12    \mu_k^*   \|  \phi \|^2_{\La}
 +\sum_{X \subset  \La}E_k^*(X )       + \sum_{  X  \# \La}   \Big(   \cT_{\La}  ( E^\#_k)_{L^{-1}} \Big) ( X  )  \\
  \end{split}
  \end{equation}
Insert these into    (\ref{poof})    and  identify  the coupling constants at the next level.  As in part I     these are: 
 \begin{equation}  
\begin{split}
\vep_{k+1}   =&  L^3 \vep_k  + \cL_1E_k   +  \vep_k^*(\la_k, \mu_k,  E_k) \\
\mu_{k+1}   =&   L^2 \mu_k  +  \cL_2E_k  + \mu_k^*(\la_k, \mu_k,  E_k)  \\
\la_{k+1}   =&  L\la_k   \\ 
E_{k+1}   =&    \cL_3 E_k  +  E^*_k(\la_k,    \mu_k,  E_k)  \\
 \end{split}
\end{equation}
Now  the terms  (\ref{poof})  can be written:
\begin{equation}  \label{nugatory}
\begin{split} &    -  \vep_{k+1} \Vol(  \La )   -   \frac12   \mu_{k+1} \| \phi \|^2_{\La} 
 -  \frac14   \la_{k+1}  \int_{\La}  \phi^4   + \sum_{X \subset  \La}E_{k+1}(X)
+   \sum_{  X  \# \La} \Big(  \cT_{\La}  (\cB E_k  +   E^\#_k)_{L^{-1}} \Big) ( X  ) \\
&=   - V_{k+1}(\La)
    + E_{k+1}(\La)+   \Big(  \cT_{\La} (\cB E_k  +   E^\#_k)_{L^{-1}} \Big) ( \La  )   \\
\end{split}    
\end{equation}
still at  $\La =  \La_{k+1}$  and    $\phi=   \phi_{k+1, \bom'}$.  
We   insert  this   back into (\ref{representation8}).

\subsection{more adjustments}

We   also  make  some changes in the  fields.    Currently we have the field $\phi_{k+1,\bom'}$.     In the active terms we  change this   to  the desired
  $ \phi_{k+1,  \bom(\La_{k+1}^*) } $ defining     tiny terms    $ R^{*,(i)}_{k+1, \bpi^+}   $  by  
\begin{equation}  \label{winter}
\begin{split}
 S^*_{k+1}( \La_{k+1}, \Phi_{k+1}, \phi_{k+1,\bom'})
= &    S^*_{k+1}( \La_{k+1}, \Phi_{k+1},   \phi_{k+1,  \bom(\La_{k+1}^*)   }  ) +    R^{*,(1)}_{k+1, \bpi^+}    \\
 V_{k+1} \big( \La_{k+1},   \phi_{k+1,\bom'}    \big)
=    &    V_{k+1} \big( \La_{k+1},    \phi_{k+1,  \bom(  \La^*_{k+1})}  \big) +     R^{*,(2)}_{k+1, \bpi^+}   \\
  E_{k+1} \big( \La_{k+1},   \phi_{k+1,\bom'}    \big)
=    &    E_{k+1} \big( \La_{k+1},    \phi_{k+1,  \bom(  \La^*_{k+1})}  \big) +       R^{*,(3)}_{k+1, \bpi^+}  \\
\end{split}
\end{equation}
In the inactive terms we  change to the field     $ \phi_{k+1,  \bom(  \La_k, \Om_{k+1},  \La_{k+1})  }  $  defining more tiny terms  by 
  \begin{equation}  \label{solstice}
\begin{split}
 S^*_{k+1}\Big(\de \La_k , \Phi_{k+1, \bom^+}, \phi_{k+1,\bom'}\Big)
=&  S^*_{k+1}\Big(\de \La_k , \Phi_{k+1, \bom^+},   \phi_{k+1,  \bom(  \La_k, \Om_{k+1},  \La_{k+1})  }   \Big )+   R^{*,(4)}_{k+1, \bpi^+}    \\ 
 V^u_{k+1} \big(\de \La_k ,   \phi_{k+1,\bom'}    \big)
=    &    V^u_{k+1} \big(\de \La_k ,    \phi_{k+1,  \bom(  \La_k, \Om_{k+1},  \La_{k+1}) }  \big) +     R^{*,(5)}_{k+1, \bpi^+}   \\
\end{split}
\end{equation}
where  $\de \La_k =  \La_k - \La_{k+1}$.
With the new arguments we identify   $ S^+_{k+1}(\La_{k+1})   =S^*_{k+1}(\La_{k+1})    + V_{k+1}(\La_{k+1})$,   and 
$ S^{+,u}_{k+1}(\de   \La_k)  =S^*_{k+1}(\de  \La_k)   + V^u_{k+1}(\de   \La_k)$,   and then  
\begin{equation}  \label{swannee}
\begin{split}
K_{k+1,  \bpi^+}   = &  [ K_{k,  L^{-1}} ]_{\bpi}   \exp  \Big(  c_{k+1}  |\Om^{c,(k)}_{k+1}|  
 -  S^{+,u}_{k+1}( \de   \La_k)    + \tilde     B_{k+1, \bpi^+}(\La_k, \La_{k+1})  \Big)\\
\end{split}
\end{equation}

Now    collect the tiny  terms  defining  $  R^{*,(0)}_{k+1, \bpi^+} =  [( R^\#_{k})_{L^{-1}}]_{\bpi^+}(\La_{k+1})$  and
\be    
R^{*}_{k+1,  \bpi^+} =   
  R^{*,(0)}_{k+1, \bpi^+}  +  \dots  +  R^{*,(5)}_{k+1, \bpi^+}
  \ee
For boundary   terms there is     $  B^{*,(0)}_{k+1, \bpi^+} =  [( B^\#_{k})_{L^{-1}}]_{\bpi^+}(\La_{k+1})$.   There is also
the part of      $  (\cB E_k)_{L^{-1}}(\La_{k}) $ left over   after we took out  $  (\cB E_k)_{L^{-1}}(\La_{k+1}) $.  This is
\be   
 B^{*,(1)}_{k+1, \bpi^+} 
\equiv   \sum_{ X  \subset  \La_k,  X \cap  \La^c_{k+1}  \neq  \emptyset}  (\cB E_k)_{L^{-1}}(X,  \phi_{k+1, \bom'} )
\ee
Finally there is     $  B^{*,(2)}_{k+1, \bpi^+} =
(  \cT_{\La_{k+1}} (\cB E_k  +   E^\#_k)_{L^{-1}} )( \La_{k+1}) $.  Altogether then  we define  
 \begin{equation}  \label{naval}
 B^{*}_{k+1,  \bpi^+}  =   B^{*,(0)}_{k+1, \bpi^+} +  \dots  +   B^{*,(2)}_{k+1, \bpi^+}  
\end{equation}
 With   these changes   and  (\ref{nugatory})   the representation  (\ref{representation8}) becomes
   \begin{equation}     \label{representation9}
\begin{split}
   \rho_{k+1}( \Phi_{k+1})     =     &Z_{k+1}  \sum_{\bpi^+}  \int d \Phi_{ \bom^{+,c}}  d W_{k+1,\bpi^+}  K_{k+1,  \bpi^+}   \cC_{k+1,\bpi^+}   \\
&  \chi_{k+1}(  \La_{k+1} )   \exp \Big(-  S^+_{k+1}(\La_{k+1})    
    + E_{k+1}(\La_{k+1}) +  R^{*}_{k+1,  \bpi^+}  +    B^{*}_{k+1, \bpi^+}      \Big) \\
    \end{split}
\end{equation}

\subsection{final  localization}

The last expression is  in   final  form   except that the terms  $ R^{*}_{k+1,  \bpi^+}$  and  $ B^{*}_{k+1, \bpi^+}   $  are not
properly localized   and we  need to establish some estimates.  These are   the problems  to which we now turn.
The proofs are very similar to the treatment in section   \ref{localization}.

 \begin{lem}     
  The function   $ R^*_{k, \bpi^+} $   can be  written    
  \begin{equation}
  \begin{split}
   R^{*}_{k+1, \bpi^+}   = \sum_{ X \subset  \La_{k+1} }    R_{k+1,\bpi^+}( X) 
  +    \sum_{  X \in \cD_{k+1}( \bmod  \Om^c_{k+1}), X  \# \La_{k+1}}     B^{*,(R)}_{k+1, \bpi^+}( X)  
  \end{split}
 \end{equation}
  The functions       $ R_{k+1,\bpi^+}( X,    \Phi_{k+1})$  and   $ B^{*,(R)}_{k+1, \bpi^+}( X,   \Phi_k,  \Phi_{k+1}) $   depend on the fields  only in $X$,     are   analytic  in  $\cP_{k+1}(\La_{k+1}, 2 \de)$  and  $\cP_{k+1, \bom^+}$ respectively,    
and on this domain they satisfy  
\begin{equation} \label{lulu}
\begin{split}
|   R_{k+1,\bpi^+}( X) |  \leq  &    \la_{k+1}^{n_0}   e^{- \ka d_{M}(X)}  \\
| B^{*,(R)}_{k+1, \bpi^+}( X) | \leq  &   \one    \la_{k+1}^{n_0}   e^{- \ka d_{M}(X,  \bmod  \Om^c_{k+1})}  \\
 \end{split}
\end{equation}
  \end{lem}
  \bigskip

\pr $ R_{k+1,\bpi^+}$   has many  pieces,  which we consider individually.   Keep in mind that   $\cP_{k+1, \bom^+}  \subset   \cP_{k+1}(\La_{k+1}, 2 \de)$.  
\bigskip

\noindent  
\textbf{The term  $R^{*(0)}_{k+1,  \bpi^+}$.}
 This has the form      $ \sum_{X \subset  \La_{k+1}} R^{*(0)}_{k+1,  \bpi^+}(X)$   with 
\be   \label{worn}
     R^{*(0)}_{k+1,  \bpi^+}(X,  \phi,  \Phi_{k+1})  =    (R^\#_k)_{L\bpi^+}(LX,  \phi_L,  \Phi_{k+1,L}  ) \hs  \textrm{ at  }  \phi  =   \phi_{k+1, \bom'}
\ee   
By lemma  \ref{fourth} we    know   $ R^\#_k( X, \phi,  \Phi_{k+1} )  $   is analytic  in  $\Phi_{k+1}  \in  \cP^0_{k+1}( \La_{k+1}, 2 \de)$ 
 and  $\phi   \in  \frac12  \cR_k$.
Hence the scaled version    $ R^{*(0)}_{k+1,  \bpi^+}(X,  \phi,  \Phi_{k+1})$
 is analytic in   $\Phi_{k+1}   \in   \cP_{k+1}( \La_{k+1}, 2 \de)$  and  $\phi_L  \in \frac12 \cR_k$.  
 Also  by      (\ref{twoone1})     $\Phi_{k+1}   \in  \cP_{k+1, \bom^+}$  implies that      $|\phi_{k+1, \bom'}|   \leq  C  \la_{k+1}^{-\frac14 - 2 \de}$  with similar bounds on the derivatives.      Then    $\phi_{k+1, \bom',L}$  and its derivatives    satisfy the same bounds  
   and since  $2\de <\ep$  these are more than sufficient to guarantee that   
$\phi_{k+1, \bom',L} \in  \frac 12 \cR_k$.  Thus  $\cP_{k+1, \bom^+}$   is a   correct analyticity domain  for   $R^{*(0)}_{k+1,  \bpi^+}(X)$.
From   (\ref{whale2})  and  $d_{LM}(LX) = d_M(X)$  we have the bound  on this domain
  \begin{equation}  \label{sluggish}
| R^{*(0)}_{k+1,  \bpi^+}(X)|  \leq    \cO(1)L^3  \la_k^{n_0}  e^{- L(\ka-6\ka_0 -6) d_M(X)}
\end{equation}
Next    localize by  introducing  the  weakened  field   $ \phi_{k+1, \bom'}(s)$ 
 as before   and   define
  \begin{equation}
 R^{*(0)}_{k+1,  \bpi^+}(X,s)=    R^{*(0)}_{k+1,  \bpi^+}(X,   \phi_{k+1, \bom'}(s),  \Phi_{k+1})
\end{equation}
which has the same bound.  Now make a decoupling expansion roughly following the  the treatment in  lemma   \ref{first}.  This yields  the strictly local expansion
in   $Z  \in  \cD_{k+1}$
\begin{equation}  \label{shoddy}
 R^{*(0)}_{k+1,  \bpi^+}  =  \sum_{ Z  \cap    \La_{k+1}  \neq  0}   ( R^{*(0)}_{k+1,  \bpi^+})'(Z,   \Phi_k,  \Phi_{k+1})
\end{equation}
with the bound 
\begin{equation}
| ( R^{*(0)}_{k+1,  \bpi^+})'(Z) |  \leq    \cO(1)L^3  \la_k^{n_0}  e^{- L(\ka -8 \ka_0-8) d_M(Z)}
\end{equation}

In the sum  (\ref{shoddy}) consider  terms with    $ Z  \subset    \La_{k+1}$.  These terms contribute to    $ R_{k+1,\bpi^+}(Z) $.
  They only depend
on  $\Phi_{k+1}$  and so are analytic  in    $\cP_{k+1}(\La_{k+1}, 2 \de)$.   We  assume   that  $\ka$  is sufficiently large such that
$ L(\ka -8 \ka_0-8) \geq   \ka $.   (It suffices for example that  $\ka  \geq  16\ka_0 +16$).      Then  the exponent is dominated by  
$ e^{- \ka d_M(Z)}$.   Furthermore     for $L$ sufficiently large and $n_0  \geq  4$ 
\begin{equation}
\cO(1)L^3  \la_k^{n_0} =  \cO(1)  L^{3-n_0}  \la_{k+1} ^{n_0}   \leq  \frac16  \la_{k+1} ^{n_0}
\end{equation}
This is why we chose  $n_0 \geq  4$.  This  is the basic   mechanism which keeps  the   tiny terms  tiny,  in spite of the growth factor  $L^3$.
Thus the bound  is  $| ( R^{*(0)}_{k+1,  \bpi^+} )'(Z) |  \leq  \frac16  \la_{k+1} ^{n_0}e^{-  \ka   d_M(Z)}$.

Now consider  terms  in  (\ref{shoddy})  with   $Z  \#  \La_{k+1}$  which   contribute to  $ B^{*,(R)}_{k+1, \bpi^+}( X)  $.
We  add   connected components of   $\Om^c_{k+1}$  connected to   $Z$,  and   
resum  to   polymers  $X \in  \cD_{k+1}(\bmod \  \Om^c_{k+1})$.   Each  term is  a contribution  to      
$ B^{*,(R)}_{k+1, \bpi^+}(X  )$  and is bounded by  say $  \one L^3  \la_{k} ^{n_0}e^{-L(\ka  - 9\ka_0 -9)  d_M(X,   \bmod  \Om^c_{k+1}) }$.
See step  (H.) in  the proof  of lemma \ref{first}  for details.  
Again    for    $\ka, L  $ sufficiently large this is dominated by  the required    $  \one  \la_{k+1} ^{n_0}   e^{-\ka  d_M(X,   \bmod  \Om^c_{k+1})} $.

 \bigskip

\noindent  
\textbf{The terms  $R^{*(1)}_{k+1,  \bpi^+}  $ }
We     have     $ R^{*(1)}_{k+1,  \bpi^+}  = \sum_{\square \subset  \La_{k+1}}  R^{*(1)}_{k+1,  \bpi^+}( \square)$.
where $\square$ is an  $M$-cube  and
\be   
R^{*(1)}_{k+1,  \bpi^+}( \square) =    S^*_{k+1}( \square  , \Phi_{k+1},  \phi_{k+1, \bom'} )
-   S^*_{k+1}( \square  , \Phi_{k+1},  \phi_{k+1,   \bom(  \La^*_{k+1})} )
\ee
Again  (\ref{twoone1}) implies     that  $ \phi_{k+1, \bom'}$  and $ \phi_{k+1,   \bom(  \La^*_{k+1})} $
are  bounded  by  $C  \la_{k+1}^{-\frac14 - 2 \de}$, also for  derivatives.    Therefore  
$|R^{*(1)}_{k+1,  \bpi^+}( \square)|  \leq  C M^3 \la_{k+1}^{-\frac12 - 4 \de}$. 

  Next introduce  the abbreviated notation 
\be
\begin{split}
\phi  = &   \phi_{k+1,  \bom(  \La^*_{k+1})}   \hs
\de \phi =     \phi_{k+1, \bom'} -  \phi_{k+1,  \bom(  \La^*_{k+1})  }\\
\end{split} 
\ee
Then     $| \de \phi|  \leq  C  \la_{k+1}^{-\frac14 - 2   \de} e^{-r_{k+1}} $  on    $\La_{k+1}$. 
This  follows  since     $G_{k+1, \bom(  \La^*_{k+1})}$ and   $G_{k+1,  \bom'}$     have random  walk expansions
which only  differ   outside  $ \La^*_{k+1}$  which  is   $[r_{k+1}]$  steps    away from  $\La_{k+1}$.  
Then the representation
\begin{equation}
\begin{split}
 R^{*,(1)}_{k+1, \bpi^+}(\square) = & S^*_{k+1}( \square  , \Phi_{k+1},  \phi + \de  \phi)
-   S^*_{k+1}( \square  , \Phi_{k+1},  \phi )\\
=  & \frac{1}{2 \pi i}   \int_{|t|  =   e^{r_{k+1}}}  \frac{dt}{t(t-1)}    S_{k+1}\Big( \square, \Phi_{k+1}, \phi    + t\de \phi \Big)\\
\end{split}
\end{equation}
yields     the bound  
\begin{equation}  \label{noisette}
| R^{*,(1)}_{k+1, \bpi^+}(\square) | 
 \leq   CM^3  \la_{k+1}^{-\frac12 - 4\de} e^{-r_{k+1}} 
  \leq   \la_{k+1}^{n_0 +1} 
\end{equation}

For decoupling we have to be a little more careful,    since   there   are    two different Green's functions to  decouple.
First  in $\de \phi$ in  $ R^{*,(1)}_{k+1, \bpi^+}(\square)$     we  replace     $ \phi_{k+1,  \bom(  \La^*_{k+1})  },  \phi_{k+1, \bom'}$    by truncated versions   $ \phi^{\tr}_{k+1,  \bom(  \La^*_{k+1})  },  \phi^{\tr}_{k+1, \bom'}$   in which   $ G_{k+1, \bom(  \La^*_{k+1})},    G_{k+1,  \bom'}$
are replaced by 
\be 
\begin{split}
 G^{\tr}_{k+1, \bom(  \La^*_{k+1})}
\equiv  & \sum_{\om:  X_{\om_0} \subset  \La_{k+1},  X_{\om}  \cap  \La_{k+1}^{*,c}  \neq  \emptyset }
G_{k+1,  \bom (  \La^*_{k+1} ),  \om }  \\
  G^{\tr}_{k+1,  \bom'}
\equiv  & \sum_{\om:  X_{\om_0} \subset  \La_{k+1},  X_{\om}  \cap  \La_{k+1}^{*,c}  \neq  \emptyset }
G_{k+1,  \bom',  \om }    \\
\end{split}
\ee
Here the random walk is based on   $  \bom(  \La^*_{k+1}) $ in the first case,  and on   $\bom'$ in the second case.
The  condition $X_{\om}  \cap  \La_{k+1}^{*,c}  \neq  \emptyset $  is  appropriate since  terms with     $X_{\om}  \subset  \La_{k+1}^*$
are the same for the two fields and so cancel.  The  fields    $ \phi^{\tr}_{k+1,  \bom(  \La^*_{k+1})  },  \phi^{\tr}_{k+1, \bom'}$  now separately  satisfy the $C  \la_{k+1}^{-\frac14 - 2   \de} e^{-r_{k+1}} $ 
bound.

Next we     weaken  the   $  \bom(  \La^*_{k+1}) $  fields   by  introducing  parameters  $s$ for  the  $\bom(  \La^*_{k+1})$ random
walk and defining
\begin{equation}
\begin{split}
& R^{*,(1)}_{k+1, \bpi^+}(\square, s) \\
=  & \frac{1}{2 \pi i}   \int_{|t|  =   e^{r_{k+1}}}  \frac{dt}{t(t-1)}    S_{k+1}\Big( \square, \Phi_{k+1}, \phi_{k+1,  \bom(  \La^*_{k+1})}(s)   
  + t(     \phi^{\tr}_{k+1,  \bom'} - \phi^{\tr}_{k+1,  \bom(  \La^*_{k+1})}(s) ) \Big)\\
\end{split}
\end{equation}
This   also satisfies   the   bound  (\ref{noisette}).  
  A decoupling expansion in $\square^c$      leads to a sum of terms  indexed by  $Y \in  \cD_{k+1,   \bom(  \La^*_{k+1})}$.
We   resume   to    terms  indexed  by  $Z   \in  \cD_{k+1}$
As  before this leads to the  representation
\begin{equation}
 R^{*(1)}_{k+1,  \bpi^+}  =  \sum_{ Z  \cap    \La_{k+1}  \neq  \emptyset,  Z \subset  \Om_{k+1}} 
  ( R^{*(1)}_{k+1,  \bpi^+} )'(Z, \Phi_{k+1},  \phi^{\tr}_{k+1,  \bom' }  )
\end{equation}
local in the indicated fields      
and    (the coefficient $3 \ka$ achieved for  $M$ sufficiently large)
\begin{equation}
|(   R^{*(1)}_{k+1,  \bpi^+})'(Z)|     \leq  \one
 \la_{k+1}^{n_0+1} e^{- 3\ka d_M(Z)}   
\end{equation}

Next   we  weaken  the  $\bom'$  field by introducing parameters $s$  for  the  $\bom'$  random walk and defining  for $X \in \cD_{k+1}$
\be  ( R^{*(1)}_{k+1,  \bpi^+} )'(X, s )  =
 ( R^{*(1)}_{k+1,  \bpi^+} )'(X, \Phi_k,  \phi^{\tr}_{k+1,  \bom' }(s)  )
\ee
 A decoupling expansion leads to a sum of terms  indexed by  $Y \in  \cD_{k+1,   \bom'}$.
We   resum   to    terms  indexed  by  $Z   \in  \cD_{k+1}$.    Then we have   
\begin{equation}
 R^{*(1)}_{k+1,  \bpi^+}  =  \sum_{ Z  \cap    \La_{k+1}  \neq  0} 
  ( R^{*(1)}_{k+1,  \bpi^+} )''(Z, \Phi_k, \Phi_{k+1}  )
\end{equation}
local in the indicated fields   and  
\begin{equation}
|(   R^{*(1)}_{k+1,  \bpi^+})''(Z)|     \leq  \one
 \la_{k+1}^{n_0+1} e^{- 2\ka d_M(Z)}   
\end{equation}

Now  split the terms  into  a  contribution to  
 $ R_{k+1,\bpi^+}( X) $ and  
$ B^{*,(R)}_{k+1, \bpi^+}( X  )$
     as in the previous case.  For the  terms  $ R_{k+1,\bpi^+}( X)$  we use  $ \one
 \la_{k+1}^{n_0+1} e^{- 2\ka d_M(X)} \leq  
 \la_{k+1}^{n_0} e^{- \ka d_M(X)}  $ 
\bigskip

\noindent  
\textbf{The terms  $R^{*(2)}_{k+1,  \bpi^+},  R^{*(3)}_{k+1,  \bpi^+} $.}
   Define   $\phi,  \de \phi$  as  before.   
   We  have the representations for $\square, X \subset  \La_{k+1}$
     \begin{equation}
    \begin{split}
 R^{*,(2)}_{k+1, \bpi^+}(\square  ) = &
   \frac{1}{2 \pi i}   \int_{|t|  =   e^{r_{k+1}}}  \frac{dt}{t(t-1)}    V_{k+1}\Big( \square, \phi    + t\de \phi \Big) \\
    R^{*,(3)}_{k+1, \bpi^+}(X  ) = &
   \frac{1}{2 \pi i}   \int_{|t|  =   e^{r_{k+1}}}  \frac{dt}{t(t-1)}    E_{k+1}\Big( X, \phi    + t\de \phi \Big) \\
\end{split}
\end{equation}
  which shows  these are tiny.    Then proceed with the localization  and split  as before. 
   \bigskip

\noindent  
\textbf{The terms  $R^{*(4)}_{k+1,  \bpi^+},     R^{*(5)}_{k+1,  \bpi^+}$ }.  The term 
$R^{*(4)}_{k+1,  \bpi^+}$  is  treated like  $R^{*(1)}_{k+1,  \bpi^+}$, and  
$R^{*(5)}_{k+1,  \bpi^+}$  is treated like $R^{*(2)}_{k+1,  \bpi^+}$.

A   difference  is that these terms are initially    localized in    $\de \La_k = \La_k - \La_{k+1}$.  In this domain 
 $\de \phi  = \phi_{k+1,\bom'}     -  \phi_{k+1,  \bom(  \La_k, \Om_{k+1},  \La_{k+1})  }$   satisfies   
 $ |  \de  \phi   |  \leq   \one  e^{-r_{k+1}}$.  This is
 because  
the  random walk expansions for    $G_{k+1,  \bom'}$  and  $ G_{k+1,  \bom(  \La_k, \Om_{k+1},  \La_{k+1}) }$   
starting in  $\de \La_k$   only  differ   outside     $\La_{k+1}^{c,*}$, i.e.  in  $\La_{k+1}^{\nat}$.  Hence  they     have at least   $[r_{k+1}]$  steps   and  this   gives the   tiny   factor  $e^{-r_{k+1}}$.  

Another point  is  that   in the expression for  say  $(R^{*(5)}_{k+1,  \bpi^+})''(Z)$,    only  paths   which  stay  in  $Z$  contribute.
If    $Z  \cap  \La_{k+1}  = \emptyset$   then   the paths   must  stay  in   $\La^c_{k+1}$  as  well as visiting   $\La_{k+1}^{\nat}$.   Hence  there are no paths contributing to     $\de \phi$  in this case,   from which  one  can deduce   $(R^{*(5)}_{k+1,  \bpi^+})'(Z)  =  0$.    Thus  we  can restrict  to  $Z \cap  \La_{k+1}  \neq  \emptyset$   and  hence   $Z \# \La_{k+1}$.  All these terms contribute to   $  B^{*,(R)}_{k+1, \bpi^+}( X)  
$.

  \bigskip

 \begin{lem}      
  The function   $ B^*_{k+1, \bpi^+} $ can be written in the form  
  \begin{equation}
   B^*_{k+1, \bpi^+}   = \sum_{  X \in \cD_{k+1}( \bmod  \Om^c_{k+1}), X  \# \La_{k+1}}   (  B^{*}_{k+1, \bpi^+})'( X)  \ + \  \tilde     B_{k+1, \bpi^+} \textrm{\   terms}
 \end{equation}
 where    $( B^{*}_{k+1, \bpi^+})'( X,   \Phi_{k+1,\bom^+},  W_{k+1,\bpi^+})  $  depends on the  fields only in $X$,     is analytic in
 \be  \label{citrus}
  \Phi_{k+1,\bom^+} \in  \cP_{k+1, \bom^+}   \hs   |W_j|  \leq    B_w \   p_j L^{\frac12(k+1-j)}   \ \ \ \    j=0,1, \dots k
  \ee
On this domain it satisfies   
\begin{equation} \label{lulu2}
 | (B^{*}_{k+1, \bpi^+})'( X) |  \leq   \frac12 B_0  \la_{k+1}^{\beta}   e^{- \ka d_M(X,\bmod \Om^c_{k+1})}  \\
\end{equation}
  \end{lem}
  \bigskip

\re
The    $\tilde  B_{k+1, \bpi^+}$ terms   are absorbed into the  $\tilde B_{k+1, \bpi^+}(\La_k, \La_{k+1})$   in   (\ref{swannee}).
\bigskip

\pr   $ B^{*}_{k+1,  \bpi^+}$  has  three    parts  which we consider separately.
\bigskip

\noindent  
\textbf{The term  $B^{(0)}_{k+1, \bpi^+}  =     [( B^\#_{k })_{L^{-1}}]_{\bpi^+}(\La_{k+1})$   }.  
This has the form     $B^{(0)}_{k+1, \bpi^+}  =  \sum_X  B^{(0)}_{k+1, \bpi^+}  (X)$
with the sum  over    $X \in  \cD_{k+1}(\bmod \  \Om^c_{k+1}), X \# \La_{k+1}$  and      
\be 
B^{(0)}_{k+1, \bpi^+}  (X, \phi,   \Phi_{k+1,\bom^+},  W_{k+1,\bpi^+}       ) 
 =      [ B^\#_{k }]_{L\bpi^+}  (LX, \phi_L,   \Phi_{k+1,\bom^+,L},  W_{k+1,\bpi^+,L})
\ee
evaluated at     $\phi = \phi_{k+1, \bom'}$.
By  lemma   \ref{third}    $ B^\#_{k }$   is analytic  in the domain   $\phi  \in  \frac12 \cR_k$  and   (\ref{500}).   Hence   $B^{(0)}_{k+1, \bpi^+} $  is analytic in    $\phi_L  \in  \frac12  \cR_k$  and (\ref{citrus})      which is the scaling of    (\ref{500}).     Furthermore (\ref{citrus}) and    (\ref{twoone1})  imply     $\phi_{k+1, \bom',L}    \in  \frac12  \cR_k$,    hence  with  $\phi = \phi_{k+1, \bom'}$    we  are in the analyticity domain for the function.
Also from lemma  \ref{third}  since     $\beta  <\frac14 - 10 \ep $ and $\la_k < \la_{k+1}$  we have the bound  
 $|B^\#_{k, \bpi^+}(X) |  \leq   \cO(1) L^3 \la_{k+1}^{\beta}  e^{ - L ( \ka- 6\ka_0-6)   d_{LM}(X,  \bmod \Om^c_{k+1})  }$.
 Since   $ d_{LM}(LX,  \bmod \   L\Om^c_{k+1}) = d_{M}(X,  \bmod \Om^c_{k+1})$ we have   on   (\ref{citrus}) 
    \begin{equation}
|   B^{(0)}_{k+1, \bpi^+}(X) |  \leq    \cO(1)L^3  \la_{k+1}^{\beta  }  e^{- L ( \ka- 6\ka_0-6)   d_M(X, \bmod \Om^c_{k+1})}
\end{equation}

To  localize we    weaken the coupling by  again    introducing  the field   $ \phi_{k+1, \bom'}(s)$  where  $s =  \{ s_{\square} \}$
is indexed by  cubes  $\square$ compatible with  $\bom' =  ( \bom(\La_k^*), \Om_{k+1})$  and not in   $X$.    This includes cubes in 
a   connected component of  $\Om^c_{k+1}$  disjoint from    $X$,  but  not   cubes in 
a  connected component of  $\Om^c_{k+1}$  contained in     $X$.    Now  define  
  \begin{equation}
 B^{*(0)}_{k+1,  \bpi^+}(X,s)=    B^{*(0)}_{k+1,  \bpi^+}(X,   \phi_{k+1, \bom'}(s),   \Phi_{k+1,\bom^+},  W_{k+1,\bpi^+})
\end{equation}
which has the  same analyticity domain and the  same bound.  Now make a decoupling expansion similar to lemma    \ref{third}.   
This generates terms indexed by    multiscale   polymers   $Y \in  \cD_{k+1, \bom'}$.    These are resumed to  give terms indexed 
by   polymers  $Z_0  \in  \cD_{k+1}$.   We  take  the union with any new  connected components of $\Om_{k+1}^c$ to get  
$Z = Z_0^+   \in   \cD_{k+1}( \bmod  \Om^c_{k+1})$.
We  then find that   
\begin{equation}
 B^{*(0)}_{k+1,  \bpi^+}  =  \sum_{ Z  \#    \La_{k+1} }   ( B^{*(0)}_{k+1,  \bpi^+} )'(Z,   \Phi_{k+1,\bom^+},  W_{k+1,\bpi^+})
\end{equation}
In  estimating this we use  the bound  
\be   |Z_0-X|_M   +  d_M( X ,  \bmod \ \Om_{k+1}^c)  \geq   d_M( Z ,  \bmod \ \Om_{k+1}^c) 
\ee
which is proved as in  (\ref{shonuff}).    Using also  (\ref{999}) we    find that 
\begin{equation}
|  ( B^{*(0)}_{k+1,  \bpi^+} )'(Z) |  \leq    \cO(1)L^3  \la_{k+1}^{\beta}  e^{- L(\ka- 8\ka_0 -8) d_M(Z, \bmod \Om^c_{k+1})} 
\leq    \frac14 B_0 \la_{k+1}^{\beta}   e^{- \ka d_M(Z, \bmod \Om^c_{k+1})} 
\end{equation}
We  have assumed     $B_0$  is sufficiently  large so that   $ \cO(1)L^3  \leq   \frac14 B_0$
\bigskip

\noindent  
\textbf{The term  $  B^{*,(1)}_{k+1, \bpi^+}$   }.     This is similar to the previous case.    After a decoupling expansion we find
\be
   B^{*,(1)}_{k+1, \bpi^+}  =  \sum_{Z_0 \cap  \La^c_{k+1} \neq \emptyset }     B^{*,(1)}_{k+1, \bpi^+} (Z_0,  \Phi_{k+1, \bom^+} )
\ee
with
\be
    |  B^{*,(1)}_{k+1, \bpi^+} (Z_0 )  |   \leq  L^3 \la_k^{\beta} e^{-{L(\ka - \ka_0 -1)}d_M(Z_0) }
\ee
Terms with  $Z_0  \subset  \La^c_{k+1}$  are  $ \tilde     B_{k+1, \bpi^+}$  terms.    For  terms      with   $Z_0 \#  \La_{k+1}$   we  adjoin 
any connected components of  $\Om^c_{k+1}$   and call  the result   $  ( B^{*,(1)}_{k+1,  \bpi^+} )'(X)  $.
\bigskip

\noindent  
\textbf{The term  $  B^{*,(2)}_{k+1, \bpi^+}$   }.   
This term depends on
\be   E_k^*(X, \phi)  \equiv  ( \cB E_k  +   E^\#_k)_{L^{-1}}(X, \phi)  = ( \cB E_k  +   E^\#_k)(LX,  \phi_L)   
\ee
For   $\phi  \in \frac12  \cR_k$   we have  
$ |E_{k}(X, \phi)|  \leq  \la_{k}^{\beta}e^{-\ka d_M(X)}$  
and    $ |E^\#_{k}(X,\phi)|  \leq  \one L^3 \la_{k}^{\frac14  - 10 \ep}e^{-L(\ka- 3 \ka_0 -3) d_{LM}(X)}$.  
Hence for   $\phi_L  \in \frac12  \cR_k$   
\be  
\label{estar}   |E^*_k(X, \phi)| \leq     \one L^3  \la_{k+1}^{\beta  }e^{-2 \ka d_M(X)}
\ee
and the  same holds in the smaller domain  $\phi  \in \cR_{k+1}$.

Now   
$
  B^{*,(2)}_{k+1, \bpi^+} =  \sum_{X} B^{*,(2)}_{k+1, \bpi^+}(X )
$
where  the sum is over  $X \in \cS$  and    $ X  \#  \La_{k+1}$  and  
\begin{equation}
\begin{split}
 B^{*,(2)}_{k+1, \bpi^+}(X,\phi_{k+1, \bom'})  =  & ( \cT_{\La_{k+1}} E^*_k)(X,\phi_{k+1, \bom'}) \\
 =   &   \al_0(E^*_k,X)\  \Vol( \La_{k+1} \cap  X ) +      \al_{2}(E_k^*,X)\  \| \phi_{k+1, \bom'} \|^2_{X \cap  \La_{k+1}} \\
 + & \sum_{\mu}      \al_{2, \mu}(E_k^*,X)  
    \int_{X  \cap  \La_{k+1}}  \phi_{k+1, \bom'} \  \pa_{\mu}   \phi_{k+1, \bom'} \\
 \end{split}   
 \end{equation}
 The bound    (\ref{estar})  implies  (see similar estimates in part I)
 \be \begin{split}
 | \al_0(E^*_k,X) |  \leq  & \cO(1)  \Vol(X)^{-1} L^3  \la_{k+1}^{\beta }e^{-2\ka d_M(X)}  \\
      |\al_{2}(E_k^*,X)|  \leq  &  \cO(1)\Vol(X)^{-1} L^3    \la_{k+1}^{\beta  + \frac12  +  6 \ep}e^{-2 \ka d_M(X)}  \\
     |\al_{2, \mu}(E_k^*,X)|  \leq &   \cO(1)\Vol(X)^{-1}  L^3   \la_{k+1}^{\beta  + \frac12 + 5 \ep}e^{-2\ka d_M(X)}  \\   
     \end{split}
     \ee
 Furthermore  $ |  \phi_{k+1, \bom'} | \leq  \one     \la_{k+1}^{-\frac14  - 2 \de} $  and hence      $ \| \phi_{k+1, \bom'}  \|^2_{X \cap  \La_{k+1}}  
\leq  \cO(1)\Vol(X \cap  \La_{k+1}) \la_{k+1}^{-\frac12  - 4 \de}$.  The same bound holds for    $ \int_{X  \cap  \La_{k+1}}  \phi_{k+1, \bom'} \  \pa_{\mu}   \phi_{k+1, \bom'} $.    Therefore  since $2 \de  <  \ep$ 
\begin{equation}
 |B^{*,(2)}_{k+1, \bpi^+}(X)|  \leq     \one   L^3 \la_{k+1}^{\beta} e^{-2\ka d_M(X)}
\end{equation}
Note that  $X \in \cS$  and    $ X  \#  \La_{k+1}$ rules out that  $X$ contains any connected components of  $\Om_{k+1}^c$ 
so we can replace the  $d_M(X)$  by  $d_M(X, \bmod \  \Om^c_{k+1})$.  
Now  localize  as  before    and  get  a   sum over     strictly localized functions  $ ( B^{*(2)}_{k+1,  \bpi^+} )'(Z, \Phi_k, \Phi_{k+1})$  satisfying    
\begin{equation}
|  ( B^{*(2)}_{k+1,  \bpi^+} )'(Z) | \leq    \frac14 B_0 \la_{k+1}^{\beta}   e^{- \ka d_M(Z, \bmod \Om^c_{k+1})} 
\end{equation}

The    lemma now holds with  $ ( B^{*}_{k+1,  \bpi^+} )'(Z)  =   ( B^{*(0)}_{k+1,  \bpi^+} )'(Z)   +  \dots   + ( B^{*(2)}_{k+1,  \bpi^+} )'(Z) $.
\bigskip

\noindent
\textbf{Conclusion:}
From the last two lemmas we can write 
\begin{equation}  \label{almonds}
 R^{*}_{k+1,  \bpi^+}  +    B^{*}_{k+1, \bpi^+}  
=      R_{k+1,\bpi^+}( \La_{k+1})   +   B_{k+1,\bpi^+}( \La_{k+1})   
\end{equation}
where \begin{equation}
  B_{k+1,\bpi^+}( X)   =
   B^{*,(R)}_{k+1, \bpi^+}( X)  +  (  B^{*}_{k+1, \bpi^+})'( X)  
\end{equation}
Then    $ B_{k+1,\bpi^+}( X)   $  is analytic in the domain   (\ref{citrus})   and by  
  (\ref{lulu})  and  (\ref{lulu2}) satisfies there
\begin{equation}
 |B_{k+1,\bpi^+}( X)|   \leq   \Big (\one  \la_{k+1}^{n_0} +     \frac12    B_0 \la_{k+1}^{\beta}\Big)   e^{- \ka d_M(X,  \bmod  \Om^c_{k+1})} 
 \leq     B_0 \la_{k+1}^{\beta}   e^{- \ka d_M(X,  \bmod  \Om^c_{k+1})} 
\end{equation}
Also note   that   $\tilde  B_{k+1, \bpi^+}(\La_k, \La_{k+1})$  is  analytic  on  (\ref{citrus}) and satisfies     for $B_0$ sufficiently large
\be 
   |\tilde  B_{k+1, \bpi^+}(\La_k, \La_{k+1})|  \leq  B_0|\La^{(k+1)}_k -  \La^{(k+1)}_{k+1}  |
\ee

Finally  substituting   (\ref{almonds})  into   (\ref{representation9})
yields
   \begin{equation}     \label{representation10}
\begin{split}
  & \rho_{k+1}( \Phi_{k+1})     =     Z_{k+1}  \sum_{\bpi^{+}}  \int d \Phi_{k+1, \bom^{+,c}}  d W_{k+1, \bpi^+}  K_{k+1,  \bpi^+}   \cC_{k+1,\bpi^+}   \\
&  \chi_{k+1}(  \La_{k+1} )   \exp \Big(-  S^+_{k+1}(\La_{k+1})    
    + E_{k+1}(\La_{k+1}) +   R_{k+1,\bpi^+}( \La_{k+1})   +   B_{k+1,\bpi^+}( \La_{k+1})    \Big) \\
    \end{split}
\end{equation}
All properties  of the various functions have been established, so  this   completes the induction  and the proof of    the main
theorem.
\bigskip

In the third and final paper   we  establish the convergence of the expansion and prove the stability bound.

\newpage
\appendix

\section{notation}
 For  various reasons we  have deviated  from the notation  employed by Balaban  in \cite{Bal95}- \cite{Bal98c}.
The  following  table  is  a dictionary   for  connecting   those papers  with the present paper.   It  is not  exact.  

\begin{table}[h]  
\hspace{3cm}
\begin{tabular}{||l r | l ||}  \hline
This work   &    & Balaban     \\  \hline  \hline
& &  \\
$\Om_k$  &   &    $\Om_k$   \\  
%& &  \\
\hline
& &  \\
$\La_k $  &   &    $\Om_k''$   or  $Z_k^c$ \\ 
%& &  \\ 
\hline
& &  \\
$\La_k^*$  &   &    $W_k$ \\ 
%& &  \\ 
\hline
& &  \\
$ \de    \Om^{(k)}_k$  &   &    $ \La_k$  \\ 
%& &  \\ 
\hline
& &  \\ 
$  \bpi $ &   &    $ \bbA $  \\ 
%& &  \\ 
\hline
& &   \\
$  \bom(\La_k^*) $&   &    $ \bbB(W_k)$  \\ 
%& &  \\ 
\hline
& &  \\ 
$  \bom'  = ( \bom(\La_k^*),   \Om_{k+1}  ) $  &   &    $ (  \bbB_k(W_k) \cap \Om^c_{k+1}, \Om_{k+1} )  $  \\ 
%& &  \\ 
\hline
& &  \\ 
$  \Phi_{k, \bom}  $   &   &    $ \psi  $  \\ 
%& &  \\ 
\hline
& &  \\
$  W_{k, \bom}$  &  &   $\psi'$   \\
%& &   \\ 
\hline
& &  \\
$ \Phi_{k, \Om_{k+1}} $  &   &    $  \theta  $  \\ 
%& &  \\ 
\hline
& &  \\
$\hat    \Phi_{k+1, \bom'} $ &   &    $\tilde  \theta  $  \\ 
%& &  \\ 
\hline
& &  \\
$\hat   \Psi_{k, \bom'} $ &   &    $\psi^{(k)} (\tilde  \theta)  $  \\ 
%& &  \\ 
\hline
\end{tabular}
\caption{comparison of notation}
\end{table}

\section{a lattice identity}  \label{zzz} 

  $\La$  be a union of cubes in  $\bbT^{-k}_{\sM + \sN -k}$  as in the text,   and let  $f,g$  be functions on a neighborhood of $\La$.
We  prove the following  identity:  

\begin{thm}   
\begin{equation}  \label{zero}
< \pa  f,  \pa g>_{*, \La}   =   <  (- \De) f,  g>_{\La}    + \frac12   \sum_{x \in \La,  x' \in \La^c  }   L^{-2k}    \pa  f( x,x' )   (g(x) + g(x'))
\end{equation}
\end{thm}

\pr   
We  have 
\begin{equation}  \label{primo}
\begin{split}   
 <\pa  f,\pa  g>  _{*, \La} =&  \sum_{ <x,x'>  \in \La } L^{-3k}  \pa f(x,x')  \pa g(x,x') 
 +\frac12   \sum_{ x \in \La,  x'\in \La^c} L^{-3k} \pa f(x,x')  \pa g(x,x')  \\
=&  \sum_{ <x,x'>  \in \La } L^{-3k}  \pa f(x,x')  \pa g(x,x') 
 +\frac12   \sum_{ x \in \La,  x'\in \La^c} L^{-2k} \pa f(x,x')  (g(x')-g(x))  \\ 
 \end{split}
\end{equation}
The first  sums  are   over  oriented bonds  $<x,x'> = <x , x + L^{-k} e_{\mu}>$.   The  first  line is the definition and the
second line follows by  $ \pa g(x,x') = L^k  (g(x') -g(x)$.

On the other hand  if  $g_{\La}$ is the restriction to $\La$  we have 
\begin{equation}
\begin{split}
&<  (-\De)  f,  g>_{\La}  =  < (-\De)  f,    g_{\La}>  =  < \pa f,  \pa  g_{\La}>    
=  \sum_{ <x,x'>  \in \La } L^{-3k}  \pa f(x,x')  \pa g(x,x') \\
 +&   \sum_{ <x,x'> : x \in \La,  x'\in \La^c} L^{-3k} \pa f(x,x')   \pa g_{\La}(x,x')  
 +   \sum_{ <x,x'> : x' \in \La,  x\in \La^c} L^{-3k} \pa f(x,x')   \pa g_{\La}(x,x')  \\
\end{split}
\end{equation}
In  the last sum  $\pa g_{\La}(x,x')  =   L^k  g(x')$  and in  the previous sum     $\pa g_{\La}(x,x')  =  - L^k  g(x)$.
Therefore   the last two sums  are  
\begin{equation}  \label{secundo}
\begin{split}
 -&  \sum_{ <x,x'> : x \in \La,  x'\in \La^c} L^{-2k} \pa f(x,x')  g(x)  
 +  \sum_{ <x,x'> : x' \in \La,  x\in \La^c} L^{-2k} \pa f(x,x')   g(x')  \\
 =   &  -      \sum_{ x \in \La,  x'\in \La^c} L^{-2k} \pa f(x,x')  g(x)  \\ 
 \end{split}
\end{equation}
Here in the  second   sum  over   we have  relabeled  $x \leftrightarrow  x'$  and    used   $ \pa f(x',x) = -   \pa f(x,x')$.  Then
this sum and the one preceding  it  are sums  of the same function over  outward bonds $x' \in \La,  x\in \La^c$.    But the first  sum comes  from   outward bonds   so  $<x,x'>$ is  oriented and the second sum  comes from outward bonds    so  $<x',x>$ is oriented.   We  combine 
them into an unrestricted sum  over all  outward bonds to get the last line.

Now  $ <\pa  f,\pa  g>  _{*, \La}-<  (-\De)  f,  g>_{\La}$    gives  the surface integral  in  (\ref{zero})   as announced.

\section{another  identity} \label{moonshine}

We   seek an  alternate   expression for   
\begin{equation}
  C_{k,\bom^+,  r}  =\Big[ \De_{k,\bom} +  \frac{a}{L^2} Q^TQ+ r\Big]^{-1}_{\Om_{k+1}}
\end{equation}
where  as in the text    $\bom^+  = ( \bom ,  \Om_{k+1}) =  ( \Om_1, \dots,  \Om_k,  \Om_{k+1})$.

\begin{lem}

\begin{equation}  \label{z4}
 C_{k,   \bom^+, r} =  \Big[ A_{k,r}  +   a_k^2  A_{k,r} Q_k  G_{k,  \bom^+,  r} Q_k^T A_{k,r}\Big]_{\Om_{k+1}}
\end{equation}
where 
\begin{equation} 
\begin{split}
A_{k,r}   =&   \frac{1}{a_k+r}  (I - Q^TQ)   +   \frac{1}{ a_k + aL^{-2}  +r}   Q^T Q \\
B_{k,r}  = &   \frac{r}{a_k+r}  (I - Q^TQ)   +   \frac{aL^{-2}  +r}{ a_k + aL^{-2}  +r}   Q^T Q \\
G_{k,  \bom^+, r}  = &  \Big[ -\De  + \bar \mu_k  
+ \left[ Q_{k, \bom}^T  \ba  Q_{k, \bom} \right]_{\Om_{k+1}^c}      +   a_k  
\left[   Q_k^T   B_{k,r} Q_k \right]_{\Om_{k+1}}  \Big]_{\Om_1}^{-1} \\
\end{split}
\end{equation}
\end{lem}
\bigskip

\pr   Start with 
\begin{equation}  \label{gumdrop1}
\exp  \Big( \frac12  <f,C_{k,\bom^+,r} f>  \Big)
=  \const \int   d\Phi   \exp \left(  <\Phi, f>  - \frac{a}{2L^2}  \|Q\Phi \|^2 - \frac{r}{2}\|\Phi\|^2 - \frac12  <\Phi,  \De_{k, \bom} \Phi>  \right)  
\end{equation}
where    $f,   \Phi:  \Om^{(k)}_{k+1}  \to  \bbR$.  In general  for  $\phi:   \Om_1 \to \bbR$
\begin{equation}
\begin{split}
&\exp \left(  - \frac12  <\Phi_{k, \bom},  \De_{k, \bom} \Phi_{k, \bom}>  \right)   \\
= &  \const   \int 
  \exp \left(   -\frac{1}{2} \|\ba^{1/2} (\Phi_{k, \bom}    -   Q_{k, \bom}   \phi ) \|^2  - \frac12  < \phi,  ( -\De  + \bar \mu_k  )  \phi  >   \right)   \ d \phi \\
\end{split}  
\end{equation}
This follows from  (\ref{tony3}), (\ref{bombshell2}), (\ref{rain})  with  $\phi_{\Om_1^c} =0$.
Specializing to   $\Phi_{k, \bom}  =   (0, \Phi)$ with $\Phi$ on  $\Om_{k+1}$    this says
\begin{equation}   \label{gumdrop2}
\begin{split}
&\exp \left(  - \frac12  <\Phi,  \De_{k, \bom} \Phi>  \right)  \\
= &  \const   \int 
  \exp \left(   -\frac{a_k}{2} \| \Phi    -   Q_{k}   \phi  \|^2_{\Om_{k+1}}  
   -\frac{1}{2} \|\ba^{1/2}   Q_{k, \bom}   \phi  \|^2_{\Om^c_{k+1}}
   - \frac12  < \phi,  ( -\De  + \bar \mu_k  )  \phi  >   \right)   d \phi  \\
\end{split}   
\end{equation}

Insert  (\ref{gumdrop2}) into  (\ref{gumdrop1})   and  do the integral  over  $\Phi$  which is 
\begin{equation}
\begin{split}
&  \int   d\Phi   \exp \left(  <\Phi, f>  - \frac{a}{2L^2}  \|Q\Phi \|^2 - \frac{r}{2}\|\Phi\|^2   -  \frac{a_k}{2} \| \Phi    -   Q_{k}   \phi  \|^2   \right) \\
=&     \int   d\Phi   \exp \left(  <\Phi, f +a_k Q_k \phi>  
-\frac12  < \Phi,  \Big(   a_k + r  +   aL^{-2}Q_k^TQ_k  \Big)  \Phi  >    -  \frac{a_k}{2}  \|  Q_k  \phi  \|^2   \right) \\
=&  \const    \exp   \Big(  \frac12  \Big<(f  + a_kQ_k),   A_{k,r}  (f  + a_kQ_k)  \Big> -  \frac{a_k}{2}  \|  Q_k  \phi  \|^2  \Big) \\
\end{split} 
\end{equation}
Here  we  used 
$
\Big(   a_k + r  +   aL^{-2}Q_k^TQ_k  \Big) ^{-1}   
 =A_{k,r}
$
which follows since $Q_k^TQ_k$ is a projection. 
 Now   we  have
\begin{equation}
\begin{split}
&\exp  \Big( \frac12  <f,C_{k,  \bom^+,r} f>  \Big)\\
=&   \const
\int   \exp   \Big(  \frac12  \Big<(f  + a_kQ_k\phi),   A_{k,r}  (f  + a_kQ_k\phi)\Big>   \\
- & \frac12  < \phi,  \Big( -\De  + \bar \mu_k  + a_k Q_k^TQ_k +\left[ Q_{k, \bom}^T  \ba  Q_{k, \bom} \right]_{\Om_{k+1}^c}  \Big)   \phi  >  \Big)   \ d \phi
 \\
 =  &   \const \exp   \Big(  \frac12  \left<f ,   A_{k,r}  f    \right>   \Big)
\int    \exp   \Big(    \left< \phi, a_k Q_k^T A_{k,r}  f \right>   \\
& - \frac12  < \phi,  \Big( -\De  + \bar \mu_k  + a_k Q_k^TB_{k,r}Q_k +\left[ Q_{k, \bom}^T  \ba  Q_{k, \bom} \right]_{\Om_{k+1}^c}  \Big)  \phi  >   \Big)
  \ d \phi
 \\
= &     
 \const \exp   \Big(  \frac12  \left<f ,   A_{k,r}  f    \right>   \Big)
\int    \exp   \Big(    \left< \phi, a_k Q_k^T A_{k,r}  f \right>    - \frac12  < \phi,  G_{k, \bom^+,r}^{-1}  \phi  >   \Big)
  \ d \phi
 \\
=& \const \exp \Big(    \frac12   \left<f ,   A_{k,r}  f    \right>  
+   \frac{a_k^2}{2}  <   f,  A_{k,r}Q_k  G_{k,\bom^+, r}  Q_k^T A_{k,r}  f>   \Big)  \\
\end{split}
  \end{equation}
which  gives the result.  Here we have used the identity
 \begin{equation}
 \begin{split}
 a_k^2 A_{k,r}   -a_k   =  &   a_k^2\Big(    \frac{1}{a_k+r}  (I - Q^TQ)   +   \frac{1}{ a_k + aL^{-2}  +r}   Q^T Q \Big)   - a_k \\
 =  &   a_k\Big(   \left( \frac{a_k}{a_k+r} - 1\right)  (I - Q^TQ)   + \left(  \frac{a_k}{ a_k + aL^{-2}  +r} - 1 \right)  Q^T Q \Big)   \\
 =  & - a_k  \Big(    \frac{r}{a_k+r}   (I - Q^TQ)   +  \frac{ aL^{-2}  +r}{ a_k + aL^{-2}  +r}    Q^T Q \Big)  \\
 =&  - a_k B_{k,r}  \\
 \end{split}
  \end{equation}

\section{connected polymer sums}   \label{sanibel}

 Let   $X  \in \cD_{k, \bom}$  be a  multiscale
polymer,   and  let    $  |X|_{\bom}$    be the number of elementary cubes in  $X$  as in section  \ref{polymersection},  except 
now we  work in arbitrary dimension  $d$.

\begin{lem}  \label{donut1}
For  $\ka_*$ sufficiently large   (  $\ka_*  \geq  \cO( \log L)$)  and any   elementary cube  in  $ \square  \subset   \cD_{k, \bom}$
\be   \label{lugnut1}
\sum_{X  \in \cD_{k, \bom},   X  \supset  \square}   e^{-  \ka_*  |X|_{\bom} } \   \leq    e^{- \frac12 \ka_*}
   \ee
\end{lem}
\bigskip

\pr   We  have
\be  
\sum_{X  \supset  \square}    e^{-  \ka_*  |X|_{\bom} }  \leq      \sum_{n \geq 1}   e^{-\ka_* n}  |\{ X \supset  \square:   |X|_{\bom}  =  n\}|
\ee
To  count  $ |\{ X \supset  \square:   |X|_{\bom}  =  n\}|$  note  that for each such  $X$  there  is   a tree  (not unique)    whose lines are  pairs  of  adjacent   cubes  in  $X$.  This  tree   will have  $n-1$ lines. 
   Distinct  polymers  give  distinct  trees  so  that number is less than the number  of   such  trees.     Each tree  can be traversed with
   a path  starting at  $\square$  and  traversing each  line exactly  twice.    Thus the number of  trees in less than the number of paths of 
   length  $2n$  starting at  $\square$.   Since  each cube  has  at most  $2d L^{d-1}$  neighbors  this is bounded by  $(2dL^{d-1})^{2n}$.
Thus the sum is bounded by 
\be       
  \sum_{n\geq 1}   e^{-\ka_* n} (2dL^{d-1})^{2n}
\leq      \sum_{n\geq1}  e^{-(\ka_*  -  2 \log  (2dL^{d-1}) )  n }    \leq   e^{- \frac12 \ka_*}
\ee
provided   $\frac 12  \ka_*   \geq       2 \log  (2dL^{d-1})   +  \log 2$.   
This completes the proof.
\bigskip

For the next result we  relax the condition that  $X$ be connected,    so   $X$  is just   a union of elementary cubes:  $L^{-(k-j)} M$ cubes in $\de \Om_j$ ($M$-cubes in $\Om_k$).     We  sum  over  $Y \supset  X$ of the same form.   A connected component of   $Y$ has the property that every  connected component of $X$ is either contained in it   or is disjoint from it.

\begin{lem}  \label{donut2}  With $\ka_*$ as above
\be  \label{lugnut2}
\sum_{ Y \supset  X }\  '    \  e^{- \ka_* |Y-X|_{\bom} }   \leq  \exp  \Big(   e^{- \frac12 \ka_*} (2dL^{d-1}+1) |X|_{\bom}\Big)
\ee
where the primed sum means every connected component of     $Y$ contains   at least one connected component of $X$ 
\end{lem}
\bigskip

\pr  Let   $\{W_{\al} \}$    be  the connected components  of  $Y-X$.   Let   $ X'$ be the  enlargement  of   $X$  formed by 
adding all cubes which have a face in common with  $X$.  Then  each  $W_{\al}$  contains some   cube   in   $ X' - X$,
(If  not      $W_{\al}$ is disjoint from $X$.   Then   $W_{\al}$ is a connected  subset of   $Y$  with no path in $Y$ to any other cube in $Y$,  since
any such path  would have to pass through  $X$.  Hence  $W_{\al}$  is a connected component of $Y$
 which contains no connected component of $X$ which is a contradiction.)
    Conversely   let  $\{ W_{\al} \}$  be  a collection of   disjoint   connected   subsets   in   $X^c$  with the property that
each  contains  a cube   in  $ X'-X$.  Then    $ Y  = X  \cup  ( \cup_{\al}  W_{\al} )$ has the property that    every  connected component of  
$Y$    contains  at  least one connected component  of  $X$.   (It contains either a connected component of $X$ or   some $W_{\al}$.  In the latter  case   the cube in     $W_{\al}$   provides a link  to some connected component of $X$ which
is therefore included.)  The upshot is  that the   sum can be written as a sum   over  the  $ \{  W_{\al}  \}$.

 We  enlarge the sum  to  a sum  over   collections $ \{  \square_{\al}  \}$ of disjoint cubes  in  $ X'-X  $  or even  $X'$,    and     connected  $W_{\al}$   so  $W_{\al}  \supset  \square_{\al}$.  
Using also the previous lemma the sum is dominated by 
\be
\begin{split}
&\sum_{  \{  \square_{\al}  \}  }   \sum_{   \{W_{\al}:   W_{\al}  \supset  \square_{\al} \}  }    e^{- \ka_*  \sum_{\al} |W_{\al }|_{\bom}}   
=    \sum_{  \{  \square_{\al}  \}  }    \prod_{\al}   \sum_{   W  \supset  \square_{\al}   }   e^{- \ka_* |W|_{\bom}}  \\
  \leq    &    \sum_{  \{  \square_{\al}  \}  }  \prod_{\al} e^{- \frac12 \ka_*}   =     \Big(1 +   e^{- \frac12 \ka_*} \Big)^{|X'|_{ \bom}}
\leq   \exp  \Big(   e^{- \frac12 \ka_*}  |X'|_{\bom}\Big)  
\\
\end{split}
\ee
Since    $|X'|_{\bom}  \leq    (2dL^{d-1}+1) |X|_{\bom}$  we have the result.
\bigskip

\re  A variation is the following.   Suppose  $X \in  \cD_k( \bmod  \Om_k)$, so  $X$ is connected and either contains a connected component
of   $\Om_k^c$ or is disjoint from it.   
Then  $X'-X  \subset  \Om_k$ so the anchors  $\{ \square_{\al}  \}$ in  $X'-X$  are  $M$-cubes.  Hence we  get
$|X-X'|_M   \leq  |X'|_M$  rather than  the  (possibly much larger) $ |X'|_{\bom}$.   The result  is 
\be     \label{999}
  \sum_{Y \in \cD_{k, \bom}:   Y \supset X}     \  e^{- \ka_* |Y-X|_{\bom} }   \leq  \exp  \Big(   e^{- \frac12 \ka_*} (2dL^{d-1}+1) |X|_{M}\Big)
\ee

\section{disconnected polymer sums}      \label{tree}

Let   $Y$  be  a   collection of $M$-blocks  $\square$  in a lattice of dimension $d$.  $Y$  is  not necessarily connected.      We   define  
various  lengths  $\ell(Y),  \ell'(Y),  \tilde  \ell(Y)$  associated with $Y$.   In the following ''tree'' means continuum tree.
\begin{enumerate}
\item    $M  \ell'_M(Y)$  is the length of  a  minimal      tree  whose vertices are the    centers of the  blocks   in   $Y$.
\item  $M \ell_M (Y)$   is   the   length of a minimal     tree   whose vertices are one point from each    block   in    $Y$.   
\item    $M \tilde   \ell_M(Y)$  is the length of  a  minimal  tree    whose vertices are one point from each    block   in    $Y$  and possibly  other  points.
\end{enumerate}
We  have trivially   $\tilde  \ell_M(Y)   \leq  \ell_M(Y)  \leq   \ell'_M(Y)$.      If  $Y$  is connected
then   $ \tilde  \ell_M(Y)$  differs slightly  from $d_M(Y)$  defined in part I,  since  the latter requires a minimal tree to lie in $Y$.  
But we   do have  $  \tilde  \ell_M(Y)  \leq  d_M(Y)$.   Recall also that  $|Y|_M$ is the number of $M$-blocks  in $Y$. 

\begin{lem}   {  \  }  \label{steamy}
\begin{enumerate}
\item   $  \ell_M(Y)   \leq  2 \tilde  \ell_M(Y)$
\item   
$  \ell'_M(Y)     \leq  \ell_M(Y)   +  |Y|_M  $
\item  
$|Y|_M  \leq  4(2^{d}+1)  ( \ell_M(Y)  +1  )$
\end{enumerate}
\end{lem}
\bigskip

\pr   We  can take $M=1$ and drop the subscript  $M$.
\begin{enumerate}
\item   \cite{DIS73}.   Let   $\tilde  \tau$ be a minimal    tree of length  $\ell(\tilde  \tau )  =  \tilde \ell(Y)$.    One can traverse  $\tilde \tau$  with a path  $\tilde \ga$   which 
passes through   every vertex  and has length  $2 \tilde  \ell_M(Y)$.   The path runs through the vertices in some order  $v_1, \dots,  v_n$.
  Replace the segment from the vertex $v_i$   to the vertex $v_{i+1}$ by a straight line.    This gives a path $\ga$  which passes thru each vertex exactly once and is shorter than $\tilde \ga$.    Hence  $\ell(Y)  \leq  \ell(\ga)  \leq    \ell(\tilde \ga)  =  2 \tilde  \ell(Y)$

\item  Let  $\tau$  be a minimal   tree on the  blocks  of    $Y$    of  length $\ell(\tau) =  \ell(Y)$.   Replace each
line  by   a line from center to center  and call the resulting tree   $\tau'$.   This increases  the  lengths by
at most  one.  (We use the metric   $|x-y| = \sup |x_{\mu}- y_{\mu}|$).    Then we have
\be
\ell'(Y)  \leq   \ell(\tau')  \leq   \ell(\tau)  +   |Y|    =  \ell(Y)    +    |Y| 
\ee
   
\item  
Given  $Y$   construct a path   $\ga$ which goes through each vertex  exactly once and has  length $\ell(\ga) \leq  2 \ell(Y)$ as in part (1.).
Let  $\ga_1$ be the first  $2^d+1$  lines,  let  $\ga_2$ be the next  $2^d+1$ lines,  etc., and let   $\ga_n$ be
the last $2^d +1$ or fewer lines, so  $\ga  =   \ga_1 \cup  \dots  \cup  \ga_n$.   For   $1 \leq  i \leq  (n-1)$   we must have
$\ell(\ga_i) \geq  1$ since at most  $2^d$ blocks be mutually touching.  Then if  $|\ga|$ is the number of lines in the path $\ga$
\be 
\begin{split}
  |Y| -1  = & |\ga|   =  \sum_{i=1}^n  |\ga_i|   \leq   n   (2^d +1) 
    \leq \Big( \sum_{i=1}^{n-1}   \ell(\ga_i)   +1\Big) 2(2^d+1)  \\
\leq  &  \Big (\ell(\ga)  +1 \Big)  2(2^d +1)    \leq  4(2^d +1) \ell(Y)   +   2(2^d +1) \\
\end{split}
\ee    
which  is sufficient.
\end{enumerate}

\begin{lem}    \label{kumquat}   \  \begin{enumerate}
\item    There  are  constants   $a', b'= \one$  such that       for any $M$-cube  $\square_0$
\begin{equation}
\sum_{Y : \  Y  \supset  \square_0}
\exp  (  -   a' \ell'_M (Y)   )   \leq  b'
\end{equation}
\item       There  are  constants   $a,  b = \one  $  such that       
\begin{equation}
\sum_{Y : \  Y  \supset  \square_0}
\exp  (  -   a \ell_M (Y)   )   \leq  b
\end{equation}
\end{enumerate}
\end{lem}
\bigskip

\re   These  generalize bounds in part I where  $Y$ was required to be connected.   
 Bounds of this type   were used   extensively in the papers of Gawedski and Kupiainen,  see for example    \cite{GaKu81}.
\bigskip

\pr  
\begin{enumerate}
\item   We take  $M=1$  and drop the subscript  $M$.
The  sum is dominated  by  
\begin{equation}
\sum_{n=0}^{\infty}\sum_{\{ \square_1, \dots,    \square_n  \} }   \exp  (  -   a'  \ell' (\square_0 \cup    \square_1\cup \dots   \cup    \square_n    )   )  
=\sum_{n=0}^{\infty}\frac{1}{n!} \sum_{ (\square_1, \dots,  \square_n)  }  \exp  (  -  a' \ell' (\square_0 \cup    \square_1\cup \dots   \cup    \square_n    )   )  
\end{equation}
where the sum is  first  over unordered    collections of   distinct  blocks
and then over ordered   collections of  distinct  blocks.

For  every  $ (\square_1, \dots,  \square_n) $ there is at least one  tree  $\tau$   on  $(0,1,2, \dots,n)$  such   that  the induced length
\be d_{\tau}  (\square_0,\square_1, \dots,  \square_n) \equiv  \sum_{\{i,j\}    \in \tau}  d'(\square_i, \square_j)
\ee  
 satisfies   
  $d_{\tau}  (\square_0,\square_1, \dots,  \square_n) 
=  \ell' (\square_0, \square_1, \dots,  \square_n) $.   Here  $d'(\square, \square')$ is the distance between centers.   Thus  our sum is dominated by   
 \begin{equation}
\begin{split}
&\sum_{n=0}^{\infty}\frac{1}{n!}    \sum_{\tau}
  \sum_{   \stackrel{ (\square_1, \dots,  \square_n): }{  d_{\tau}  (\square_0,\square_1, \dots,  \square_n) 
=  \ell' (\square_0, \square_1, \dots,  \square_n)  }  }
  \exp  \Big(  -   a'   d_{\tau} (\square_0, \square_1, \dots,  \square_n  )   \Big)\\
\leq  &  \sum_{n=0}^{\infty}\frac{1}{n!}    \sum_{\tau}  \sum_{ (\square_1, \dots,  \square_n)   }  \prod_{\{i,j\}    \in  \tau} 
 \exp \Big (  -   a' d'(\square_i, \square_j  ) \Big) \\
\end{split}  
\end{equation}
In   the second step we  dropped the restriction of the sum over   $(\square_1, \dots,  \square_n) $.
 Now  we   sum  over  the outer leaves of the tree  working our way back to the root   at  $0$,  using   the bound
\be
 \sum_{\square':  \square' \neq   \square}   \exp \Big (  -   a' d'(\square,\square') \Big)
\leq    \one  e^{-  a'}
\ee
    Then  the expression is dominated by 
\begin{equation}
  \sum_{n=0}^{\infty}\frac{ 1}{n!}   \Big ( \one e^{-a'}\Big )^n     \sum_{\tau}  1  \leq    \sum_{n=0}^{\infty}   (\one e^{-a'} )^n   \leq    \one  
\end{equation}
for $a'$ sufficiently large.   Here  we  use  Cayley's formula  that  there  are  $n^{n-2}$ tree graphs on  $n$ vertices.  Here this is 
$(n+1)^{n-1}  \leq  \one^n n!$. 

\item
Using the  second and third  bound  in lemma \ref{steamy}    we have  for $a,b$  sufficiently large
\begin{equation}
\begin{split}
\sum_{Y : \  Y  \supset  \square_0}
\exp  (  -   a  \ell(Y)   ) \leq  &   \sum_{Y : \  Y  \supset  \square_0}
\exp \Big (  -  ( a -a')   \ell(Y) +a'   |Y| - a'    \ell'(Y)  \Big)  \\
\leq  & \one\sum_{Y : \  Y  \supset  \square_0}
\exp \Big (  -  ( a -\one ) \ell(Y)    \Big ) \exp (   -  a' \ell'(Y)  )  \\
\leq  & \one  \sum_{Y : \  Y  \supset  \square_0}
 \exp (   -  a'  \ell'(Y)  )  
\leq       b\\
\end{split}
\end{equation}
\end{enumerate}
\bigskip

\begin{lem}  \label{summit}  Let  $\Om$ be a union of $M$-cubes.
 For  $\square \subset  \Om$  and constants  $\ka_0,  K_0 = \one$  
\begin{equation}   \label{sumsum}
\sum_{X  \in   \cD_k(\bmod  \Om^c),  X \supset  \square}
\exp \Big(   - \ka_0  d_M (X,  \bmod  \Om^c)  \Big )  \leq   K_0
\end{equation}
\end{lem}
\bigskip

\pr   We have     $ d_M (X,  \bmod  \Om^c)  \geq \tilde    \ell_M(X \cap  \Om) \geq  \frac 12   \ell_M( X  \cap  \Om) $.
   Thus it suffices to show   
 \be
    \label{sumsum2}
\sum_{X  \in   \cD_k(\bmod  \Om^c),  X \supset  \square}
\exp \Big (   -  \frac12 \ka_0   \ell_M( X  \cap  \Om) \Big )  \leq   K_0
\ee
 We  classify  the terms in the sum   by  the $Y =  X \cap  \Om$  they generate.   Although $X$ is connected, $Y$ need not be.
The sum can then be written
\begin{equation}
 \sum_{Y:  Y \supset \square}    \exp  \Big(   - \frac12 \ka_0  \ell_M (Y)  \Big ) 
\Big| \{ X  \in   \cD_k(\bmod  \Om):   X \cap \Om =   Y\}    \Big |  
\end{equation}    
Thus we  must estimate the number  of   polymers $X \in   \cD_k(\bmod  \Om)$  such that  $  X \cap \Om =   Y$.
If   $\{  \Om_{\al}^c  \}$  are the connected components of  $\Om^c$,   then  any such  $X$  can be written
\begin{equation}
X  =  Y  \cup  \Big(   \cup_{\al}   (X \cap  \Om^c_{\al}  )  \Big)
\end{equation}  
By our assumptions  either  $X \cap  \Om^c_{\al} = \emptyset   $  or    
 $X \cap  \Om^c_{\al} = \Om^c_{\al} $.
Since   $X$ is connected  each    non-empty   $ X \cap  \Om^c_{\al} $  must have  a 
block  sharing a face with  a   block  in  $Y$.     Thus  the number of non-empty  
$ X \cap  \Om^c_{\al} $   is at most   $2^d |Y|_M$.  Counting the number of $X$'s
generating  a particular $Y$  means  choosing a subset  of   this set.  Thus  there  are less than  $2^{2^d|Y|_M}$     such  $X$.  
Now   our   sum is bounded  by   
\begin{equation}
  \sum_{Y:  Y \supset \square}    \exp \Big(   -\frac12  \ka_0 \ell_M (Y) +   \cO(1)|Y|_M     \Big   ) 
\end{equation}    
But  $ |Y|_M   \leq  \cO(1) \ell_M(Y) + \cO(1)$  by  lemma  \ref{steamy}    and  so  if $\ka_0$ is large enough  
the sum is bounded by 
\begin{equation} \cO(1)  \sum_{Y:  Y \supset \square}    \exp \Big(   - \frac 14 \ka_0 \ell_M (Y)  \Big   ) 
\end{equation}    
The result now follows by lemma \ref{kumquat}  provided  $\ka_0  > 4a$.
\bigskip

\section{cluster expansion with holes}  \label{fancycluster}
 
 We   quote   a  special    version of the cluster expansion   in which  there are holes  for the large field region.  See also  \cite{BIJB84},  \cite{Bry86}.

  On a unit lattice   consider     subsets  $\Om,  \La  $ which    are unions    of  $M$-cubes,  and satisfy  $\La  \subset  \Om$.
we   consider  integrals  of the form   
\begin{equation}  \label{gog1}
\Xi =  \int \exp     \left(  \sum_{X \in \cD_k( \bmod  \Om^c),  X  \cap  \La  \neq  \emptyset }
   H( X, \Phi',    \Phi  )  \right)    d\mu_{\La}(\Phi)        
\end{equation}
Here  $\mu_{\La} $  is   an  ultralocal    probability measure on  the fields  $\Phi: \La  \to  \bbR$  rendering them  independent random variables.
The  $\Phi'$ are any other fields,    
and      $H(X, \Phi',  \Phi)$    depends on    $\Phi', \Phi$   only  in  $X$.

\begin{thm}  \label{cluster3}      Let  $c_0= \one$ be sufficiently small,  let  $H_0 \leq  c_0$,  let  $\ka \geq  3\ka_0 + 3$, and 
suppose      
\begin{equation}
|H(X,\Phi',  \Phi  )  |   \leq    H_0  e^{- \kappa   d_M( X,  \bmod  \Om^c)  }   \hs    X \in \cD_k( \bmod  \Om^c)
\end{equation}
on the support of  $\mu_{\La}$.   Then     
\begin{equation}   \label{gog2}
\Xi   = \exp  \left(      \sum_{Y \in \cD_k( \bmod\  \Om^c), Y \cap \La  \neq  \emptyset }      H^\#(Y, \Phi')  \right)
\end{equation}
where  $H^\#(Y, \Phi') $    depends on  $\Phi'$  only  in  $Y$  and  satisfies
\begin{equation}  \label{sunshine} 
|H^\#(Y, \Phi')  |   \leq  \cO(1)      H_0  e^{ - (\kappa-  3 \ka_0 - 3)   d_M( Y,  \bmod  \Om^c)  }
\end{equation}
\end{thm}
\bigskip

\re   In   addition  $H^\#(Y)$  only  depends  on $H(X)$  for  $X \subset  Y$.  We  call this the \textit{local influence} property of the cluster expansion.
\bigskip

\pr   This  closely   follows   the   proof  of the standard  cluster  expansion.  It   is  exposed in Appendix B  in   part  I,  to which we  refer for more details.
The differences  are that  instead of  general   polymers  $X$,   we  have  polymers  $X \in      \cD_k( \bmod\  \Om^c) $ with holes  $\Om^c$,
and  instead  of   decay rates  $e^{- \ka d_M(X) }$  we  have  decay  rates  $e^{- \kappa   d_k( X,  \bmod  \Om^c)  }$   outside the holes.
Key ingredients    are   the   bound   
\be    \label{keykey1}
  \sum_{Y:  Y  \cap  Y'  \cap  \Om \neq  \emptyset  }  e^{  - \ka_0  d_M(Y,  \bmod \Om^c)}     \leq    \one  |Y' \cap  \Om|_M  
\ee
from (\ref{sumsum})   and the bound
\be    \label{keykey2}
  |Y'  \cap  \Om|_M   \leq    \one   (d_M(Y', \bmod  \Om^c)  +1)  
\ee

We have stated these estimates in a form that takes into account that 
  there is     a  modified notion of connectedness  between   polymers.     
  Now we    say  that  $X_1,X_2  \in    \cD_k( \bmod\  \Om^c) $   are  
$\Om$-connected if  $X_1\cap  \Om$  and  $X_2 \cap \Om$  have non-empty intersection  (i.e  if   $X_1 \cap  X_2 \cap  \Om  \neq  \emptyset$).
Otherwise    $X_1\cap  \Om$  and  $X_2 \cap \Om$  have  empty intersection and they are called $\Om$-disjoint.   Note that $\Om$-connected implies 
connected,  but  $\Om$-disjoint  does not imply disjoint.

We  sketch some details of the proof.   First  we  make  a Mayer  expansion and write  
\be  \exp     \left(    \sum_{X  }
   H( X )  \right) =  \sum_{ \{  Y_j\} }  \prod_j   K(Y_j)   
\ee
where the  $ Y_j$  are  $\Om$-disjoint,   and where  
where  for  $Y  \subset    \cD_k( \bmod\  \Om^c)$   and $Y \cap  \La \neq  \emptyset$.
\be   
K(Y)  =  \sum_{  \{X_i \} :  \cup_i X_i  =Y}  \prod_i  (e^{H(X_i)}  -1)
\ee
The latter   sum  is  restricted by the condition that  the $X_i$  are $\Om$-connected,  i.e.    cannot  be  divided into  $\Om$-disjoint sets. 
$K(Y) =  K(Y, \Phi' ,\Phi)$  only depends on  $\Phi'$ in   $Y$   and $\Phi$ in  $Y \cap  \La$.

 To  estimate   $K(Y)$   we  need to  show  that   if   $ \{X_i \}$ has  $n$  elements
\begin{equation}  \label{otter}
d_M(  Y, \bmod\ \Om^c  )   \leq   \sum_{i}   d_M(  X_i, \bmod\ \Om^c  ) + (n-1)
\end{equation}
To   see  this  let  $\tau_i$   be minimal graphs   on  the cubes  in    $X_i \cap  \Om$   of length
$\ell( \tau_i)   =    M d_M(  X_i, \bmod\  \Om^c  ) $.    Stitch together   the   $\tau_i$   to get  a graph  $\tau$  on    
the   cubes   in  $\cup_i  (X_i \cap   \Om)  =  Y \cap  \Om $   with   $\ell(\tau)   \leq  \sum_i \ell(\tau_i)+M(n-1)$
Since   $Md_M(  Y, \bmod\   \Om^c  )   \leq     \ell(\tau)   $ this gives the result.   

With some further analysis the bounds    (\ref{keykey1}),   (\ref{keykey2}), (\ref{otter})       lead to  the bound on the support of  $\mu_{\La}$
\be \label{simplon}
  |K(Y)|  \leq   \one H_0    e^{-( \ka -  \ka_0 -2)   d_M( Y,  \bmod  \Om^c)  }
\ee

Because the $Y_j \cap  \Om$  are disjoint,  the  $Y_j \cap  \La$  are disjoint.    Since   $K(Y, \Phi' \Phi)$  depends on  $\Phi$ only  in   $Y_i \cap  \La$,
  and  because  fields  at  different sites  are independent  random variables  
  \begin{equation}
   \int   \Big(    \sum_{   \{Y_j\}  } \prod_j    K(Y_j,\Phi', \Phi  )  \Big)     d\mu_{\La} (\Phi)   
= \sum_{   \{Y_j\}  } \prod_j    K^\#(Y_j , \Phi'  ) 
\end{equation}
where  
\begin{equation}
 K^\#( Y, \Phi' )=   \int    K(Y,\Phi', \Phi)    d\mu_{\La}(\Phi)   
\end{equation}
again satisfies the bound  (\ref{simplon}).

Next  we  exponentiate the sum and get
\begin{equation}
 \sum_{   \{Y_j\}  } \prod_j    K^\#(Y_j  )   =  \exp \Big(     \sum_Y   H^\#(Y)     \Big)
 \end{equation}
 where   for   $Y  \subset    \cD_k( \bmod\  \Om^c)$   and $Y \cap  \La \neq  \emptyset$.
 \begin{equation}    \label{hstar}
   H^\#(Y)  
 =    \sum_{n=1}^{\infty}
 \frac{1}{n!}  \sum_{(Y_1, \dots,  Y_n): \cup_i Y_i  =Y}  \rho^T(Y_1, \dots,  Y_n)  \prod_i K^\#( Y_i )
 \end{equation}
and   $\rho^T(Y_1, \dots,  Y_n) $  is    now defined so it    vanishes  if  the    $Y_j $   are  not   $\Om$-connected.

Finally,   with some further analysis,    the bound  on    $K^\#( Y )$   and  the  estimates   (\ref{keykey1}),  (\ref{keykey2}),  (\ref{otter}),  (\ref{simplon})      lead to
the convergence of the series and    the estimate  (\ref{sunshine}).

\end{document}